\documentclass[pre,aps,amsmath,amssymb,onecolumn,nofootinbib]{revtex4}
\usepackage{graphicx}% Include figure files
\usepackage{dcolumn}% Align table columns on decimal point
\usepackage{subfigure}
\usepackage{epsfig}
\usepackage{color}
\usepackage{bm}% bold math

\newcommand\beq{\begin{equation}}
\newcommand\eeq{\end{equation}}
\newcommand\bea{\begin{eqnarray}}
\newcommand\eea{\end{eqnarray}}
\newcommand\noi{\noindent}
\newcommand\non{\nonumber}
\newcommand{\bra}[1]{\langle #1|}
\newcommand{\ket}[1]{|#1\rangle}
\newcommand\al{\alpha}
\newcommand\ga{\gamma}
\newcommand\de{\delta}
\newcommand\De{\Delta}
\newcommand\ep{\epsilon}
\newcommand\la{\lambda}
\newcommand\si{\sigma}
\newcommand\om{\omega}
\newcommand\ct{\cite}
\newcommand\dg{\dagger}

\begin{document}
%---------------------------------------------------------------------
\title{Quantum phase transitions in transverse field spin models: From Statistical Physics to Quantum
Information}
\author{Amit Dutta}
\email{dutta@iitk.ac.in}
\affiliation{Department of Physics, Indian Institute of Technology, Kanpur
208 016, India}
\author{Uma Divakaran}
\email{udiva@lusi.uni-sb.de}
\affiliation{Theoretische Physik, Universit\"at des Saarlandes, 66041 
Saarbr\"ucken, Germany}
\author{Diptiman Sen}
\email{diptiman@cts.iisc.ernet.in}
\affiliation{Centre for High Energy Physics, Indian Institute of Science, 
Bangalore 560 012, India}
\author{Bikas K. Chakrabarti}
\email{bikask.chakrabarti@saha.ac.in}
\affiliation{Theoretical Condensed Matter Physics Division and Centre for 
Applied Mathematics and Computational Science, Saha Institute of Nuclear 
Physics, 1/AF Bidhan Nagar, Kolkata 700 064, India}
\author{Thomas F. Rosenbaum}
\email{tfr@uchicago.edu}
\affiliation{Department of Physics and the James Franck Institute, 
University of Chicago, Chicago, IL 60637}
\author{Gabriel Aeppli}
\email{gabriel.aeppli@ucl.ac.uk}
\affiliation{London Centre for Nanotechnology and Department of Physics 
and Astronomy, UCL, London, WC1E 6BT, United Kingdom}
%---------------------------------------------------------------------

\begin{abstract}

We review quantum phase transitions of spin systems in transverse magnetic 
fields
taking the examples of some paradigmatic models, namely, the spin-1/2 Ising
and $XY$ models in a transverse field in one and higher spatial dimensions.
Beginning with a brief overview of quantum phase transitions, we introduce
the model Hamiltonians and discuss the equivalence between the quantum
phase transition in such a model and the finite temperature phase transition
in a higher dimensional classical model. We then provide exact solutions in
one spatial dimension connecting them to conformal field theoretical studies
when possible. We also discuss Kitaev models, {and some other 
exactly solvable spin systems in this context.} Studies of quantum phase
transitions in the presence of quenched randomness and with frustrating
interactions are presented in details. {We discuss 
novel phenomena like Griffiths-McCoy singularities associated with
quantum phase transitions of low-dimensional transverse Ising models with 
random interactions and transverse fields}. We then turn to more recent
topics like information theoretic measures of the quantum phase transitions
in these models; {we elaborate on the scaling behaviors of 
concurrence, entanglement entropy (for both pure and random models), quantum 
discord and quantum fidelity close to a quantum critical point. At the same 
time, we discuss the decoherence (or Loschmidt echo) of a qubit coupled to a 
quantum critical spin chain. We then focus on non-equilibrium dynamics 
of a variety of transverse field systems across quantum critical points 
and lines. After briefly mentioning rapid quenching studies, we 
dwell on slow dynamics and discuss the Kibble-Zurek scaling for the defect 
density following a quench across critical points and its modifications for 
quenching across critical lines, gapless regions and multicritical points. 
We present a brief discussion on adiabatic perturbation theory indicating the 
limits of slow and sudden quenching; we introduce the concept of a 
generalized fidelity susceptibility in this context. Topics like the role 
of different quenching schemes, local quenching, quenching of models with 
random interactions and quenching of a spin chain coupled to a heat bath are 
touched upon. The connection between
non-equilibrium dynamics and quantum information theoretic measures (introduced in the previous section) is presented at some length. We indicate the 
connection between Kibble-Zurek scaling and adiabatic evolution of a state as
well as the application of adiabatic dynamics as a tool of a quantum 
optimization technique known as quantum annealing (or adiabatic quantum 
computation). The final section is dedicated to a detailed discussion} 
on recent experimental studies of transverse Ising-like systems.
\end{abstract}
%---------------------------------------------------------------------
\maketitle
\tableofcontents
%---------------------------------------------------------------------
\section{Introduction and the aim of the review}

A plethora of systems exhibit phase transitions as the temperature or
some other parameter is changed. Examples range from the 
ice-water phase transition observed in our daily life to the loss of 
ferromagnetism in iron or to the more sophisticated Mott insulator-superfluid 
phase transition observed in optical lattices \ct{greiner02}. The last five 
decades have witnessed a tremendous upsurge in the studies of phase 
transitions at finite temperature 
\ct{stanley87,chaikin95,goldenfeld92,cardy96,ma76,mussardo10,nishimori10}. 
The success of Landau-Ginzburg theories and the concepts of spontaneous 
symmetry breaking and the renormalization group 
\ct{amit84,zinn-justin89,wilson74,parisi88} in explaining many of the 
finite temperature phase transitions occurring in nature has been spectacular. 

In this review, we will consider only a subclass of phase transitions called 
quantum phase transitions (QPTs) 
\ct{chakrabarti96,sachdev99,sondhi97,vojta03,continentino01,belitz02,
belitz00R,belitz05,carr10}; these are zero temperature phase transitions 
which are driven by quantum fluctuations 
% arising due to the Heisenberg uncertainty principle 
and are usually associated with a non-analyticity in the ground state
energy density of a quantum many-body Hamiltonian. We will focus on 
continuous QPTs where the order parameter vanishes 
continuously at the quantum critical point (QCP) at some value of the 
parameters which characterize the Hamiltonian. We will not discuss first order 
quantum phase transitions associated with an abrupt change in the order
parameter.
Usually, a first order phase transition is characterized by a finite
discontinuity in the first derivative of the ground state energy density. A 
continuous QPT is similarly characterized by a finite discontinuity, or 
divergence, in the second derivative of the ground state energy density, 
assuming that the first derivative is continuous. This is of course the 
classical definition; we will later mention some
QPTs where the ground state energy density is not necessarily singular.

{The central theme of the review is quantum phase 
transitions in transverse field models, (namely, Ising and $XY$ models in a 
transverse magnetic field) which are paradigmatic models exhibiting zero 
temperature continuous QPTs, and the recent studies (both experimental and 
theoretical) involving these models.} To the best of our knowledge, the 
one-dimensional version of the transverse Ising model (TIM) first appeared 
in the context of the exact solution of a two-dimensional
nearest-neighbor ferromagnetic classical Ising model; the row-to-row transfer
matrix of the two-dimensional model can be mapped to a transverse Ising chain 
in some limit \ct{lieb61} and the exact solution of this one-dimensional 
version of the model soon followed \ct{katsura62}. Shortly, the model was 
invoked to mimic the order-disorder transitions in KDP ferroelectrics 
\ct{degennes63}. Our review will attempt to highlight the aspects of the 
TIMs for which these models and their variants continue to be enormously 
useful, even more than fifty years 
after their first appearance, in understanding QPTs, non-equilibrium dynamics
of quantum critical systems, and connections between QPTs, dynamics and quantum
information,{and in possible experimental and theoretical 
realizations of quantum annealing}. Moreover, experimental realizations of 
these transverse Ising models have opened up new vistas of research.

 {Before presenting the section-wise plan of the review, let 
us first discuss in what sense the review will be useful to the community. 
Here, we place the most recent theoretical as well as 
experimental studies of QPTs in transverse field models under the same roof 
pointing to the open problems wherever possible.
%to the best of our knowledge, this is the first attempt in this direction.
For example, a very recent experimental study of the exotic low lying spectrum 
of the transverse Ising systems in a longitudinal field \ct{coldea10} 
has been discussed here at some length mentioning the corresponding 
theoretical prediction of the same made decades ago (Sec.~\ref{expcorr}).
Even when discussing conventional topics, also reviewed elsewhere, we emphasize recent developments and debates; for example, 
we allude to recent studies of matrix product states and their 
connection to quantum information and dynamics. The correspondence between
QPTs and classical finite-temperature phase transitions
(in a model with one added dimension) is well studied for these models, but 
the possible breakdown of this established scenario 
in the context of the spin-boson model is relatively less familiar; we 
include this in the review. Although the novel features associated with random 
quantum Ising transitions have been reviewed previously, we mention the most 
generalized model introduced in connection to quantum information theory 
(more precisely, to the entanglement entropy, to be defined in 
Sec.~\ref{e_entropy}) and point to the recent debate on the possibility of a 
generalized $c$-theorem. Similarly, the long-standing debate concerning
the width of the floating phase in the phase diagram of a one-dimensional 
transverse Ising model with regular frustrations is highlighted.
At the same time, it needs to be emphasized that in Secs. 5 and 6, we have 
attempted to provide a general perspective of the related physical problems;
the general theory is illustrated using the example of transverse field 
models. For example, we discuss the generic scaling of quantum fidelity in 
all limits (Sec.~\ref{fidelity}) and verify these scaling relations using 
one-dimensional transverse field models. It is to be noted that the section 
on regularly frustrated system (Sec.~\ref{sec_ANNNI}) is model specific and 
in Sec. 3 we elaborate the features of quantum phase transitions in random 
systems which are the most prominent in TIMs.}

 {In the first two sections we briefly present the 
phenomenology 
of the models and discuss at length their exact solution in one dimension. 
These discussions are intended to help a beginner to navigate through the more 
involved topics discussed in subsequent sections. Although the main emphasis 
is on the transverse Ising and $XY$ models, we also briefly dwell on other 
one-dimensional exactly solvable models (closely related to the Ising and $XY$ 
models) and the two-dimensional Kitaev model, and discuss their applications. 
In that sense, this review covers the entire gamut of Jordan-Wigner solvable 
models (to be explained below) especially from the viewpoint of quantum 
information and dynamics. Moreover, we use Appendix \ref{jorwigbos} to provide 
a discussion on the method of bosonization and QPTs in Tomonaga-Luttinger 
liquids which go beyond Jordan-Wigner solvable models.}

{With this preamble, let us elaborate on the section-wise plan 
of this review. In Sec.} 1, we will briefly introduce 
QPTs and critical exponents. We then introduce the transverse Ising models in 
one and higher dimensions. This is followed by a discussion on a $XY$ spin 
chain which is a 
generalized version of the Ising chain with $XY$-like interactions. 
The anisotropic version of the $XY$ spin model has the same symmetry ($Z_2$) as 
the transverse Ising model and both the models are exactly solvable in one 
spatial dimension. However, the transverse XY chain exhibits a richer 
phase diagram with critical, multicritical points and gapless critical lines. 
Transverse Ising models have been extended to $n$-component quantum rotor 
models which are in fact quantum generalizations of classical $n$-vector 
models; we also briefly touch upon quantum rotor models. There are some 
variants of transverse Ising and $XY$ spin chains {(Sec. 1.3)} 
which can be exactly solved and show rich phase diagrams ({and 
are being studied extensively}), e.g., the transverse $XY$ spin chain with 
alternating interactions and fields, and a transverse Ising chain with 
three-spin interactions. The latter model is known to have a matrix product 
ground state and {undergoes a quantum phase transition which 
has some unconventional properties as explained in Sec. 1.1.} 

Another useful aspect of the TIMs is their connection to classical 
Ising models in the sense of the universality class which we briefly mention 
in Sec. 1.4; {we summarize the essential idea here}. 
The zero-temperature transition of a $d$-dimensional TIM with 
nearest-neighbor ferromagnetic interactions belongs to the same universality
class as the thermal phase transition of a classical Ising model with similar 
interactions in $(d+1)$-dimensions; this equivalence can be established using 
a Suzuki-Trotter formalism and through the imaginary time path-integral 
formalism of the quantum model where the additional dimension is the Trotter 
direction or imaginary time direction. Similarly, starting from the transfer 
matrix of the two-dimensional Ising model, one can arrive at the quantum Ising 
chain in a limit which is known as the $\tau$-continuum limit. The underlying 
idea in both cases is that the extreme anisotropy in interactions along 
different spatial directions does not alter the universality class, i.e., the 
critical exponents. The quantum-classical correspondence holds even when 
the interactions in the quantum model are frustrated or random; in this case 
interactions along $d$ spatial dimensions of the higher-dimensional classical 
models become frustrated or random as in the quantum model but the 
interactions in the Trotter (time) direction remain ferromagnetic and
nearest-neighbor. Therefore, one ends up with a classical model with 
randomness or frustration correlated in the Trotter direction. The
quantum-classical correspondence turns out to be useful for several 
reasons: (i) since classical models are well studied, one can infer 
some of the critical exponents of the equivalent quantum model 
beforehand. For example, one can immediately conclude that the upper critical 
dimension for a short-range interacting quantum Ising model would be three 
since this is equivalent to the thermal transition of a four-dimensional 
classical Ising model. (ii) Secondly, this mapping renders quantum Ising 
models ideal candidates for quantum Monte Carlo studies, and to understand 
the quantum transitions one has to study a classical Ising model with one 
additional dimension. Although an equivalent classical model exists for many 
quantum spin models (though not all), one usually ends up with a classical 
model with complicated interactions. Therefore, this equivalence is a unique
property of TIMs which has been exploited intensively to gain a better 
understanding of QPTs in random models. In a similar spirit, one can show that
the QPT in a quantum rotor model is equivalent to the thermal transition in a 
classical $n$-vector model with one additional dimension. 
{However, this traditional
notion has been challenged in the context of the spin-boson model (Sec. 1.5)}.

The integrability of the transverse Ising and XY chains with nearest-neighbor 
interactions has possibly attracted the attention of the physicists the 
most\footnote{{We will discuss later what integrability 
means; here the expression implies
that all the eigenfunctions and energies can be exactly obtained.}}.
In Sec. 2, we will see how these spin models can be solved exactly 
using the Jordan-Wigner (JW) transformation. The reduced Hamiltonian in the 
Fourier space gets decoupled into $2 \times 2$ Hamiltonians; this two-level 
form eventually turns out to be extremely useful in calculating the defect 
density following a sweep of the spin chain across a QCP as we shall discuss 
in Sec. 5. Moreover, the ground state takes a direct product form of 
qubits; this simple form of the ground state wave function is extremely helpful
in understanding the behavior and scaling of information theoretic measures 
close to a QCP which we shall discuss in Sec. 5. The Kitaev model on a 
honeycomb lattice is one of the very models which are exactly solvable in 
two dimensions; the solution employs a mapping of spin$-1/2$'s to Majorana
fermions through a JW transformation. Hence the model has been widely used to 
understand non-equilibrium dynamics of quantum critical systems as well as 
in quantum information studies which we shall refer to in subsequent sections.
We shall introduce the Kitaev model, the appropriate JW transformation and the
phase diagram of the model in Sec. 2. Finally, we will briefly discuss the 
connection of various one-dimensional models to conformal field theory and 
show how the scaling dimensions of different operators can be
extracted from the conformal properties of the equivalent action at the QCP. 
This information turns out to be useful, e.g., for the transverse Ising
chain in the presence of a longitudinal field that destroys the integrability 
of the model. The recent study of the QPT of a model that challenges this 
quantum-classical equivalence will also be briefly mentioned.

It is well known that the presence of disorder or randomness in interactions 
or fields deeply influences the phase transition occurring in a classical
system, often wiping it out completely. A modified Harris criterion points to 
a stronger dominance of randomness in the case of a zero-temperature quantum 
transition. Although the quantum versions of classical Ising and $n$-vector 
spin glasses have been studied in the last four decades and their quantum 
counterparts were introduced in the early eighties, the experimental results 
for $\rm {LiHoF_4}$ and its disordered version $\rm {LiHo_xY_{1-x}F_4}$ which 
are ideal realizations of TIMs with long-range dipolar interactions led to a 
recent upsurge in exploring the QPT in quantum Ising spin-glasses which are 
briefly mentioned in Sec. 4. The most exciting feature associated with 
low-dimensional random quantum critical systems is the prominence of 
Griffiths-McCoy (GM) singular regions close to the QCP and the activated 
quantum 
dynamical scaling right at the QCP. We shall argue that these off-critical 
singularities are more prominent in transverse Ising models than in the 
rotor models due to the Ising symmetry in the former. These singular regions 
arise due to locally ordered large ``rare'' clusters, which can be viewed as 
a giant spin. If the spins are Ising-like, this giant spin can flip only
through the barrier tunneling which is an activated process thereby leading to 
prominent GM regions. These issues and the 
contrast with the $O(n)$ symmetric quantum rotor models will be discussed 
in Sec. 3 where we allude to the effect of a random longitudinal 
field on the quantum Ising transitions. Experimental signatures of this 
GM phase are also discussed later. {The role of 
these GM singularities in the quantum information theoretic measures and 
dynamics of disordered spin chains will be discussed in appropriate sections}.

In Sec. 4, we shall introduce a version of the classical Ising 
model with competing interactions in which in addition to the 
nearest-neighbor ferromagnetic interactions along all the spatial 
directions, there is a next-nearest neighbor antiferromagnetic interactions 
along a particular direction, called the axial direction. This model is known 
as the axially anisotropic next-nearest-neighbor Ising (ANNNI) model which is 
the simplest model with non-trivial frustration and exhibits a rich phase 
diagram that has been
studied for the last three decades. We focus on the bosonization study of the
equivalent quantum model in one dimension. We show that different numerical 
studies and the bosonization studies contradict each other and whether a 
floating phase of finite width exists in the phase diagram of the 
one-dimensional transverse ANNNI is not clearly established till today. 
To make a connection with bosonization studies, we present the main ideas 
of bosonization and its applications to one-dimensional spin systems in 
Appendix \ref{jorwigbos}. {In this context, we refer to the 
quantum rotor models with ANNNI-like frustrations and mention theoretical and 
experimental studies of the quantum Lifshitz point which appears in the phase 
diagram of these rotor models.} 

In Sec. 5, we will discuss some connections between quantum information 
theoretic measures and QPTs. It is interesting that these quantum 
information theoretic measures can capture the ground state singularities 
associated with a QPT. The quantum correlations of a state can be quantified
in terms of bipartite entanglements. While the entanglement is a measure of 
the correlation based on the separability of two subsystems of the composite 
system, the quantum discord is based on the measurement on one of 
the subsystems. Both the concurrence (which is one of the measures of 
bipartite entanglement) and quantum discord show distinctive behaviors 
close to the QCP of a one-dimensional transverse $XY$ spin model and 
interesting scaling relations which incorporate the information about the 
universality of the associated QPT. We also discuss the generic scaling 
for the entanglement entropy close to a QCP and how these scaling 
relations are verified for transverse field spin models. Subsequently, we 
review the notion of quantum fidelity which is the modulus of the overlap of 
two ground state wave functions with different values of the parameters of the
Hamiltonian. Although the fidelity vanishes in the thermodynamic limit, for 
a finite system it shows a dip at the QCP and hence is an ideal indicator of 
a QCP. We discuss the concept of fidelity susceptibility valid 
in the limit when two parameters under consideration are infinitesimally 
close to each other; this fidelity susceptibility shows a very interesting 
scaling behavior close to a QCP which can be verified analytically for 
transverse field spin chains. Moreover, one finds surprising results close 
to a multicritical point. In the process,
we comment on the quantum geometric tensor defined on a 
multi-dimensional parameter space pointing to the close connection between the
fidelity susceptibility and the geometric phase. Very recently, the fidelity 
has been studied for the case when the difference in the parameters 
characterizing two ground state wave functions is finite; this is referred to 
as the fidelity in the thermodynamic limit. A scaling of fidelity has been 
proposed in this limit also, which has again been verified for transverse field
spin chains. Finally, we comment on the decoherence of a qubit (spin-$1/2$) 
or a central spin coupled to an environment which is conveniently chosen to 
be a transverse Ising/$XY$ spin chain. Due to the interaction between the qubit
and the environment, the initial state (chosen to be the ground state) of the 
spin chain evolves with two different Hamiltonians and the corresponding 
Loschmidt echo is measured. The echo shows a sharp dip close to the QCP and 
shows a collapse and revival there which can be taken to be an 
indicator of quantum criticality. The sharp dip in the echo at the same time 
indicates a complete loss of coherence of the qubit which reduces from the 
initial pure state to a mixed state. The connection between the static 
fidelity and the dynamic Loschmidt echo is also mentioned. 

In Sec. 6, we study in detail many aspects of the quenching dynamics of 
quantum systems across quantum critical points or lines; this is the longest 
section in this review. When a quantum system is driven across a quantum 
critical point by changing a parameter of the Hamiltonian following some 
protocol, defects are generated in the final state. This is due to the 
diverging 
relaxation time close to the QCP; no matter how slowly the system is quenched 
the dynamics is not adiabatic. The scaling of the density of defects in the 
final state is given by the rate of change of the parameter and some of the 
exponents of the QCP across which the system is driven. What is exciting 
about this problem is that the defects are generated through a non-equilibrium
dynamics but the scaling of the defect density is dictated by the equilibrium 
quantum critical exponents. This scaling is known as the Kibble-Zurek scaling 
which can be readily verified using the integrability of the transverse field
spin chains and the Landau-Zener transition formula for the non-adiabatic 
transition probability. We discuss in some detail adiabatic 
perturbation theory in order to arrive at the Kibble-Zurek scaling law; 
we also discuss the counterpart of this law 
in the limit when a parameter of the Hamiltonian is suddenly changed by 
a small amount starting from the QCP. We discuss how the scaling of the 
defect density following a sudden or adiabatic quench can be related to a 
generalized fidelity susceptibility. We indicate how to calculate the defect 
density following a quench across the critical points for the transverse $XY$ 
spin chain. When the system is quenched across a multicritical point or a 
gapless critical line of the transverse $XY$ spin chain or an extended 
gapless region of the Kitaev model, the traditional Kibble-Zurek scaling is 
non-trivially modified as we review at some length. Interesting results are 
obtained for non-monotonic variations of parameters and also when the spin 
chain is coupled to a bath. The effect of randomness in the interactions 
on the quenching is also discussed {to indicate the role of 
GM singular
regions (discussed in Sec. 3) in modifying the Kibble-Zurek scaling relations}.
%We also discuss about recent studies of quantum phase transition in space
It is natural to seek an optimized rate of quenching that minimizes defects in 
the final state within a fixed time interval (-$T$ to $T$). This knowledge 
would be extremely useful for the 
adiabatic evolution of a quantum state; we take up a discussion of related 
studies. One may wonder what happens when a parameter of the quantum 
Hamiltonian is changed locally; it so happens that in that case, one finds 
an interesting time evolution of the correlation function following the 
quench. Finally, we embark on establishing the connection between 
the non-equilibrium quantum dynamics and quantum information 
{discussed in Sec. 5}. As mentioned 
above, a sweep across a QCP generates defects in the final state which 
satisfy a scaling relation; interestingly the scalings of the concurrence, 
negativity, entanglement entropy and other measures of entanglement are also 
dictated by the scaling of the defect density although for the time evolution 
of the entanglement entropy one observes a clear signature of the Loschmidt 
echo for integrable models. In the last section, we comment on some recent 
studies of the possibilities of quantum annealing and adiabatic quantum 
computation. {We reiterate that all the generic discussions 
presented in this section are illustrated using transverse field models and 
their variants.}

A discussion of experimental realizations of Ising systems in transverse 
fields will be taken up in Sec. 7. As mentioned already, the model was 
introduced to study the order-disorder transition in KDP type ferroelectrics 
which we discuss at some length. We present a discussion of singlet ground 
state magnets where the ground state is well separated from higher lying spin 
multiplets, e.g., ${\rm LiTbF_4}$ which was previously employed as an example 
of an ideal Ising 
dipolar coupled ferromagnet. The systems for which the most data are available 
is $\rm {LiHoF_4}$ for which the ground state is a non-Kramers doublet with 
strong anisotropy effectively producing an Ising spin along the tetragonal 
$c$-axis. When a laboratory field is applied perpendicular to the $c$-axis,
the ground state degeneracy is lifted and a transverse field is generated due 
to the mixing between the states which essentially leads, at the lowest 
order, to an Ising model representation of the system with an 
effective transverse field. The disordered version of the model (when the
magnetic Ho is substituted by non-magnetic Y), $\rm {LiHo_xY_{1-x}F_4}$, is an
ideal realization of a quantum Ising spin glass. The phase diagram in the 
laboratory field and temperature plane shows a ferromagnetic phase for 
$x > x_c \simeq 0.2$ and a spin glass phase for $x <x_c$. This spin-glass 
transition and related experimental studies and observations which are not 
yet explained theoretically are discussed in detail in Secs. \ref{expt4}
and \ref{expt5}. 
%The systems $\rm {LiHoF_4}$ and 
%its disordered version $\rm {LiHo_xY_{1-x}F_4}$ are ideal realization of TIMs
%with long-range dipolar interactions to be discussed in these sections.

%into two terms with $\langle J_z \rangle=0$.
%{The splitting of the ground state doublet in a transverse field, 
%$\Gamma$,} is even in the external field because it is generated via mixing, 
One of the most exciting experiments on the one-dimensional 
transverse field Ising model is discussed in Sec.~\ref{expt3}, the 
material for the study being CoNb$_2$O$_6$. The low-lying spectrum in the 
presence of a longitudinal field is described by the Lie algebra of the 
exceptional group $E_8$ which was theoretically predicted using a combination 
of CFT and an exact scattering matrix analysis for the low-energy particles 
that is discussed in Sec.~\ref{expcorr}. The experimental study discussed 
in Sec.~\ref{expt3} in fact corroborates this finding. We shall make concluding remarks in Sec. 8. 

%In short, the aim 
%of this review is to provide an overview of the transverse field spin models 
%starting from their exact solution in one dimension which has turned out to 
%be extremely useful in recent studies of
%non-equilibrium quantum dynamics and quantum information.
%We also review version of the models with disorder and frustration and 
%refer to their experimental realizations. 
%Our focus is however not only limited to transverse field models; we also
%mention briefly the quantum rotor models and the two-dimensional Kitaev model 
%at some length to emphasize the great importance of spin models in recent 
%studies of dynamics and information. so that 
%the reader obtains a complete picture of the recent developments in the 
%studies of spin models, especially in low dimensions.
% whose crystal structure is illustrated in Fig.~\ref{GA4}. 
%The building blocks are ferromagnetic zig-zag chains of edge sharing CoO$_6$ 
%octahedra; { this is in contrast to the linear chains} of apex 
%sharing NiO$_6$ octahedra in the model $S=1$ antiferromagnet, YBaNiO$_5$ 
%\ct{guangyong07} which are 
%arranged so as to be coupled antiferromagnetically. In an external field 
%transverse to the ``easy" spin axis, the ordering along the easy axis is 
%destroyed at a QPT due to delocalized domain walls, 
%sometimes referred to as kinks or solitons.

% and the 
%expressions for Landau-Zener tunneling in two-level systems in Appendix B.

{Finally, we have strategically incorporated eight Appendices 
which are complementary to the material presented in the body of the review. 
In Appendix \ref{app_largespin}, we discuss QPTs of a transverse $XY$ spin 
chain in the limit when the spin magnitude $S$ is large, known as the $S \to 
\infty$ limit; 
the results obtained here can be compared to the solution in the extreme 
quantum case (spin-$1/2$) discussed in Sec.~\ref{jorwigtr}. At the same time, 
this section provides a thorough discussion of an interesting (mean field) 
method. In Appendix \ref{app_mps}, we have shown how a Hamiltonian 
(which is a variant of the transverse Ising spin chain Hamiltonian) with a 
given matrix product ground state can be constructed. We provide a brief 
overview of Tomonaga-Luttinger liquids in Appendix \ref{jorwigbos} while 
the corresponding studies of quantum dynamics and information are presented 
in Appendix \ref{app_ll}. The intention here is to enable the reader to get 
an idea of recent developments in these low-dimensional models. The discussion 
of Appendix \ref{jorwigbos} would also be useful in following the bosonization 
results used in Sec.~\ref{sec_ANNNI}. In Appendix \ref{sec_gp_scaling}, we 
discuss the scaling of the geometric phase close to a QCP which is in fact 
closely related to the geometric tensor discussed in Sec. \ref{fidelity}. A 
derivation of the LZ transition formula that has been used extensively in 
estimating the defect density following a quantum quench across QCPs 
(discussed in Sec. 6) is given in Appendix \ref{app_lz}. Finally, in 
Appendices \ref{kz_space} and \ref{sec_topology}, we briefly dwell on two 
exciting but involved topics, namely, the Kibble-Zurek mechanism in space and 
the effect of topology on quantum quenching; these studies in fact complement 
those presented in Sec. 6.}

\subsection{Quantum phase transitions}
\label{qpt}

Let us first recall some features of classical phase transitions driven 
by thermal fluctuations before considering QPTs in detail.
Suppose that a translation invariant classical system in $d$ spatial 
dimensions has a critical point at some finite temperature $T=T_c$. Such a 
critical point is associated with a number of critical exponents which are 
defined as follows \ct{amit84,cardy96}. Let ${\cal O} ({\vec x})$ 
denote an order parameter so that the thermal expectation value $\langle
{\cal O} ({\vec x}) \rangle = 0$ for $T \ge T_c$ and $\ne 0$ for $T < T_c$.
If $T>T_c$, the two-point correlation function $\langle {\cal O} ({\vec x}_1) 
{\cal O} ({\vec x}_2) \rangle$ falls off at large distances as 
$\exp (-|{\vec x}_1 - {\vec x}_2|/\xi)$ 
%as
%$|{\vec x}_1 - {\vec x}_2| \to \infty$. 
This defines a correlation length
$\xi$ which is a function of $T$. Then $\xi$ diverges when $T \to T_c^+$
as $\xi \sim (T - T_c)^{-\nu}$, where $\nu$ is called the correlation
length exponent. Another exponent is $\beta$ which is defined by
the way the order parameter approaches zero as $T \to T_c^-$, namely,
$\langle {\cal O} ({\vec x}) \rangle \sim (T_c - T)^{\beta}$. The specific
heat diverges as $T \to T_c^+$ as $C \sim (T - T_c)^{-\al}$. Next,
if $h$ denotes the field which couples to the order parameter in the
Hamiltonian of the system, then the zero-field susceptibility $\chi \equiv
(d\langle {\cal O} ({\vec x}) \rangle/dh)_{h=0}$ diverges as $T \to T_c^+$
as $\chi \sim (T - T_c)^{-\ga}$. Exactly at $T=T_c$, the order
parameter scales with $h$ as $\langle {\cal O} ({\vec x}) \rangle \sim |h|^{1/
\de}$ as $h \to 0$, while the two-point correlation function falls off at large
distances as $|{\vec x}_1 - {\vec x}_2|^{-(d-2+\eta)}$. Finally, there
is a dynamical critical exponent, $z$, which determines the response of
the system to time-dependent fields; there is a response time $\tau$ which
diverges as $\tau \sim \xi^z$ as $T \to T_c^+$. The connection between a 
finite temperature classical phase transition (CPT) and the corresponding 
QPT in transverse field models will be discussed below.

{The critical exponents defined above are not independent of 
each other; they are connected through scaling relations like $\al+2\beta
+\gamma=2$, $\gamma=(2-\eta)\nu$, $2 \beta + \ga = \beta(\de+1)$, etc., so 
that only two of the six exponents are independent \ct{stanley87,chaikin95}. 
These scaling
relations stem from the fact that close to a critical point, quantities like 
the free energy density, correlation functions, etc., can be expressed as 
homogeneous functions of dimensionless variables scaled with the diverging 
correlation length $\xi$. Let us consider a correlation volume in a 
$d$-dimensional hyper-space of size $\xi^{d}$ so that the free energy density 
($f$) scales as $f \sim \xi^{-d }$. Noting that $f \sim |T-T_c|^{2-\al}$ (as 
$C \sim |T-T_c|^{-\al}$) and $\xi \sim |T-T_c|^{-\nu}$, one readily arrives at 
one of the scaling relations $2 -\al = \nu d$; this is known as the 
hyperscaling relation. This hyperscaling relation is valid at or below the 
upper-critical dimension $d_u^c$ of the model considered. For $d>d_u^c$, the
critical exponents are determined by the mean field
theory. For classical Ising and $n$-vector models $d_u^c=4$. }

The striking feature of any continuous phase transitions is the concept of 
$``universality"$, which means that the critical exponents depend only 
upon the dimensionality of the system, the symmetry of the order 
parameter, and the nature of the fixed point,
irrespective of the microscopic details of the Hamiltonian. Thus, the critical
exponents of the phase transitions occurring in nature can in principle be 
studied by exploring a simpler model Hamiltonian belonging to the same 
universality class.

We now present a brief overview of quantum phase transitions 
\ct{sachdev99,sondhi97,vojta03}. QPTs are phase
transitions which occur at zero temperature by driving a non-thermal
parameter of a quantum system. At zero temperature, the system will 
be in the ground state dictated by the values of the parameters of
its Hamiltonian. Each term of the Hamiltonian,
when dominant, corresponds to a specific ground state and determines a
phase in the phase diagram. The presence of non-commuting terms in
the Hamiltonian produces a superposition between various states.
% the amount of
%mixing depends upon the strength of the non-commuting terms. 
By varying 
different terms of the Hamiltonian, one can access the entire phase diagram, 
crossing different phase boundaries involving second order QPTs. In QPTs, it 
is the quantum fluctuations (due to non-commuting terms in the Hamiltonian) 
% mandated by Heisenberg's uncertainty principle 
which are responsible for
taking the system from one phase to another, in contrast to the thermal
fluctuations in finite temperature phase transitions. One might think that 
such phase transitions are not relevant to the real world due to the 
unattainability of zero temperature. However, it has been found that many 
finite temperature properties of a system can be explained by understanding
its quantum critical point (QCP).

We now discuss various critical exponents which characterize a QCP 
\ct{sachdev99,vojta03}. A QCP is a point at which the ground state energy of 
the system is a non-analytic function of some parameter which is different 
from temperature,
such as the pressure, magnetic field or interaction strength. At this point, 
the energy difference $\De$ between the ground state and the first excited
state vanishes. As a parameter $g$ in the Hamiltonian is increased from zero, 
$\De$ decreases till it vanishes at the quantum critical point as
\beq \De \propto |g-g_c|^{\nu z}, \label{intro_nuz} \eeq
where $\nu$ and $z$ are the critical exponents associated with the critical 
point $g_c$ and are defined as follows. In a continuous phase transition, one 
can associate a length scale $\xi$ which determines the exponential decay of 
the equal-time two-point correlation in the ground state. For example, the 
equal-time connected correlation function $G(r)$ of the order parameter
between two points separated by a distance $r$ is given by
\beq G(r)=\langle O(0,t)O(r,t)\rangle - \langle O(0,t)\rangle \langle O(r,t)
\rangle \propto \frac{e^{-r/\xi}}{r^{d-2+\eta}}, \label{intro_equaltime} \eeq
where $d$ is the dimensionality of the system and $\eta$ is called the Fisher 
exponent associated with the critical point. As the critical point is 
approached, the length scale $\xi$ diverges with a critical exponent $\nu$ as
\beq \xi \propto |g-g_c|^{-\nu}. \label{intro_onlynu} \eeq
At the critical point, $\xi \to \infty$ and the correlation function $G(r)$
decays as a power law. Similarly, one can 
define a time scale $\xi_{\tau}$ for the decay of equal-space correlations
analogous to (\ref{intro_equaltime}), which diverges as
\beq \xi_{\tau} \sim \De^{-1} \propto \xi^z \propto 
|g-g_c|^{-\nu z}. \label{intro_onlyz} \eeq
The above equation shows that, unlike CPTs, space and time are connected 
with each other in QPTs. This complex interplay between space and time in 
QPTs makes it a challenging area of research.

At the critical point, the correlation length and time are both
infinite; fluctuations occur on all length and time scales, and the system 
is said to be scale invariant. As a result, all observables show
power-law behaviors near the critical point. For example, the order 
parameter $\cal O$, characterizing order in the system, is given by
\bea {\cal O} & \propto & (g_c-g)^{\beta}~~{\rm ~ for~~} g<g_c, 
\non \\
&=& 0~{\rm ~~for~~} g\geq g_c. \eea

The specific heat $C$ in QPTs is defined as the change in the ground state
energy due to a small change in the varying parameter.
Although there are seven critical exponents 
{ (as shown in Table I for transverse Ising models)}, not all 
are independent. As in the classical continuous phase transitions,
one can obtain four scaling relations connecting these exponents and hence 
only three of the exponents are independent \ct{chaikin95,goldenfeld92}.

Till now, we have discussed everything at zero temperature where 
there are no thermal fluctuations. For the sake of completeness, we briefly 
describe the effect of finite temperature on the QPTs.
Let us introduce the frequency associated with the diverging relaxation time 
by $\om_c$. As discussed before, the energy associated
with this frequency vanishes as $$\hbar \om_c \propto |g-g_c|^{\nu z}$$
when the critical point is approached. Quantum mechanics is important
as long as this typical energy scale is larger than the thermal
energy $k_B T$. In the regions of the phase diagram where $k_B T$ is larger
than $\hbar \om_c$, the classical description of the phase transition can 
be applied. Namely, quantum mechanics can still be important
on microscopic scales but it is the thermal fluctuations which
dominate the macroscopic scales and is more relevant
in determining the critical behavior.
\begin{figure}
\begin{center} \includegraphics[height=3.0in]{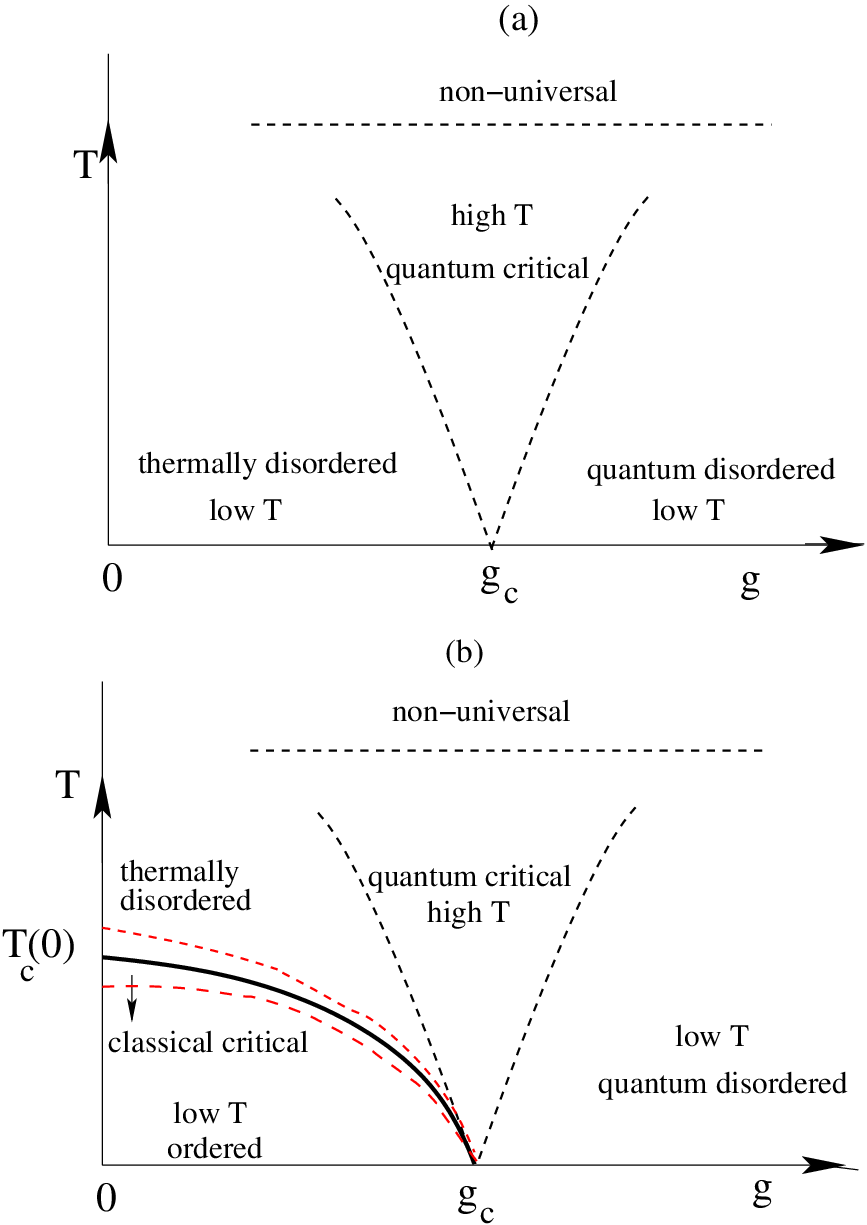}
\caption{(Color online) 
Schematic phase diagram in the vicinity of a quantum critical point. 
Fig. (a) corresponds to the case where order exists only at $T=0$, for example,
the one-dimensional transverse Ising model (TIM) with $g$ denoting the 
transverse field. Fig. (b) refers to the case
where order can exist even at finite temperatures, for example, the 
two-dimensional TIM. We discuss Fig. (b) with respect to the two-dimensional 
TIM. The phase transition at $T=0$ and $g=g_c$ has the critical exponents of 
a two-dimensional TIM or the thermal phase transitions of the three-dimensional
classical Ising model as obtained by the Suzuki-Trotter formalism (to be 
discussed in Sec.~\ref{sec_class+quant}). The 
classical critical point at $g=0$ and $T=T_c(0)$ denotes the thermal phase 
transition of a two-dimensional classical Ising model. The solid line marks 
the phase boundary between ordered and disordered phases. Phase transitions 
at any finite temperature belong to the universality class of the 
two-dimensional classical Ising transition governed by $T_c(0)$. The region 
between the two dashed red lines is dominated by classical fluctuations. The 
crossover from a low $T$ quantum disordered region to a high $T$ quantum 
critical region occurs when $\hbar \omega \simeq k_B T$, where $\omega$ is the
characteristic frequency scale (see Sec.~\ref{sec_class+quant}).
(After Sachdev, 1999).} \label{fig_qpt} \end{center} \end{figure}

The phase diagram of a quantum system can be of two types depending on 
whether a finite temperature phase transition exists or not in the system.
Figure \ref{fig_qpt} (a) describes the situation where the order exists only 
at $T=0$, for example, the one-dimensional transverse Ising 
model (discussed later) with the critical temperature $T_c=0$ at $g=g_c$.
In this case, there will be no true phase transition in any real experiment 
carried out at finite temperatures. However, the finite temperature region in 
the same figure is characterized by three different regions, separated by 
crossovers, depending on whether the behavior is dominated by thermal or 
quantum fluctuations. In the thermally disordered region, the long-range order
is destroyed by thermal fluctuations
whereas in the quantum disordered state, the physics is dominated by quantum
fluctuations. In the quantum critical region, both types of fluctuations
are important. In the other type of systems undergoing phase transitions at 
finite temperatures, for example, the two-dimensional transverse Ising model,
the phase diagram is richer as shown in Fig.~\ref{fig_qpt} (b). There is an 
additional line of finite temperature transitions, the regions surrounding 
which are referred to as ``classical critical'' where classical fluctuations 
dominate. Each of these regions has some special properties \ct{sachdev99} 
which will be discussed later.

There have been many experimental demonstrations of QPTs.
The cuprate superconductors which can be tuned
from a Mott insulating phase to a $d-$wave superconducting phase by carrier
doping are a paradigmatic example of QPTs. $\rm {LiHoF_4}$, $\rm{La_2CuO_4}$ 
and the heavy fermion $\rm {CeCu_{6-x}Au_x}$ 
are some other examples of materials which show phase transitions at zero 
temperature when a particular parameter, say, the transverse field or the
concentration $x$ of a component is varied. The systems $\rm {LiHoF_4}$ and 
its disordered version $\rm {LiHo_xY_{1-x}F_4}$ are ideal realizations of 
the transverse Ising model (TIM) with long-range dipolar interactions (to be 
discussed in Secs. \ref{expt4} and \ref{expt5}) which exhibit QPTs. 
%For example, the low-lying 
%magnetic excitations of the insulator LiHoF$_4$ consist of fluctuations of
%the Ho ions between two spin states that are aligned parallel or antiparallel
%to a particular axis. These states can be represented by a two state ``Ising"
%spin variable on each Ho ion. At $T=0$, the magnetic dipolar interactions
%between the Ho ions cause all the Ising spins to align in the same orientation,
%and so the ground state is a ferromagnet. Bitko, Rosenbaum and Aeppli 
%experimentally demonstrated the corresponding phase transition in LiHoF$_4$
%by placing the material in a magnetic field transverse to the magnetic axis 
%\ct{bitko96}. Varying the transverse field introduces quantum tunneling 
%between the two Ising states of each Ho ion and can take the system from the 
%FM phase to the PM phase, with the 
%ferromagnetic moment vanishing continuously as the field is increased.

{So far we have discussed QPTs which are associated with a 
singularity in the ground state of the quantum Hamiltonian.}
Before ending this section, let us mention a different class of QPTs in 
which the ground state has a matrix product form \ct{wolf06}. In 
these QPTs, the ground state may be completely analytic near $g=g_c$, there 
may not be any spontaneous symmetry breaking on either side of $g=g_c$, and 
the two-point correlation function may not approach zero as a power of the 
separation between the two points at $g=g_c$; in these respects, these QPTs 
are quite different from
those discussed above. However, as $g \to g_c$, the correlation length 
diverges and the energy gap approaches zero as powers of $|g-g_c|$ as in 
conventional QPTs. In this review, we will mainly discuss 
conventional QPTs. However, the concept of a matrix product wave function 
is useful in many problems, including the exact ground state of the 
Affleck-Kennedy-Lieb-Tasaki model which describes a spin-1 chain with 
isotropic nearest-neighbor interactions \ct{affleck88}.
A matrix product wave function has the form \ct{wolf06,cozzini071}
\beq | \psi \rangle ~=~ \sum_{i_1,\cdots,i_N=1}^d ~tr (A_{i_1} \cdots
A_{i_N}) ~|i_1,\cdots,i_N \rangle, \label{eq_mpstate}\eeq
where $N$ denotes the number of sites, $d$ is the Hilbert space dimension
at each site of the system, and $\{A_i\}$ is a set of $d$ $D$-dimensional
matrices {where $D$ is an appropriately chosen integer known 
as the bond dimension}\footnote{{Typically,
variational states of the matrix product form become more accurate as
$D$ is increased.}}. For such a wave function, correlation functions can be
calculated by transfer matrix methods; they typically decay exponentially at 
large distances, where the correlation length is related to the ratio of the 
second largest eigenvalue to the largest eigenvalue of the transfer matrix. 
Further, given a matrix product wave function, one can usually construct a 
Hamiltonian for which it is the ground state \ct{wolf06} {(see
Sec.~\ref{sec_relatedmodels} and Appendix \ref{app_mps})}.

{Finally, we note that there have been many studies of QPTs 
which occur between a topological phase and a conventional phase.
It is to be noted that a phase with topological order cannot be characterized
by any local order parameter; rather it is characterized in other ways such as 
the ground state degeneracy and the nature of the low-lying excitations. The 
toric code model is a microscopic lattice model that has been used in this 
context of topological order and quantum criticality \ct{kitaev03}. We shall
not discuss this issue here and refer interested readers to a recent review 
article \ct{castelnovo10}.}

\subsection{Transverse Ising and $XY$ models}
\label{models}

One of the prototypical examples capturing the essence of QPTs is the 
transverse Ising model (TIM) \ct{katsura62,pfeuty70}. On a $d$-dimensional 
hypercubic lattice, the model consisting of $N$ spins is described by the 
Hamiltonian
\bea H ~=~- ~\sum_{<{\bf ij}>} ~J_{\bf ij} ~\si_{\bf i}^x 
\si_{\bf j}^x ~-~ h~ \sum_{\bf i} ~\si_{\bf i}^z, \label{eq_timgen} \eea
where $\si^a_{\bf i}$, $a = x, y, z$, denote the Pauli spin matrices which 
satisfy the commutation relations $[\si_{\bf i}^a,\si_{\bf i}^b]=2i \ep_{abc} 
\si_{\bf i}^c$ at the same site and commute at two different sites (we will 
generally set $\hbar=1$). The number of spins $N=(L/a)^d$, where 
$L$ is the linear dimension of the system, and $a$ is the lattice 
spacing; we will usually set $a=1$ in subsequent
sections if it is not explicitly mentioned. Here,
$\langle {\bf ij} \rangle$ denotes nearest-neighbor interactions. 
This Hamiltonian is invariant under the $Z_2$ symmetry $\si_{\bf i}^x \to 
-\si_{\bf i}^x$, $\si_{\bf i}^y \to - \si_{\bf i}^y$, and $\si_{\bf i}^z \to 
 \si_{\bf i}^z$. In this section, we assume that $J_{\bf ij} = J_x \ge 0$, 
i.e., a ferromagnetic (FM) interaction between the nearest-neighbor spins.
% which simplifies the Hamiltonian to the form
%\bea H ~=~ - ~J ~\sum_{<{\bf ij}>}^N ~\si_{\bf i}^z
%\si_{\bf j}^z ~-~ h~ \sum_{\bf i} ~\si_{\bf i}^x. \label{eq_tim} \eea
We take $h \ge 0$ without loss of generality.
% since we can always resort to the unitary transformation $\si_{\bf i}^x \to
%\si_{\bf i}^x$, $\si_{\bf i}^y \to - \si_{\bf i}^y$ and $\si_{\bf i}^z \to 
%-\si_{\bf i}^z$, which flips the sign of
%$h$ but leaves $J$ unchanged. 
%The operator $\si_{\bf i}^x$ acting on the 
%eigenstate of $\si_{\bf i}^z$ with eigenvalue $+1$ (up state) changes it to 
%the down state with eigenvalue $-1$. 
The non-commuting transverse field term introduces quantum fluctuations in 
the model causing a QPT from an ordered phase ($m_x= <\si^x> \neq 0$, where 
$< \cdots>$ represents the ground state average of an operator) to a 
disordered paramagnetic (PM) phase ($m_x=0$) at a critical value of $h=h_c$.
For $h<h_c$, there are two degenerate ground states with FM ordering in 
which the $Z_2$ symmetry is broken. For a finite chain, there is a finite 
rate of tunneling between the two states which is exponentially small in the 
system size (due to a large barrier); the tunneling leads to a breaking of 
the exact degeneracy which exists for an infinite system and produces an 
exponentially small energy gap ($\sim \exp(-cL^d)$) between the two lowest 
energy states. The long-range order persists up to $h_c$. For $h>h_c$, the 
field term $h$ wins over the cooperative interaction $J_x$, leading to the 
vanishing of the order parameter, and the system is in a PM phase. The TIM
Hamiltonian can be further generalized incorporating a longitudinal
magnetic field $h_L$ such that
\bea H ~=~- ~\sum_{<{\bf ij}>} ~J_{\bf ij} ~\si_{\bf i}^x \si_{\bf j}^x ~-~ 
h~ \sum_{\bf i} ~\si_{\bf i}^z-h_L \sum_i \sigma_i^x. \label{eq_timlongi} \eea
%This model undergoes a phase transition at $T=0$ from a ferromagnetic phase
%to a paramagnetic phase at $g = g_c$.

In Table I, we present definitions of all the major critical 
%\ref{\intro_tab1}
exponents taking the example of TIM with magnetization $m_x$ as the
order parameter $\cal O$, $h_L$ as the conjugate field or the longitudinal
field, and $h$ (analogous to $g$ in (\ref{intro_nuz})) as the non-commuting
transverse field which leads to quantum fluctuations resulting in a QPT.

%\begin{widetext} 
\begin{table}[h]
\caption{Definition of critical exponents associated with a quantum critical 
point occurring in magnetic system (see text) \ct{vojta03}.}
\begin{tabular}{c|c|c|c}
\hline
Quantity & Exponent & Definition & Conditions \\
\hline \hline
Correlation length $\xi$ & $\nu$ & $\xi \propto |h-h_c|^{-\nu}$ & $h \to h_c$ 
and $h_L = 0$\\[1ex]
Order parameter $m_x$ & $\beta$ & $m_x \propto (h_c-h)^\beta $ &
$~h \to h_c$ from below and $h_L=0$\\ [1ex]
Specific heat $C$ & $\al$ & $C \propto |h-h_c|^{-\al}$
& $h \to h_c$ and $h_L=0$\\ [1ex]
Susceptibility $\chi$ & $\ga$ & $\chi \propto |h-h_c|^{-\ga}$ &
$h \to h_c$ and $h_L=0$\\[1ex]
Critical isotherm & $\de$ & $h_L \propto |m_x|^{\de} {\rm sign}~ m_x$ &
$h_L \to 0$ and $h=h_c$\\[1ex]
Correlation function $G~$ & $\eta$ & $~G(r) \propto |r|^{-d+2-\eta}~$ &
$h=h_c$ and $h_L=0$\\[1ex]
Correlation time $\xi_{\tau}$ & $z$ & $\xi_{\tau} \propto \xi^z$ & $h \to h_c$
and $h_L=0$ \\
\hline
%\label{intro_tab1} 
\end{tabular} \end{table} 
%\end{widetext}

{We mention here some of the very early studies of the phase 
transitions in TIMs. This system was studied by de Gennes \ct{degennes63}}, 
to theoretically model the order-disorder transition in some double well 
ferroelectric systems such as potassium dihydrogen phosphate (KDP or 
KH$_2$PO$_4$) crystals \ct{blinc60}; we shall provide a detailed discussion 
on this in Sec.~\ref{expt2}. The connection to KDP ferroelectrics led to 
studies of the mean field phase diagram \ct{brout66,stinchcombe73}, series 
studies \ct{elliott61} of the model, and the exact solution in one dimension 
\ct{katsura62,pfeuty70}; perturbative calculations of transverse 
susceptibilities in one dimension and on the Bethe lattice were also 
carried out \ct{fisher63}. The major experimental systems which are 
modeled in terms of TIMs are summarized in Table II. {In 
addition, an experimental study of quantum simulations of a 
transverse Ising Hamiltonian of three spins with frustrating interactions 
between them has been reported recently; this is realized in 
a system of three trapped atomic ions whose interactions can be precisely 
controlled using optical forces \ct{kim10}.}

{For the sake of completeness, we mention here some mean field 
studies of the model in higher dimensions (or for higher spin values).}
The TIM in a general dimension was studied by Brout $et~al.$ \ct{brout66}, 
within mean 
field theory by reducing it to an effective single-site problem; this method 
has been generalized to study the elementary (spin-wave) excitations around
the mean field state for pure \ct{brout66} as well as a dilute TIM \ct{lage76} {(for details see \ct{chakrabarti96,das03})}.
Quantum spin models can also be solved in the large spin ($S \to \infty$) 
limit when one incorporates quantum fluctuations over the classical ground 
state to first order in $1/S$ by applying the Holstein-Primakoff 
transformation \ct{holstein40,anderson52,kubo52} {(see 
Appendix \ref{app_largespin})}.
 %appropriately and diagonalizing the
%Hamiltonian using Fourier transform and Bogoliubov transformation to obtain 
%decoupled harmonic oscillators in wave vector space. This method has been 
%useful in studying transverse Ising spin chains and their variants 
%\ct{sen90,dsen91} and also for a Kitaev model \ct{baskaran08}.

% In the ordered phase, each proton of the hydrogen bond of KDP can occupy 
% one of the two minima of the potential, created by the oxygen atoms 
%\ct{degennes63,blinc60}, which is modeled in terms of the cooperative
%interaction term in Eq.~(\ref{eq_timgen}) with $J_{ij}$ being the electrostatic
%interaction between the neighboring protons. The transverse field term 
%denotes the tunneling integral which determines the rate of tunneling from 
%one minimum to the other. For small transverse fields, protons will choose
%one of the minima leading to a spontaneous symmetry breaking. On the other 
%hand, if the transverse field exceeds a critical value, the symmetry is 
%restored at the cost of the ferroelectric long-range order. The strength of 
%the tunneling term can be tuned by deuteration (i.e., replacing the hydrogen 
%by deuteron) or by applying pressure which increases the double well overlap 
%integral. 

The one-dimensional version of the model (\ref{eq_timgen}) with 
nearest-neighbor FM interactions can be written as 
\beq H ~=~ - J_x \sum_{i=1}^{N-1} \si_i^x \si_{i+1}^x ~-~ h \sum_{i=1}^N 
\si_i^z. \label{eq_tim1d} \eeq
A duality transformation between low field and high field, which is a quantum 
generalization of the duality relation between the low and high temperature 
phases of the two-dimensional classical Ising model \ct{kramers41}, enables 
one to locate the QCP exactly \ct{fradkin78,kogut79}. Defining Pauli spin 
operators on a dual lattice (using for example, $\mu^x_i = \si_{i+1}^z\si_i^z$
and $\mu^z_i = \prod_{j<i} \si_j^x$) and 
exploiting the self-duality of the QCP, one finds $h_c=J_x$ (see Kogut, 1979, 
for a review) which turns to be the exact value (see Sec.~\ref{jorwigtr}).
In subsequent sections, we will often set $J_x=1$ so that $h_c=1$. 
{In Sec.~\ref{jorwigtr}, we shall present the exact 
diagonalization of (\ref{eq_tim1d}) using the Jordan-Wigner transformation.}
%Recently, it has been shown that the model shows a first-order phase 
%transition under a hyperbolic deformation \ct{ueda10}.

% We consider a new \emph{dual lattice} 
%in which the sites of the original lattice are replaced by the bonds in the 
%dual lattice and vice versa. The new operators residing on the sites of
%the dual lattice defined by the relations 
%\bea \mu^x_i &=& \si_{i+1}^z\si_i^z, \non \\
%\mu^z_i &=& \prod_{j<i}\si_j^x, \label{eq_duality} \eea
%also satisfy the Pauli spin algebra. The original Hamiltonian in 
%Eq.~(\ref{eq_tim1d}) can be rewritten as follows:
%\bea H_D=-\sum_i \mu^z_i\mu^z_{i+1} - \la \sum_i \mu^x_i,
%\eea
%which implies $$H(\si,\la)=\la H_D(\mu,\la^{-1}).$$
%Clearly, in the dual lattice, $\la > 1$ corresponds to the disordered 
%phase and $\la<1$ is the ordered phase, in contrast to the original $H$ 
%in the $\si$ representation, and one gets $E(\la)=\la E(\la^{-1})$. At the 
%critical point, the energy gap between the ground state and the first excited 
%state vanishes. Assuming a unique $\la_c$ for this model, we find that 
%the self-duality of the models leads to $\la_c=1$ which%
%in fact will be shown later to be the exact value.

%Apart from the experimental realizations of the TIM to be discussed in 
%Sec.~\ref{expt}, we note that it can be mapped to the problem of loading 
%one-dimensional hard core bosons or non-interacting fermions onto a
%commensurate optical lattice potential \ct{grandi08}, and it describes the 
%Toulouse point in the sine-Gordon model where this model can be mapped to a 
%system of spinless fermions \ct{giamarchi04}. 

%\begin{widetext}
\begin{center}
\begin{table}
\begin{tabular}{|p{4cm}|p{3cm}|p{6cm}|}
\hline
Category & Some specific examples & References \\
\hline
Order disorder ferroelectrics & KH$_2$PO$_4$ &\ct{kobayashi68}\\
& KD$_2$PO$_4$ &\ct{kaminow68,cochran69,samara71}\\
\hline
Jahn-Teller systems & DyVO$_4$ & \ct{gehring71}\\
& TbVO$_4$ & \ct{elliott71b}\\
\hline
Mixed hydrogen bonded ferroelectrics (proton glasses) & 
Rd$_{1-x}$(NH$_4$)$_x$H$_2$PO$_4$ & \ct{pirc85}\\
\hline
Dipolar magnets (quantum spin-glass) & LiHo$_x$Y$_{1-x}$F$_4$ & 
\ct{wu91,wu93}\\
\hline
Quasi-one-dimensional Ising systems & CoNb$_2$O$_6$ & \ct{coldea10}\\
\hline
\end{tabular}
\caption{Systems described by to Ising model in a transverse field; for a 
more exhaustive list we refer to \ct{stinchcombe73} }
%\label{timsystems}
\end{table}
\end{center}
%\end{widetext}
\begin{figure}[htb]
\begin{center}
%\hspace*{-2cm}
\includegraphics[height=2.8in,width=3.0in]{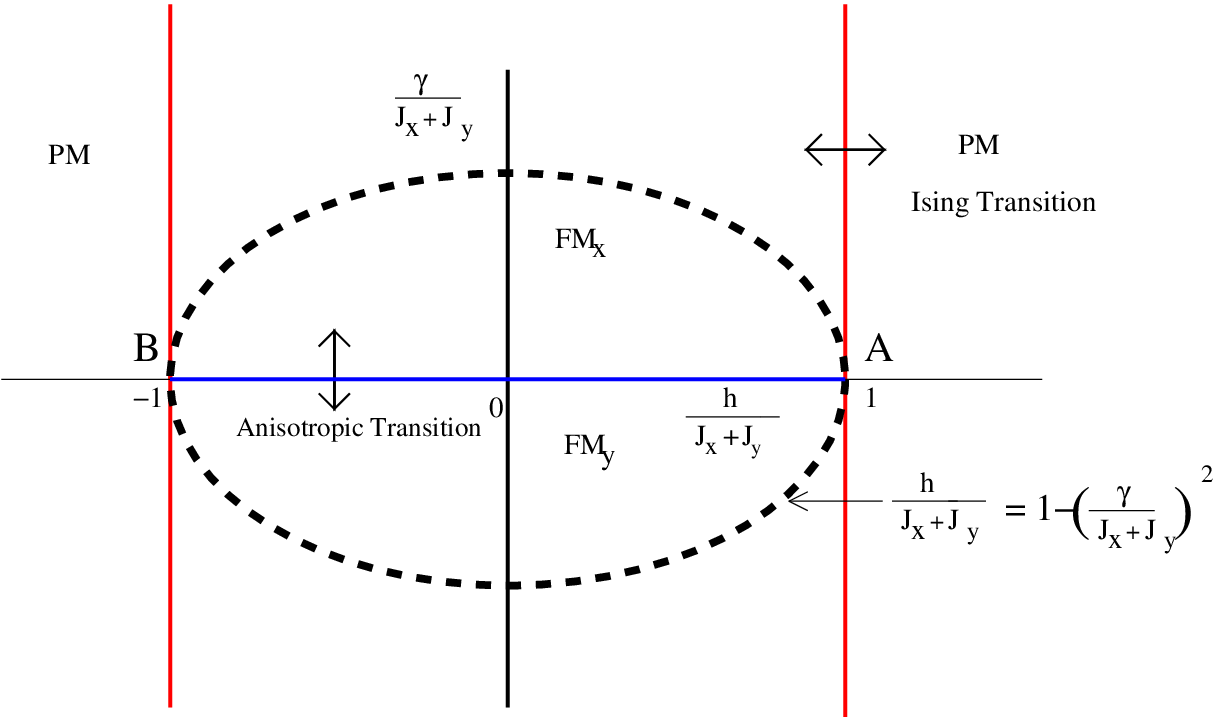}
\caption{(Color online)
The phase diagram of a spin-1/2 transverse $XY$ chain. The vertical 
red lines denote the Ising transitions between the FM phase and the PM phase. 
The horizontal bold line ($\ga=J_x -J_y=0, -(J_x + J_y) < h <J_x+J_y$) stands 
for the anisotropic phase transition between the ferromagnetic phases 
${\rm FM_x}$ and ${\rm FM_y}$ with magnetic ordering along $\hat x$ and 
$\hat y$ directions, respectively. The multicritical points where 
the Ising and anisotropic transitions line meet ($\ga=0, h = \pm (J_x+J_y)$) 
are denoted by $A$ and $B$ in the figure. The dotted line given by the equation
$h/(J_x + J_y) =1- \ga^2/ (J_x + J_y)^2$, denotes the boundary between the 
commensurate and incommensurate ferromagnetic phases. For the Ising and the 
anisotropic transitions critical exponents are $\nu=z=1$ while for the 
transition at the MCPs, $\nu_{mc}=1/2$ and $z_{mc}=2$. (After Bunder and 
McKenzie, 1999).} \label{fig_xyphase} \end{center} \end{figure}

The TIM can be generalized to the $XY$ Hamiltonian in a transverse field 
incorporating additional interactions defined by the following Hamiltonian in 
one dimension \ct{katsura62,barouch70,barouch71,barouch711,suzuki71},
\bea H~=~ - ~\sum_i ~[J_x \si_i^x \si_{i+1}^x +J_y \si_i^y \si_{i+1}^y +
h \si_i^z]. \label{ham_xy} \label{eq_txy}\eea
Without loss of generality, we
can assume that $J_x \ge |J_y| \ge 0$ and therefore that $J_x + J_y > 0$.
The higher dimensional version of the model appears in the pseudo-spin 
representation of the BCS Hamiltonian,
and its mean field treatment yields exactly the BCS gap equation 
\ct{anderson58}. The spin chain defined in (\ref{eq_txy}) is exactly 
solvable as will be shown in Sec.~\ref{exactsol}. The phase diagram of the 
one-dimensional model is shown in Fig.~\ref{fig_xyphase}
with the anisotropy parameter $\ga = J_x -J_y$.
%and $J_x + J_y=1$. 
The system undergoes transitions from PM to 
FM phase at $h=\pm (J_x+J_y)$; these transitions belong to the universality 
class of the transverse Ising chain and are hence called Ising transitions. The
ferromagnetic order in the $XY$ plane is in the $\hat x$ direction ($\hat y$
direction) if $J_x > J_y$ ($J_x < J_y$) and $|h| < J_x+J_y$; there is a QPT 
between these two phases at $J_x=J_y$ referred to as anisotropic transition. 
On this anisotropic transition line ($J_x=J_y$) between two FM phases
(referred to as ${\rm FM_x}$ and ${\rm FM_y}$ phases) the energy gap vanishes
for a non-commensurate wave vector $k_0$; $k_0=0$ or $\pi$ when $h \to 
-(J_x+J_y)$ or $h \to J_x +J_y$. There are two multicritical points (MCP)
(at $h=\pm (J_x +J_y)$ and $\ga=0$) denoted by A and B in the phase diagram 
(\ref{fig_xyphase}) where the Ising and anisotropic transition lines meet 
\ct{damle96}. In subsequent sections, we will often set $J_x + J_y = J =1$.
{Due to its exact solvability and rich phase diagram with 
critical and multicritical points as well as gapless critical lines, the 
model in (\ref{eq_txy}) has been used throughout Secs. 5 and 6.}

%The transverse $XY$ model can be obtained using the simple Bose-Hubbard 
%model for one-component (spinless) bosons with nearest-neighbor hopping which 
%can be realized in optical lattices \ct{lewenstein06}. This is given by the 
%Hamiltonian
%\beq H=-t \sum_{<ij>} \left ( b^{\dg}_i b_j + H.c. \right) + \frac{U}{2} 
%\sum_i n_i(n_i-1) - \mu \sum_i n_i, \eeq
%where $b_i$'s are bosonic operators and $n_i=b_i^\dg b_i$; $t$, $U$ and 
%$\mu$ are the tunneling amplitude, on-site repulsive interaction and 
%chemical potential, respectively. In the limit, $U \gg t,\mu$, we have at most
%one particle per site and consequently the Bose-Hubbard Hamiltonian can be 
%mapped to the transverse $XY$ Hamiltonian with interactions $J_x=J_y=J$ and 
%the chemical potential $\mu$ playing the role of the transverse field. 

\subsection{Some exactly solvable models related to transverse Ising and 
$XY$ models}
\label{sec_relatedmodels}

In this section, we shall briefly introduce a few spin 
models which are closely related to the transverse Ising and $XY$ spin chains. 
These models are exactly solvable 
and hence have turned out to be extremely useful in recent studies of quantum 
information and dynamics as discussed later.
% However, we suggest that this 
%section may be skipped in the first reading because of its technical nature.}

Let us first introduce an extended transverse Ising spin chain whose ground 
state is given exactly by a finite rank matrix product state (see the 
discussion at the end of Sec. \ref{qpt}). Referring to
Eq. (\ref{eq_mpstate}), with $D=d=2$, and defining $A_1 = (I-\si^z)/2 + \si^-$
and $A_2 = (I + \si^z)/2 + g \si^+$, where $I$ is $2 \times 2$ identity 
matrix and $\si^{\pm} = \si^x \pm i \si^y$, we obtain a $Z_2$ symmetric 
Hamiltonian \ct{wolf06}
\beq H ~=~ \sum_i ~[2(g^2 -1) \si_i^z \si_{i+1}^z ~-~ (1+g)^2 \si_i^x ~+~
(g-1)^2 \si_i^z \si_{i+1}^x \si_{i+2}^z] \label{eq_mps_hamil} \eeq
with periodic boundary conditions; {see Appendix \ref{app_mps} 
for the derivation}. This Hamiltonian undergoes a QPT at $g_c=0$ with a 
diverging correlation 
length with critical exponents $\nu=1$ and $z=2$; the state at the QCP is a 
Greenberger-Horner-Zeillinger state \ct{greenberger89}. The QCP at $g_c =0$ 
is unconventional in the sense that
% neither the entanglement entropy nor 
the ground state energy is analytic at this point.

We can define a duality transformation to a spin-1/2 chain whose site labels 
run over $i + 1/2$, with the mappings
\beq \tau_{i+1/2}^z ~=~ \sigma_i^z \si_{i+1}^z, ~~~\tau_{i-1/2}^x 
\tau_{i+1/2}^x ~=~ \si_i^x, ~~~\tau_{i-1/2}^y \tau_{i+1/2}^y ~=~ 
- ~\si_{i-1}^z \si_i^x \sigma_{i+1}^z, \label{map} \eeq
where $\tau_{i+1/2}^a$ denote Pauli matrices for the dual spin-1/2 chain.
Under this transformation, Hamiltonian in Eq.~(\ref{eq_mps_hamil}) gets mapped
to the transverse $XY$ model in Eq.~(\ref{eq_txy}), with $h/(J_x + J_y)= 
(1-g^2)/(1+g^2)$ and $ \gamma/(J_x + J_y) =2g/(1+g^2)$. We then see that
the QPT at $g=0$ corresponds to the multicritical point lying at
$h/(J_x + J_y)=1$ and $ \gamma/(J_x + J_y) = 0$ in Fig.~\ref{fig_xyphase}
and $g$ parametrizes a path along the dotted line. {The notion 
of a matrix product state and the Hamiltonian in (\ref{eq_mps_hamil}) have 
been useful in recent quantum information theoretic studies 
\ct{tagliacozzo08,pollmann09,zhou08,cozzini071} (see Secs. \ref{e_entropy}
and \ref{fidelity}). }
%in the phase diagram in Fig.~\ref{fig_xyphase}.

We note that a transverse Ising spin chain incorporating three-spin 
interactions, given by the Hamiltonian 
\bea H ~=~ - ~\sum_i \si_i^z [h + J_3 \si_{i-1}^x \si_{i+1}^x] ~+~ J_x ~\sum_i 
\si_i^x \si_{i+1}^x \label{eq_threespin} \eea
which is dual to the transverse $XY$ spin chain in Eq.~(\ref{eq_txy}), has 
already been studied \ct{kopp05}.\footnote{The interaction $J_3$ which is 
generated in the first step of a real space renormalization group (RSRG) study
of transverse Ising models \ct{hirsch79} has been found to be irrelevant in 
determining the critical behavior of the system.} {The model 
is exactly solvable and hence has been useful in studying quantum dynamics 
across its QCPs \ct{divakaran07}.}

Let us also introduce an anisotropic spin-1/2 $XY$ spin chain Hamiltonian in 
which the strength of the transverse field alternates between $h+\de$ and 
$h-\de$ at odd and even sites, respectively, \ct{perk75,okamoto90,deng08} 
given by
%\bea H &=& - ~\sum_j ~[{\frac{(J_x + J_y)}{4}} (\si^x_j 
%\si^x_{j+1} + \si^y_j \si^y_{j+1}) \non \\ 
%& & ~~~~~~~~~+~ {\frac{(J_x-J_y)}{4}} (\si^x_j \si^x_{j+1} - \si^y_j 
%\si^y_{j+1}) + \frac {(h-(-1)^j\de )}{2} \si^z_j]. \label{eq_dimerham} \eea
\bea H ~=~ - ~ \sum_j ~\left[{\frac{(J_x + J_y)}{4}} (\si^x_j \si^x_{j+1} + 
\si^y_j \si^y_{j+1}) ~+~ {\frac{(J_x-J_y)}{4}} (\si^x_j \si^x_{j+1} - \si^y_j 
\si^y_{j+1}) + \frac {(h-(-1)^j\de )}{2} \si^z_j \right]. 
\label{eq_dimerham} \eea
%where we have set $J_x+J_y =J$, and $\ga= J_x-J_y$ measures the anisotropy.
%The presence of two underlying sub-lattices necessitates the introduction of 
%a pair of fermion operators $a$ and $b$ \ct{perk75,okamoto90,deng08} 
%for even and odd sites. 

\begin{figure}
\begin{center} \includegraphics[height=1.8in,width=3.0in]{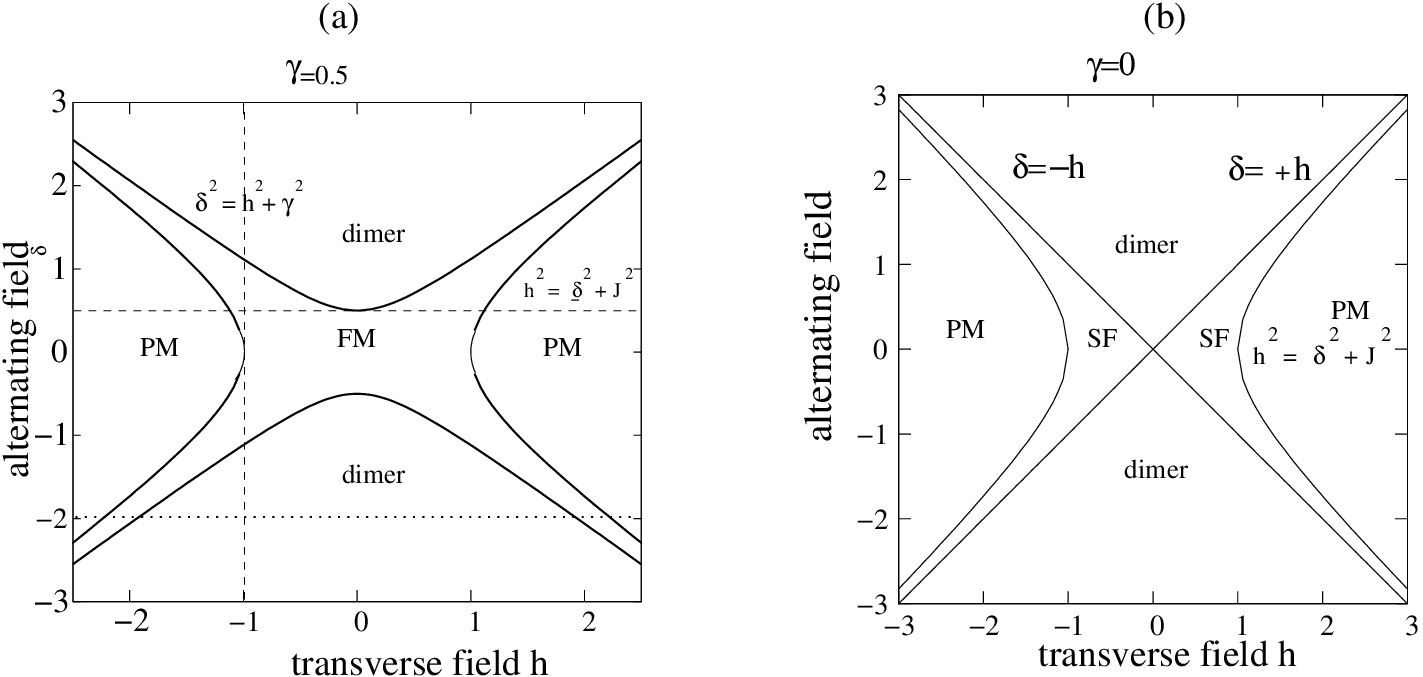}
\caption{Phase diagram of the $XY$ chain in an alternating transverse field for
$\ga=0.5$ (Fig. (a)) and $\ga=0$ (Fig. (b)). In (a), the PM $\leftrightarrow$ 
FM and FM $\leftrightarrow$ dimer phase transitions with critical exponents 
$\nu=z=1$ are shown by the phase boundaries $h^2 = \de^2 + J^2$ and $\de^2 = 
h^2 + \ga^2$, respectively. There is also a QPT with $\nu =2$ and $z=1$ when 
two special points ($h=\pm 1$, 
$\de = 0$) and ($h=0$, $\de = \pm \ga$) points are approached along the dashed 
line. In (b), the gapless $U(1)$-symmetric superfluid (SF) $\leftrightarrow$ PM phase boundary is given 
by $h^2 = \de^2 + J^2$ with ordering wave vector $\cos k_0 = \sqrt{h^2 - 
\de^2}/J$ for $\de^2 < h^2 < \de^2 + J^2$; this transition belongs to the 
Lifshitz universality class with critical exponents $\nu=1/2$, $z=2$. A 
different critical behavior occurs at $h=\pm 1$, $\de \to 0$ where $\nu=1$, 
$z=2$. Ising exponents are recovered while approaching the point $h=0$, $\de =
0$ along every path other than $\de=0$, $h\to 0$ when the spin chain is on 
the anisotropic
critical line of Fig.~\ref{fig_xyphase}. (After Deng $et~al.$, 2008).
%The dotted line in Fig. (a) shows the direction of 
%the transverse quenching. We quench the system along a gapless
%line parallel to the $\ga$-axis (perpendicular to the plane of this paper)
%passing along the phase boundary $h^2 = \de^2 + J^2$ at all times.
} 
\label{fig_dimerphase} \end{center} \end{figure}
{The rich phase diagram of the model in (\ref{eq_dimerham}), 
obtained by exact solution using the Jordan-Wigner transformation, is shown in 
Fig.~\ref{fig_dimerphase} (with $\gamma=J_x-J_y$ and $J_x+J_y =J$) for 
$\ga=0$ and $\ga=0.5$. We list below the associated
QCPs and corresponding critical exponents.}
\medskip

{{\noi For $\ga > 0:~$} The PM $\leftrightarrow$ FM and FM 
$\leftrightarrow$ dimer phase transitions 
are given by the phase boundaries
$h^2 = \de^2 + J^2$ and $\de^2 = h^2 + \ga^2$, respectively. (The dimer
phase is one in which $<\si_j^z>$ has a staggered order). \\
\noi (i) In general, for $\ga > 0$, the QCPs belong to the $d=2$ Ising 
universality class with critical exponents $\nu=z=1$. \\
\noi (ii) A different critical behavior is observed for
$h \to 0, \de=\pm \ga$ and $h=\pm 1, \de \to 0$ with $\nu=2,z=1$.}
\medskip

{{\noi For $\ga = 0:~$} \\
(i) Superfluid (SF) $\leftrightarrow$ dimer
and SF $\leftrightarrow$ PM phase transitions.
The ordering wave vector in the gapless SF phase is $k_0$ given by
$\cos k_0 = \sqrt{h^2 - \de^2}/J$ for $\de^2 < h^2 < \de^2 + J^2$. \\
\noi (ii) In general, the QCPs on the boundary line belongs to
the Lifshitz universality class with critical exponents $\nu=1/2,z=2$.\\ 
\noi (iii) Once again, a different critical behavior occurs at $h=\pm 1, 
\de \to 0$ where $\nu=1,z=2$. \\
\noi (iv) The Ising exponents are recovered while approaching the point 
$h=0,\de =0$ along every path other than $\de=0,h\to 0$ where there is 
no critical behavior. \\}

{Because of the exact solvability and the possibility of exploring different 
QCPs, this model has turned out to be extremely useful
% from the viewpoint of non-equilibrium quantum dynamics 
\ct{deng08,divakaran08} (see Sec. \ref{sec_gapless}).}

{Finally, let us briefly mention an infinite range interacting 
$XY$ Hamiltonian known as the Lipkin-Meshkov-Glick (LMG) model \ct{lipkin65}; 
this is an exactly solvable model with infinite coordination number 
\ct{botet82,ribeiro07}. Although this model is not discussed in detail
in this review, it has been studied extensively in recent years, particularly 
from the point of view of quantum information 
\ct{dusuel04,dusuel05,vidal04a,vidal04b,barthel06,wichterich10,kwok081} and 
dynamics \ct{caneva08}.} The model consists of $N$ spin-1/2 objects and given 
by 
\beq H ~=~ - ~\frac{J}{N} ~\sum_{i<j} ~(\si_i^x \si_j^x + \ga \si_i^y 
\si_{i+1}^y) ~-~ h ~\sum_i ~ \si_i^z, \label{haminfinite} \eeq
where $N$ is the number of spins and $\ga \leq 1$ is the anisotropy parameter.
%We note that this model differs 
%from the one studied in \oct{chakrabarti06a,chakrabarti06b}, where the 
%spins were taken to be living on two sub-lattices, with Ising interactions 
%only between spins on different sub-lattices. 
Introducing $\vec S_a = \sum_i {\vec \si}_i^a, a=x,y,z$,
the Hamiltonian in (\ref{haminfinite}) can be written up to a constant as
$-(1/2N) (S_x^2 + \ga S_y^2) - h S_z$. The Hamiltonian satisfies the 
commutation relations $[H, {\vec S}^2]=0$ and also $[H, \prod_i \si_i^z]=0$ 
which implies that the parity of the number of spins pointing in the 
direction of the field is conserved. Moreover, $[H, S_z]=0$ for $\ga=1$. 

In the $N \to \infty$ limit, the LMG model shows a QPT as $2h/J \to 1^-$ 
characterized by mean field exponents for all values of $\ga$ \ct{botet83}; 
the magnetization in the $\hat x$ direction (or in the $xy$-plane for the 
isotropic case $\ga=0$) is given by $m = \sqrt{1 - 
4 h^2/J^2}$ for $h \leq J/2 $ and vanishes for $h > J/2$. 
For $h>J/2$, the ground state is non-degenerate for any $\ga$; for $h<J/2$, 
it is doubly degenerate for any $\ga \neq 0$, indicating the breakdown of Ising
symmetry. The energy gap vanishes at the QCP as $\sqrt{(h-J/2)(h-\ga)}$ for 
$h \geq J/2$. For finite $N$ (which also means finite 
$S$ for this model), the model can be diagonalized using Holstein-Primakoff 
transformation \ct{holstein40,anderson52,kubo52} both for $\ga=0$ \ct{das06} 
and $\ga \neq 0$ \ct{dusuel05}. The finite temperature phase diagram in the 
$N \to \infty$ limit can be found by 
%To derive the phase boundary, let us denote the mean field value with 
%magnetization $m_z$ given by $m_z={\sum_i \langle S_i^z \rangle}/N$ so that 
%the effective field on a single spin is given
%The Hamiltonian governing any one of the spins is 
%\beq H_s ~=~ - ~4Jm_zS_{tot}^z ~-~ 2h S_{tot}^x. \eeq
using the self-consistent method for the equivalent single particle 
Hamiltonian \ct{das06}; see Sec. 6.1.

\subsection{Quantum-classical correspondence and scaling}
\label{sec_class+quant}

The mapping from temperature to imaginary time ($\bar \tau$), given by 
${\bar \tau} = \beta$ where $\beta=1/(k_B T)$,
enables us to make use of an imaginary time path 
integral formalism of a quantum Hamiltonian \ct{feynman72} which states that 
the transition amplitude between two states can be calculated by summing over 
the amplitudes of all possible paths going between them. To formulate the 
corresponding path integral, one divides the temperature interval $L_{\tau}=
1/(k_B T)$, where $k_B$ is the Boltzmann constant, into $M$ small intervals 
$\de \bar \tau$ such that $M \de \bar \tau = \beta$ in the limit of $\de 
\bar \tau \to 0$ and $M \to \infty$ with $\beta$ remaining finite. The path
integral formalism of the partition function of a 
$d$-dimensional quantum system effectively gives us an equivalent action of a 
$(d+1)$-dimensional classical system, with the additional time dimension being
of finite size $(=1/(k_B T))$. In the limit $T \to 0$, the size of the 
temporal direction diverges and one gets a truly $(d+1)$-dimensional classical 
action. For $T \neq 0$, the time direction $L_{\tau}$ is finite and does not 
contribute close to the critical point when $\xi \to \infty$ and one arrives
at a $d$-dimensional classical action \ct{sondhi97}. A QPT at $T=0$ is 
therefore associated with a diverging correlation length in the temporal 
direction (or the relaxation time) which scales as $\xi_\tau \sim 1/\De$, 
where $\De$ is the energy gap between the ground and the first excited 
state \ct{kogut79}. Although 
as mentioned above, such a correspondence exists for all models discussed
before, we shall illustrate this classical quantum correspondence
using the example of the transverse Ising models.

For a TIM in $d$ dimensions, the classical-quantum correspondence was 
established \ct{young75,hertz76} using the imaginary time path integral in the
Matsubara representation with Matsubara frequencies (MFs) $\om_m = 2\pi m k_B 
T$, $m=0,1,2,\cdots$ \ct{mahan00}. At $T=0$, these MFs become continuous and 
one arrives at the Landau-Ginzburg-Wilson (LGW) action of a classical model 
with an additional temporal direction. This correspondence establishes that 
the upper critical dimension ($d_u^c$) for the QPT of TIM is $(3+1)$; one can 
calculate the associated critical exponents using the $\ep$-expansion 
technique \ct{amit84} around the upper critical dimension \ct{sachdev99}. 
For $T \neq 0$, the MFs become discrete and only the mass 
term corresponding to $m=0$ vanishes at the critical point, whereas the 
mass terms for $m \neq 0$ become irrelevant under renormalization.
This effectively results in a LGW action of a classical
Ising model in $d$ dimensions. 

\begin{figure}
\begin{center} \includegraphics[height=2.8in]{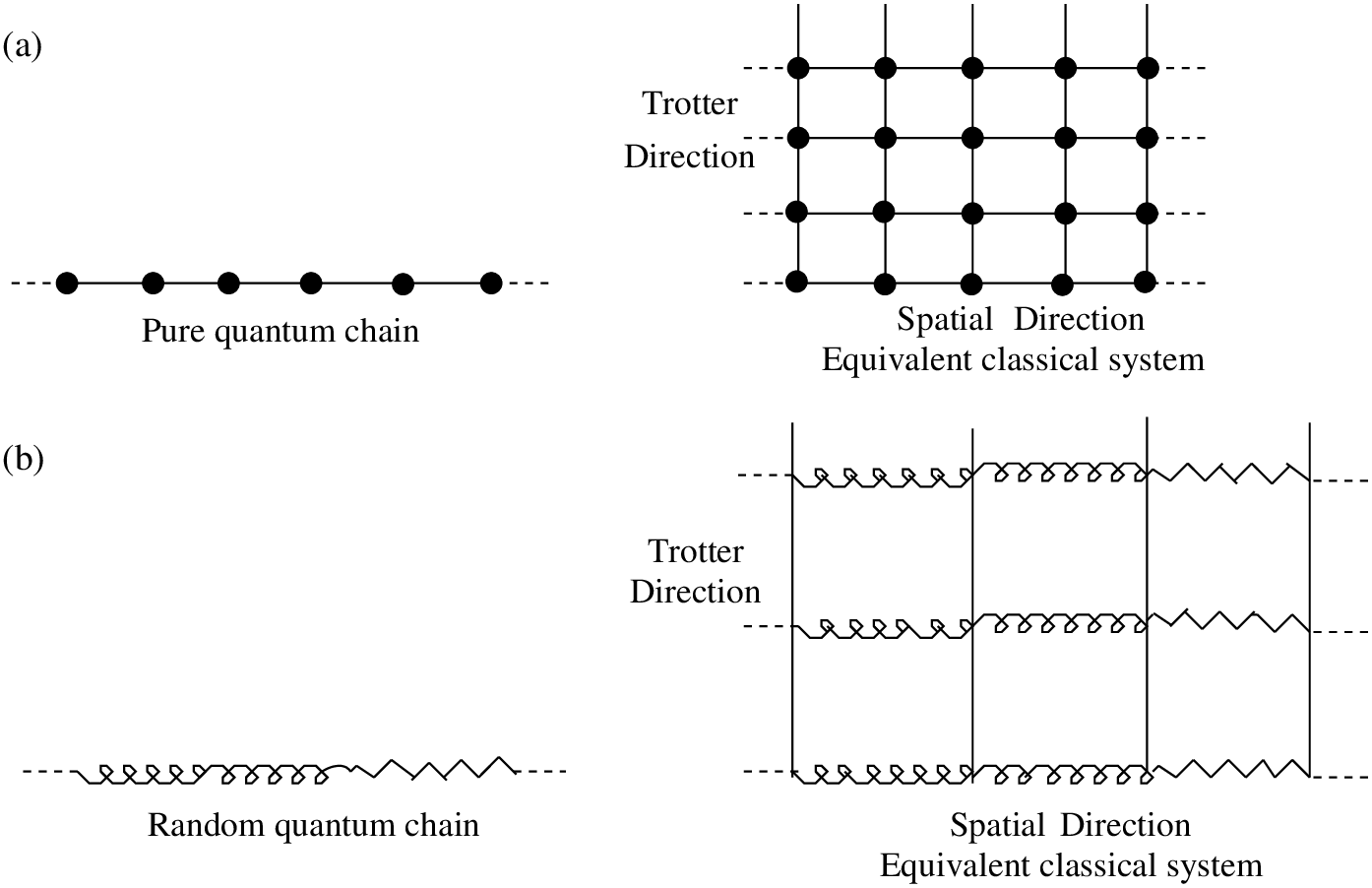}
\caption{The equivalent classical model: (a) for a pure transverse Ising 
chain, and (b) for a chain with random interactions but uniform transverse 
field. Here continuous lines indicate ferromagnetic interactions and zig-zag 
lines denote random interactions. Figure (b) shows that the randomness is 
infinitely correlated along the Trotter direction (i.e., the same pattern is 
replicated) of the equivalent classical random model. This represents a 
version of the McCoy-Wu model \ct{mccoy68}.} \label{fig_mapping1}
\end{center} \end{figure}

We can use the Suzuki-Trotter (ST) \ct{suzuki76,suzuki86} formalism 
to show that the QPT of a $d$-dimensional TIM is equivalent to a 
$(d+1)$-dimensional classical Ising model \ct{elliott70}. The ST formalism 
relies on the Trotter formula $\exp(\hat A_1+ \hat A_2) = \lim_{M \to \infty} 
[\exp (\hat A_1/M) \exp(\hat A_2/M)]^M$, where $\hat A_1$ and $\hat A_2$ are 
quantum mechanical bounded operators which may not commute with each other. 
Starting from a transverse Ising chain (\ref{eq_tim1d}) of length $L$ with 
interaction term $K =\beta J_x$, we use the Trotter formula; inserting 
$M$ complete set of eigenstates of the operator $\si_x$, we arrive at 
%We
%consider the canonical partition function of a one-dimensional TIM with NN FM 
%interactions as in Eq.~(\ref{eq_tim1d}) with periodic boundary conditions; 
%the partition function is given by
%\beq {\cal Z} = {\rm Tr} ~\exp(-\beta \hat H) = {\rm Tr} ~ \exp \left [ 
%\sum_{j=1}^N (K \hat \si_j^z \hat \si_{j+1}^z + \beta h \hat \si_j^x) \right],
%\label{eq_suzuki1} \eeq
%where $K =\beta J$. 
%using the Trotter formula and inserting a complete set of 
%eigenvectors $|\si \rangle$ of the operator $\hat \si_i^z$ such that $\hat 
%$\si^z|\si \rangle= \si |\si \rangle$ (where $\si=\pm 1$), one generates 
%$M$ copies of the original quantum spin along an additional direction (called 
%the Trotter direction). In the process
the partition function of a classical Ising model on a $L\times M$ square 
lattice, where $M$ denotes the size of the Trotter direction, with anisotropic 
interactions $K/M$ and $K_M$ $=(1/2) \ln \coth (\beta h/M)$ in the spatial and
Trotter directions, respectively (for details, see Ref. \ct{dutta96}).
We note that in the limit $M \to \infty$, 
$K/M$ vanishes while $K_M$ diverges logarithmically unless $T = 0$. We take 
$M \to \infty$ and $\beta \to \infty$ simultaneously so that $\beta/M$ is
finite; then the model becomes equivalent to a two-dimensional classical 
Ising model. The above calculation can be generalized to higher dimensions. 
%in order to 
%establish the $d_{\rm quantum} \to (d+1)_{\rm classical}$ correspondence. When
If the interactions of the spins are random, the ST mapping leads to a higher 
dimensional classical Ising Hamiltonian with randomness being striped or 
infinitely correlated along the Trotter direction (see 
Fig.~\ref{fig_mapping1}). The ST formalism and imaginary time path integral 
formalism are at the root of quantum Monte Carlo methods; for example, 
Rieger and Kawashima used continuous imaginary time cluster Monte Carlo 
algorithms for a transverse Ising model in $d =2$ \ct{rieger991}.

In a similar spirit, one can derive an equivalent quantum Hamiltonian starting
from the partition function of a classical Hamiltonian in an extreme 
anisotropic limit; this is known as the $\tau$-continuum formulation 
\ct{fradkin78,kogut79}. Let $\hat T$ denote the transfer matrix of a 
$d$-dimensional classical Hamiltonian $H_{\rm cl}$. The equivalent
quantum Hamiltonian $\hat H$ is defined through the relation $\hat T = 1 -\de 
\bar \tau \hat H + O( {\de \bar \tau}^2)$ where $\de \bar \tau$ ($\to 0$) is 
the infinitesimal lattice spacing along one of the spatial directions.
For example, one considers the row-to-row transfer matrix of the 
two-dimensional 
anisotropic classical Ising model at temperature $T$ \ct{schultz64}.
%The free energy of the classical system, given by the largest 
%eigenvalue of the transfer matrix, is now related to the ground state energy 
%of $\hat H$. For example, one considers with NN FM interactions 
%$J_1$ ($J_{\tau}$) along the rows (columns) with the Hamiltonian
%\beq H_{\rm cl} = -J \sum_{i,k} \si_{i,k} \si_{i+1,k} - J_{\tau} \sum_{i,k}
%\si_{i,k}\si_{i,k+1}, \label{eq_suzuki4} \eeq 
%is given by \ct{schultz64}
%$\hat T = \hat T_2^{1/2} \hat T_1 \hat T_2^{1/2}$, where 
%\beq \hat T_1 = \exp \left[ \sum_i {\tilde K} \si_i^z \si_{i+1}^z \right], ~~~
%\hat T_2 = \exp \left[\sum_i {\tilde K}_\tau \si_i^x \right], 
%\label{eq_suzuki5} \eeq
%where ${\tilde K} = - (1/2) \ln [\tanh (\beta J)]$ and ${\tilde K}_\tau =
%\beta J_{\tau}$. Clearly, $[\hat T_1, \hat T_2] \neq 0$. To cast the transfer 
%matrix into the form in Eq.~(\ref{eq_suzuki3}), we take 
In the extreme anisotropic limit, in which the interaction between spins of 
neighboring rows vanishes as $\de \bar \tau$ and between spins of the same 
row as $g\de \bar \tau$ so that the ratio of these two interactions ($=g$) 
remains finite, 
%in which 
%$\de \bar \tau \to 0$, ${\tilde K} \sim \la \bar \de \tau$ and ${\tilde K}_\de
%\bar \tau \sim \de \bar \tau$, so that $\la = {\tilde K}/{\tilde K}_\tau$ is 
%finite. In this limit, we can neglect the 
%non-commutativity between $\hat T_1$ and $\hat T_2$. One can then recast 
%$\hat T$ into the form of (\ref{eq_suzuki3}) 
one gets the equivalent one-dimensional quantum Hamiltonian (defined along the 
row) given in (\ref{eq_tim1d}). The temperature $T$ in the classical model maps
to the transverse field $h$ in the quantum Hamiltonian.

The scaling relations associated with QCP are discussed in detail in 
Ref. \ct{sachdev99}, and in Ref. \ct{sondhi97}; 
we present here a brief recapitulation. 
The inherent quantum dynamics and the diverging correlation length as well as
diverging relaxation time $\xi_{\tau} \sim \xi^z$ modify the scaling relation 
when compared to a classical phase transition. 
% At $T=0$ and close to the QCP ($ \la =g-g_c=0$), an operator $\cal O$ 
%scales as ${\cal O}(k,\om, \la)\sim \xi^{d_{\cal O}}f(k\xi,\om \xi^z, \la)$; 
%here $k$ and $\om$ are the wave vector and the frequency, $f$ is 
%the scaling function and $d_{\cal O}$ is the scaling dimension 
%of the variable $\cal O$ \ct{amit84}. At the QCP, on the other hand,
%${\cal O}(k,\om)\sim k^{-d_{\cal O}} \tilde f(\om/k^z)$, where $\tilde f$
%is another scaling function. 
For example, close to the QCP, the 
energy gap between the ground state and the first excited state scales as 
$\la^{z \nu}$, where $\la=g -g_c$ is the measure of the deviation from the 
critical point. Exactly at $\la = 0$, the energy gap vanishes at some 
momentum ${\vec k}_c$ as $|{\vec k} - {\vec k}_c|^z$; if $z=1$, the QCP is
Lorentz invariant. The ground state energy density therefore scales 
as $E_g\sim \xi^{-(d+z)} f(k\xi, \om \xi^z, \la \xi^{1/\nu})$, where we have used the fact
that the correlation volume is $\xi^{d+z}$. Using the scaling relation $\xi 
\sim \la^{-\nu}$ and the definition of the exponent $\al$ (see Sec.~\ref{qpt})
which yields $E_g \sim \la^{2-\al}$, we get the modified hyperscaling 
relation for a QPT given by 
\beq 2-\al = \nu(d+z), \label{eq_hyperscaling}\eeq which is valid below the 
upper critical dimension. When compared with the hyperscaling relation $2 - 
\al_{\rm cl} = \nu_{\rm cl} d$ of a classical system \ct{stanley87} where 
$\al_{\rm cl}$ and $\nu_{\rm cl}$ are critical exponents of the equivalent 
CPT, the modified hyperscaling relation once again implies 
{the following: the QPT is equivalent to the finite 
temperature transition of an equivalent classical model
with one additional dimension, i.e., the Trotter dimension, along which the 
correlation length $\xi_{\tau}\sim \xi^z$, and the exponent $z$ manifests in 
(\ref{eq_hyperscaling}); if $z=1$, one gets a $d_{quantum} \to 
(d+1)_{classical}$ correspondence. Therefore the upper critical dimension of 
the QPT of the TIM ($\nu=z=1$) is $d_u^c=3$ in comparison to $d_u^c=4$ for 
the thermal transition of the classical Ising model \ct{chaikin95}.}
%In both the cases, the hyperscaling relation is valid. We have already shown 
%in Sec.~\ref{jorwigtr} that $\nu=z=1$ for the
%one-dimensional TIM, which gives $\al=0$ from Eq.~(\ref{eq_hyperscaling}). 
%This confirms that the exponents are the same as those of the two-dimensional 
%classical Ising model \ct{stanley87,chaikin95}.
%In a similar fashion, one can derive other 
%scaling relations associated with a QPT.

%The scaling of the equal-time correlation function is given by $G(r, \tau =0) 
%\sim 1/r^{(d-2+\eta)} f(r/\xi)$, where $\eta$ is the anomalous dimension of 
%the order parameter field. Using the fluctuation-dissipation theorem 
%which connects the static 
%susceptibility to the correlation function, we get one more scaling
%relation $(2 - \eta)\nu = \ga$. 

At a finite temperature, the size of the temporal direction is finite and
given by $L_{\tau} = \beta$. Using the finite size scaling \ct{barber83}, 
%we note that a classical system has an effective critical 
%temperature $T_c(L)$, where the correlation length $\xi$ equals the linear 
%dimension $L$. The finite size scaling of the critical temperature is given by
%$T_c(L)= T_c(\infty) + B L^{-1/\nu_{\rm cl}}$, where $B$ is a non-universal 
%constant (which may depend on the boundary condition \ct{cardy96}), and 
%$\nu_{\rm cl}$ is the correlation length exponent of the associated classical 
%phase transition at $T_c(\infty)$ in the thermodynamic limit. A quantum system
%at a finite temperature will similarly show an effective critical point when 
%$\xi_{\tau} \sim L_{\tau} \sim L^{1/z}$; note that $g_c(T)=g_c(L_{\tau})$, 
%while the QCP is at $g_c(T=0)= g_c(L_{\tau} \to \infty)$. Generalizing the 
%finite size scaling argument for a classical system, 
one gets the relation
\beq g_c(T) = g_c(T=0) + B L_{\tau}^{-1/\nu z}=g_c(T=0)+B T^{1/\nu z}, \eeq
%where we have used the facts that $L_{\tau} \sim L^z$ and $L_{\tau} = \beta$.
where $B$ is a constant,
implying that the quantum critical exponents determine the phase boundary 
at a finite temperature close to the QCP (see Fig.~\ref{fig_qpt}).
%We can write down the finite temperature scaling relation for a quantity $X$ 
%diverging as $\de^y$ along the $T=0$ line as
%\beq X(\de,T,\om)=|\de|^{-y} F_{\pm} \left( \frac{(|\de|)^{\nu z}}{k_B T}, 
%\frac{\om}{k_B T} \right), \eeq
%where we have assumed different scaling functions $F_{\pm}$ depending on the 
%sign of $\de$.
Referring to Fig.~\ref{fig_qpt} (a), 
%we note that there is a 
%singularity in the zero-frequency scaling at the critical line 
%which occurs at $T_c \sim \de^{\nu z}$ corresponding to the phase boundary in 
%the $g-T$ plane. For $\om \gg k_B T$, the dominating fluctuations are 
%quantum in nature. For $k_B T \gg \om$, we get quantum critical regions 
%where the critical ground state and thermal fluctuations dominate. 
the crossover from the quantum paramagnetic region to the quantum critical 
region occurs when $\hbar \om \sim k_B T$, where $\om$ is the characteristic 
frequency scale. In the vicinity of the phase boundary, the phase transition 
is classical in nature due to the critical slowing down as mentioned already. 
Now consider a TIM in $d=1$ which exhibits long-range order only at $T=0$. In 
this case, the ordered region is replaced by a ``renormalized'' classical 
state (see Fig.~\ref{fig_qpt} (a)) \ct{sachdev99}. In this region, the 
fluctuations are classical because $\xi$ diverges as $T \to 0$ for $g < g_c$.
For the one-dimensional classical Ising model, $\xi \sim \exp(\beta)$ 
\ct{chaikin95} grows much faster than the de Broglie wavelength and 
hence the fluctuations are classical. 

\subsection{Quantum spin chains coupled to a bath}

The quantum-classical correspondence discussed in the previous section has
been critically looked at and also challenged in recent years in the context of
the quantum phase transition in a two-level system (or a single spin-1/2) 
coupled to an infinite number of bosonic degrees of freedom which are
characterized by a spectral function $J(\om)$; this is known as the 
spin-boson model \ct{leggett87,weiss99,prokofev00}. The Hamiltonian is
\beq H= \De \frac{\si^x}{2} + \frac{\si^z}{2} \sum_i \la_i (a_i + 
a_i^{\dagger}) + \sum_i \om_i a_i^{\dagger} a_i, \label{eq_spinboson} \eeq
where $\si^a, ~a=x,z$, are Pauli matrices, $a_i$, $a_i^\dagger$ are 
bosonic annihilation and creation operators, $\De$ 
represents the tunneling matrix element, and $\om_i$ denote the oscillator 
frequencies of the bosonic degrees of freedom. The coupling between the spin 
and the oscillator bath via $\la_i$ is determined by the spectral function of 
the bath, 
\beq J(\om) =~ \pi \sum_i \la_i^2 \de(\om -\om_i).
% ~=~ 2\pi \al \om_c^{1-s} \om^s, s> -1 
\eeq
This is parametrized as $2\pi \al \om_c^{1-s} \om^s$ for $0<\om<\om_c$, 
where $\om_c$ is the cut-off frequency and $\al$ represents the coupling 
strength; we take $s > -1$. {When the exponent $s$ lies in 
the range $0<s<1$, 
the model is said to be in the sub-ohmic range. The case $s=1$ refers to 
the ohmic situation; then $J(\om) \sim \om$ and is independent of $\om_c$. 
The case $s>1$ is called super-ohmic.}

{
The quantum critical behavior of the model in (\ref{eq_spinboson}) depends 
crucially on the value of the exponent $s$.} For $s<1$, the spin-boson model 
shows a QPT from a delocalized phase (when tunneling dominates) to a 
localized phase at a critical value of $\al = \al_c (s,\De,\om_c)$. In view 
of the quantum-classical mapping discussed earlier, this QPT is expected to 
be in the same universality class as that of the finite-temperature
transition of a classical Ising chain with long-range interactions 
\ct{fisher72,dutta01}. The case $s=1$ corresponds to a classical Ising chain 
with inverse-square interactions \ct{kosterlitz76,bhattacharjee81} which 
exhibits a {Berezinskii-Kosterlitz-Thouless (BKT) transition}
\ct{kosterlitz73} from a power-law correlated phase below $T_c$ to a 
disordered phase with exponentially decaying correlations above $T_c$.

Studies using numerical RG techniques \ct{bulla03} have 
established the existence of the aforementioned QPT for both the ohmic ($s=1$)
and sub-ohmic ($s<1$) situations \ct{bulla03}. Moreover, the possibility of 
a breakdown of the quantum-classical mapping for $0 <s <1/2$ has been pointed 
out in \ct{vojta051}. In this range of $s$, the equivalent classical 
Ising chain shows a mean field behavior \ct{fisher72}, whereas 
$\ep$-expansion studies indicate that the QPT is governed by the interacting 
fixed point {which implies a non-mean field critical 
behavior} \ct{vojta051}. 
However, the critical exponents obtained by an accurate quantum Monte Carlo 
method with a continuous imaginary time cluster algorithm including finite 
temperature corrections indicated the correctness of the quantum-classical 
mapping for all values of $s$ \ct{winter09} and {predicted 
mean field behavior 
for the QPT for this range of $s$} (see also \ct{alvermann09})\footnote{The 
failure of the numerical RG technique in the localized phase 
for $s<1/2$ is argued to be due to the presence of a dangerously irrelevant 
operator \ct{winter09} so that the exponents are determined by the Gaussian 
fixed point, as happens in Landau theory for $d>d_u^c$ \ct{chaikin95}.}. 
{A very recent study \ct{kirchner12}, on the other hand, again
points to the breakdown of the quantum-classical correspondence\footnote{This 
breakdown is attributed to the presence of a Berry-phase 
term in a continuum path-integral representation of the model.}
for the QPT with $s<1/2$ implying that the debate on this issue is far from
being settled.} It should be mentioned here that dissipative spin dynamics 
in this model has also been studied using the Majorana representation 
and numerical RG techniques \ct{narayanan10,narayanan11}.

{Let us note that the quantum Ising spin chain coupled to a 
heat bath has been studied in recent years.} The Hamiltonian is given by 
\ct{werner05}
%\bea H &=& -J \sum_i \si_i^x \si_{i+1}^x ~-~ h \sum_i \si_i^z \non \\
%&& +~ \sum_{i,k} \la_k (a_{ik} + a_{ik}^{\dagger})\si_i^x ~+~ 
%\sum_{ik} \om_{ik} a_{ik}^{\dagger} a_{ik}, \eea
\bea H ~=~ -J \sum_i \si_i^x \si_{i+1}^x ~-~ h \sum_i \si_i^z ~+~ \sum_{i,k} 
\la_k (a_{ik} + a_{ik}^{\dagger})\si_i^x ~+~ \sum_{ik} \om_{ik} 
a_{ik}^{\dagger} a_{ik}, \eea
where the spin at each site $i$ is linearly coupled via $\la_k$ to an 
independent set of bosons denoted by the operators $a_{ik}$ and
$a_{ik}^{\dagger}$ with frequencies $\om_{ik}$. Using the Suzuki-Trotter
formalism, the above chain can be mapped to the two-dimensional classical 
Ising Hamiltonian with a long-range interaction in the temporal direction. 
The phase boundary between the FM and PM phases in the 
presence of dissipation (denoted by $\al$ as above) has been obtained using 
extensive Monte Carlo studies for the ohmic case $s=1$ \ct{werner05}. These 
studies also established a new quantum critical behavior in the presence of 
dissipation with critical exponents independent of the value of $\al$. 

The QCP mentioned above has been experimentally studied in
\ct{ronnow05} for the system $\rm LiHoF_4$ where the electron spins 
are effectively coupled to a spin bath consisting of the nuclear spins (see 
also Sec.~\ref{expt4}). This coupling modifies the QCP of the transverse 
Ising model describing the electron spins, so that the energy gap, defined 
by following the most prominent feature of the magnetic exciton dispersion, 
does not go to zero but reaches a minimum (and the corresponding coherence 
length reaches a maximum) at some value of the 
transverse field. This work demonstrates an intrinsic limitation to the 
observation of a QCP for an electronic system which is coupled to a nuclear
spin bath even at very low temperatures where the coupling to a phonon bath 
becomes unimportant.

\subsection{Quantum rotor models}

Finally, we provide a brief note on the quantum lattice rotor 
models because of their close connection to the TIMs. These models were 
introduced \ct{chakravarty89} as a version
of the non-linear $\si$-models to study the low-temperature properties of 
two-dimensional quantum Heisenberg antiferromagnetic systems. The Hamiltonian
of a quantum rotor (QR) model on a $d$-dimensional regular lattice is given by
\beq H_R= \frac{g}{2} \sum_{\bf i} \hat L_{\bf i}^2 - J \sum_{<{\bf ij}>} 
{\hat x}_{\bf i} \cdot {\hat x}_{\bf j}. \label{eq_hamrotor} \eeq
The $n$-component unit-length vectors ${\hat x}_{\bf i}$'s denote the 
orientation of the rotors on the surface of a sphere in the $n$-dimensional 
rotor space, where $n \ge 2$. For simplicity, we have assumed only 
nearest-neighbor ferromagnetic interactions between the rotors. The operators 
$\hat L_{{\bf i}\al \beta}$ ($\al,\beta=1,2,...,n$) are the $n(n-1)/2$ 
components of the angular momentum generator operator in the rotor space 
which can be put in the differential form as
$$L_{{\bf i}\al\beta} = -i \hbar \left(x_{{\bf i}\al} \frac{\partial}{\partial
x_{{\bf i}\beta}} -x_{{\bf i}\beta} \frac{\partial}{\partial x_{{\bf i}\al}}
\right).$$

The non-commutativity between the operators ${\hat x}_{\bf i}$ and 
${\hat L}_{\bf i}$ 
introduces quantum fluctuations in the model which can be tuned by changing 
the kinetic term $g$. In the limit $g=0$, the rotors at different sites 
choose a particular orientation (ferromagnetic phase) thereby breaking the 
underlying $O(n)$ symmetry, while for $g \gg J$ the symmetry is
restored. We therefore expect a QPT from a symmetric PM phase to 
a symmetry-broken FM phase at a critical value $g=g_c$ at $T=0$.

 {For example, consider the case for $n=2$ and $d=2$, 
when the Hamiltonian in (\ref{eq_hamrotor}) describes an array of Josephson 
junctions on a $d$-dimensional lattice which is in fact the quantum 
counterpart of the $d$-dimensional classical $XY$ model 
\ct{sondhi97,nagaosa99}. The Hamiltonian in Eq. (\ref{eq_hamrotor}) 
can be written in the form
\beq H = \frac{2 e^2}{C} \sum_i \frac{\partial^2}{\partial \theta _i^2} - 
E_J \sum_{<ij>} \cos( \hat \theta_i - \hat \theta_j), \label{eq_path7} \eeq 
where we have set $\hat x_i= \cos \hat \theta_i$, $g/2 =2e^2/C $, $J_{ij}=J =
E_J$, and $\hat L_i^z = -(i \hbar) \partial/ \partial \theta_i$. Here, the 
operator $\hat \theta_j$ represents
the phase of the superconducting order parameter on the $j$-th grain, while 
the term $-i \hbar (2e/C) \partial/ \partial \theta_i$ is canonically 
conjugate to the phase and denotes the voltage on the $j$-th junction. The 
Cooper pair charge is $2e$ and $C$ is the capacitance of each grain. 
The interaction term $E_J$ represents the Josephson coupling energy between 
the superconducting grains. In the Hamiltonian in (\ref{eq_path7}), the 
kinetic term tends to destroy any long-range order in the phases 
$\theta_i$; hence the model exhibits a QPT from a 
superconducting to an insulating phase when the ratio $E_C/E_J$ exceeds a 
critical value, where $E_C (= 2e^2/C)$ is called the charging energy. One 
can construct an imaginary time path integral \ct{wallin94} which is in fact 
identical to that of the TIM with the order parameter field being 
two-component. As argued in Sec.~\ref{sec_class+quant}, the finite 
temperature transition (when the non-commuting kinetic term is irrelevant) 
belongs to the universality class of the $d$-dimensional classical $XY$ model 
given by the $XY$ term $E_J \sum_{<ij>} \cos( \hat \theta_i - \hat \theta_j)$ 
in (\ref{eq_path7}). On the other hand, the QPT corresponds to a $XY$ model 
with one added dimension coming from the kinetic term. The finite temperature 
transition of a two-dimensional array (or the equivalent zero-temperature 
transition of a one-dimensional array) is therefore a BKT transition 
\ct{kosterlitz73}.}\footnote{{A similar scenario is shown
in a $1d$ TIM with FM interactions falling as the inverse square of the 
distance between the spins \ct{dutta01}.}}

The discussion above shows that the QR Hamiltonian is in fact 
a $n$-component generalization of the transverse Ising Hamiltonian 
from the viewpoint of universality of QPTs. 
%This similarity can be visualized 
%in terms of the time action \ct{sachdev99}. Noting 
%as follows \ct{sachdev99}. 
Noting that the components of the operators $\hat {x}_{\bf i}$ commute with 
each other (unlike the Pauli spin operators), one can show that the zero 
temperature phase transition of a $d$-dimensional 
QR model belongs to the universality class of the finite temperature 
transition of a $(d+1)$-dimensional classical $n$-vector model 
{as shown above for $n=2$}. On the 
other hand, the finite temperature transition for the QR corresponds to the 
finite temperature transition of a $d$-dimensional classical $n$-vector 
model. In the limit $n \to \infty$, the Hamiltonian reduces to a quantum 
version of a spherical model \ct{vojta96} with the quantum critical behavior 
belonging to the universality class of the finite temperature transition of 
a classical spherical model \ct{berlin52} with one dimension added. 
The QR model has been extensively studied in the presence of random 
interactions \ct{ye93,read95,sachdev95} and random longitudinal fields 
\ct{vojta961,dutta98} in order to investigate the zero temperature and finite 
temperature properties of quantum spin glass models of rotors. We refer 
to Ref. \ct{sachdev99}, for exhaustive discussions on quantum rotor models;
{however, we shall briefly refer to these models in the 
context of GM singularities (Sec. \ref{sec_griff+act}) and the 
quantum Lifshitz point (Sec. \ref{sec_ANNNI})}.

\section{Exact solutions in one and two dimensions}
\label{exactsol}

\subsection{Jordan-Wigner transformation}
\label{jorwigtr}

Let us consider the spin-1/2 $XY$ model placed in a magnetic field pointing 
in the $\hat z$ direction. The Hamiltonian is given by 
\beq H ~=~ - ~ \sum_{n=1}^{N-1} ~[~ J_x \si_n^x \si_{n+1}^x ~+~ J_y \si_n^y 
\si_{n+1}^y] ~-~ \sum_{n=1}^N ~h \si_n^z , \label{xyh} \eeq
with $J_x + J_y >0$, where $\si_n^a$ denote the Pauli matrices at site $n$, 
and we will assume periodic boundary conditions so that ${\vec \si}_{N+1} 
\equiv {\vec \si}_1$. 
% if
%necessary, this can be ensured by performing a unitary transformation which 
%flips the signs of $\si_n^x$ and $\si_n^y$ on alternate sites or interchanges 
%$\si_n^x$ and $\si_n^y$. 
Clearly, for $J_y=0$, the Hamiltonian in (\ref{xyh}) reduces to that of the 
transverse Ising chain. { A discussion of QPTs in this model 
in the $S \to \infty$ limit is presented in Appendix \ref{app_largespin}.}

The above system can be mapped to a model of 
spinless fermions using the Jordan-Wigner (JW) transformation \ct{lieb61}.
We map an $\uparrow$ spin or a $\downarrow$ spin at any 
site to the presence or absence of a spinless fermion at that site. This can 
be done by introducing a fermion annihilation operator $c_n$ at each site, 
and writing the spin at that site as
\bea \si_n^z &=& 2 c_n^\dg c_n -1 = 2 \rho_n - 1, \non \\
\si_n^- &=& \frac{1}{2} (\si_n^x -i \si_n^y)= c_n ~e^{i\pi \sum_{j=1}^{n-1} 
\rho_j}, \label{jw1} \eea
where $\rho_n = c_n^\dg c_n = 0$ or $1$ is the fermion occupation number at 
site $n$. The 
expression for $\si_n^+$ can be obtained by taking the Hermitian conjugate 
of $\si_n^-$. The string factor in the definition of $\si_n^-$ is necessary 
to ensure the correct anticommutation relations between the fermionic 
operators, namely, $\{ c_m, c_n^\dg \} = \de_{mn}$ and $\{ c_m, c_n \} = 0$.

Following the JW transformation, (\ref{xyh}) takes the form 
\bea H &=& \sum_{n=1}^{N-1} ~\left[ - (J_x + J_y) ~(c_n^\dg c_{n+1} + 
c_{n+1}^\dg c_n) ~+~ (J_x - J_y) ~(c_{n+1}^\dg c_n^\dg + c_n c_{n+1}) \right] 
\non \\
& & -~ (-1)^{N_f} ~[- (J_x + J_y) ~(c_N^\dg c_1 + c_1^\dg c_N) ~+~ 
(J_x - J_y)~ (c_1^\dg c_N^\dg + c_N c_1)] \non \\
& & -~ \sum_{n=1}^N ~h ~(2 c_n^\dg c_n -1), \label{jorwig2} \eea
where $N_f$ denotes the total number of fermions: $N_f = \sum_{n=1}^N c_n^\dg 
c_n$. (Note that $N_f$ commutes with $H$, hence it is a good quantum number).
We now Fourier transform to the operators $c_k = \sum_{n=1}^N c_n 
e^{-ikna}/\sqrt{N}$, where $a$ is the lattice spacing and the momentum $k$ 
lies in the range $[-\pi /a, \pi /a]$ and is quantized in units of 
$2\pi/(Na)$. [The factor of $-(-1)^{N_f}$ in Eq. (\ref{jorwig2})
implies that $k=2j\pi/(Na)$ if $N_f$ is odd, while $k=(2j+1)\pi/(Na)$ if $N_f$ 
is even; here $j$ runs over a total of $N$ integer values. The chain length 
$L=Na$; in subsequent sections we will usually set $a=1$]. We then obtain
\bea H &=& \sum_{k>0} ~\left( \begin{array}{cc}
c_k^\dg & c_{-k} \end{array} \right)~ H_k ~
\left( \begin{array}{c}
c_k \\
c_{-k}^\dg \end{array} \right), \non \\
H_k &=& 2 \left( \begin{array}{cc}
- (J_x + J_y) \cos (ka) - h & i (J_x - J_y) \sin (ka) \\
-i(J_x - J_y) \sin (ka) & (J_x +J_y) \cos (ka) +h \end{array} \right), \non \\
& & \label{eq_ham2by2}\eea
where $k$ now lies in the range $[0,\pi/a]$. Since this Hamiltonian is 
quadratic in the fermion operators, it can be diagonalized using a fermionic 
Bogoliubov transformation,
\bea d_k^\dg &=& \sin \theta_k ~c_k ~+~ i \cos \theta_k ~c_{-k}^\dg, \non \\
d_{-k}^\dg &=& \sin \theta_k ~c_{-k} ~-~ i \cos \theta_k ~c_k^\dg, \eea
where $\theta_k$ is fixed by the condition
\beq \tan (2\theta_k) ~=~ -~\frac{(J_x - J_y) \sin (ka)}{(J_x + J_y) \cos (ka)
+ h}.
\label{eq_def_theta} \eeq
We then arrive at a Hamiltonian given by \ct{lieb61,bunder99}
\beq H ~=~ \sum_{k>0} \om_k ~(d_k^\dg d_k + d_{-k}^\dg d_{-k} - 1),
\label{ham3} \eeq
where $\om_k = 2 ~[h^2 + J_x^2 + J_y^2 + 2 h (J_x + J_y) \cos (ka) + 2 J_x J_y
\cos (2ka) ]^{1/2}$. The ground state $|GS\rangle$ of the Hamiltonian in 
(\ref{ham3}) satisfies $d_k |GS\rangle = d_{-k} |GS\rangle=0$ for all $k>0$.
If we define the vacuum state $|\phi \rangle$ to be the state satisfying $c_k |
\phi \rangle = c_{-k} |\phi \rangle$ for all $k>0$, the ground state of 
(\ref{ham3}) can be written as
\bea |GS\rangle &=& \bigotimes_{k>0} ~(\cos \theta_k ~+~ i \sin \theta_k ~
c_k^\dg c_{-k}^\dg ) ~|\phi \rangle \non \\
&\equiv& \bigotimes_{k>0} ~(\cos \theta_k ~|0\rangle ~+~ i \sin \theta_k ~ 
|k,-k \rangle). \label{eq_direct_product} \eea
In the thermodynamic limit $N \to \infty$, the ground state energy is given 
by $E_0 = -N \int_0^{\pi/a} dk/(2\pi) \om_k$. 

As a function of $k$, $\om_k$ has extrema at the points $0$, $\pi/a$ and 
$k_0$ which is given by 
\beq \cos (k_0a) ~=~ - ~\frac{h(J_x + J_y)}{4 J_x J_y}, \label{coska} \eeq
provided that $h(J_x + J_y)/(4 J_x J_y)$ lies in the range $[-1,1]$. The 
values of $\om_k$ at these three points are given by $2|J_x + J_y + h|$,
$2|J_x + J_y - h|$ and $2|J_x - J_y| \sqrt{1 - h^2/(4J_x J_y)}$ respectively.
The system is therefore gapless for the following three cases: (i) $J_x + J_y
= -h$, (ii) $J_x + J_y = h$, and (iii) $J_x = J_y$ and $|h/J_x| \le 2$. In
all three cases, $\om_k$ vanishes linearly as $k \to 0$, $\pi/a$ or $k_0$;
hence the dynamical critical exponent is given by $z=1$. Further $\nu z =1$, hence 
$\nu =1$. Similarly, one can show that for a transition across the MCP in 
Fig.~\ref{fig_xyphase}, $\nu_{mc}=1/2$ and $z_{mc}=2$. At the MCP, $\om_k 
\sim k^2$ yielding $z_{mc}=2$, and $\om_k \sim (h+2J_x)$ for $k=0$, as the 
MCP B is approached along the anisotropic transition line, 
i.e., $h \to -(J_x + J_y)=-2J_x$, so that $\nu_{mc} z_{mc}=1$. 
%The phase diagram defined
%by these critical lines is shown in Fig.~\ref{fig_xyphase}.

The QPTs in cases (i) and (ii) belong to the 
universality class of the critical transverse Ising model in one dimension 
\ct{pfeuty70,bunder99} which is also related to the classical Ising model 
in two dimensions at the critical temperature. We will therefore refer to 
these cases as the ``Ising transition". The quantum phase transition
in case (iii) will be referred to as the ``anisotropic transition" since this 
line is crossed when the $XY$ couplings $J_x$ and $J_y$ are made unequal 
giving rise to an anisotropic $XY$ model.

The generally incommensurate value of $k_0$ given by (\ref{coska})
becomes equal to the commensurate value of $\pi/a$ if $h/(J_x+J_y) = 1 -
\ga^2/(J_x + J_y)^2$, where $\ga = (J_x - J_y)$. This corresponds
to the dotted line shown in Fig.~\ref{fig_xyphase} where there is a 
transition from an incommensurate to a commensurate phase. On that line,
the model can be mapped by a duality transformation (\ref{map}) to the model 
given in (\ref{eq_mps_hamil}).

The gaplessness of the three cases discussed above suggests that it may be 
possible to describe them in terms of quantum field theories (at length 
scales which are much larger than $a$) which are conformally invariant. 
We will now identify the appropriate conformal field theories.

Cases (i) and (ii) with $J_x + J_y = \pm h$ are similar, so we will consider
only one of them, say, $J_x + J_y = -h$. Eq.~(\ref{ham3}) shows that 
there are two modes, $d_k$ and $d_{-k}$, whose energies vanish linearly as 
$\om_k = 2a |J_x - J_y|k$ as $k \to 0^+$. The velocity of these modes is 
given by $v=(d\om_k/dk)_{k=0} = 2a |J_x - J_y|$. Further, the operators
$d_k$ and $d_{-k}$ describe right- and left-moving modes whose wave functions 
are given by $e^{ik(x-vt)}$ and $e^{-ik(x+vt)}$ respectively, where $k > 0$.
These are precisely the modes of a massless Majorana fermion described in 
Sec.~\ref{expcorr}.

%\begin{figure}[h] \begin{center}
%\includegraphics[height=2.0in]{rmp_xyspectrum.eps}
%\caption{Excitation spectrum $\ep_k$ of the $XY$ model in a transverse 
%field for $h=0.9,~J_x=0.8$ and $J_y=0.2$.} \end{center} \end{figure}

Case (iii) with $J_x = J_y$ and $|h/J_x| \le 2$ has $\om_k$ vanishing
linearly as $k \to k_0$ from both above and below. The velocity is given by 
$v=(d\om_k/dk)_{k=k_0} = 2a \sqrt{4J_x^2 - h^2}$. Let us redefine the
operators $d_k \to d_k^\dg$ and $d_{-k} \to d_{-k}^\dg$ for $k < k_0$.
Since this transforms $d_k^\dg d_k \to d_k d_k^\dg = - d_k^\dg 
d_k$ plus 1 (and similarly for $d_{-k}$), we see that the energy 
$\om_k$ becomes negative for $k < k_0$. Further, the wave functions for 
$d_k$ are given by $e^{i[kx-(k-k_0)vt]}$ and $e^{-i[kx-(k-k_0)vt]}$ for $k > 
k_0$ and $k < k_0$ respectively, while the wave functions for $d_{-k}$ are 
given by $e^{-i[kx+(k-k_0)vt]}$ and $e^{i[kx+(k-k_0)vt]}$ for $k > k_0$ and 
$k < k_0$ respectively. Upon noting that $k-k_0$ can take both positive and 
negative values, and redefining the momentum from $k-k_0$ to $k$, we see that 
these are the modes of a massless Dirac fermion, with $d_k$ and $d_{-k}$ 
describing right- and left-moving modes respectively.

We see that the velocities $v$ of the above field theories vanish at the 
two points given by $J_x = J_y$ and $h=\pm 2J_x$. These correspond to 
multicritical points. At these points, $\om_k$ vanishes quadratically, rather
than linearly, as $k \to 0$ or $\pi$; this implies that $z=2$. Hence, the 
corresponding field theories are not Lorentz invariant and therefore not 
conformally invariant.

We can consider small perturbations from the gapless theories discussed above,
namely, making $J_x + J_y$ slightly different from $|h|$ in cases (i) and 
(ii), and making $J_x$ slightly different from $J_y$ in case (iii). 
In the quantum field theoretic language, such a perturbation gives rise to
a Majorana mass term for cases (i) and (ii), and to a Dirac mass term for 
case (iii). The forms of these mass terms are given in Sec.~\ref{expcorr}.

Finally, we consider a transverse Ising chain in a longitudinal 
field given by the Hamiltonian
\beq H = -J_x\sum_{<ij>} \si_i^x \si_i^x - h\sum_i \si_i^z - h_L\sum_i 
\si_i^x. \label{eq_t1d_long} \eeq
Due to the presence of a longitudinal field, the model is no longer exactly solvable; we
will discuss the effect of this field in a particular limit below.

\subsection{Connection to conformal field theory}
\label{expcorr}

Near a QCP, we can generally use a continuum field theory to
describe the system since the correlation length $\xi$ is much larger than
microscopic length scales such as the lattice spacing (if the quantum
model is defined on a lattice) or the distance between nearest-neighbor 
particles. Exactly at $g = g_c$, the field theory will be 
scale invariant since $\xi = \infty$. Further, if the dynamical critical 
exponent $z=1$, time and space will be on the same footing (apart from a 
factor given by the velocity $v$ of the low-energy excitations), and the field 
theory would be expected to have a Lorentz invariant form. A combination of 
Lorentz invariance and scale invariance turns out to put rather powerful 
constraints on the correlation functions of the quantum theory, particularly 
in two space-time dimensions, i.e., if the original classical system is in 
two spatial dimensions. We will now discuss this case a little more 
\ct{itzykson88}.

Consider a quantum field theory in two-dimensional Minkowski space-time 
with coordinates $(x,t)$. It is convenient to transform to two-dimensional 
Euclidean space-time with coordinates $(x,\bar \tau)$, by making the 
substitution $t = -i\bar \tau$; when we analytically continue from real $t$ 
to imaginary $t$, $\bar \tau$ becomes real. 
% In the path integral language, this corresponds to rewriting
% the partition function for some field $\phi$ as
% \beq Z ~=~ \int ~{\cal D} \phi ~e^{i S_M [\phi]} \to \int ~{\cal D} 
% \phi ~e^{- S_E [\phi]}, \eeq
% where $S_M$, $S_E$ denote the actions in Minkowski and Euclidean space-times.
Let us introduce the complex coordinates $z=\bar \tau - ix/v$ and $z^* = \bar 
\tau + ix/v$, where $v$ denotes the velocity. It turns out that a theory in two
Euclidean dimensions which is invariant under translations, rotations and 
scaling is also invariant under the full conformal group corresponding to 
transformations given by all possible analytical functions, $z \to z' = f(z)$;
such a theory is called a conformal field theory (CFT). The Lie algebra for 
the conformal group in two dimensions is infinite-dimensional since a 
transformation close to the identity can be written as a Laurent series 
\beq z' ~=~ z ~-~ \sum_{m=-\infty}^\infty ~\ep_m z^{m+1}, \eeq
where the $\ep_m$'s are infinitesimal quantities. The generators of
the transformations are given by $l_m = - z^{m+1} \partial_z$ which satisfy 
the Lie algebra $[l_m,l_n] = (m-n)L_{m+n}$. It turns out that in 
the quantum theory, the Lie algebra of the generators, now denoted by $L_m$,
contains an extra term if $m=-n$; namely, we find that
\beq [L_m,L_n] ~=~ (m-n) ~L_{m+n} ~+~ \frac{c}{12} ~(m^3 - m) ~\de_{m+n,0}.
\eeq
This is called the Virasoro algebra and $c$ is called the central charge of
the theory. The number $c$ plays an important role in several properties
of the theory, such as finite size corrections to the free energy of the 
theory at finite temperature \ct{blote86,affleck86}, the entropy of a 
finite region of the system \ct{holzhey94}, and the entropy of entanglement 
between two parts of the system \ct{vidal03} (see Sec.~\ref{e_entropy}).

In a CFT, the correlation functions of various field
operators fall off as powers of their space-time separations. 
Consider an operator ${\cal O} (z,z^*)$ whose two-point correlation function 
takes the form
\beq \langle {\cal O} (z_1,z_1^*) {\cal O} (z_2,z_2^*) \rangle ~=~ 
\frac{C}{(z_1 - z_2)^{2x_{\cal O}} (z_1^* - z_2^*)^{2{\bar x}_{\cal O}}}, \eeq
where $C$ is a constant which depends on $\cal O$. Then $d_{\cal O} =
x_{\cal O}+ {\bar x}_{\cal O}$ is called the scaling dimension and 
$x_{\cal O}-{\bar x}_{\cal O}$ is called the spin of the operator $\cal O$.

We now present three simple examples of CFTs. The first 
example is a massless Dirac fermion. In Minkowski space-time, the action
for this is given by
\beq S ~=~ i~ \int ~dt dx ~[\psi_R^\dg (\partial_t + v \partial_x) \psi_R ~
+~ \psi_L^\dg (\partial_t - v \partial_x) \psi_L]. \eeq
The Euler-Lagrange equations of motion imply that $\psi_R$ and $\psi_L$ are 
functions of $x-vt$ and $x+vt$ respectively. We find that the central charge 
of this theory is $c=1$. The scaling dimension and spin of the fields $\psi_R$
and $\psi_L$ are given by $(1/2,1/2)$ and $(1/2,-1/2)$ respectively. The
second example is a massless Majorana fermion; such a fermion is described
by a field operator which is equal to its own Hermitian conjugate. Given 
a Dirac field $\psi_R$, we can form two Majorana fields from it, namely,
$\chi_R = \psi_R + \psi_R^\dg$ and $\xi = i(\psi_R - \psi_R^\dg)$. 
Using only one of these fields, say, $\chi_R$ and a similar field $\chi_L$
formed from $\psi_L$, we can write down the action
\beq S ~=~ i~ \int ~dt dx ~[\chi_R (\partial_t + v \partial_x) \chi_R ~+~ 
\chi_L (\partial_t - v \partial_x) \chi_L]. \eeq
As before, the equations of motion show that $\chi_R$ and $\chi_L$ are 
functions of $x-vt$ and $x+vt$ respectively. The central charge of this 
theory is given by $c= 1/2$. The third example of a CFT is a massless boson 
whose action is given by
\beq S ~=~ \int ~dt dx ~[ \frac{1}{2v} ~(\partial_t \phi)^2 ~-~ \frac{v}{2}~
(\partial_x \phi)^2 ]. \eeq
The equations of motion show that $\phi$ can be written as $\phi= \phi_R + 
\phi_L$, where $\phi_R$ and $\phi_L$ are functions of $x-vt$ and $x+vt$ 
respectively. The central charge of this theory is also given by $c=1$. The 
fields $\partial_z \phi_R$, $e^{i2 \sqrt{\pi} \phi_R}$ and $e^{-i2 \sqrt{\pi}
\phi_R}$ have scaling dimension and spin equal to $(1,1)$, $(1/2,1/2)$ and
$(1/2,1/2)$ respectively. Similarly, we can use $\phi_L$ to construct three 
fields with scaling dimension and spin equal to $(1,-1)$, $(1/2,-1/2)$ 
and $(1/2,-1/2)$. We will discuss later how these fermionic and bosonic 
theories can be related to each other by bosonization.

The three CFTs mentioned above are massless; indeed, masslessness is a 
necessary condition for a theory to be scale invariant. However, one can 
perturb a CFT to give the particles a small mass. The corresponding mass 
terms in the action are given by $\int dt dx ~\mu (\psi_R^\dg \psi_L
+ \psi_L^\dg \psi_R)$ for the Dirac fermion, $\int dt dx ~i \mu \chi_R 
\chi_L$ for the Majorana fermion, and $\int dt dx ~\mu^2 \cos (2 \sqrt{\pi} 
\phi)$ for the boson respectively. The last term assumes that the bosonic 
theory is invariant under $\phi \to \phi + \sqrt{\pi}$.

The effect of a perturbation of a CFT is governed by the scaling dimension of 
the corresponding operator $\cal O$ as follows. 
%Let $d_{\cal O}$ be the scaling dimension of $\cal O$,
Under a perturbation of the action of the 
CFT by an amount $\de S = \la \int dt dx {\cal O}$, it turns out 
that the parameter $\la$ effectively becomes a function of the length 
scale $L$, and it satisfies the renormalization group (RG) equation 
\beq \frac{d\la}{d \ln L} ~=~ (2-d_{\cal O}) ~\la \eeq
to first order in $\la$. The perturbation given by $\cal O$ is relevant, 
irrelevant and marginal for $d_{\cal O} < 2$, $>2$ and $=2$ respectively. If
$d_{\cal O} <2$, a small value of $\la$ at the microscopic length scale 
(say, the lattice spacing $a$ if the underlying model is defined on a lattice)
will grow to be of 
order 1 at a length scale $L_0$ given by $L_0/a \sim 1/(\la (a))^{1/(2-
d_{\cal O})}$. This implies that the perturbation gives rise to a finite
correlation length $\xi$ of the order of $L_0$, or, equivalently, produces an
energy gap (sometimes also called a mass) which scales with $\la (a)$ as
\beq \De E ~\sim ~v ~(\la(a))^{1/(2-d_{\cal O})}, \eeq
since energy scales as $v/{\rm length}$.
If the correlation or energy gap can be experimentally measured as a function
of $\la (a)$, one can deduce the value of $d_{\cal O}$. Conversely,
if the value of $d_{\cal O}$ can be found using analytical or numerical
methods, one can predict the power-law for $\xi$ or $\De E$. If $d_{\cal O}
> 2$, the perturbation is irrelevant and the correlation length (energy gap)
will remain infinite (zero) respectively.

We now apply these ideas about scaling dimensions and the effects of
perturbations to the Ising transition mentioned in Sec.~\ref{jorwigtr}. To 
be specific, let us consider the Hamiltonian given in (\ref{eq_tim1d}).
Close to the critical point $h=J_x$, the system is described by a 
CFT which is the theory of a massive Majorana fermion; from (\ref{ham3}),
we see that the energy-momentum dispersion near $k=\pi/a$ is given by 
$\om_k = \sqrt{v^2 \de k^2 + m^2}$, where $\de k = k -\pi/a$, the
velocity is $v=2ha$ and the mass is $m=2|h-J_x|$. The exact values of the 
various critical exponents in this theory are known and are given by $z=1$ 
(since the energy gap scales as $|\de k|$ for $h=J_x$), $\nu =1$ (since 
$z=1$ and the energy gap scales as $|h-J_x|$ for $\de k =0$), $\beta = 1/8$,
$\al = 0$, $\ga = 7/4$, $\de = 15$, and $\eta = 1/4$. The value $\nu =1$ and 
the equivalence of the temperature in the two-dimensional classical Ising 
model and the transverse field $h$ in (\ref{eq_tim1d}) implies that near the 
critical point, the correlation length scales as $(T-T_c)^{-1}$ or $|h - 
J_x|^{-1}$. Next, the combination $\nu =1$ and $\beta = 1/8$ implies that the 
scaling dimension of the longitudinal magnetization ($<\si_n^x>$) is given by 
$1/8$. (This implies that at criticality, the two-point correlation function 
$<\si_n^x \si_0^x> \sim 1/n^{1/4}$, and hence that $\eta = 1/4$ as noted 
above). At the critical point, the addition of a small longitudinal magnetic 
field $h_L$ (pointing in the $\hat x$ direction as in (\ref{eq_t1d_long})) 
will produce an energy gap scaling as 
$h_L^{1/(2-1/8)} = h_L^{8/15}$, and a longitudinal magnetization scaling as 
$({\rm gap})^{1/8} \sim h_L^{1/15}$; the latter implies that $\de = 15$. We 
note that the value of critical exponent $\nu$ depends upon the scaling 
dimension of the perturbing operator that takes the system away from the 
gapless QCP and thereby generates a gap. As shown above, if the perturbation 
is due to the transverse field $h$, $\nu=1$. In contrast, if the gap is 
generated by applying a longitudinal field in the $\hat x$ direction,
the resultant $\nu$ would be different (see (\ref{eq_gen_mag_field}) and 
discussions preceding that).

It turns out that the addition of a small longitudinal magnetic field 
at criticality produces a rather intricate pattern of energy gaps for the 
low-lying excitation spectrum. This was shown using a combination of CFT and 
an exact scattering matrix analysis for the low-energy particles in 
Ref. \ct{zamolodchikov89,delfino95,delfino96,fonseca06}. This was confirmed 
experimentally by Coldea $et~al.$ \ct{coldea10} (see also Ref. \ct{affleck10}, 
and Sec.~\ref{expt3}); for a recent theoretical analysis, see 
Ref. \ct{kjall10}. The low-lying spectrum is described by the Lie algebra 
of the exceptional group $E_8$, and the scattering matrix approach gives the 
exact ratios of the energy gaps of the 8 states; in particular, the ratio of 
the energies of the second and first excited states is predicted to be the 
golden ratio $(1+\sqrt{5})/2 \simeq 1.618$. This was found to be the case in 
the quasi-one-dimensional Ising ferromagnet CoNb$_2$O$_6$; the Co$^{2+}$ ions 
form spin-1/2 chains which are coupled very weakly 
to each other. A small longitudinal field $h_L$ is believed to be present 
intrinsically in this system due to the interchain couplings. The ratio of 
the energies of the two lowest-lying states was found to approach the golden 
ratio when the transverse field was tuned to be close to the critical value 
of about $5.5$ T \ct{coldea10} (see Fig.~\ref{GA6}). 

Before ending this section, we note that a huge literature exists on the 
dynamic correlation function of a transverse Ising chain 
\ct{barouch70,barouch71,barouch711,mccoy71,wu66,mccoy73}. We will not
discuss this in this review and refer to Ref. \ct{sachdev99} 
(see also \ct{perk09}, and references therein).

\subsection{Exact solution of the Kitaev model}
\label{sec_kitaev}

\begin{figure} \begin{center}
\includegraphics[height=2.5in,width=2.5in]{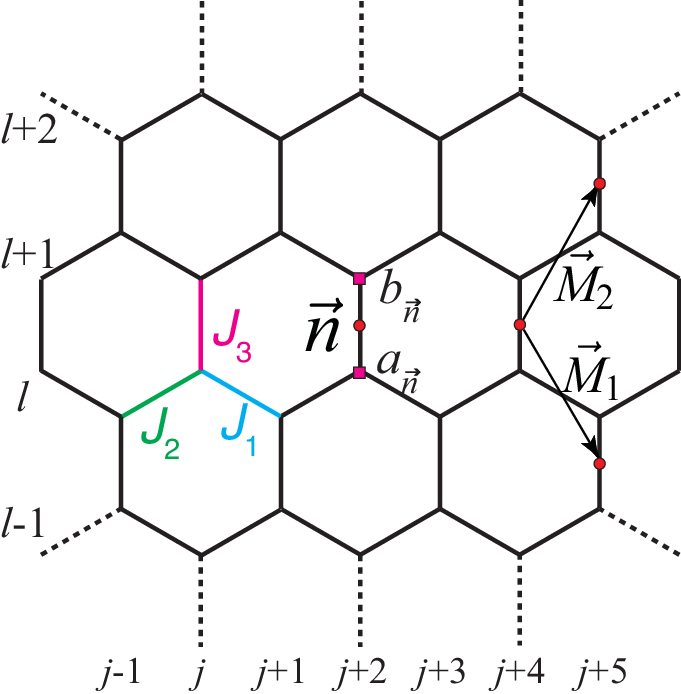}
\caption{(Color online) Schematic representation of a Kitaev model on a 
honeycomb lattice 
% in (a) and a brick-wall lattice in (b) 
showing the interactions $J_1$, $J_2$ and $J_3$ between $x$, $y$ and $z$ 
components of the spins respectively. $\vec n$ represents the position 
vector of each vertical bond (unit cell). The vectors $M_1$ and $M_2$ are 
spanning vectors of the lattice. In the fermionic representation
of the model, the Majorana fermions $a_{\vec n}$ and $b_{\vec n}$ sit at 
the bottom and top sites of the vertical bond with
center coordinate $\vec n$ as shown. (After \ct{hikichi10}.)}
\label{fig_reverse2d} \end{center} \end{figure}

The two-dimensional Kitaev model defined on a honeycomb lattice is 
described by the Hamiltonian \ct{kitaev06}
\beq H_{2d} ~=~ \sum_{j+l={\rm even}} ~(J_1\si_{j,l}^x \si_{j+1,l}^x ~+~ 
J_2\si_{j-1,l}^y \si_{j,l}^y ~+~ J_3\si_{j,l}^z\si_{j,l+1}^z), 
\label{rmp_kitaev2d} \eeq
where $j$ and $l$ defines the column and row indices of the lattice shown
in Fig.~\ref{fig_reverse2d}. One of the main properties of the Kitaev model 
which makes it theoretically attractive is that, even in two dimensions, it 
can be mapped onto a non-interacting fermionic model by a JW
transformation \ct{kitaev06,feng07,nussinov08,chen08,lee07}. 
%The rich phase diagram of this model consists of a gapless region in the
%parameter range $|J_1-J_2|\le J_3 \le |J_1+J_2|$ where the energy gap
%$\De_{\vec k}$ vanishes for special values of $\vec k$. 

%We first discuss the one-dimensional version of the Kitaev model which 
%corresponds to $J_3=0$,. 
The one-dimensional Kitaev model, namely, just one row of interacting spins 
in Fig.~\ref{fig_reverse2d}, is described by the Hamiltonian 
%\ct{kitaev06,chen08,lee07,feng07}
\bea H_{1d} ~=~ \sum_{n=1}^{N/2} ~(J_1 \si_{2n}^x\si_{2n+1}^x ~+~ J_2
\si_{2n-1}^y\si_{2n}^y), \label{rmp_kitaev1d} \eea
%where $n$ refers to the site index and $N$ is the total number of sites. 
The summation label $n$ takes $N/2$ values where $N$ is the total number 
of sites and the number of unit cells is $N/2$. 

% {\it Phase diagram of one-dimensional Kitaev model}

%\subsection{}

As before, the Hamiltonian in (\ref{rmp_kitaev1d}) can be diagonalized
using the JW transformation 
\ct{lieb61} (for details, see \ct{mondal08,divakaran09}), defined as
\bea a_n=(\prod_{j=-\infty}^{2n-1} \si_j^z) ~\si_{2n}^y~~~~{\rm and}~~~~
b_n=(\prod_{j=-\infty}^{2n} \si_j^z) ~\si_{2n+1}^x, \label{jw2} \eea
where $a_n$ and $b_n$ are independent Majorana fermions at site
$n$, satisfying the relations $ a_n^\dag = a_n,~ b_n^\dag = b_n,~ 
\{a_m,a_n\} = 2\de_{m,n}, \{b_m,b_n\} = 2\de_{m,n},~ \{a_m,b_n\} = 0.$
%Substituting $\si_n^x$ and $\si_n^y$ in terms of Majorana fermions,
%the Hamiltonian takes the form
%\bea H_{1d} = i\sum_n (J_1 b_n a_n + J_2 b_n a_{n+1}).
%\label{reverse_ham} \eea
[Eq.~(\ref{jw2}) has the same content as Eq.~(\ref{jw1}) but is more useful 
here because we want to work with Majorana fermions $a_n$ and $b_n$ rather 
than Dirac fermions $c_n$]. We perform a Fourier transformation of $a_n$ 
and $b_n$, 
%where the Fourier component $a_k$ is defined as follows
%\bea a_n= \sqrt{\frac{4}{N}}\sum_{k=0}^{\pi}[a_k e^{ikn} + 
%a_k^\dag e^{-ikn}], \label{reverse_fourierak} \eea
%and taking the limit $N \to \infty$, Eq.~(\ref{reverse_ham}) can 
%be rewritten as
%\bea H_{1d} &=& 2i\sum_{k=0}^{\pi}[b_k^\dag a_k(J_1+J_2 e^{ik})
%+a_k^{\dag}b_k(-J_1-J_2e^{-ik})]. \non \\
%& & \label{reverse_kitaev_map} \eea
%It can be verified that these Fourier components
%satisfy the standard anticommutation relations $\{a_k,a_{k'}^\dag\}=
%\de_{k,k'}$ and $\{a_k,a_{k'}\}=0$ as required.
to $a_k$ and $b_k$, and note that the sum over $k$ 
%in Eq.~(\ref{reverse_fourierak}), or (\ref{reverse_kitaev_map}) 
goes only over half the Brillouin zone as the $a_n$'s
are Majorana fermions. One can check that
% It can be checked that 
the number of modes lying in the
range $0\leq k \leq \pi$ is $N/4$ so that $k$ ranges from $0$ to $\pi$. 
%which confirms the limits of the summation.
By defining $\psi_k=(a_k,b_k)$, the Hamiltonian in (\ref{rmp_kitaev1d})
%(\ref{reverse_kitaev_map})
can be rewritten as
$H_{1d} = \sum_{k=0}^{\pi}\psi_k^\dag H_k \psi_k$
where
\bea H_k= 2 i\left[ \begin{array}{cc} 0 & -J_1-J_2e^{-ik} \\
J_1+J_2 e^{ik} & 0 \end{array} \right]. \non \\
& & \label{reverse_kitmat1} \eea

%\begin{figure}
%\includegraphics[height=2.4in]{rmpkit_groundstate.eps}
%\caption{The second derivative of ground state energy $E_0$ diverges at the 
%critical point $J_1=J_2$ as discussed in the text. Figure taken from 
%\oct{feng07}.} \label{kit_groundstate} \end{figure}

\noi The above Hamiltonian can be diagonalized by a Bogoliubov 
transformation, and the eigenvalues are given by
$E_k^{\pm}=\pm2\sqrt{J_1^2+J_2^2+2J_1J_2 \cos k}$. In the
ground state, one of the bands is fully occupied while the other is empty. The
gap in the spectrum $\De_k = E_k^+ - E_k^-$ vanishes at $J_1=\pm J_2$ 
for $k=\pi$ and $0$ respectively. The critical exponents $\nu$ and $z$ can 
be calculated using the definition
$\De_{k=k_c}=(J_1-J_c)^{\nu z} ~~~{\rm and}~~~ 
\De_{J_1=J_c} \sim k^z,$
where $J_c$ is the value of the coupling
$J_1$ at which the gap closes at the critical mode $k_c$. In this case, we 
have $J_c=\pm J_2$ for $k_c=\pi$ and $0$, respectively, and $\nu=z=1$.
%By varying $J_1/J_2$, we find that the second derivative of the ground 
%state energy $E_0$ diverges logarithmically at $J_1=J_2$ 
%as shown in Fig.~\ref{kit_groundstate}.
%The ground state energy is obtained by integrating over
%$\ep_k$ for all $k$ as in the ground state, the states with energy
%$-\ep_k$ are filled for all $k$. 
%The point where the gap vanishes corresponds to a quantum phase transition. 
It can be shown by
introducing a duality transformation that the transition between the two gapped 
phases at $J_1/J_2=1$ does not involve any change of symmetry but
there is a change of topological order \ct{feng07,perk771,perk772,chen08}
\footnote{{We reiterate that states with different topological 
order cannot be distinguished from each other by any local order parameter, 
but they differ in other ways such as the ground state degeneracy and the 
nature of the low-lying excitations.}}.

%\subsubsection{Two-dimensional Kitaev model}
% {\it Two-dimensional Kitaev model}

For $J_3 \ne 0$, 
%the Kitaev model described by Eq.
%(\ref{rmp_kitaev2d}) describes a spin model on a hexagonal lattice.
we define the JW transformation as
\bea a_{j,l} &=& \left( \prod_{i = - \infty}^{j-1} \si_{i,l}^z \right) 
\si_{j,l}^y~~~\text{for even} ~~j+l, \non \\
a'_{j,l} &=& \left( \prod_{i = - \infty}^{j-1} \si_{i,l}^z \right) 
\si_{j,l}^x~~~\text{for even} ~~j+l, \non \\
b_{j,l} &=& \left( \prod_{i = - \infty}^{j-1} \si_{i,l}^z \right) 
\si_{j,l}^x~~~\text{for odd} ~~j+l, \non \\
b'_{j,l} &=& \left( \prod_{i = - \infty}^{j-1} \si_{i,l}^z \right) 
\si_{j,l}^y~~~\text{for odd} ~~j+l, \label{maj2d} \eea
where $a_{j,l}$, $a'_{j,l}$, $b_{j,l}$ and $b'_{j,l}$ are all Majorana 
fermions.
%i.e., they are Hermitian, their square is equal to 1, and they
%anticommute with each other. Under
%this transformation $xx$ and $yy$ interactions become local and quadratic in 
%Majorana fermions; although the $zz$ interaction usually becomes non-local 
%and quartic, in this model this remains local and only couples fermions on 
%nearest-neighbor sites due to large number of conserved quantities. 
The spin Hamiltonian in (\ref{rmp_kitaev2d})
gets mapped to a fermionic Hamiltonian given by 

\bea H_{\rm 2d} = i ~\sum_{\vec n} [J_1 b_{\vec n} a_{{\vec n}
- {\vec M}_1} + J_2 b_{\vec n} a_{{\vec n} + {\vec M}_2} 
+J_3 D_{\vec n} b_{\vec n} a_{\vec n}], \label{kitham2d1} 
\eea
where $D_{\vec n} = i ~b'_{\vec n} a'_{\vec n}$ and $\vec n = {\sqrt 3} 
{\hat i} ~n_1 + (\frac{\sqrt 3}{2} {\hat i} +
\frac{3}{2} {\hat j} ) ~n_2$ denote the midpoints of the vertical
bonds. Here $n_1, n_2$ run over all integers so that
the vectors $\vec n$ form a triangular lattice whose vertices
lie at the centers of the vertical bonds of the underlying honeycomb
lattice; the Majorana fermions $a_{\vec n}$ and $b_{\vec n}$ sit at the
bottom and top sites respectively of the bond labeled $\vec n$. The
vectors ${\vec M}_1 = \frac{\sqrt 3}{2} {\hat i} -
\frac{3}{2} {\hat j}$ and ${\vec M}_2 = \frac{\sqrt 3}{2} {\hat i} +
\frac{3}{2} {\hat j}$ are spanning vectors for the reciprocal lattice. We note 
that the operators $D_{\vec n}$ have eigenvalues $\pm 1$, and commute with 
each other and with $H_{2d}$; hence all the eigenstates of $H_{2d}$ can be 
labeled by specific values of $D_{\vec n}$. 
%(We observe that 
%the Hamiltonian $H^{\prime}$ gives dynamics to the fermions $a_{\vec n}$
%and $b_{\vec n}$, but the fermions $a'_{\vec n}$ and $b'_{\vec n}$ have no 
%dynamics since $i b'_{\vec n} a'_{\vec n}$ is fixed). 
The ground state can be shown to correspond to $D_{\vec n} = 1$ for all 
${\vec n}$ for any value of the interaction parameter \ct{kitaev06}.
% Each of the $D_{\vec n}$'s is a Hermitian operator whose square is one; they
%commute with each other so that their eigenvalues can be equal t
%\pm 1$ independently for each $\vec n$. The crucial point that makes 
%the solution of the Kitaev model feasible is that the $D_{\vec n}$'s also 
%commute with $H_{\rm 2d}$, so that all the eigenstates of
%$H_{\rm 2d}$ can be labeled by specific values of $D_{\vec n}$. It has been
%shown that for any value of the parameters $J_i$, the ground state of the
%model always corresponds to $D_{\vec n}=1$ on all the bonds. Since 
%$D_{\vec n}$ is a constant of motion, the dynamics of the model starting from 
%any ground state never takes the system outside the manifold of states with 
%$D_{\vec n}=1$. Also, there are $2^{N/2}$ decoupled sectors corresponding to 
%the values of $D_{\vec{n}} = \pm 1$ in the $N/2$ different hexagons.
\begin{figure}[htb]
\begin{center} \includegraphics[height=2.4in,width=3.1in]{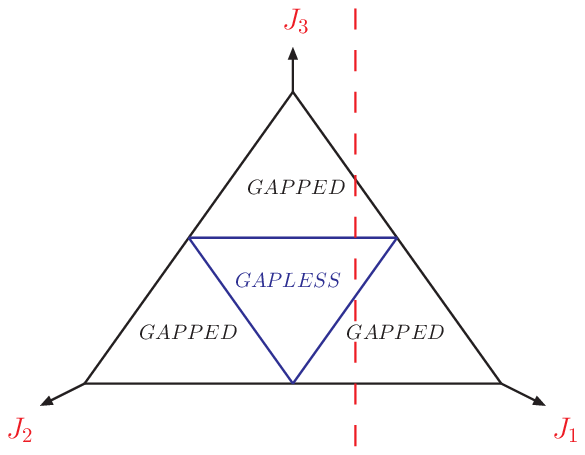}
\caption{(Color online) Phase diagram of the Kitaev model showing one gapless
and three gapped phases inside the equilateral triangle in which $J_1 + J_2 + 
J_3$ is a constant. (The symbols $J_i$ denote that the values of $J_i$ are 
measured from the opposite side of the triangle). The red dashed line 
indicates the quenching of $J_3$ from $-\infty$ to $\infty$ to be discussed 
in Sec.~\ref{sec_gapless}. (After Sengupta $et~al.$, 2008).} 
\label{kitaev_phase} \end{center} \end{figure} 
For $D_{\vec n}=1$, it is straightforward to diagonalize $H_{\rm 2d}$ in 
momentum space (for details, see \ct{mukherjee113}). 
%We define Fourier transforms of the Majorana operators $a_{\vec n}$ as
%\beq a_{\vec n} ~=~ \sqrt{\frac{4}{N}} ~\sum_{\vec k} ~[~ 
%a_{\vec k} ~e^{i\vec k \cdot \vec n} ~+~ a_{\vec k}^\dg ~ e^{-i\vec k 
%\cdot \vec n} ~], \label{fourier2} \eeq
%(and similarly for $b_{\vec n}$), where $N$ is the number of sites (assumed to
%be even, so that the number of unit cells $N/2$ is an integer), and the sum 
%over $\vec k$ extends over half the Brillouin zone of the hexagonal lattice. 
We then obtain 
$H_{\rm 2d} = \sum_{\vec k} \psi_{\vec k}^\dg H_{\vec k} \psi_{\vec k}$, 
where $\psi_{\vec k}^\dg =(a_{\vec k}^\dg, b_{\vec k}^\dg)$, and
$H_{\vec k}$ can be expressed in terms of Pauli matrices
\bea H_{\vec k} ~=~ 2 ~[J_1 \sin ({\vec k} \cdot {\vec M}_1) ~-~ J_2
\sin ({\vec k} \cdot {\vec M}_2)] ~\si^x ~+~ 2 ~[J_3 ~+~ J_1 \cos ({\vec k} 
\cdot {\vec M}_1) ~+~ J_2 \cos ({\vec k} \cdot {\vec M}_2)] ~\si^y.
\label{eq_kitaev2by2} \eea
The energy spectrum of $H_{\rm 2d}$ therefore consists of two bands 
with energies
\bea E_{\vec k}^\pm ~=~ \pm ~2 ~[(J_1 \sin ({\vec k} \cdot {\vec M}_1) - J_2 
\sin ({\vec k} \cdot {\vec M}_2))^2 ~+~ (J_3 + J_1 \cos ({\vec k} \cdot 
{\vec M}_1) + J_2 \cos ({\vec k} \cdot {\vec M}_2))^2 ]^{1/2}. \label{hk1} \eea
We note for $|J_1-J_2|\le J_3 \le J_1+J_2$, these bands touch each other so 
that the energy gap $\De_{\vec k} = E_{\vec k}^+ - E_{\vec k}^-$ vanishes 
for special values of $\vec k$ for which $J_1 \sin ({\vec k} \cdot {\vec M}_1)
- J_2 \sin ({\vec k} \cdot {\vec M}_2)=0$ and $J_3 + J_1 \cos ({\vec k} \cdot 
{\vec M}_1) + J_2 \cos ({\vec k} \cdot {\vec M}_2)= 0$. The
gapless phase in the region $|J_1-J_2|\leq J_3 \leq J_1+J_2$ is shown in 
Fig.~\ref{kitaev_phase} \ct{kitaev06,feng07,lee07,chen08,nussinov08}. It is 
straightforward to show that the transition on the critical line $J_3 = J_1 + 
J_2$ that separates one of the gapped phase from the gapless phase in 
Fig.~(\ref{kitaev_phase}) belongs to the universality class of a anisotropic 
quantum critical point with exponents $\nu_1=1/2, z_1=2 $ and $\nu_2=1, z_2= 
2$, respectively \ct{hikichi10,mukherjee113}. {Both the
one-dimensional and two-dimensional versions of the Kitaev model have been 
used extensively in studies of quantum information and non-equilibrium
dynamics (see e.g., Secs. \ref{fidelity}, \ref{sec_gapless}, Appendix 
\ref{sec_topology}, etc.).}

%\section{Experimental realizations: ({\it Aeppli $\sim$ 4pp})}

%\subsection{Singlet ground state magnetic systems ~~(\it {Aeppli})}

%\subsection{KDP-ferroelectrics: order-disorder transitions {(\it AD)}}

%\ct{degennes63,blinc60}

%\subsection {One-dimensional systems ~~(\it {Aeppli})}

%\subsection{Dipolar transverse Ising systems: ${\rm LiHo_x Y_{1-x}F_4}$ ~
%(\it {Aeppli})}

%\subsection{Superconducting arrays ~(\it {AD})}

\section{Role of quenched disorder}

\label{sec_qdis}

In this section, we will discuss quantum phase transitions in TIMs in the 
presence of random interactions or fields. We will provide a brief overview 
and refer to some review articles 
\ct{igloi05,vojta06,bhatt98,rieger97,refael09} for extensive discussions. 
Randomness has a drastic effect on QPTs, especially in low dimensions which 
has been a subject of extensive research both theoretically and experimentally.

In a seminal paper \ct{harris74} Harris addressed the question of the 
relevance of disorder close to a finite temperature classical transition 
using the following argument. For a model with a random local critical 
temperature \ct{grinstein76}, one can use the central limit theorem to argue
that the fluctuation in $T_c$ for a domain of size $\xi$ is of the order of 
$\Delta T_c \sim \xi^{-d/2} \sim (T-T_c)^{\nu_{\rm pure} d/2}$, where 
$\nu_{\rm pure}$ is the correlation length exponent corresponding to the 
critical point of the non-random version of the model. This shows that 
$\Delta T_c/|T - T_c|$ 
grows as $T \to T_c$, if ${\nu_{\rm pure}< 2/d}$ (or $\al_{\rm pure}>0$ 
because of the hyperscaling relation $2 - \al_{\rm pure}=\nu_{\rm pure}d$ 
valid for $ d< d_u^c$), and hence randomness is a relevant perturbation.
If ${\nu_{\rm pure}> 2/d}$, randomness is irrelevant in determining the 
critical behavior which remains that of the pure system. When disorder is 
relevant, it has been shown that the correlation length critical exponent 
$\nu_{\rm random}$ associated with the random critical fixed point satisfies 
$\nu_{\rm random} \geq d/2$; {this is known as the Chayes 
criterion \ct{chayes86}. However, this criterion is not universally 
valid;} there exist a number of systems, including the metal-insulator 
transition in uncompensated, doped semiconductors and helium in aerogel, that 
have experimental results in conflict with this (see Ref. \ct{pazmandi97} and 
references therein)\footnote{{The Chayes criterion is 
obtained using finite size scaling and disorder
averaging of a particular type. This leads to an additional length scale in
the problem, as well as noise. It has been posited that self-averaging is
not the appropriate treatment of the disorder in all cases, circumventing the 
strict bound on the critical exponent at $2/d$ in such instances 
\ct{pazmandi97}.}}.

Let us consider a TIM in $d$-dimensions with randomness only in the 
interactions {(assumed to be nearest neighbor for the time 
being)} given by the Hamiltonian 
\beq H = -\sum_{<ij>} J_{ij} \si_i^x \si_i^x - h\sum_i \si_i^z, 
\label{eq_random1} \eeq
where the $J_{ij}$'s are chosen from a random distribution. Similarly one can 
define QR models (\ref{eq_hamrotor}) with random interactions \ct{ye93}. As 
discussed already, the QPT of the Hamiltonian in (\ref{eq_random1}) is 
equivalent to the finite temperature CPT in a $(d+1)$-dimensional classical 
Ising model with randomness infinitely correlated in the Trotter direction (see
Fig.~\ref{fig_random1}). Due to this correlated nature of randomness, this 
additional dimension does not influence the 
fluctuation $\Delta T_c$ of the $(d+1)$-dimensional model which is still
given by $\Delta T_c \sim (T-T_c)^{\nu_{\rm pure}^q d/2}$, where $\nu_{\rm 
pure}^q$ is the correlation length exponent for the QPT in the corresponding 
pure system. Therefore, following similar arguments as
given above, we find that
disorder is relevant when ${\nu_{\rm pure}^q< 2/d}$ or $\al_{\rm pure}^q
+\nu_{\rm pure}^q >0$ \ct{harris741,boyanovski83} where we have used the 
hyperscaling relation in (\ref{eq_hyperscaling}) with $z=1$. Since 
$\nu_{\rm pure}^q >0$, randomness is usually relevant for a QPT.

\subsection{Quantum Ising spin glass (QISG)}

The free energy ${\cal F}$ of a random system is a self-averaging quantity 
which means that in the $N \to \infty$ limit, ${\cal F} =- 1/(\beta N) 
\overline{\ln{\cal Z}(J)}= {\cal F}_{\infty}(\beta)$ where ${\cal Z}$ is the 
partition function for a given configuration of disorder. The overline bar 
denotes the average over disorder such that $\bar A = \int dJ P(J) A(J)$ where
$P(J)$ is distribution of disorder. For a self-averaging quantity,
the fluctuation falls off as $1/N$.
To calculate $ \overline{\ln{\cal Z}(J)}$, the method of replicas 
\ct{edwards75} turns out to be extremely useful. This involves using the 
mathematical formula $\ln {\cal Z} = \lim_{n \to 0} ({{\cal Z}^n -1})/{n}$, 
which means that {we are in fact considering $n$ identical 
copies of the system for a given disorder configuration. As a result of 
disorder averaging, i.e., in evaluating the disorder-averaged 
$\overline{\ln{\cal Z}(J)}=\int dJP(J) \ln Z$ (using its replicated form),
the different replicas get coupled; eventually one takes the limit $n \to 0$}.
%However, in the process of disorder 
%averaging different replicas get coupled.

This method has been used to study classical Ising spin glasses (for reviews 
see \ct{binder86,chowdhury86,fisher91,castelloni05,dedominicis06}) given by 
the Hamiltonian 
\beq H ~=~ - ~\sum_{ij} J_{ij} S_i S_j, \label{eq_random_cl}\eeq
where the $S_i$'s are classical binary variables ($S_i=\pm 1$). In the 
Edwards-Anderson (EA) version of Hamiltonian (\ref{eq_random_cl}) 
\ct{edwards75}, the interactions are Gaussian distributed but are restricted 
to nearest neighbors only. In the mean field Sherrington-Kirkpatrick (SK) 
version \ct{sherrington75}, all the spins in
 (\ref{eq_random1}), interact with each other, and the interactions 
$J_{ij}$'s are chosen from a Gaussian distribution,
\beq P(J_{ij}) = \left(\frac{N}{2\pi {\tilde J}^2} \right) \exp \left(-
\frac{NJ_{ij}^2}{2{\tilde J}^2} \right),
\label{eq_SK} \eeq
where we have set $\overline {J_{ij}}=0$ and 
{${\tilde J/\sqrt N}$} is the variance of the distribution. 
{The presence of $1/N$ in the expression of the variance is 
essential for the extensivity of the model, i.e., to obtain finite 
thermodynamic quantities in the thermodynamic limit ($N \to \infty$).} 

The natural choice for the spin-glass order parameter is 
the EA parameter defined as $q= 1/N \sum_{i=1}^N \overline{(\langle S_i 
\rangle)^2}$, {which is the configuration averaged 
mean-squared local magnetization. Note that one applies a conjugate 
longitudinal field ${\tilde H}$ (that breaks the spin-reversal symmetry of 
(\ref{eq_SK})), which is set equal to zero after the thermodynamic limit is 
taken; the precise definition of $q$ is therefore given by $q=
\lim_{{\tilde H}\to 0} \lim_{N\to \infty}(1/N)\sum_{i=1}^N 
\overline{(\langle S_i \rangle)^2}$. Otherwise, $\langle S_i \rangle=0$ due to 
the symmetry.} On the other hand, the overlap between different replicas 
defined by $q_{\al\beta} = 1/N \sum_{i=1}^N \langle S_i^{\al} \rangle \langle 
S_i^{\beta} \rangle$, is also a measure of spin glass order. In the saddle 
point solution {(obtained in the $N \to \infty$ limit) 
\ct{sherrington75}\footnote{{One in fact interchanges the 
limits $n\to 0$ and $N\to \infty$ to arrive at the saddle point solution; 
first the $N \to \infty$ limit is taken keeping $n$ finite, and then the limit
$n\to 0$ is applied.}}} of the classical SK model, one assumes a replica 
symmetric (RS)
ansatz $q_{\al\beta} = q_0$ for $\al\neq \beta$ and $q_{\al\al}=0$. However, 
the stability analysis of the RS saddle point solution shows that the spin 
glass phase with $q \neq 0$ is unstable at low temperature 
{because of the emergence of a negative entropy. In a magnetic
field (${\tilde H}$), the RS solution is unstable below a line in the 
$T-{\tilde H}$ plane, this is known as the Almeida-Thouless (AT) line 
\ct{almeida78}}.

To cure this problem, the concept of replica symmetry breaking (RSB) was 
introduced \ct{parisi79,parisi791,parisi80,parisi83} (see also \ct{binder86}). 
In this ansatz, the $n \times n$ order-parameter matrix $q_{\al\beta}$ is 
divided into $n/m_1 \times n/m_1$ blocks
of size $m_1 \times m_1$. In the off-diagonal blocks, nothing is changed while
in the diagonal blocks $q_0$ is replaced by $q_1$. The method is repeated for 
each of the blocks along the diagonal which are split into $m_1/m_2 \times 
m_1/m_2$ subblocks each of size $m_2 \times m_2$ and along the diagonal of 
the sub-block $q_1$ is replaced by $q_2$;{the procedure is 
repeated $k$ times such that $n \geq m_1 \geq m_2 ...>m_k \geq 1$. While the 
procedure is meaningful for positive integer $n$ and finite $k$, one assumes 
that an analytical continuation to the limit $n \to 0$ and $k \to \infty$ 
is possible such that $0 \leq m_i \leq 1$ as $k \to \infty$.}
In this limit $m_i$ becomes continuous,
$m_i \to x, 0<x<1$ and $q_i \to q(x)$; the EA order parameter is given by 
$q = q(x=1)$. The RSB is related to the rugged free energy landscape of the 
spin glass phase when the system may get trapped in a local minima leading to 
the breaking of ergodicity. The ground state of SK model is expected to be 
infinitely replica symmetry broken and there are thermodynamically large 
number of local minima. Later works showed the RSB solution is stable 
\ct{dedominicis83} and exact for SK spin glass \ct{talagrand80}. We note that 
an alternative picture known as the ``droplet picture'' \ct{bray84,fisher88} 
in fact rules out the possibility of an AT line and RSB for any finite 
dimensional spin-glass. Whether the AT line exists for a real short-range 
interacting spin glass is still not clear \ct{young04}; a recent study 
\ct{katzgraber09} shows that it occurs only for $d >d_u^c$ ($=6$ for 
short-range spin class \ct{binder86}), i.e., in the mean field region. However,
we shall comment below that in a quantum spin glasses there is a 
possibility of restoration of replica symmetry due to
quantum tunneling at zero-temperature.

The quantum counterpart of classical Ising spin glasses given by Hamiltonian 
(\ref{eq_random1}) was introduced in the early 1980's \ct{chakrabarti81}.
% In the Sherrington-Kirkpatrick (SK) version of 
%the quantum Ising spin glass (QISG) 
%where $N$ is the number of spins and
The SK version (with $P_{ij}$'s given by (\ref{eq_SK})) has been studied 
extensively \ct{ishii85,ray89,thirumalai89,goldschmidt90,buttner90}; in 
particular, the possibility of the restoration of replica symmetry of the 
replica symmetry broken classical ground state was addressed \ct{ray89}. 
There is a possibility that quantum fluctuations due to the transverse field 
may help tunneling across macroscopically large but narrow barriers which is 
not possible through thermal activated dynamics.

In the $N \to \infty$ limit, the classical SK model can be reduced to a single 
spin problem in a Gaussian distributed random field $h_{\rm eff}^r$, with zero 
mean and variance $J^2 q$, where $q$ is the replica symmetric version of the 
spin-glass order parameter $q_{\al\beta}$. For the quantum case, one can 
intuitively write an effective single spin Hamiltonian ($H_{\rm SP}$) 
\ct{kopec881,kopec882} given by
$H_{\rm SP} = - h_{\rm eff}^r\si^x - h \si^z$. The Hamiltonian $H_{\rm SP}$
can then be solved using 
%by generalizing the method discussed in Sec.~\ref{brout} to the random case
a self-consistent method which necessitates random averaging in addition to 
the thermal averaging. The phase boundary between the spin-glass phase 
($q= 1/N \sum_{i=1}^N \overline { \langle \si_{i\al}^x \si_{i\beta}^x\rangle} 
\neq 0$) and the PM phase ($q=0$) 
is given by the condition $h_c(T)/J = \tanh (h_c(T)/(k_B T))$. 

The random averaged quantum SK model can be mapped to a single spin problem
due to the infinite range of interactions. However, a non-local interaction
in imaginary time $\cal R(\bar\tau -\bar\tau')$ is generated due to the 
infinite correlation of disorder in the temporal direction as happens also 
for a vector quantum spin glass \ct{bray80}; here the replica symmetric case 
is assumed. Within the static approximation which ignores the time dependence 
of $\cal R$, the phase diagram of the model has been studied 
\ct{goldschmidt90,usadel87}. Miller and Huse \ct{miller93} 
used a non-perturbative 
argument that goes beyond the static approximation to estimate the critical 
value of the transverse field $h_c$ at $T=0$, starting from the replica 
symmetric PM phase.
% the spin autocorrelation function was found to decay as $1/t^2$ at the QCP.

Following the experimental realization of a QISG with long-range dipolar 
interaction in the material ${\rm LiHo_xY_{1-x}F_4}$ \ct{wu91,wu93,aeppli05}
(see Sec. 9 for details),
a number of studies have been directed towards understanding the QPT in the 
EA QISG. The equivalent classical spin-glass model with correlated randomness 
(Fig.~{\ref{fig_mapping1}) has been studied using quantum Monte Carlo 
methods for spatial dimension $d=2$ \ct{rieger94} and $d=3$ \ct{guo94}. The 
critical temperature and the exponents are obtained by a finite-size scaling 
analysis of the Binder cumulant ${\tilde g}$ \ct{binder86}, given by 
\beq {\tilde g}= \frac{1}{2} \left(3 - \frac{\langle q^4 \rangle}{\langle q^2 
\rangle^2} \right) = \bar g_0 \left(L^{1/\nu}(T-T_c), \frac{M}{L^z} \right),
\label{eq_random3} \eeq
where $\bar g_0$ is the scaling function and $L$ ($M$) is the size of the 
spatial (Trotter) direction. It is found that the Binder cumulant 
shows a maximum as a function of $M$, and the value of this maximum becomes 
independent of the spatial size $L$ at $T=T_c$; knowing $T_c$, one can find the 
dynamical exponent $z$ using the scaling relation in (\ref{eq_random3}). 
Similarly, by studying the variation of $\tilde g$ as a function of 
temperature for fixed $M/L^z$, one can estimate $T_c$ (at which $\tilde g$ 
becomes independent of $L$)
and also the critical exponent $\nu$. These studies show that $\nu \approx 
1.0$ and $z \approx 1.7$ for $d=2$ \ct{rieger94}, and $\nu \approx 0.8$ and 
$z \approx 1.3 $ for $d=3$ \ct{guo94}. Both the linear and non-linear 
susceptibilities were found to diverge for $d=3$, while for $d=2$ only the 
non-linear susceptibility was shown to diverge. 

The LGW actions in terms of the spin glass order parameter $q_{\al\beta}$ for 
an EA QISG and also a QR were proposed in \ct{read95} where it 
was shown that the quantum fluctuations are dangerously irrelevant (see 
\ct{chaikin95}, for a discussion of dangerously irrelevant variables) close 
to the QCP. Below the upper critical dimension $d_u^c$ ($=8$ for this model, 
consistent with the hyperscaling relation in (\ref{eq_hyperscaling}) since 
the mean field $z=2$, $\nu=1/4$ for QISG and $d_u^c =6$ for classical spin 
glass), the renormalization group calculations fail to locate any stable weak 
coupling fixed point, and a run-away to strong coupling was observed. This was
attributed to the existence of an infinite randomness fixed point (see below) 
for $d < d_u^c$. A similar result was obtained \ct{dutta02} for the quantum 
version of the classical spin-glass where the random interaction decays 
algebraically with the distance between the spins 
\ct{kotliar83,bray86,katzgraber09}.
%i.e. $J_{ij} \to J_{ij}/{r_{ij}^{(d+p)/2}$ with $p>0$ 
A quantum version of the droplet model has also been proposed \ct{thill95}.

%magnetic excitations \ct{rosnow071}
%Ising ferroglass \ct{kao01}
%Ising Spin Liquid \ct{silevitch071}
%Recent theory of transverse field ising, especially with random field effects
%etc. 
%single ion anisotropy: 
%tunable random field; experimental study \ct{silevitch07}

\subsection{Griffiths singularities and activated dynamics} 
\label{sec_griff+act}

The free energy of a dilute Ising ferromagnet is a non-analytic function of 
the external field below the critical temperature of the corresponding pure 
model; this is known as a Griffiths-McCoy (GM) singularity \ct{griffiths69}. 
More generally speaking, a random magnetic system is in its GM phase 
if it is above its ordering temperature but below the highest ordering 
temperature allowed by the distribution \ct{bray82}. The GM singular regions
occur due to locally ordered ``rare regions'' 
\ct{griffiths69,mccoy69,mccoy691} as seen for example in two-dimensional 
Ising models with frustration as well as randomness 
\ct{shankar871,shankar872}, and it strongly influences the dynamical response 
of the system \ct{randeria85,bray87,dhar88}.

In a QPT, statistics and dynamics are mingled and hence these ``rare regions'' 
have a more prominent effect on the QPT of a low-dimensional system as we 
shall indicate below using the example of a dilute transverse Ising chain in 
the ferromagnetic phase; {the existence of these regions 
results in fascinating features associated with random QPTs in low-dimensional 
systems. These features are the activated quantum dynamical scaling ($z \to 
\infty$) at the QCP and the existence of GM singular regions where the 
susceptibility diverges even away from the critical point 
\ct{fisher92,fisher94,fisher95}.} {The role of these GM 
singular regions in quantum information and dynamics will be discussed in 
Secs. \ref{e_entropy}, \ref{sec_quench_disorder}, etc.} For a discussion on
experimental signature of such singularities see the discussion around 
Eq.~(\ref{eq_ga1}).

Let us consider a random one-dimensional transverse Ising chain given by the 
Hamiltonian
\beq H ~=~ - ~\sum_{i=1}^{N-1} J_{i,i+1} \si_i^x \si_{i+1}^x ~-~ \sum_{i=1}^N 
h_i \si_i^z, \label{eq_random4} \eeq 
which includes randomness in the transverse field also. For $d=1$, one can 
perform a gauge transformation to make all the $J_{i,i+1}$ and $h_i$ positive.
Moreover, using the duality transformation \ct{kogut79} (see also 
Sec.~\ref{models}), one can argue that the critical point is expected when the 
distribution of bonds and fields are identical. Hence, defining variables 
$\Delta_h = \overline {\ln h}$ and $\Delta_J = \overline {\ln J}$, the QCP 
occurs when $\Delta_h =\Delta_J$. The deviation from the QCP is measured as 
$\la = (\Delta_h -\Delta_J)/ ({\rm var} [h] + {\rm var} [J])$; 
here ${\rm var}$ stands for the variance of the distribution. 

The model in (\ref{eq_random4}) has been studied by Fisher 
\ct{fisher92,fisher95,fisher99,fisher98} using a strong disorder 
renormalization group (SDRG) technique introduced in \ct{dasgupta80}. 
% which is exact in the asymptotic limit.
In this approach, we choose the strongest bond or the strongest field and 
minimize the corresponding term in the Hamiltonian; therefore the degrees of 
freedom associated with the maximum energy scale ($\Omega_0$) are frozen at the 
lowest energy scale. If the strongest coupling is a field $h_i$ at the site 
$i$, then the spin variable at this site is fixed in the direction of the 
field (i.e., decimated) and an effective interaction generated due to quantum 
fluctuations between neighboring spins $\tilde J_{i-1,i+1}$ (much smaller 
than $\Omega$) is calculated using second-order perturbation theory. 
%\simeq$ max$(J_{i-1,i+1}, J_{i-1,i} J_{i,i+1}$.
In contrast, if the interaction between two sites $J_{i,i+1}$ are the 
strongest, we set the spins at those two sites $i$ and $i+1$ to point in the 
same direction, i.e., they form a ferromagnetic cluster which can be viewed 
as a single spin with higher magnetic moment in an effective transverse field
again calculated by second-order perturbation.
%and the cluster interacts the neighboring spins with interaction 
%$\tilde J_{i-1,i}$ chosen as the maximum of the interactions 
%$J_{i-1,i}$ and $J_{i-1,i+1}$. 
In the process, the maximum energy scale decreases to $\Omega < \Omega_0$. 
The process is iterated and the renormalization group flow equations for the
distributions of $h_i$ and $J_{i,i+1}$ as a function of $\Lambda= \ln 
(\Omega_0/\Omega)$ have a fixed point solution which is an attractor for all 
initial distribution of randomness. This is the infinite 
randomness fixed point (IRFP) distribution, so called because disorder
grows beyond limit under renormalization as $\Omega \to 0$; in this sense,
SDRG is asymptotically
exact in describing the low-energy properties near a random QCP ($\la=0$).
We see that the clusters grow in the ordered phases while bonds become 
disconnected in the disordered phase. {Qualitatively one 
can argue that at the QCP 
%can therefore be viewed as a novel percolation point where 
the annihilation and aggregation of clusters 
compete with each other at each energy scale. }
% Moreover, and hence the critical fixed point is an 
%with a continuously varying dynamical exponent which diverges at QCP.
% established. The results of the RSRG show that . Under the RSRG, in which 
%the only the stronger of the bond strength and the field strength is 
%retained at every step of iteration, .
%One can argue that disorder broadens without limit at the QCP when 
%the energy scale is lowered,

An IRFP is characterized by three critical exponents $\psi$, $\phi$ and $\nu$.
For a rare large cluster at energy scale $\Omega$, the linear dimension $L$ 
of the cluster is related to $\Omega$ through the exponent $\psi$ as 
$\ln(\Omega/\Omega_0) \sim L^{\psi}$.
%where $\Omega_0$ is the basic energy scale. 
This logarithmic dependence is the signature of the activated quantum 
dynamics. Similarly, the magnetization of the cluster scales as $L^{\phi 
\psi}$ and the correlation length exponent $\nu$ determines the decay of the 
average correlation function \ct{fisher92}. Fisher showed that 
\ct{fisher92,fisher95} for the one-dimensional chain (\ref{eq_random4}), 
$\psi=1/2$, $\phi=(\sqrt 5 + 1)/2$ and $\nu=2$. Close to the QCP ($\la$ 
non-zero but small), the spin chain is in the GM phase, where the low-energy 
behavior is dominated by gapless but well-localized
excitations which lead to off-critical singularities; at early stages the 
chain obeys the critical scaling. However, when the typical sizes and bond
length is of the order of the correlation length $\xi \sim \la^{-2}$ and 
$\Lambda= \ln (\Omega_0/\Omega) \sim \la^{-1}$, a crossover to ordered 
($\la <0$) or disordered ($\la >0$) phase takes places. For small $\la$, in 
both phases $\Omega \sim L^{-z(\la)}$ with the effective dynamical exponent 
$z(\la) \sim |\la|^{-1}$ diverging at the QCP where dynamics is activated.
%The typical correlation length is however determined by a different exponent 
%$\bar \nu=1$ as typical correlation function defined by 
We mention that the SDRG method has been extended to $d=2$ 
\ct{motrunich00,lin00}.

Numerical diagonalization of a one-dimensional random Ising chain using a 
JW mapping to fermions \ct{young96} confirmed the SDRG predictions; the 
dynamical properties of a random chain has also been studied 
\ct{kisker98,young97}. Long-range spatial correlations of disorder
\ct{weinrib83} were found to enhance the GM singularities \ct{rieger99}. 
Similar results have been obtained for a random transverse $XY$ chain by 
mapping to a Dirac equation with random mass \ct{bunder99} and also exploiting 
the analogy between a one-dimensional random TIM and a random walk 
\ct{igloi97,rieger98}.

Quantum Monte Carlo studies of EA QISG 
\ct{rieger96,guo96} as well as of two-dimensional random bond Ising 
models \ct{pich99} show that critical points in all these cases happen 
to be an IRFP and hence one can conclude that at an IRFP frustration is 
irrelevant. However, the situation of QISG needs more attention at this point.
For QISG in spatial dimension two \ct{rieger96}, signatures of strong GM 
singularities are observed in the disordered phase near the quantum 
transition where one finds power-law distributions of the local susceptibility 
and local non-linear susceptibility characterized by a smoothly varying 
dynamical exponent. The local non-linear susceptibility diverges in the 
GM phase though the local susceptibility does not; it diverges only at 
the QCP. Approaching from the disordered phase, the limiting value of the 
dynamical exponent apparently tends to its value ($z \simeq 1.7$) at the QCP 
quoted in the previous section. This is in contrast to the one-dimensional 
situation where $z \to \infty$ at the QCP and if one assumes that $z$ is 
finite, the dynamical scaling should be conventional. However, one can not 
rule out the possibility of increasing $z$ for larger system size and hence 
an activated quantum critical dynamics.

The SDRG studies of the random spin-1/2 Heisenberg chain \ct{fisher94} has 
revealed that the ground state is a random singlet phase which is another
example of IRFP. This is true also for higher spin values \ct{damle02}. We 
note that the existence of GM phases and an infinite randomness fixed 
point has been reported for the finite temperature transition of a layered 
Heisenberg ferromagnet \ct{mohan09}.

The effect of GM singularities are less prominent for QR models. This can be 
argued \ct{read95} taking the example of a random Ising (rotor) model with a 
uniform transverse field (kinetic term). A locally ordered region of size 
$L^d$ can be viewed as a single spin with a large magnetic moment $L^d$ coupled
ferromagnetically with similar blocks along the Trotter direction, so that we 
have an equivalent one-dimensional chain of these giant moments. If the 
moments are Ising-like, then the correlation time goes as $\xi_{\tau} \sim 
\exp(L^d)$ which is an activated behavior, while for the $n$-vector case 
$\xi_{\tau} \sim L^d$ \ct{chaikin95}. The latter suggests a power-law quantum 
critical dynamics and hence less prominence of GM singularities for $n>1$.
Therefore, it is the Ising nature of interactions that lead to activated 
quantum dynamical scaling which in turn results in exotic GM singular
phases associated with low-dimensional random quantum Ising transitions. 

\subsection{A generalized random transverse field Ising spin Chain}

We now briefly mention a generalized random transverse field Ising chain with 
spins having $q$ states (labeled by $\ket{s_i} = \ket{0_i}, \ket {1_i}, \cdots,
\ket{(q-1)_i}$). The spin chain described by the Hamiltonian \ct{santacharia06}
\beq H_q ~=~ - ~\sum_i J_i \sum_{n=1}^{q-1} \al_{n} (S_i^{z \dagger} S_i^z)^n ~
-~ \sum_i h_i \sum_{n=1}^{q-1} \al_n \Gamma_i^n . \label{eq_random_gen} \eeq
{is useful in studying the entanglement entropy of a random 
system to be discussed in Sec.~\ref{e_entropy}}.
Here, $S^z$ is a $q \times q$ matrix with $(S^z)_{lm} = \exp[(2i \pi l)/q] 
\de_{lm}$ (with $l,m = 0, \cdots , q-1$), $\Gamma= |s\rangle \langle s+1|$ is 
the spin-raising operator, i.e, $\Gamma \ket{s}= \ket{s+1}$; the variables 
$\alpha_n$ are disorder-free and satisfy $\alpha_n = \alpha_{q-n}$ for 
Hermiticity. For each set of $\al_n$, there exist a duality transformation 
$S_i^{z\dagger} S_i^z = \Gamma_i^{*}$ and $\prod_{j<i} \Gamma_j= S_i^{z*}$ 
which interchanges $h$ and $J$. For $\alpha_n=1$ for all $n$, the Hamiltonian 
describes a $q$-state random quantum Potts chain. The disorder free
ferromagnetic quantum Potts chain has a first order quantum transition for 
$q>4$ and a continuous transition for $q \leq 4$ \ct{wu82}. For $\al_n = 
\de_{1,n}$, model (\ref{eq_random_gen}) represents a random $Z_q$ clock model. 
We note that the ferromagnetic clock chains have, for $q > 4$, a 
quasi-long-range ordered phase sandwiched between a FM phase and a PM phase 
with the corresponding transitions being of BKT type; for $q \leq 4$,
the model has a second order phase transition \ct{jose77}. SDRG studies of 
random $q$-state Potts model and $Z_q$ clock model show that the critical 
fixed point is in fact the IRFP of the model (\ref{eq_random4})
\ct{senthil961}. In particular, for the $Z_q$ clock model with $q >4$, 
the existence of an unstable intermediate-disorder fixed point separating a 
low disorder fixed point and the IRFP has been found both analytically
and using density matrix renormalization group \ct{senthil961,carlon01}.

An important realization of (\ref{eq_random_gen}) is obtained by choosing 
$\al_n = \sin(\pi/q)/\sin(\pi n/q)$. The pure version of the model 
is critical along the self-dual line $J=h$, (unlike Potts and clock models) 
for each $q$ \ct{alcaraz87} and its fluctuations are governed by the $Z_q$ 
parafermionic field theories with central charge $c_q^{\rm pure} = 2(q -1)/
(q +2)$ \ct{fateev85}; $q=2$ for Ising model and $q=3$ for $3$-state 
Potts model. Reference {\ct{santacharia06} suggests that 
even for infinitesimal disorder the Hamiltonian flows to an IRFP for all 
values of $q$.} However, under SDRG, the set $\al_n$ renormalizes to $\al_n = 
\de_{1,n}$ which coincides with the random clock model discussed in the
previous paragraph {for which there is an intermediate fixed 
point. Therefore, whether the Hamiltonian (\ref{eq_random_gen}) with $q >4$ 
flows to the IRFP upon introduction of a weak disorder (as indicated in 
\ct{santacharia06}) is still an open question although it happens for $q <4$}.

\subsection{Higher dimensional realization of IRFP}

The ideal higher-dimensional realization of an IRFP is provided by a random 
TIM on a $d-$ dimensional dilute lattice which is again given by (\ref{eq_random1}) where 
the bonds $J_{ij}$ are present with a probability $p$, i.e.,
\beq P(J_{ij}) = p\de(J_{ij}-J) + (1-p) \de(J_{ij}).
\label{eq_random_dilute} \eeq
%where $p$ is the concentration probability of bonds.
This model connects the QPT of the TIM (from a FM to a PM phase) to a 
geometrical phase transition, 
namely, the percolation transition \ct{stauffer92,aharony86}. For $h=0$ (see 
Fig.~\ref{fig_random1}), if $p >p_c$ (the percolation threshold), at least 
one system spanning cluster exists, while for $p<p_c$, only finite clusters 
appear. The geometrical transition at $p=p_c$ belongs to the universality class
of the thermal transition of a $q$-state Potts model in the limit $q \to 1$ 
\ct{kasteleyn69}. The critical line at $p_c$ in Fig.~\ref{fig_random1} is 
vertical due to the fact that even at the percolation threshold there exists 
an infinitely connected cluster with a fractal dimension $d_f$ ($<d$), and 
hence a finite amount of $h$ is necessary to destroy the long-range order 
\ct{harris741}. This prediction was verified \ct{stinchcombe81} by extending 
the real space renormalization group applicable to classical dilute magnets 
\ct{stinchcombe83,young76}. 

The exponents associated with quantum transitions across this vertical line 
below the multicritical point ($h <h_0$) (see Fig.~\ref{fig_random1}) are the 
exponents of the percolation transition \ct{stauffer92}; in this case the 
fixed point at $p=p_c,h=0$, plays the role of an IRFP \ct{senthil96}. This 
has been verified numerically \ct{ikegami98}. These 
exponents can be found in the following way: the energy of a cluster of 
diameter $L$ scales as $L^{d_f}$ showing that the exponent $\psi=d_f$, while 
the exponent $\nu= \nu_p$ and $\phi = (d -\beta_p/\nu_p)/d_f =1$. Here 
the subscript $p$ denotes the corresponding percolation exponents which are 
%exactly known for $d=2$ and 
well known from numerical studies, e.g., for $d=2$, $\nu_p = 4/3$, $d_f=91/48$
and $\beta_p=5/36$. \ct{stauffer92}. Senthil and Sachdev \ct{senthil96},
also showed that either side of the percolation 
threshold $p_c$ is flanked with off-critical GM singular phases. For 
example, in the disordered phase ($p<p_c$), the disorder averaged imaginary
time part of the local dynamical susceptibility satisfies a power-law scaling 
$\overline{\chi_L^{''}}(\om) \sim \om^{d/z-1}$. This power-law behavior shows
that the PM phase is gapless, and the average local susceptibility 
diverges as $T^{d/z -1}$ as $T \to 0$. The 
dynamical exponent is related to the percolation correlation length $\xi_p$ 
as $z\sim \xi_p^{d_f} \sim (p_c - p)^{-\nu d_f}$ and diverges as $p \to p_c$. 

Similar results are obtained in \ct{dutta07,divakaran07} for 
a TIM on dilute lattices with long-range connection probabilities 
\ct{schulman83,newman86}, and for contact processes \ct{hinrichsen10} on
dilute lattices \ct{vojta07}. For quantum rotors on dilute lattices, the 
quantum dynamics at the percolation threshold is of the conventional power-law 
type with $z=d_f$ which is consistent with the argument given above 
\ct{vojta05}. Let us also mention some interesting studies that connect 
geometrical and quantum phase transitions of magnetic models
\ct{monthus97,sandvik01,sknepnek04,vajk02,vajk021}.

\begin{figure}
\begin{center} \includegraphics[height=2.4in]{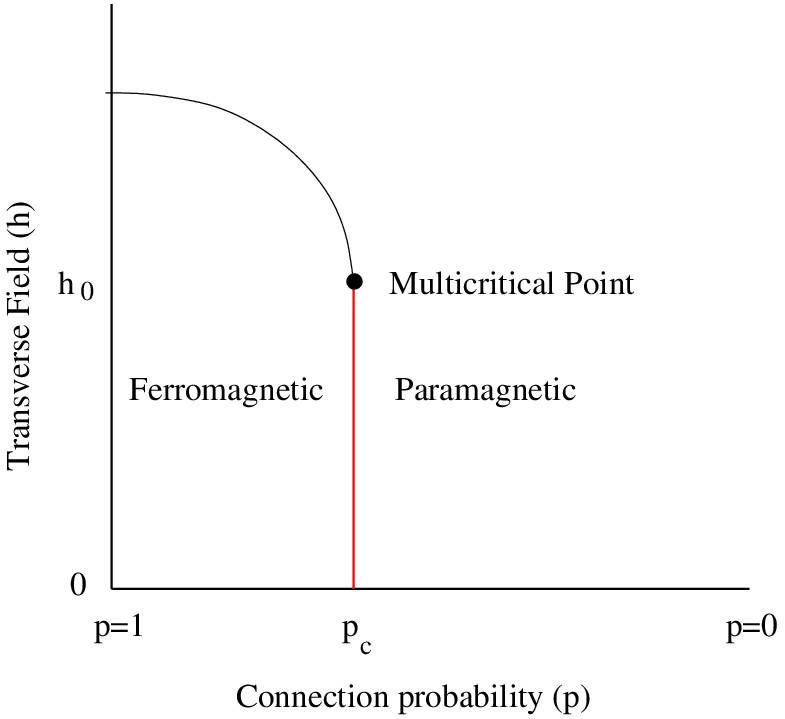}
\caption{The phase diagram of a dilute TIM for $d \geq 2$. The quantum phase 
transition across the vertical 
line at the percolation threshold $p_c$ is governed by the percolation fixed 
point. Here, $h_0$ denotes the value of $h$ corresponding to the multicritical 
point (After Harris, 1974b).} \label{fig_random1} \end{center} \end{figure}

Finally, let us comment on the reason behind the greater prominence of GM 
regions in a QPT compared to a CPT using the example of a dilute Ising model 
in the FM phase \ct{rieger97}. The probability of a rare cluster of size 
$L^d$ is exponentially small and goes as $p^{L^d} \sim \exp (-cL^d)$, where 
$c$ is a constant. However, for the classical system, the spins within this 
cluster have a relaxation time that is exponentially large (because of the 
activation energy that is needed to pull a domain wall through the cluster to 
turn it over), where the relaxation time $\sim \exp(\si L^{d-1} )$, and $\si$ 
is the surface tension.
These two together render the final decay of the autocorrelation function 
faster than algebraic. {Let us now consider the dynamics of 
a similar rare cluster of size $L$ in the case of the dilute quantum Ising 
system in (\ref{eq_random_dilute}) which again occurs with a probability 
$\sim \exp (-cL^d)$. The quantum dynamics in this model is essentially due to 
barrier tunneling between two nearly degenerate minima of the cluster; for a 
small transverse field $h$, the energy gap between these two minima scales 
as $\exp(-\si' L^d)$. Consequently, the inverse of the tunneling rate (or the 
relaxation time) of the cluster scales as $\exp(\si'L^d)$; this (together 
with the probability $\sim \exp (-cL^d)$) in turn leads to an algebraic decay 
of the autocorrelation function. This algebraic decay is in fact a signature 
of gaplessness which signals the existence of GM singularities.} We 
emphasize that it is the difference in the nature of dynamics (activated for 
a random CPT and barrier tunneling for a random QPT in transverse Ising 
models) which is at the root of the stronger GM singularities in a QPT.

\subsection{Quantum Ising model in a random longitudinal field}

The classical Ising model in a site-dependent random longitudinal field with 
zero mean and non-zero variance is given by the Hamiltonian 
\ct{nattermann98,belanger91,rieger95,belanger98}
\beq H = -J \sum_{<ij>} S_i S_j -\sum_i h_i S_i. \label{eq_class_rf} \eeq
This model shows an order-disorder transition from a FM phase ($m= \overline 
{\langle S_i\rangle} \neq 0$) to PM phase ($m=0$) at a critical temperature in 
dimensions greater than the lower critical dimension ($d_l^c$); the domain 
wall argument by Imry and MA \ct{imry75} 
and also calculations on hierarchical lattices 
\ct{bricmont87} showed that $d_l^c =2$. Moreover, from LGW calculations it has 
been shown that $d_u^c=6$ and thermal fluctuations are dangerously irrelevant 
for $d < d_u^c$ \ct{grinstein761,aharony76}. The phase transition of a random 
field Ising model (RFIM) has been shown to be of the same universality class as that 
of a site-dilute Ising antiferromagnet in a longitudinal field 
\ct{cardy84,fishman79}.

The quantum version of the model (\ref{eq_class_rf}) is given as
\beq H = -J_x\sum_{<ij>} \si_i^x \si_i^x - h\sum_i \si_i^z -\sum_i h_L^i 
\si_i^x, \label{eq_random5} \eeq
where $\overline{h_L^i}=0$ and $\overline {h_L^i h_L^j} = \Delta \de_{ij}$.
The order-disorder transition in model (\ref{eq_random5}) has been studied 
\ct{aharony82,boyanovski83,senthil98,dutta00,dutta96,dutta98} within
mean field theory and $\ep$-expansion calculations around the upper critical 
dimension. It has been shown that a TIM model in a random longitudinal field 
(\ref{eq_random5}) is equivalent to a site-dilute Ising antiferromagnet in a 
transverse field \ct{dutta96}. 

The model in (\ref{eq_random5}) has been studied using the imaginary time path 
integral formalism \ct{dutta98,senthil98,boyanovski83}; it can be shown that 
the random field couples only to the static ($\om=0$) part of the order 
parameter, and hence the lower and upper critical dimensions of the quantum 
models remain the same as those of the classical counterpart (see also 
\ct{greenblatt09}). It has also been shown that the quantum fluctuations are 
dangerously irrelevant. The critical exponents for the QPT are therefore 
identical to the classical model. In short, the random field fluctuations mask 
the quantum fluctuations and random field fixed point at ($ T=h=0$) 
is the critical fixed point. This has been established 
using a phenomenological scaling for quantum systems \ct{dutta00} which is an 
extension of the real space renormalization group (RSRG) studies of the classical model in (\ref{eq_class_rf}) 
\ct{bray85,stinchcombe96}. 
The classical RFIM shows an activated dynamical behavior 
\ct{fisher86} due to the presence of free energy barriers which scale with
the cluster size close to the critical point; it has been argued that in a 
quantum RFIM this effect is expected to be stronger as the domains are 
correlated in the Trotter direction \ct{dutta00}.

Finally, let us point out some recent theoretical studies of the system 
${\rm LiHoF_4}$. The magnetic phase diagram has been obtained by quantum 
Monte Carlo studies which go beyond the mean field theory and include the 
on-site hyperfine interaction through a renormalization of the transverse 
magnetic field \ct{chakraborty04}, and also by employing a perturbative 
quantum Monte Carlo technique \ct{tabei081}. Although domain wall formation 
in these systems is favored due to the combination of strong Ising 
anisotropy and long-range forces, the long-range forces have been found to 
destroy the roughening transition \ct{mias05}. The role of strong single-ion 
anisotropy in random field dipolar spin glasses has been studied in 
\ct{tabei08}. We conclude by mentioning some recent theoretical 
studies which especially include the effects of a random field 
\ct{schechter08,schechter07,schechter05,tabei06}.

We conclude this section by noting that disordered itinerant magnetic systems 
and related GM singularities have been studied in several interesting 
papers \ct{vojta97,belitz00,narayanan99,belitz99};
we refer to the review article \ct{vojta06} for relevant discussions.

\section{Related models with frustration}

In this section, we shall study variants of TIMs in the presence of regular 
frustration, for example, models with competing nearest-neighbor and 
next-nearest-neighbor interaction or long-range antiferromagnetic interactions.
The discussion will point to the fact that even the phase diagram of the model
with minimal frustration is not yet fully understood. 

\subsection{Quantum ANNNI model}
\label{sec_ANNNI}

The classical axial next-nearest-neighbor Ising (ANNNI) model was introduced 
by Fisher and Selke \ct{fisher80}
to simulate the spatially modulated structures observed 
in magnetic and ferroelectric systems (for reviews see \ct{selke88,selke92}).
The model is described by a system of Ising spins with nearest-neighbor FM 
interactions in all directions as well as a competing next-nearest-neighbor 
antiferromagnetic interaction along one axis which leads to frustration in 
the model. The Hamiltonian for a two-dimensional system is 
\beq H ~=~ - ~{\tilde J}_0 ~\sum_{i,j} ~S_{i,j}^z S_{i,j+1}^z ~-~ 
{\tilde J}_1 ~\sum_{i,j} ~S_{i,j}^z S_{i+1,j}^z ~+~
\frac{{\tilde J}_2}{2} ~\sum_{i,j} S_{i,j}^z S_{i+2,j}^z ~,
\label{eq_classical_annni} \eeq
where the $S_{i,j}$'s are binary classical variables with values $\pm1$, 
$i$ labels the layers perpendicular to the axial direction, ${\tilde J}_0$ 
denotes nearest-neighbor FM interactions within a layer, and ${\tilde J}_1$ 
and ${\tilde J}_2$ respectively denote ferromagnetic nearest-neighbor and 
antiferromagnetic next-nearest-neighbor interactions along the axial 
direction. The model given by (\ref{eq_classical_annni}) has been 
studied extensively over nearly
three decades \ct{muller-hartmann77,villain81,selke88,selke92,yeomans87,sen89}.
The phase diagram of the model obtained via analytical and numerical studies 
is presented in Fig.~\ref{fig_annni_phase} with ${\tilde J}_0={\tilde J}_1$.
It consists of a FM phase, a PM phase and an antiphase (with a modulated
$++--$ structure with period 4) and a floating phase (FP) with incommensurate 
modulation where the spin-spin correlations decay algebraically separating 
the PM
phase from the antiphase. The multicritical point at $\kappa= |{\tilde J}_2|/
{\tilde J}_1=1$ is infinitely degenerate with the degeneracy being equal to 
$[(1 + \sqrt 5)/2]^N$ for a system of $N$ spins. A disorder line starts
from this point and touches the temperature axis asymptotically; it divides 
the PM phase in two regions such that on the large $\kappa$ side the 
exponential decay of correlations is modulated by periodic oscillations. The 
antiphase and the modulated phase are separated by a line corresponding to the 
Pokrovsky-Talapov (PT) transition \ct{dzhaparidze78,pokrovsky79,nijs88} 
between a commensurate and an incommensurate phase, while the transition from 
the gapless modulated phase to the gapped PM phase is a BKT transition 
\ct{kosterlitz73,berezinskii70}. There is probably a Lifshitz 
point at which the three phases, modulated, para and antiphase, meet.

We will review here some recent studies of the one-dimensional quantum ANNNI 
model \ct{allen01,dutta03,sen89,sen88,umin96} in a transverse field which is 
related to the two-dimensional classical ANNNI model through the 
$\tau$-continuum formalism \ct{barber81,barber821,barber822,rujan81}; hence 
they belong to the same universality class and a similar phase diagram 
is expected. The Hamiltonian is given by
\beq H_A ~=~ \sum_i~[~ - \frac{J_1}{2} \mu_i^x \mu_{i+1}^x + \frac{J_2}{4} 
\mu_i^x \mu_{i+2}^x + \frac{ \Gamma}{4} \mu_i^y ~], \label{annni1} \eeq
where the $\mu_i^\al$ are Pauli matrices, and $J_1$ and $J_2$ are 
nearest-neighbor FM and next-nearest-neighbor antiferromagnetic Ising 
interactions, respectively, and $ \Gamma$ is the transverse magnetic field. 
The ground state is exactly solvable on the disorder line $\Gamma/J_1 = 
\kappa - 1/4 \kappa$, 
%which separates the PM phase into a commensurate 
%phase on the left and an incommensurate phase on the right 
%The disorder line has the special property that the ground state is exactly 
%solvable on it 
\ct{peschel81,kurmann82,muller85}, and is given by a direct product of certain
spin configurations on each of the sites, i.e., is a matrix product state.

\begin{figure}
\begin{center}
\includegraphics[height=2.3in]{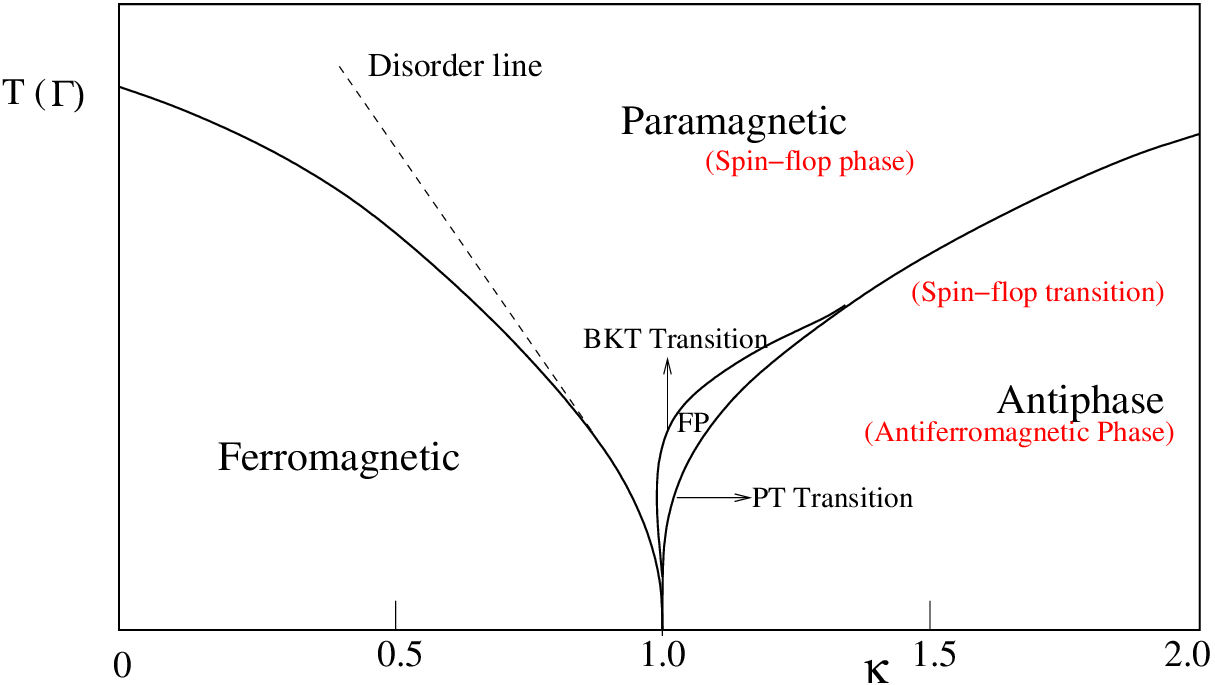}
\caption{(Color online) The phase diagram of a two-dimensional classical 
ANNNI model ($\tilde {J}_0 =\tilde J_1$) (\ref{eq_classical_annni}) and 
one-dimensional transverse ANNNI model (\ref{annni1}) with $x$-axis as 
$\kappa = |\tilde J_2/|\tilde J_1$ or $\kappa = | J_2|/J _1$, respectively. 
The $y$ axis denotes thermal fluctuations ($T$) for the $2d$ classical model 
and quantum fluctuations induced by the transverse field $\Gamma$ for the 
$1d$ quantum model. As discussed in the text, the dual model 
(\ref{annni2}) is expected to show an identical phase diagram but 
the corresponding phases are different and are shown in parenthesis in red. 
A bosonization study of (\ref{annni2}) suggests that the quantum transition 
from the antiferromagnetic to the spin-flop phase far away from the 
multicritical point ($\kappa =1$) is a spin-flop transition and there is no 
intermediate gapless floating phase.} \label{fig_annni_phase} \end{center} 
\end{figure}

Although large parts of the phase diagram of this model are well established, 
the width of the floating phase shown in the above phase diagram for a quantum 
ANNNI chain (\ref{annni1}) (and equivalently for the $2d$ classical ANNNI 
model in (\ref{eq_classical_annni})) 
has been a subject of serious debate over last three decades. 
Analytical \ct{villain81,nijs88,selke92,yeomans87,chakrabarti96,allen01} and 
numerical \ct{sato99,shirahata01,beccaria07,chandra07a,chandra07b} studies
of the phase diagrams of the two-dimensional classical ANNNI model and the 
one-dimensional quantum ANNNI model provide contradictory results about the 
extent of the floating phase. While some of these studies indicate that the 
floating phase, if it exists at all, is restricted to only a line 
\ct{shirahata01,dutta03}, other studies showed that the floating phase has 
a finite width \ct{allen01,beccaria07,chandra07a}. We will address 
this particular issue in the present section.

The Hamiltonian in (\ref{annni1}) has been studied using bosonization and 
the RG in recent years \ct{allen01,dutta03}. Dutta and Sen \ct{dutta03}
 studied the dual Hamiltonian 
\beq H_D ~=~\sum_i ~[~\frac{J_2}{4} \si_i^x \si_{i+1}^x ~+~\frac{\Gamma}{4} 
\si_i^y \si_{i+1}^y -\frac{J_1}{2} \si_i^x ~], \label{annni2} \eeq
to address the issue of the width of the floating phase.
Here, the $\si_i^{\al}$ are Pauli matrices dual to $\mu_i^{\al}$ (for 
instance, $\si_i^x = \mu_i^x \mu_{i+1}^x$ and $\mu_i^y = \si_{i-1}^y \si_i^y$).
Scaling this Hamiltonian by an appropriate factor gives
\bea H ~=~ \sum_i ~\left[ \frac{1+a}{4} ~\si_i^x \si_{i+1}^x ~+~ \frac{1-a}{4}~
\si_i^y \si_{i+1}^y ~+~ \frac{J_z}{4} ~\si_i^z \si_{i+1}^z ~-~ \frac{h}{2} ~
\si_i^x ~\right], \label{ham1} \eea
where the parameters in Eqs.~(\ref{annni2}) and (\ref{ham1}) are related as
$a ={J_2 - \Gamma}/{J_2 + \Gamma}$, $h = {2J_1}/{J_2 + \Gamma}$, and $J_z 
= 0$. The Hamiltonian in (\ref{ham1}) describes 
a spin-1/2 $XYZ$ chain with a magnetic field applied along one of the three
directions; a non-zero value of $J_z$ allows for the study of a more 
general model than the one in (\ref{annni2}). We will assume that
% We will assume 
%that the $XY$ anisotropy $a$ and the $zz$ coupling $\De$ satisfy 
$-1 \le a, J_z \le 1$, $|J_z| < 1 \pm a$, and $h \ge 0$; it then turns out 
that the critical behavior of the Hamiltonian in (\ref{ham1}) does not 
depend on $J_z$ because, as discussed below, the Luttinger parameter $K$ 
(which is a function of $J_z$) changes very little under the RG flows.
%We also assume without loss of generality
%that the magnitude of the $zz$ coupling is smaller than the $yy$ coupling
%(i.e., $|\De| < 1 - a$), and that the magnetic field strength $h \ge 0$.
%We note that the Hamiltonian in Eq.~(\ref{ham1}) is invariant under the global
%$Z_2$ transformation $\si_n^x \to \si_n^x, \si_n^y \to - \si_n^y, \si_n^z 
%\to - \si_n^z$.

%We will use bosonization and an RG analysis 
%\ct{amit84,cardy96} to study
%the phase diagram of the model defined in Eq.~(\ref{ham1}).
The model in (\ref{ham1}) can be bosonized as described in Appendix 
\ref{jorwigbos}. For $a=h=0$, the bosonic theory is governed by a parameter 
$K$ given by (\ref{kpara}), namely, $1/K = 1 + (2/\pi) \arcsin (J_z)$.
%The bosonization
%formulae given in Sec.~\ref{jorwigbos} show that in the continuum limit
%(i.e., on length scales much larger than the lattice spacing), the spin 
%component $\si^x_n$ and the $XY$ anisotropy are proportional to the bosonic
%operators ${\cal O}_1$ and ${\cal O}_2$ respectively, where
%\bea {\cal O}_1 &=& \cos (2 {\sqrt \pi} \phi) ~\cos ({\sqrt \pi} \theta), 
%\non \\
%{\cal O}_2 &=& \cos (2 {\sqrt \pi} \theta). \eea
%It is convenient to introduce a third operator 
%\beq {\cal O}_3 ~=~ \cos (4 {\sqrt \pi} \phi). \eeq
%The three operators ${\cal O}_1$, ${\cal O}_2$ and ${\cal O}_3$ appear in 
%Eq.~(\ref{ham1}) with the coefficients $h$, $a$ and $b$ respectively, where
%$b=0$ at the scale of the lattice spacing; however we will see below that $b$ 
%is generated by the RG flows as we go to larger length scales. 
Up to second order, one finds the RG
%that these coefficients and the parameter $K$ satisfy the following RG
equations \ct{dutta03,giamarchi88,nersesyan93,yakovenko92},
\bea \frac{dh}{dl} &=& (2 - K - \frac{1}{4K}) h ~-~ \frac{1}{K} ah ~-~ 4Kbh, 
\non \\
\frac{da}{dl} &=& (2 -\frac{1}{K}) a ~-~ (2K -\frac{1}{2K}) h^2, \non \\
\frac{db}{dl} &=& (2 - 4K) b ~+~ (2K -\frac{1}{2K}) h^2, \non \\
\frac{dK}{dl} &=& \frac{a^2}{4} ~-~ K^2 b^2, \label{rg} \eea
where $l$ denotes the logarithm of the length scale, and $b$ is 
a coupling generated in the process of renormalization. 
%It is interesting to observe that Eqs.~(\ref{rg}) are invariant under the 
%duality transformation $K \leftrightarrow 1/(4K)$ and $a \leftrightarrow b$.
Noting that $K$ renormalizes very little, one
% regime of RG flows 
%that we will be interested in. For any value of $K=K^*$, it turns out,
finds remarkably that Eqs.~(\ref{rg}) have a non-trivial fixed point (FP) 
given by
\bea h^* &=& \frac{\sqrt {2K^*(2-K^*-1/(4K^*))}}{2K^*+1}, \non \\
a^* &=& (K^*+\frac{1}{2}) ~h^{*2}, \quad {\rm and} \quad b^* ~=~ 
\frac{a^*}{2K^*}. \label{fp} \eea
%One might object that Eqs.~(\ref{rg}) can only be trusted if $a$, $b$ and $h$ 
%are not too large, otherwise one should go to higher orders. We note, however,
%that the FP values $(a^*,b^*,h^*)$ approach zero as $K^* \to 1 + 
%{\sqrt 3}/2 \simeq 1.866$. Thus the RG equations can certainly be trusted 
%for $K^*$ close to $1.866$. 
For $K^*=1$ (corresponding to $J_z = 0$), the 
FP is at $(a^*,b^*,h^*) = (1/4,1/8,1/{\sqrt 6})$.
Fig.~\ref{annni_fig1} shows the numerically obtained RG flows in 
(\ref{rg}) projected on to the $a-h$ plane with $b=0$, 
$K=1$ and $a$, $h$ chosen to be very small initially; the flow of $K$ has
been ignored \ct{dutta03}. 
%where we have shown, and study which set 
%of values flows to a non-trivial FP. We find that there is a line of points 
%which flow to a FP at $(K^*, a^*, b^*, h^*) = (1.020, 0.246, 0.122, 0.404)$.
%This line projected on to the $(a,h)$ plane is shown in Fig.~\ref{annni_fig1}.
%We see that $K$ changes very little during this flow; it is therefore not a 
%bad approximation to ignore the flow of $K$ completely.

%\begin{figure}[htb]
%\begin{center} \hspace*{-0.5cm}
%\epsfig{figure=annni_fig1.ps,width=6cm,height=7cm} \end{center}
%\vspace*{-0.2cm}
%\caption{RG flow diagram in the $(a,h)$ plane. The solid line shows the set of
%points which flow to the FP at $a^*=0.246$, $h^*=0.404$ marked by an asterisk.
%The dotted lines show the RG flows in the gapped phases A and B (see text).}
%\label{annni_fig1} \end{figure}

\begin{figure}
\begin{center}
\includegraphics[height=2.1in]{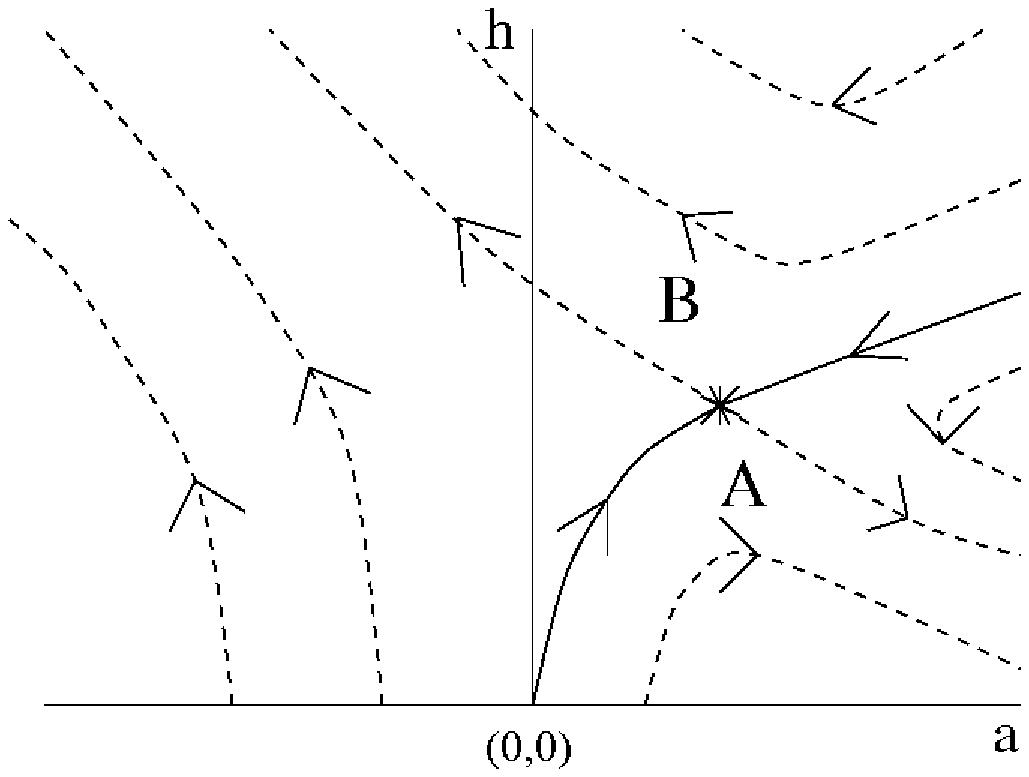}
\caption{RG flow diagram in the $(a,h)$ plane of a $XYZ$ model where $a$ is the 
anisotropy and $h$ is the longitudinal field. The solid line shows the set of
points which flow to the FP at $a^*=0.246$, $h^*=0.404$ marked by an asterisk.
The dotted lines show the RG flows in the gapped phases A and B (see text). 
The dual version of the ANNNI model shows a similar RG flow in the 
bosonization picture. (After Dutta and Sen, 2003).} \end{center}
\label{annni_fig1}\end{figure}

%We now examine the stability of small perturbations away from the non-trivial
%FP. (We will not discuss the stability of the trivial FP lying at $a=b=h=0$).
The non-trivial FP has two stable directions, one unstable direction and one
marginal direction; the last one corresponds to changing $K^*$ and 
simultaneously $a^*$, $b^*$
and $h^*$ to maintain the relations in Eqs.~(\ref{fp}). The presence of two
stable directions implies that there is a two-dimensional surface of
points, in the space of parameters $(a,b,h)$, which flows to this FP;
the system is gapless on that surface. A perturbation in the unstable
direction produces a gap in the spectrum.
% For instance, at the FP with $(K^*,a^*,b^*,h^*)=(1,1/4,1/8,1/{\sqrt 6})$,
%the four RG eigenvalues are given by 1.273 (unstable), 0 (marginal), and
%$-1.152 \pm 1.067 i$ (both stable). The positive eigenvalue corresponds to
%an unstable direction given by $(\de K, \de a, \de b, \de h)=
%\de a (0.113, 1, -0.092, -0.239)$. 
A small perturbation of size $\de a$ at the FP with $(K^*,a^*,b^*,h^*)=(1,
1/4,1/8,1/{\sqrt 6})$ in that direction will produce a gap in the spectrum 
which scales as $\De E \sim |\de a|^{1/1.273} = |\de a|^{0.786}$; the 
correlation length is then given by $\xi \sim 1/\De E \sim |\de 
a|^{-0.786}$. In the gapped regions A ($\de a > 0$) and B ($\de a <0$), 
the RG flows are towards $a=\infty$ and $a,h \to -\infty$, respectively.

%Figure \ref{annni_fig1} shows that the set of points which do not flow to the 
%non-trivial FP belong to either region A or region B. These regions can be
%reached from the non-trivial FP by moving in the unstable direction, with
%$\de a > 0$ for region A, and $\de a < 0$ for region B. In region A,
%the points flow to $a=\infty$; this corresponds to a gapped phase in which
%the $xx$ coupling is larger than the $yy$ and $zz$ couplings. In region
%B, both $a$ and $h$ flow to $-\infty$; this is a gapped phase in which the
%$yy$ coupling is larger than the $xx$ and $zz$ couplings. We will now see that
%the difference between these two phases lies in the way in which the $Z_2$
%symmetry of the Hamiltonian is realized.
To distinguish between the phases A and B we introduce an order 
parameter, namely, the staggered magnetization in the $\hat y$ direction, 
defined in terms of a ground state expectation value as $m_y ~=~ [~ 
\lim_{i \to \infty} ~(-1)^i \langle S_0^y S_i^y \rangle ~]^{1/2}$.
This is zero in phase A; hence the $Z_2$ symmetry is unbroken. In phase B,
$m_y$ is non-zero, and the $Z_2$ symmetry is broken. The scaling of $m_y$ 
with the perturbation $\de a$ can be found as follows 
\ct{dmitriev02a,dmitriev02b}. At $a=h=0$, the leading term in the 
long-distance equal-time correlation function of $\si^y$ is given by
$ \langle \si_0^y ~\si_i^y \rangle ~\sim ~ {(-1)^n}/{|i|^{1/2K}}$ giving
the scaling dimension of $\si_i^y$ to be $1/(4K)$. In a gapped phase in 
which the correlation length is much larger than the lattice spacing, $m_y$ 
will therefore scale with the gap as $m_y \sim (\De E)^{1/(4K)}$. If we 
assume that the scaling dimension of $\si_i^y$ at the non-trivial FP remains 
close to $1/(4K)$, then the numerical result quoted above for $K=1$ implies 
that $m_y \sim |\de a|^{0.196}$ for $\de a$ small and negative.

% While it is not clear how to reconcile all these different results, 
%we would like to make some comments which may be relevant \ct{dutta03}.
%Many of the earlier studies which found a floating phase with a finite width 
%were carried out at values of $J_2 /J_1$ which are close to 1 (here we are 
%using the notation of Eq.~(\ref{annni1}), while our RG results are expected 
The bosonization study is valid only if $a,h$ are small, i.e., if $J_2 /J_1$ is
large, and it predicts a vanishingly small width of the FP in the phase diagram
in (\ref{fig_annni_phase}). At $J_2/J_1 \simeq 1$, the situation is quite 
different for the following reason. Exactly at $\kappa =1$ and $\Gamma=0$, 
the Hamiltonian in (\ref{annni2}) is 
$H_{MC} ~=~ (J_2/4) ~\sum_i ~(\si_i^x ~-~ 1)~(\si_{i+1}^x ~-~1).$
This is a multicritical point with a ground state degeneracy growing
exponentially with the system size, since any state in which every pair
of neighboring sites $(i,i+1)$ has at least one site with $\si^x =1$ is a
ground state. 
%Slightly away from this point To lowest order, this involves doing 
%perturbation theory. 
For $J_2 - J_1$ and $\Gamma$ non-zero but small, one can argue, using lowest 
order perturbation theory within the large space of degenerate states
\ct{villain81,nijs88,selke92,yeomans87,sen89}, that the low-energy 
properties of the model in (\ref{annni2}) do not change if $\Gamma \si_i^y 
\si_{i+1}^y$ is replaced by $(\Gamma/2) (\si_i^y \si_{i+1}^y + \si_i^z 
\si_{i+1}^z)$.\footnote{This is because the difference between the two kinds 
of terms is given by operators which, acting on one of the degenerate ground 
states, take it out of the degenerate space to an excited
state in which a pair of neighboring sites have $\si^x = -1$.}
Thus the fully anisotropic model in (\ref{annni2}) becomes equivalent to a 
different model which is invariant under the $U(1)$ symmetry of rotations in 
the $y-z$ plane. The $U(1)$ symmetric model has been studied earlier using 
bosonization which shows a finite width of the gapless region separating 
two gapped phases \ct{cabra98,giamarchi88}; in the bosonization study by 
Dutta and Sen \ct{dutta03}, the symmetry away from the MCP is $Z_2$ which 
reduces this region to a line.

%it has a gapless phase of finite width which lies between two gapped phases 
%\ct{allen01}. Thus the difference between the study in \oct{dutta03}, 
%(in which $J_2 -\Gamma$ and $J_1$ are small) and the earlier studies (in which
%$J_2 - J_1$ and $\Gamma$ are small) is that they have different symmetries
%away from the transition line, namely, $Z_2$ and $U(1)$ respectively. These
%sets of studies are therefore complimentary to each other; a combination of 
%the two should lead to a complete understanding of the model over the entire 
%parameter range.

To summarize, the transition from phase A to phase B can occur either through 
a gapless line (if $a,h$ are small), or through a gapless phase of finite 
width (if $a,h$ are large). The complete phase diagram of the ground state of 
the dual model in (\ref{annni2}) is shown in Fig.~\ref{fig_annni_phase}
incorporating all the possibilities discussed above 
where different phases are shown in parenthesis.
%\ct{villain81,nijs88,selke92,yeomans87,chakrabarti96}.
The three major phases are distinguished by the following properties of 
the expectation values of the different spin components. In the 
antiferromagnetic phase, the spins point alternately along the ${\hat x}$ 
and $-{\hat x}$ directions. In the spin-flop phase, they point alternately 
along the ${\hat y}$ and $-{\hat y}$ directions, with a uniform tilt towards 
the ${\hat x}$ direction. In the FM phase, all the spins point 
predominantly in the ${\hat x}$ direction. The antiferromagnetic and spin-flop
phases are separated by a floating phase of finite width for $\kappa$ close 
to $1$, and by a spin-flop transition line for large values of $\kappa$. It 
has been conjectured that the floating phase and the spin-flop transition 
line are separated by a Lifshitz point as indicated in 
Fig.~\ref{fig_annni_phase}. 
We note that in terms of the original quantum Hamiltonian in (\ref{annni1}),
the antiphase is in fact the antiferromagnetic phase while the PM phase is 
the spin-flop phase in the dual model in (\ref{annni2}).

Taking the classical limit ($S \to \infty$) \ct{sen90,sen91} of (\ref{ham1})
%one can show 
%\bea H_{S1} ~=~ \sum_n ~[~ & & (1+a) ~S_n^x S_{n+1}^x ~+~ (1-a) ~S_n^y 
%S_{n+1}^y \non \\
%& & +~ \De ~S_n^z S_{n+1}^z ~-~ 2 S h ~S_n^x ~], \label{ham2} \eea
%where the spins satisfy ${\bf S}_n^2 = S(S+1)$.
% and we are interested in the
%classical limit $S \to \infty$ \ct{sen90,sen91}. (We have multiplied the 
%magnetic field by a factor of $2S$ in Eq.~(\ref{ham2}) so that we recover 
%Eq.~(\ref{ham1}) for spin-1/2). and the $zz$ coupling is assumed to be 
%smaller in magnitude than the $yy$ coupling. The classical ground state 
%of Eq.~(\ref{ham2}) is given by a configuration in which all the spins lie in 
%the $x-y$ plane, with the spins on odd and even numbered sites pointing 
%respectively at an angle of $\al_1$ and $-\al_2$ with respect to the $x$-axis.
% The ground state energy per site is given by
%\bea e (\al_1, \al_2 ) ~=~ S^2 [ & & - ~h ~(\cos \al_1 + \cos \al_2 ) + 
%\cos (\al_1 + \al_2 ) \non \\
%& & + ~a ~\cos (\al_1 - \al_2 ) ]. \label{solns} \eea
and minimizing the ground state energy, 
%with respect to $\al_1$ and $\al_2$, one 
%discovers that there is a special line given by $h^2 ~=~ 4 a$ on which all 
%solutions of the equation
%$ h ~\cos (\frac{\al_1 - \al_2}{2} ) ~=~ 2 ~\cos (\frac{\al_1 + \al_2}{2})$
%give the same ground state energy per site, namely, $e_0 =-(1+a)S^2$ and the 
%solutions of Eq.~(\ref{gnst}) range from $\al_1 = \al_2 = \cos^{-1} (h/2)$ to
%$\al_1 = \pi, \al_2 = 0$ (or vice versa); in the ground state phase
%diagram of the ANNNI model, these two configurations correspond respectively
one finds a phase transition line $h^2 = 4a$. This denotes a spin-flop 
transition from a phase with an antiferromagnetic alignment of the spins with 
respect to the $y$-axis (with a small tilt towards the $x$-axis if $h$ is 
small) to an antiferromagnetic alignment of the spins with respect to the 
$x$-axis.
%The curve $h^2 = 4a$ is therefore a phase transition line, and the form of
%the ground states on the two sides shows that there is a spin-flop transition
%across that line. For $h^2 =4a$, there is a one-parameter
%set of classical ground states (characterized by, say, the value of
%$\al_1$ which can go all the way from 0 to $2\pi$ in the solutions of
%$h ~\cos (\frac{\al_1 - \al_2}{2} ) ~=~ 2 ~\cos (\frac{\al_1 + \al_2}{2})$
% which are all degenerate. 
Moreover the symmetry is enhanced
from a $Z_2$ symmetry away from the line to a $U(1)$ symmetry (of rotations
in the $x-y$ plane) on the line. One therefore expects a gapless mode in the
excitation spectrum corresponding to the Goldstone mode of the broken
continuous symmetry which can be found explicitly by going to the next order 
in a $1/S$ expansion \ct{sen91}. It may be mentioned here that spin-flop 
transitions in one-dimensional spin-1/2 systems have been studied earlier
\ct{karadamoglou99,wang00,sakai99}.

The corresponding numerical studies need to be mentioned here. A study based 
on the cluster heat bath algorithms \ct{sato99} points to a finite width of 
the gapless phase in the finite temperature phase diagram of the 
two-dimensional classical ANNNI model. In contrast, a subsequent study based on
the non-equilibrium relaxation method \ct{shirahata01} showed that its width 
is infinitesimally small and the correlation length diverges in a power-law 
%(exponential) 
fashion as this line is approached from the antiphase side. The 
one-dimensional quantum ANNNI model has been studied using the density matrix 
renormalization group technique \ct{white93}; while no evidence of
the floating phase is found for $\kappa <1$ \ct{beccaria06}, the floating 
phase with a finite width seems to exist for $\kappa >1$ and extends for very
large $\kappa \simeq 5$ \ct{beccaria07}. A very recent study \ct{nagy11} which
uses finite entanglement scaling of the matrix product state in the
one-dimensional quantum ANNNI system seems to corroborate the findings of 
Beccaria $et~al.$ \ct{beccaria07}. The
discussion above shows that even after thirty years of exploration, the phase 
diagram of a quantum ANNNI chain is not yet fully understood.

The model in (\ref{ham1}) with $a=0$ has been studied theoretically 
in \ct{dmitriev02a,dmitriev02b,caux03,hagemans05}. A number 
of phases have been found as a function of $J_z$ and $h$, including Neel 
and FM order along the $z$ axis (depending on the magnitude and 
sign of $J_z$), Neel order along the $y$ axis (the spin-flop phase), and 
a phase with no long-range order. Experimentally, a system with $a=0$ and 
magnetic fields along both $\hat x$ and $\hat z$ directions has been studied 
in \ct{kenzelmann02}, where the two Neel phases and the spin-flop 
phase have been found.

An $n$-component quantum rotor model (see (\ref{eq_hamrotor})) with 
ANNNI-like regular frustration in some of the spatial directions given by 
the Hamiltonian
\beq H_R ~=~ \frac{g}{2} ~\sum_i \hat L_i^2 ~-~ J_1 ~\sum_{<ij>} {\hat x}_i 
\cdot {\hat x}_j ~+~ J_2 ~\sum_{<<ij>>} {\hat x}_i \cdot {\hat x}_j, 
\label{eq_annnirotor}\eeq
where $<<...>>$ denotes next-nearest-neighbor interaction,
has been studied in recent years \ct{dutta97,dutta98a}. In this model, in 
addition to the nearest-neighbor FM interactions $J_1$, there 
exists a next-nearest-neighbor antiferromagnetic interaction $J_2$ along $m$ 
of the $d$ spatial directions. The zero-temperature phase diagram of the 
model shows the existence of a $(d,m)$ quantum Lifshitz point (QLP) 
\ct{dutta97,ramazashvilli99,nishiyama07} where the FM, PM
and helically ordered phases meet; this is a quantum generalization of a 
classical Lifshitz point \ct{hornreich75,hornreich75a,hornreich77}. The 
QPT at a QLP is an example of an anisotropic QCP with 
correlation time $\xi_{\tau} \sim \xi_{||}^{z_{||}} \sim 
\xi_{\perp}^{z_{\perp}}$ where $\xi_{||}$ ($\xi_{\perp})$ is the spatial 
correlation length in $m$ ($d-m$) directions with (without) frustration.
Using the Gaussian propagator, it can be shown \ct{dutta97} that $z_{||}=2 
z_{\perp}=2$, $2 \nu_{||}= \nu_{\perp}= 1/2$ for an anisotropic QLP with 
$m<d$, with the upper critical dimension being $d_u^c=3+m/2$. We note that 
the scaling valid close to a QLP has been
observed experimentally. The magnetic 
fluctuations in ${\rm CeCu_{6-x}Au_x}$, which has a QCP at $x=0.1$ separating 
a heavy fermion paramagnet from an antiferromagnet, has been studied using 
inelastic neutron scattering \ct{schroeder98}. The form of the susceptibility 
so obtained has been explained in terms of a QLP and the value of $2/z_{||}$ 
is found to be $0.8$ in contrast to the mean field prediction $2/z_{||}=1$.

\subsection{Models with long-range antiferromagnetic interactions}

In this section, we discuss the properties of the long-range transverse Ising 
antiferromagnet (LRTIAF) with a disorder in the cooperative interactions
superposed on it 
\ct{chakrabarti06a,chakrabarti06b,chandra09,chandra10,ganguli09}.
The general model one studies is given by the Hamiltonian
\bea H ~=~ - ~\frac{1}{N} ~\sum_{ij (j > i)} J_{ij} \si_i^x \si_j^x ~-~ 
h ~\sum_i\si_i^z, \label{eq:Hamiltonian} \eea
{where one assumes that the pairwise interactions 
$J_{ij}$'s are chosen from a Gaussian distribution, $P(J_{ij}) = \exp
[-{(J_{ij}-J_0)^2} /{2\tilde{J}^2}]/{\sqrt{2\pi}\tilde{J}}$, and $J_0 (<0)$ is 
the parameter controlling the strength of the antiferromagnetic bias.}
We thus recover the `pure' antiferromagnetic Ising model with infinite
range interactions when we consider the disorder-free limit $\tilde{J} \to 0$.
The model with $J_0 > 0$ and $h =0$ is identical to the 
classical SK model \ct{sherrington75}.
%and with $J_0 < 0$.
% and $h =0$ it is the classical (LRIAF) model.

% To check how 
%this `liquid'-like antiferromagnetic phase of the pure LRTIAF gets `frozen' 
%into a spin glass phase when a little disorder is added, we study the general 
%LRTIAF Hamiltonian with a coupling with the Sherrington-Kirkpatrick (SK) 
%spin glass Hamiltonian \ct{sherrington75,binder86}; we will examine the phase 
%transition behavior
%induced by both finite temperature and a transverse field. Indeed, a stable 
%SK-like spin glass phase is observed for both thermal and quantum fluctuations
%below some finite critical values of the temperature and transverse field.

In the mean field limit, one can derive an effective single site Hamiltonian 
\ct{binder86} 
\beq H ~=~ \vec{h}_{\rm eff} \cdot \sum_{i=1}^N \vec{\si}_i ~~~{\rm where}~~~
\vec{\si}_i = \si^z_i \hat{z} + \si^x_i \hat{x}, \eeq
and $\vec{h}_{\rm eff} = \left( {J_0} m + \tilde{J}\sqrt{q}y \right) \hat{z} + 
h \hat{x}$; we can then calculate the magnetization $m \equiv N^{-1} 
\sum_i\langle 
\vec \si_i^x \rangle$ and spin-glass order parameter $q \equiv N^{-1}\sum_i
\langle \si_i^x \rangle^2$ in a self-consistent manner. In the disorder-free 
case with $h=0$, one observes indications of a very unstable quantum
antiferromagnetic (AF) phase ($50\%$ spins up and $50\%$ spins down without 
any sub-lattice structure) which gets destabilized by both infinitesimal 
thermal (classical) as well as quantum fluctuations induced by $h$. 
In the presence of spin glass-like disorder the AF phase is destabilized 
and a spin glass order sets in; this eventually gets destroyed as the thermal 
or quantum fluctuations increase beyond their threshold values and a 
transition to a para phase occurs. This mean field calculations are
supported by numerical simulations although the replica 
symmetric solution indicates some corrections.

\begin{figure}
\begin{center}
\includegraphics[width=8.5cm,height=6.5cm]{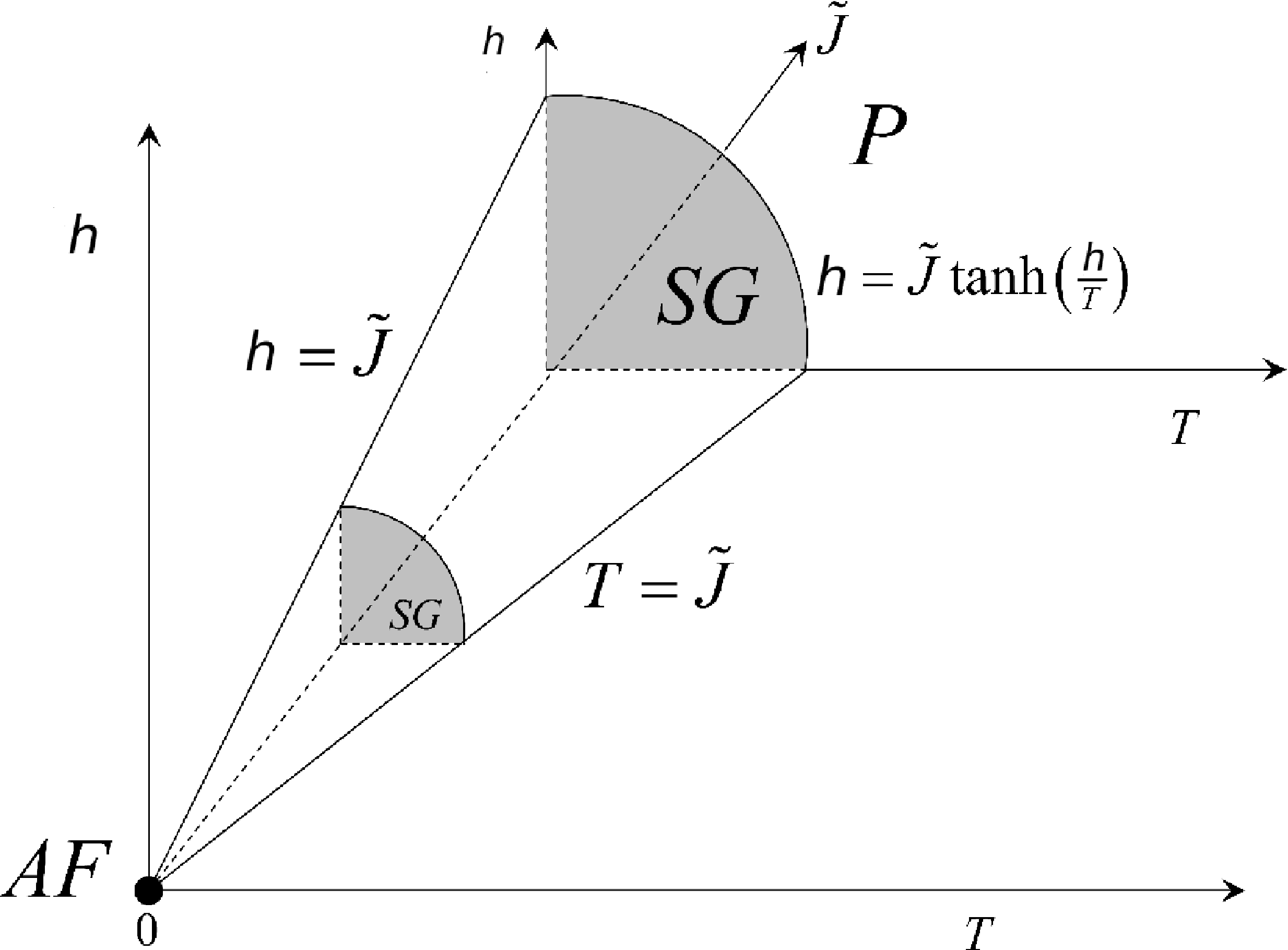} \end{center}
\caption{\footnotesize\label{fig:fg2}
Schematic phase diagram for the quantum system given in 
(\ref{eq:Hamiltonian}). The antiferromagnetic order 
(AF) exists if and only if we set $h=T=\tilde{J} =0$. As $\tilde{J}$
decreases, the spin glass (SG) phase gradually shrinks to zero and eventually 
ends up at an antiferromagnetic phase at its vertex (for $h=T=\tilde{J}=0$). 
The phase boundary between the spin-glass and paramagnetic
(P) phase is $h = \tilde J \tanh (h/T)$ as obtained from mean field theory. 
(After Chandra $et~al.$, 2010).} \end{figure}

%\begin{figure}[ht] \begin{center} \includegraphics[width=8cm]{rmpad_fig11.eps}
%\caption{\footnotesize\label{fig:fg2}
%Schematic phase diagram for the quantum system. The antiferromagnetic order 
%exists if and only if we set $T=\Gamma=0$ and $\tilde{J} = 0$. As $\tilde{J}$
%decreases, the spin glass phase gradually shrinks to zero and eventually ends 
%up at an antiferromagnetic phase at its vertex (for $h=0=T=\tilde{J}$)
%\end{document}
%\end{figure}

\noi The resulting mean field phase diagrams for the quantum model is 
presented in Fig.~\ref{fig:fg2};
% over the above mean field results. Numerical results 
%however support \ct{chandra10} the phase diagrams indicated in Fig. 
%\ref{fig:fg1} (at zero temperature) and Fig.~\ref{fig:fg2} (for finite 
%temperature or transverse field).
%In summary, Chandra $et~al.$, 2010, obtained the finite temperature 
%free energy expression for this model and studied analytically the
%magnetization, spin glass order and the correlation (using Trotter replicas).
%In summary, for the pure case (i.e., no disorder), the antiferromagnetic 
%order is seen to immediately break down as soon as thermal or quantum 
%fluctuations are added. However, when one adds disorder 
%as in the SK Hamiltonian to the LRTIAF as a perturbation, 
we find that an infinitesimal disorder is enough to induce a stable glass 
order which eventually gets destroyed due to thermal or quantum fluctuations
%as the thermal or quantum fluctuations
%increase beyond their threshold values and a 
leading to a transition to the para phase. As shown in Fig.~\ref{fig:fg2}, the 
antiferromagnetic phase of the LRTIAF (occurring only at $h = T = \tilde{J} 
= 0$) can get `frozen' into a spin glass phase if a little SK-type 
disorder is added ($\tilde{J} \neq 0$). 
%this is the only missing element 
%which is required to induce stable order (freezing of random spin 
%orientations) in a LRTIAF (which is fully frustrated, but lacks disorder).
%These results have been confirmed by Monte Carlo simulations. 
It is noteworthy that the degeneracy factor of $e^{0.693N}$ for the ground 
state of the LRTIAF with $h=0$ is much larger than the factor of $e^{0.199N}$ 
for the classical SK model. Hence, because 
of the presence of full frustration, the LRTIAF possesses a surrogate 
incubation property of a stable spin glass phase which can be induced by 
the addition of a small amount of disorder. However, the effect of replica 
symmetry breaking on these phases is yet to be explored.

%\begin{figure}[htb]
%\includegraphics[height=2.5in,width=2.5in, angle = 0]
%{rmp_longrangeantiferro.eps}
%\caption{Phase diagram for the long-range transverse Ising antiferromagnet
%with disorder ($\tilde J$). The antiferromagnetic order (half of the spins 
%up and half of the spins down) exists if and only if we set 
%$T = \Gamma = 0$ and $\tilde J = 0$. As the $\tilde J$ decreases, the spin 
%glass phase gradually shrinks to zero and eventually ends up at an 
%antiferromagnetic phase at its vertex (for $\Gamma = 0 = T = \tilde J$).}
%\end{figure}

%\section{Quantum hysteresis in transverse field models (+{\it Aeppli 
% $\sim$ 2 pp})}

%\ct{miyashita09,das09}

%\begin{figure}[htb]
%\begin{center}
%\includegraphics[height=2.8in,width=3.3in, angle =0]{rmp_hysteresis1.eps}
%\caption{The phase diagram of transverse $XY$ model. The vertical bold lines 
%denote Ising transitions, whereas the horizontal bold line stands for 
%anisotropic phase transition. The multicritical points are denoted by $A$ 
%and $B$ in the figure.} \end{center}
%\end{figure}

\section{Theoretical studies of quantum phase transition and information}

In this section, our main focus is on the connection between QPTs in 
transverse spin models and quantum information theory 
\ct{werner89,bennett96,horodecki01,nielsen09,wootters01}. Quantum 
information theoretic measures like the concurrence, fidelity, entanglement 
entropy, etc., are able to capture the
ground state singularities associated with a QPT. 
More importantly, as we shall discuss below, these measures show 
interesting scaling behavior close to the QCP which is given
in term of some of the quantum critical exponents. We shall also show how 
the integrability of the transverse Ising/$XY$ chains and the qubit form of 
the ground state as given in (\ref{eq_direct_product}) turn out to be 
extremely useful for studying quantum information theoretic measures
\ct{amico08,ghosh03,bose10}.
%the hysteresis behavior in quantum spin models and 

%\subsection{Experimental studies of entanglement: (+{\it Aeppli $\sim$1/2 p})}

\subsection{Entanglement}

We start our discussion with a brief review on the behavior of concurrence 
close to the QCP of transverse spin models.

\subsubsection{Concurrence}

The knowledge of pairwise entanglement in a 
quantum many-body system can provide a 
%deeper 
characterization of the ground 
state wave function close to a QCP. In fact, non-analyticities characterizing 
a QPT are in general directly related to the bipartite entanglement measures 
\ct{wu04}.
%One of the ways of quantifying the 
%pairwise entanglement in a mixed state is through the entanglement of 
%formation $E_F$ which, for a state described by a density matrix $\rho$
%\ct{bennett96}, is defined as $E_F(\rho) = {\rm min} \sum_j p_j S
%(\rho_{A,j})$; here the minimum is taken over all realizations of the state 
%$\rho_{AB} = \sum_j p_j |\psi_j\rangle \langle\psi_j|$, and $S(\rho_{A,j})$ is 
%the von Neumann entropy of the reduced density matrix $\rho_{A,j}= {\rm tr}_B 
%|\psi_j\rangle \langle\psi_j|$. For a two-qubit system one gets the analytical
%expression 
%\beq E_F(\rho) = - \sum_{\si=\pm 1} \frac{\sqrt{1+\si C}^2}}{2} ~ \ln ~
%\frac{\sqrt{1+\si C^2}}{2}, \eeq
%where $C^2$ is the concurrence 
%which is one of the measures of the entanglement.
The concurrence, which is one of the measures of pairwise entanglement
between two spins separated by $n$ lattice spacings in a spin-1/2 chain, 
is defined as \ct{hill97,wootters98,wootters01}
\beq { C}^n ~=~ {\rm max} \{\sqrt{\la^n_1} - \sqrt{\la^n_2} - 
\sqrt{\la^n_3} - \sqrt{\la^n_4}, 0\}, \label{conc1} \eeq
where the $\la^n_i$'s are the eigenvalues of the 
positive Hermitian matrix $\sqrt {\rho^n} \tilde {\rho^n} \sqrt {\rho^n}=
\sqrt {\rho^n}(\si^y \bigotimes \si^y )\rho^{n*} (\si^y \bigotimes \si^y )
\sqrt{\rho^n}$ in decreasing order \ct{hill97,wootters98}, and ${\rho^n}$ is 
the two-spin density matrix for two sites $i$ and $j= i+n$, 
{which is obtained by integrating the total density matrix
over all the other spins}; the eigenvalues 
$\lambda_i$ are invariant under unitary transformations of the two qubits 
which implies that they are independent of the basis. The case $C^n =1$ 
corresponds to maximum entanglement between the two spins while $C^n=0$ 
denotes the absence of entanglement. For the transverse field spin chains, 
the density matrix $\rho^n$ can be expressed in terms of different 
correlators \ct{syljuasen03}:
\bea \rho^n =\left( \begin{array}{cccc}
a_+^n & 0 & 0 & b_1^n \\
0 & a_0^n & b_2^n & 0 \\
0 & b_2^{n*} & a_0^n & 0 \\
b_1^{n*} & 0 & 0 & a_-^n \end{array}\right), \label{rhon} \eea
where the matrix elements are given in terms of the two-spin 
correlation functions as follows:
\bea a_{\pm} &=& \left \langle \frac{1}{4} (1 \pm \si_i^z)(1 \pm 
\si_{i+n}^z) \right \rangle, \\
a_0^n &=& \left \langle \frac{1}{4} (1 \pm \si_i^z)(1 \mp \si_{i+n}^z) 
\right \rangle, \\
b_{1(2)}^n &=& \langle \si_i^- \si_{i+n}^{-(+)} \rangle. 
\label{matel} \eea
The other correlators such as $\langle \si_i^{\pm}(1 \mp \si_{i+n}^z) 
\rangle$, and hence the other matrix elements of $\rho^n$, vanish if there
is a symmetry under $\si_i^x \to - \si_i^x$, $\si_i^y \to - \si_i^y$ and 
$\si_i^z \to \si_i^z$, as is true for the transverse $XY$ Hamiltonian. Using 
the available analytical results \ct{pfeuty70,lieb61,barouch71}, the 
above correlators and hence the entanglement of the ground state can be 
obtained exactly.

\begin{figure}[htb]
\begin{center}
\includegraphics [height=2.6in,width=2.6in, angle = 0]{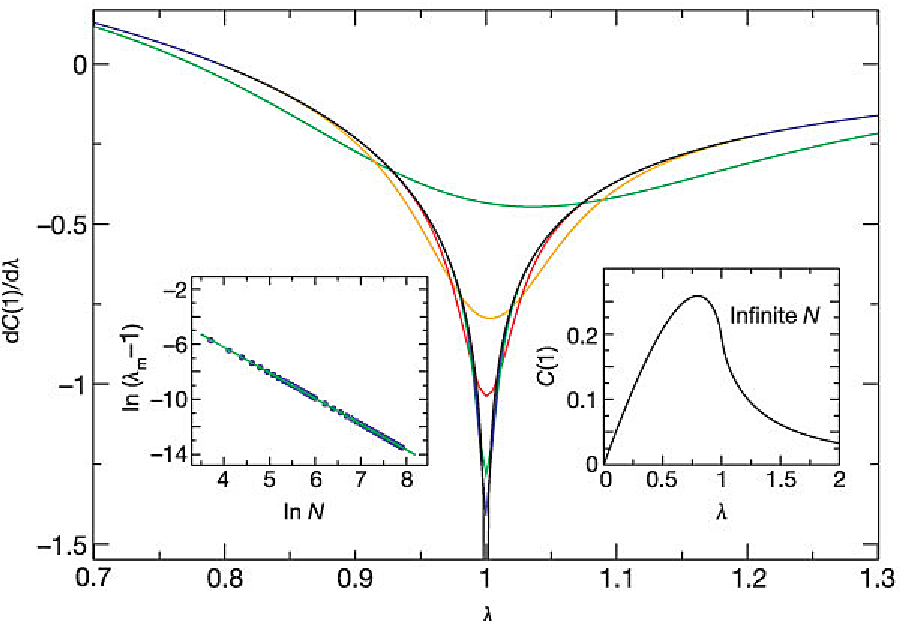}
\caption{(Color online) In the figure $C(1)=C^1$ and $\la= h$. For a 
transverse Ising chain, the minima gets pronounced with increasing system 
size (top to bottom) and also the position of 
the minima approaches the value $\la_c=h_c=1$ (left inset). The
right inset shows the variation of $C(1)$ as a function of $\la$ for a 
thermodynamic system (After Osterloh $et~al.$, 2002).} \label{fig_conc1} 
\end{center}
\end{figure}

The pairwise entanglement can be used to indicate the proximity to a QCP as 
has been shown for the transverse Ising chain in (\ref{eq_tim1d}) 
\ct{osborne02,osterloh02}. The concurrence tends to vanish in the limits when 
the transverse field $h \to 0$ and $h \gg h_c=1$ (choosing $J_x=1$) since the 
ground of the system is then fully polarized along the $\hat x$ and $\hat z$ 
directions, respectively. It has 
also been observed that the concurrence is practically zero unless two sites 
are at most next-nearest neighbors even at the QCP where the correlation 
length diverges in the thermodynamic limit; the next-nearest-neighbor 
concurrence is smaller by one order of magnitude than the nearest-neighbor 
value \ct{osborne02}. This observation signifies that the non-local
quantum part of the two-point correlations is indeed short-ranged. 

The nearest-neighbor concurrence $C^1$ is a smooth function of the field
$h$ showing a maximum close to QCP (see Fig.~\ref{fig_conc1}), while the 
next-nearest-neighbor concurrence is maximum at the QCP. The derivative of 
$C^1$ captures the signature of criticality and diverges as
\beq \frac{\partial C^1}{\partial h} \sim \frac{8}{3 \pi^2} \ln 
|h-h_c|,~~~h_c=1, \label{eq_entang1} \eeq
in the thermodynamic limit ($L \to \infty$) on approaching the QCP.
Eq.~(\ref{eq_entang1}) yields a quantitative measure of the 
non-local correlations in the quantum critical region.
For a finite system, one can use the finite size 
scaling theory \ct{barber83} which predicts that in the vicinity of the 
transition point, the concurrence scales with the combination $L^{1/\nu}
(h -h_m)$, where $h_m(L)$ is the value of $h$ where the derivative 
$\partial C^1/ \partial h$ is minimum for a system of size $L$. For a 
logarithmic divergence, the scaling ansatz is of the form
\beq \partial_h C^1 (L,h) ~-~ \partial_h C^1 (L,h_0) ~\sim ~Q[L^{1/\nu} 
(h -h_m)] ~-~ Q[L^{1/\nu} (h_0 -h_m)], \eeq
where $\partial_h \equiv \partial /\partial h$ and $h_0$ is some non-critical 
value. The scaling function goes as 
$Q(x)\sim Q(\infty) \ln x$ for large $x$. The critical value scales as $h_m (L)
\sim h_c + L^{-1.86}$, while the derivative satisfies the scaling form
\beq \partial_h C^1 (L,h_m) ~=~ -0.2702 ~\ln L ~+~ {\rm constant}. 
\label{eq_entang2} \eeq
The ratio of prefactors of the logarithms in Eqs.~(\ref{eq_entang1}) and 
(\ref{eq_entang2}) yields the critical exponent $\nu(=1)$ of the transverse 
Ising chain; this suggests that although the concurrence describes
short-range properties, it is able to capture the scaling behavior of a QCP.

For a transverse $XY$ chain, the above scaling relation has been verified 
\ct{osterloh02} for the Ising transition with 
$0<\ga <1$ (Fig.~\ref{fig_conc2}). However, the range of entanglement is 
not universal and tends to infinity as $\ga \to 0$. The total entanglement
$\sum_n C^n$ is found to be weakly dependent on $\ga$. Studies
of the entanglement in the transverse $XY$ chain in the presence of a single 
defect \ct{osenda03} and many defects \ct{huang04} show that the finite size 
scaling of the concurrence is drastically modified in the presence of disorder,
and strong disorder eventually kills the critical behavior.

\begin{figure}[htb]
\begin{center}
\includegraphics[height=2.6in,width=2.6in, angle = 0]{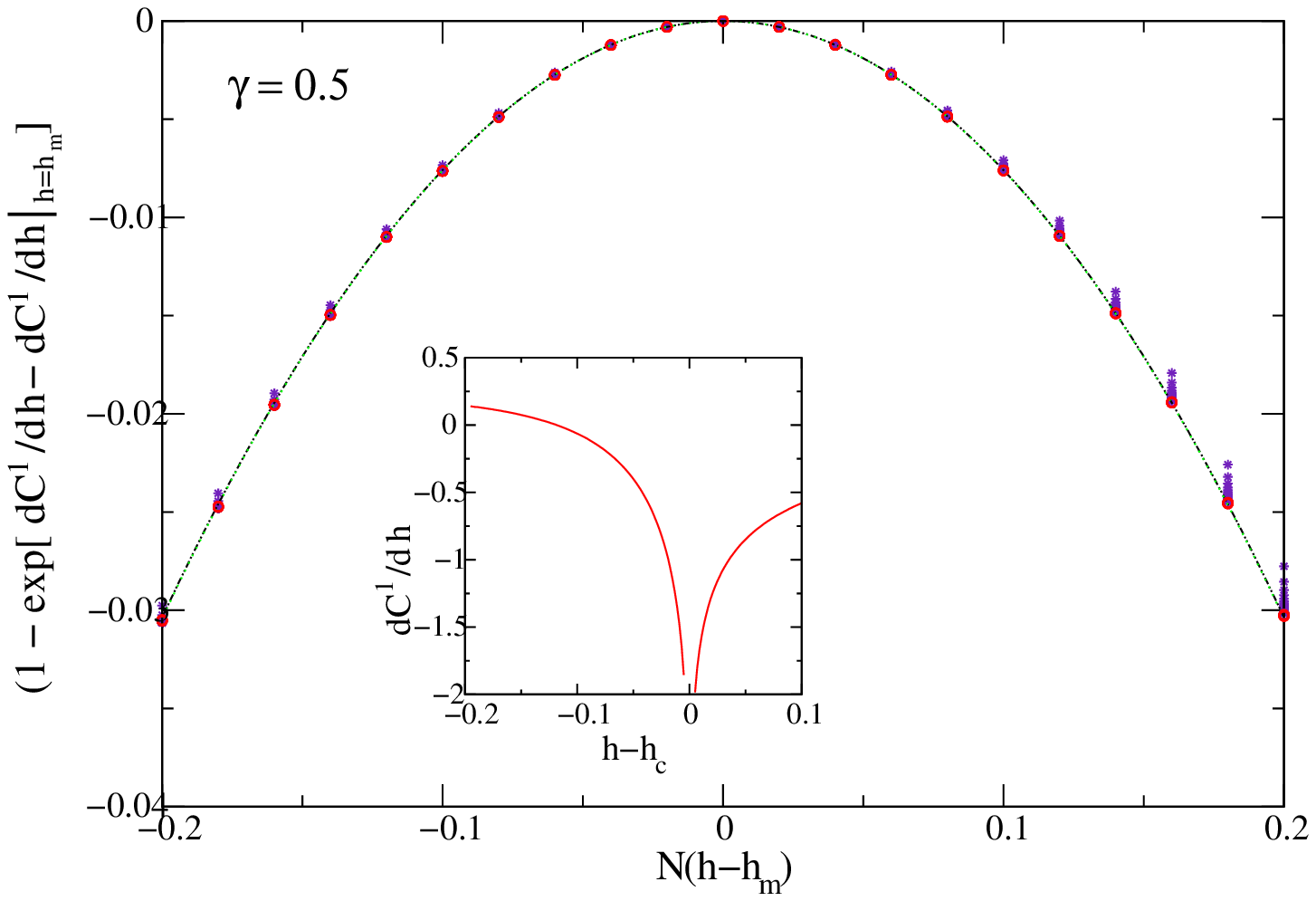}
\caption{(Color online) The universality hypothesis for the entanglement is 
verified in transverse $XY$ chain with $\ga=0.5$ and $N$ ranging from 41 up 
to 401. Data collapse shows that $\nu=1$ for the Ising transition. The inset 
shows the divergence at the critical point for the infinite system. (After 
Osterloh $et~al.$, 2002).} \label{fig_conc2} 
\end{center}
\end{figure}

The transverse Ising and $XY$ chains have been extremely useful for studying 
thermal entanglement \ct{osborne02}, localizable entanglement 
\ct{verstraete04}, macroscopic multi-species entanglement \ct{subrahmanyam11},
and the propagation and dynamics of entanglement 
\ct{amico04,subrahmanyam04,sen05}. We refer the reader to \ct{amico08},
for an extensive discussion. We note that usually a continuous QPT is 
associated with a singularity in 
the derivative of the ground state concurrence while a first order QPT shows 
discontinuities in the ground state concurrence \ct{vidal04c,bose02,alcaraz04}.

In conclusion, we would like to point out that a quantum critical environment 
can also induce entanglement between two qubits which are coupled to 
it. Coupling two spin-1/2's initially in a separable state to a transverse $XY$
chain which is initially in the ground state, one can calculate the reduced 
density matrix and hence the concurrence of the central system as a function 
of time. At a fixed time, one observes a sharp decay of the concurrence on the
anisotropic critical line (see Fig.~\ref{fig_xyphase}) as $\ga=J_x -J_y 
\to 0$ and $- (J_x + J_y) \leq h \leq J_x + J_y$, whereas the concurrence can 
become nearly unity away from the QCP \ct{yi07}.

Another commonly used measure of entanglement between two spin-1/2 objects
is called negativity ${\cal N}(\rho)$ \ct{peres96,vidal02}. This will be 
discussed in Sec.~\ref{conneg}; see (\ref{nega1}). To calculate 
negativity, one takes a partial transpose ($\rho^T$) of a bipartite mixed 
state described by $\rho$; the negativity ${\cal N}(\rho)$ is then given by 
the sum of the absolute values of the negative eigenvalues of $\rho^T$.
If the two objects are entangled ${\cal N} >0$.

\subsubsection{Entanglement entropy}
\label{e_entropy}

Motivated by the works of Fiola $et~al.$ \ct{fiola94}, and Holzhey $et~al.$ 
\ct{holzhey94}, in the context of black hole physics, another 
important measure of entanglement was introduced in which a composite system 
(which is in a pure state) is partitioned into two distinct sub-systems $A$ 
and $B$. The entanglement between $A$ and $B$ is measured by the von Neumann 
entropy associated with the reduced density matrix of one of the subsystems; 
this is called the entanglement entropy \ct{vidal03,audenaert02}; for a 
review, see \ct{latorre09}. The entropy is invariant under local 
unitary transformations, and is continuous and additive in the sense that for 
a direct product state $\ket {\psi} \bigotimes \ket {\phi}$, $S(\ket {\psi} 
\bigotimes \ket {\phi})= S(\ket {\psi}) + S( \ket {\phi})$.
For a quantum spin chain with $N$ spins with the 
ground state $|\psi_0 \rangle$, the reduced density matrix of a block of $l$ 
neighboring spins is $\rho_l ={\rm tr}_{N-l} |\psi_0 \rangle \langle \psi_0|$
(i.e., $N-l$ spins of the environment are integrated out). Then the 
entanglement entropy of the block with the rest of the system is given by 
$S_l = -{\rm tr} (\rho_l\log_2 \rho_l)$ \ct{vidal03}. For a translationally 
invariant ground state, $S_l$ is a concave function of $l$, i.e., $S_l \geq 
(S_{l-m}+ S_{l+m})/2$ where $l=0,...,N,$ and $m=0, ..., {\rm min} \{N-l,l\}$.

According to the area law \ct{srednicki93}, the entropy $S_l$ depends only 
on the surface of separation between the two regions $A$ and $B$. In a 
$d$-dimensional system, this implies that $S_l \sim l^{d-1}$ 
\ct{latorre09,calabrese09}. Therefore, the entanglement entropy for 
systems with one spatial dimension should be
independent of the block size which is indeed the case when the system is 
off-critical (gapped). However, in the gapless (critical) case, $S_l$ diverges
logarithmically with the block size, with the prefactor of the logarithmic term
being universal and related to the central charge $c$ of the underlying 
conformal field theory \ct{holzhey94}, i.e., $S_l \sim (c/3) \log_2 (l/a)$, 
where $a$ is the lattice spacing. Away from the critical point, when the 
system has a finite correlation length $\xi >>a$, the block entropy saturates 
to a value $S_l \sim (c/3) \log_2 (\xi/a)$ for $l \to \infty$. The above 
result was extended at the critical point {of the 
thermodynamic system}; for example, with periodic boundary 
conditions, one gets \ct{holzhey94} $S_l \sim (c/3) \log_2 [(N/\pi a) 
\sin (\pi l/N)]$. At finite temperature, one gets \ct{calabrese04,korepin04} 
$$S_l ~=~ \frac{c}{3} ~\log_2 [\frac{\beta}{\pi a} \sinh (\frac{\pi \beta}{l})]
~+~ A,$$ where $\beta = 1/(k_B T)$ and $A$ is some non-universal constant. Of 
course, a finite temperature block entropy is not unexpected because the 
initial state is necessarily mixed.

The entanglement between two halves of an infinitely long 
chain close to a QCP that is described by a conformal field 
theory has also been studied \ct{calabrese04}; it is given by $S \sim (c/6) 
\log_2 \xi$ and $\sim (c/3) \log_2 \xi$ for open and periodic boundary 
conditions, respectively; this is because in the periodic boundary condition, 
the contribution comes from
two boundaries. This result can be generalized as follows. If an 
infinitely long system consists of a finite subsystem A which has $n$ 
disjoint blocks, each of which has a length much larger than $\xi$, the 
entanglement entropy of A with the rest of the system is given by 
$(c/3) n \log_2 \xi$ \ct{calabrese04}. In the opposite case of a critical
system ($\xi = \infty$) with $n$ disjoint blocks, the form of the entanglement
entropy is much more complicated and has been studied in 
\ct{alba10} and \ct{fagotti10}.

\bigskip
\noi {(i) Pure systems:}
\bigskip

The entanglement entropy of transverse Ising/$XY$ spin chains has been studied
extensively in recent years \ct{latorre04,vidal03,jin04,its05,franchini06}. 
Using the mapping to Majorana fermions, it has been shown that on the 
anisotropic critical line with $\ga=0$ (Fig.~\ref{fig_xyphase}), $S_l = 1/3
\log_2 l +b$, where $b$ is a constant close to $\pi/3$ \ct{vidal03,latorre04}.
This is in agreement with the conformal field theory prediction noting that 
the central charge for the underlying CFT is $c=1$ (see Sec.~\ref{expcorr}). 
%However, for $\ga=0$ and $h > 2J_x$, 
%the entanglement entropy vanishes because in the ground state, the spins are 
%all oriented in the direction of the field so that the wave function has a 
%fully factorized form. 
%The same holds true even at the multicritical point ($\ga=0,h=2J_x$). 
On the Ising critical line, on the other hand, one gets a logarithmic scaling 
with the coefficient $1/6$ which is expected since the central charge $c=1/2$.
Hence, for a one-dimensional TIM ($J_y=0$), one gets $S_l = 1/6 \log_2 l +b$.

We mention here some interesting work on the scaling of the entanglement 
entropy near the critical point of the transverse $XY$ chain 
\ct{tagliacozzo08,pollmann09}. Let us ask how well the matrix product 
approximation for the ground state of a critical system can reproduce the 
entanglement entropy between two halves of a system. If the matrices $A_i$ in 
(\ref{eq_mpstate}) are 
$D$-dimensional, it is clear that their information content can increase
with increasing $D$. We therefore expect that the matrix product form 
may be able to correctly reproduce the properties of a critical 
system (whose correlation length $\xi$ is infinitely large) if we take the 
limit $D \to \infty$. Let us therefore make the ansatz that that there is
a matrix product state with $D$-dimensional matrices which has the 
same properties as the ground state of a system with a large correlation 
length $\xi$ if these scale with each other as $\xi \sim D^\kappa$ for 
$\xi \to \infty$ \ct{tagliacozzo08,pollmann09}. It is then found that this 
ansatz reproduces the entanglement entropy of two halves of an infinitely long
system, $S = (c/6) \log_2 \xi = (c\kappa/6) \log_2 D$, if $\kappa$ is
related to the central charge $c$ of the critical system as
\beq \kappa ~=~ \frac{6}{c ~(1 ~+~ \sqrt{12/c})}. \eeq

A generalized version of the entanglement entropy between a part A of a 
system with the rest of the system is given by the Renyi entropies 
\ct{calabrese10,calabrese08}
\beq S_\al ~=~ \frac{1}{1-\al} ~\log_2 Tr (\rho_A)^\al. \eeq
Note that the limit of $S_\al$ as $\al \to 1$ gives the von Neumann entropy;
in principle, a knowledge of $S_\al$ for all values of $\al$ allows us
to infer the complete spectrum of eigenvalues of $\rho_A$. In a critical
one-dimensional system with central charge $c$, the Renyi entropies
of a block of length $l$ lying in the middle of the system are given by
\beq S_\al ~=~ \frac{c}{6} ~(1 ~+~ \frac{1}{\al}) ~\log_2 l \label{renyi} \eeq
plus non-universal constant terms. For a Tomonaga-Luttinger liquid (see Appendix \ref{jorwigbos}) with Fermi 
momentum $k_F$ and Luttinger parameter $K$, there are subleading corrections 
to (\ref{renyi}) which are of the form $\cos (2k_F l)/ (2l\sin k_F)^{2K/\al}$ 
with non-universal coefficients \ct{calabrese10}. We note that the 
entanglement entropy has also been studied for a two-dimensional quantum 
model belonging to the Lifshitz universality class with $z=2$ \ct{fradkin09}.
The experimental observation of entangled states of magnetic dipoles will be
discussed in Sec. 7 \ct{ghosh03}.

\bigskip
\noi {(ii) Random systems:}
\bigskip

In Sec.~\ref{sec_griff+act}, we have discussed that the QCPs associated with a 
random transverse Ising spin chain is an IRFP. In the presence of 
infinitesimal randomness,
the RG flow is from the pure fixed point to the IRFP. According to
the $c$-theorem \ct{zamolodchikov86} if the RG flow is from the critical 
point $A$ to $B$, then the associated central charges satisfy the relation, 
$c_A \geq c_B$. The possibility of a pure-random $c$-theorem in random Ising 
spin chains has been explored using the entanglement entropy as a measure. In 
\ct{refael04,refael09}, the entanglement entropy of a segment of $l$ 
spins of such a chain given in (\ref{eq_random4}) has been studied at the 
QCP. Exploiting the properties of the IRFP, it has been shown that the 
universal logarithmic scaling still holds though conformal invariance is lost.
%The ground state of these models present clusters forming over an 
%average length $\sim \Lambda^2$ where $\Lambda = ln (\Omega_0/\Omega)$.
% where the $\Omega_0$ ($\Omega$) is initial (reduced) energy scale. 
%Considering the 
%entanglement of subsystem of $r$ spins in a frozen (decimated by field) 
%cluster of $s$ spins, it can be shown that the entanglement of the subsystem 
%with the rest $s-r$ spins is $\log 2$ for each $r$ and $s$. Therefore,
One can argue that for a segment of length $l$ embedded in an infinite chain, 
the frozen clusters which are totally inside or outside the segment $l$ do not 
affect the total entanglement. 
The only contribution comes from the number of decimated clusters which cross 
the two edges. Each of such clusters contribute $\ln 2$ to the entropy 
resulting in $S_L \sim 2 p \ln 2 \langle D \rangle_l$, where $\langle D 
\rangle_l$ is the average number of decimations that occur at the edge, and 
$p$ is the probability that a field is decimated instead of a coupling; the 
self-duality of the model yields $p=1/2$, and using the properties of the IRFP 
in a random Ising chain it can be shown that $\langle D \rangle_l \sim (1/3) 
\log_2 l$. We eventually get $S_l^{\rm dis} \sim \frac {\ln 2}{6}\log_2 l +
\kappa_1$; for a random $XY$ chain with $\ga=0$, one gets $S_L^{\rm dis} \sim 
\frac {\ln 2} {3} \log_2 l+\kappa_2$, where $\kappa_1$ and $\kappa_2$ are 
non-universal constants. Comparison with the pure system leads to following 
observations: (i) there is a loss of entropy due to randomness, and (ii) the 
effective central charges of the random chain are related to those of the pure 
chain as $c^{\rm dis} = \ln 2 \times c^{\rm pure};~ c^{\rm pure}=1/2$ for the
Ising transition. These charges determine the universality class of the 
associated IRFP. For this spin chain, disorder is a relevant perturbation and 
the RG flow is from the pure fixed point to the IRFP. Hence, the observation 
$c^{\rm pure} > c^{\rm dis}$ is consistent with 
the $c$-theorem \ct{zamolodchikov86}.

The above study has been extended to generalized random chains with $Z_q$ 
symmetry \ct{santacharia06} given in (\ref{eq_random_gen})
with $\al_n = \sin(\pi/q)/\sin(\pi n/q)$. The entanglement entropy is found 
to be $S_l = \frac {\ln q}{6} \log_2 l$ so that the effective central charge is
$c_q^{\rm dis} = \frac {\ln q}{6}$; comparing with the central charge of the 
pure chain $c_q^{\rm pure} = 2(q -1)/(q +2)$, one finds an increase of 
the effective central charge at the IRFP for $q > 41$ which apparently rules 
out the possibility of a pure-to-random $c$-theorem mentioned above. However, 
an intermediate disorder critical line separating weak and strong disorder 
phase may exist in these models, as discussed in Sec.~\ref{sec_griff+act} 
and one therefore, can not conclusively rule out the possibility of a 
pure-random $c$-theorem \ct{refael09}. 

The logarithmic dependence of $S_l^{\rm dis}$ has been verified 
numerically for random $XY$ (with $\ga=0$) \ct{laflorencie05} and Heisenberg 
chains \ct{chiara06}. Generalizing the strong disorder RG to $d=2$, for a 
random transverse Ising model, a double logarithmic correction to the area 
law was pointed out in contrast to the $d=1$ case \ct{lin07,kovacs10}. 

\subsection{Quantum discord}
\label{sec_discord}

A different and significant measure of quantum correlations between two spins 
other than entanglement, namely ``quantum discord" was introduced in
\ct{olliver01}, (see also \ct{zurek03}) which exploits the fact that 
different quantum analogs of equivalent classical expressions can be obtained 
because of the fact that a measurement perturbs a quantum system. While the 
concurrence is a measure of quantum correlations based on separability of 
states, the discord, on the other hand, is a measurement based measure of 
the same. There are quantum states which are completely separable but the 
discord (and hence the quantum correlation) is not necessarily zero.

The information associated with a classical system is quantified in terms of 
the Shannon entropy $H(p)$ where $p$ is the probability distribution of 
the system. If the system comprises two 
subsystems $A$ and $B$, the mutual information is defined as $I(p)=H(p^A) + 
H(p^B) - H(p)$, where $H(p^i)$, $i=A,B$, stands for the entropy associated 
with the subsystem $i$; this can alternatively be expressed as $J(p)=H(p^A)-
H(p|p^B)$, where $H(p|p^B)= H(p)-H(p^B)$ is the conditional entropy. In the
classical context, $I(\rho)=J(\rho)$.

In the quantum context, the classical Shannon entropy functional gets 
replaced by the quantum von Neumann entropy expressed in terms of the 
density matrix $\rho$ acting on the composite Hilbert space.
The natural quantum extension of the mutual information is given by 
\beq I(\rho)=s(\rho^A)+s(\rho^B)-s(\rho). \label{eq_MI} \eeq
The conditional entropy derived through local measurements however alters the 
system. {A local measurement is of the von Neumann type defined
by a set of one-dimensional projectors $\{ \hat{B_k} \}$ which sum up to 
identity.} Following a local measurement only on the subsystem $B$, the final 
state $\rho_k$ of the composite system
%which is the generalization of the classical conditional probability, 
is given by $\rho_k=(1/p_k) (\hat{I} \otimes \hat{B_k})\rho 
(\hat{I} \otimes \hat{B_k}),$ with the probability $p_k={\rm tr}(\hat{I}
\otimes \hat{B_k})\rho(\hat{I} \otimes \hat{B_k})$ where $\hat{I}$ is the 
identity operator for the subsystem $A$. The quantum conditional entropy can 
be defined as $s(\rho|\{\hat{B_k}\})=\sum_k p_k s(\rho_k)$, such that the 
measurement based quantum mutual information takes the form $J(\rho|
\{\hat{B_k}\})=s(\rho^A)-s(\rho|\{ \hat{B_k}\}).$ This expression maximized 
based on the local measurement gives the classical correlation 
\ct{henderson01,vedral03}. Hence we have 
\beq C(\rho)=max _ {\{\hat{B_k}\}} J(\rho| \{\hat{B_k}\}). \label{classical}
\eeq
We arrive at two quantum analogs of the classical mutual information: the 
original quantum mutual information $I(\rho)$ (\ref{eq_MI}) and the 
measurement induced classical correlation (\ref{classical}). As 
introduced by Olliver and Zurek \ct{olliver01}, 
the difference between these two, i.e.,
\beq Q(\rho)= I(\rho)-C(\rho) \label{quantum} \eeq 
is the quantum discord which, being the difference between total information 
(correlation) $I$ and the classical correlation $C$, measures the amount of 
quantumness in the state; a non-zero $Q$ implies that all the information about
the correlation between $A$ and $B$ cannot be extracted by local measurement on $B$.

The studies of quantum discord for spin systems close to QCPs have 
established a natural connection between QPTs and quantum information theories
\ct{olliver01,luo08,dillen08,sarandy09,pal11,maziero11,li11,tomasello11}. 
The $Z_2$ symmetry of the one-dimensional transverse Ising chain in
(\ref{eq_tim1d}) 
%with $J_x=1$]
enables us to write the density matrix for two spins separated by a lattice 
spacing $n$ given in (\ref{rhon}) and thereby to calculate quantum 
discord. The total correlation $I$ is calculated using the eigenvalues of the 
reduced density matrices for subsystems $A$ and $B$, while the classical 
correlations are derived by introducing a set of projection operators for part 
B parametrized on the Bloch sphere and optimizing for the polar and azimuthal 
angles \ct{luo08,dillen08,sarandy09}. 

Quantum discord is also able to capture the QPT occurring at transverse field 
$h=h_c =1$ ($J_x=1$). For a chain of length $L$, one finds the following.

\noi (i) The derivative of the nearest-neighbor classical correlation with 
respect to $h$ shows a pronounced minimum at $h_m$ which approaches $h_c=1$ 
as $L \to \infty$. The derivative at $h_m$ shows a logarithmically divergence 
with $L$. This behavior surprisingly resembles that of the behavior of 
nearest-neighbor concurrence shown in Fig.~\ref{fig_conc1}. 

\noi (ii) In contrast, the first derivative of the nearest-neighbor quantum 
discord, $\partial Q/\partial h$, has a point of inflection at $h_m$ while 
the second derivative has a pronounced maximum that shows a quadratic 
logarithmic divergence in the thermodynamic limit. This observation leads to 
the conclusion that near a QCP, $Q$ and the concurrence show quite different 
scaling behaviors. Although the study of discord in quantum critical 
systems is in its infancy, an experimental study to measure quantum discord 
using an NMR set up has already been reported \ct{auccaise11}.

\subsection{Quantum fidelity} 
\label{fidelity}

In this section, we will study an information theoretic measure known as 
the quantum fidelity which is able to capture ground state singularities 
associated with a QPT. In the present context, fidelity is defined as the 
overlap between the ground states of a quantum Hamiltonian calculated for 
two different values of some external tunable parameter, e.g., the transverse 
field in TIMs. The study of quantum fidelity dates back to 1967 when P. W. 
Anderson showed that the ground state fidelity for a system of $N$ fermions 
vanishes in the thermodynamic limit $N \to \infty$ \ct{anderson67}. This 
phenomenon is shared by many other quantum many body systems (see e.g., 
\ct{bettelheim06,sachdev01}) and is known as the Anderson orthogonality 
catastrophe. Quantum fidelity is currently being studied extensively as an 
indicator of a QPT without making reference to an order parameter 
\ct{zanardi06}; for a review see \ct{gu10}.

The ground state fidelity between two ground states 
of a $d$-dimensional quantum mechanical Hamiltonian $H(\la)$ described 
by parameters $\la$ and $\la+ \de$ is defined as 
\beq F (\la, \de )= |\langle \psi_0(\la)|\psi_0(\la + \de) \rangle |. \eeq
For small system sizes, the overlap is usually expected to be close to 
unity if the two states are very close to each other in the parameter space. 
In contrast, the fidelity should vanish in the limit in which the states are 
orthogonal which is the case close to a QCP where the fidelity shows a sharp 
dip (see Fig.~\ref{fig_fidelity1}). We emphasize that in the limit $N \to 
\infty$, the fidelity vanishes irrespective of the proximity to a QCP even 
though a sharper decay in the fidelity is expected close to a QCP 
\ct{zanardi06}.

%The overlap between two neighboring ground states of a quantum many-body 
%Hamiltonian when a parameter is varied is called the 
%fidelity; this is a measure of the distance between the two 
%states in the parameter space \ct{zanardi06}. 
Let us consider a Hamiltonian of the form
\bea H\left(\la \right) = H_0 + \la H_I \eea
which satisfies the eigenvalue equation
\bea H\left(\la \right)|\psi_m\left(\la \right)\rangle = E_m (\la) |\psi_m
\left(\la\right)\rangle, \eea
where $m = 0,1,2,\cdots$. Let $H_0$ describe a QCP while $H_I$ is a 
driving term which does not commute with $H_0$.
%\equiv \partial_{\la} H|_{\la=0}$ is the perturbation not commuting with $H_0$.

\begin{figure}[htb]
\begin{center}
\includegraphics[height=2.0in,width=2.8in, angle = 0]{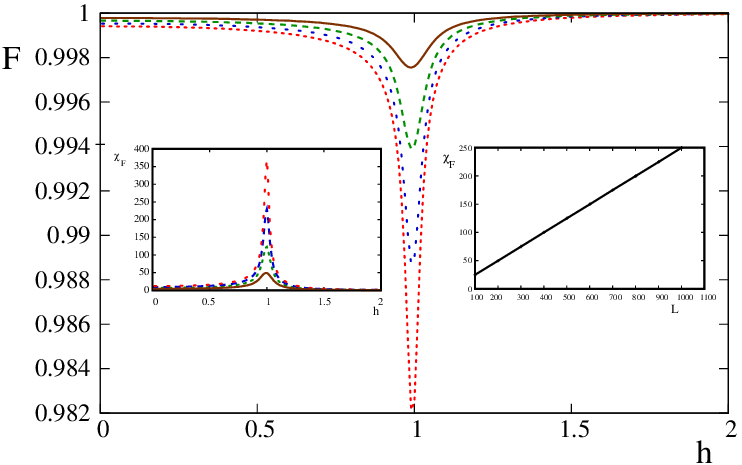}
\caption{(Color online) The fidelity between two ground states of a transverse
Ising chain with transverse fields $h$ and $h+\de$, with $\de=0.1$, is plotted 
as a function of $h$ for small system sizes. The fidelity shows a dip around 
the QCP at $h=1$ ($J_x=1$). One observes a sharper decay at the QCP when the system size increases. Left 
inset shows the divergence of the fidelity susceptibility which again becomes prominent with increasing system size.
Right inset shows the scaling $\chi_F \sim L$ at the QCP.} 
\label{fig_fidelity1} 
\end{center}
\end{figure}

When $\de \to 0$, the overlap $\langle \psi_0(\la)|\psi_0(\la + \de 
)\rangle$ between two ground states infinitesimally separated in the 
parameter space can be expanded using a Taylor expansion,
\beq \langle \psi_0(\la)|\psi_0(\la + \de )\rangle = 1 + \de \langle 
{\psi_0(\la)} |{\frac{\partial}{\partial \la} |\psi_0(\la)} \rangle
+ \frac{\de^2}{2} \langle {\psi_0(\la)} |{\frac{\partial^2}{\partial 
\la^2} |\psi_0(\la)} \rangle + \cdots. \eeq
The term linear in $\de$ is the Berry connection term which gives 
rise to the Pancharatnam-Berry phase \ct{pancharatnam56,berry84}.

\bigskip

\noi {(i) Fidelity susceptibility}\\

We now introduce the concept of fidelity susceptibility valid in the limit 
$\de \to 0$
\ct{you07,gu08,gurev08,venuti07,abasto08,venuti09,cozzini071,cozzini072,
zanardi07,venuti08,gu091,li09,schwandt09}.
We can use the expansion
$$|\psi_0(\la + \de ) \rangle \simeq |\psi_0 (\la)\rangle + \de 
\sum_{n\neq 0} \frac{\langle \psi_n(\la)|H_I|\psi_0(\la)\rangle |\psi_n(\la)
\rangle}{E_n -E_0} + \cdots$$
to express the fidelity in the form
\beq F(\la, \de) ~=~ 1 ~-~ \frac{1}{2}\de^2 ~L^d ~\chi_{ F} (\la) + ...
\label{def_fid}, \eeq
where the term linear in $\de$ vanishes due to the normalization condition
of the wave function, and $L$ is the linear dimension of the system. The 
quantity $\chi_F = -(2/L^d) \lim_{\de \to 0} (\ln F)/\de^2= - 
(1/L^d) ({\partial^2 F}/{\partial \de ^2})$ is called the fidelity 
susceptibility (FS) and is given by
\bea \chi_F\left( \la \right) = \frac{1}{L^d} \sum_{m \ne 0} {\frac{|
\langle\psi_m\left(\la \right) |H_I |\psi_0 \left(\la \right)|^2}{\left[E_m
\left(\la \right) - E_0 \left(\la \right) \right]^2}}. \label{eq_fidsus}\eea 
It is useful to compare the form of $\chi_F$ with the ground state 
specific heat density given by
\bea \chi_E &=& - \frac{1}{L^d} \frac{\partial^2 E_0}{\partial \la ^2}= - 
\frac{2}{L^d} \sum_{m \ne 0} {\frac{|\langle\psi_m\left(\la \right) |H_I | 
\psi_0 \left(\la \right)|^2}{ E_m\left(\la \right) - E_0 \left(\la \right)}},
\non \\
& & \label{sp_heat} \eea
which diverges at the QCP following the scaling relation $|\la|^{-\al}$.
We note that Eq.~(\ref{def_fid}), defined for a fixed system size 
$N=L^d$, describes a situation where the fidelity is close to unity even at 
the critical point; however, as we will show below, the scaling of the FS 
is dictated by some of the quantum critical exponents.

Using the scaling relation $\chi_E$ $\sim |\la|^{-\al}$ along with the 
hyperscaling relation $2-\al = \nu(d+z)$ valid below the upper critical 
dimension, the scaling relation of the FS close to the QCP can be readily 
obtained \ct{gritsev10,albuquerque10}. One finds that $\chi_F \sim |\la|^{\nu 
d - 2}$ away from the QCP ($|\la|^{-\nu} \ll L$), while close to the QCP
($|\la|^{-\nu} \gg L$), finite 
size scaling \ct{barber83} yields a modified scaling relation $\chi_F 
\sim L^{2/\nu - d}$. Therefore, we 
conclude that $\chi_F$ diverges when $\nu d <2$. For $\nu d > 2$, although the
FS appears to vanish at $\la=0$ in the thermodynamic
%the low-energy modes no longer play a dominant role and the scaling 
%saturates to the perturbative quadratic form $\chi_F \sim \la^2$ while at 
%$d \nu=2$ there are additional logarithmic singularities.
limit, a non-zero value of the FS is expected due to non-universal 
high-energy modes which are insensitive to the presence of a QCP 
\ct{gritsev10}.

These scaling relations have been verified for the transverse Ising model on 
a square lattice using extensive Monte Carlo simulations \ct{albuquerque10}.
Very recently, the above scaling has been generalized to the case of an 
anisotropic QCP \ct{mukherjee112} as is seen at a semi-Dirac band crossing 
point \ct{banerjee09}. At an anisotropic QCP, characterized by critical 
exponents $\nu_{1}$ in $m$ spatial directions and $\nu_2$ in the other
$(d-m)$ directions, $\chi_F \sim L^{2/\nu_1 - \nu_1/\nu_2 (d-m)-m}$ which 
reduces to the conventional scaling
when $\nu_1=\nu_2$. Also the finite size scaling of the bipartite 
entanglement and the ground state fidelity have been used to study
the QCP in the one-dimensional Bose-Hubbard model \ct{bounsante07}.
 
The FS can be related to the connected imaginary time 
($\bar \tau$) correlation function of the perturbation $H_I(\bar \tau)$ 
using the relation \ct{you07,venuti07} 
\beq \chi_F(\la) = \int_0^\infty d\bar\tau ~\bar \tau ~\langle H_I(\bar \tau)
H_I(0)\rangle_c, \label{eq_fidcorr} \eeq
where we define $H_I(\bar \tau) =\exp(H\bar \tau) H_I \exp(-H_I \bar \tau)$, 
$\bar\tau$ being the imaginary time, and 
$\langle H_I(\bar \tau) H_I(0)\rangle_c = \langle H_I(\bar \tau) H_I(0)\rangle
-\langle H_I(\bar \tau) \rangle \langle H_I(0)\rangle$. 
Assuming $H_I= \sum_{\vec r} V(\vec r)$ and using the scaling $r'=b r, \bar 
\tau'=b^z \bar \tau$ and $V(r') = b^{-\De_V} V(r)$ for a change of length 
scale by a factor $b$, one finds from (\ref{eq_fidcorr}) that the 
scaling dimension of $\chi_F$ is given by ${\rm dim}[\chi_F]=2\De_V-2 
z + d$, i.e., $\chi_F \sim L^{2z+d-2\De_V}$ at the QCP; away from the QCP 
($L > \xi \sim \la^{-\nu}$), $\chi_F \sim \la^{-\nu(2z+d-2\De_V)}$.
A marginal or relevant perturbation $H_I$ (so that $\la H_I$ scales as 
the energy, $\la H_I \sim \la^{\nu z}$) allows us to make an additional 
simplification coming from the fact that the scaling dimension of $H_I$ is 
$\De_{H_I} = z-1/\nu$ \ct{schwandt09,gritsev10}, so that 
$\chi_F\sim L^{2/\nu-d}$ at $\la=0$ as derived previously.

For the transverse $XY$ chain in (\ref{eq_txy}), the ground state can be 
written in a direct product form for a chain of $L$ sites (spins) (see 
(\ref{eq_direct_product})). 
% in terms of a parameter $\theta_k$
%\beq |\psi_0 (\ga,h)\rangle = \bigotimes_{k>0} ( \cos \frac{\theta_k}{2} 
%|0\rangle_k|0\rangle_{-k} + i \sin \frac{\theta_k}{2}|1\rangle_k|1
%\rangle_{-k}, \eeq
%where $\cos \theta_k = ((J_x+J_y)\cos k - h)/ \Lambda_k$ with $\Lambda_k = 
%\sqrt{(h + (J_x+J_y)\cos k)^2 + \ga^2 \sin ^2 k}$; we have used
%a periodic boundary condition so that the momentum $k$ is quantized as $k = 
%2n \pi/L$ where $n=0,1,2,..(L-1)$. The overlap between two neighboring states 
One then finds that the fidelity between neighboring states is
given by $F= \prod_{k >0} \cos( {\theta_k - \theta'_k})$ \ct{zanardi06}. 
%We note that the absence of the $k=0$ mode for a finite size
%system leads to a non-zero fidelity. 
Now, under a change in the transverse field from $h$ to $h+dh$, the FS can be 
derived by writing the fidelity in the form $F \simeq \exp(-\de^2 L^d 
\chi_F/2)$ as
\bea \chi_F &=& \frac{1}{L} \sum_{k>0} (\frac{\partial \theta_k}{\partial h})^2
= \frac{1}{L} \sum_{k>0}\frac{\ga^2 \sin^2 k}{[(h +\cos k)^2 + \ga^2 \sin^2 
k]^2}, \non \\
& & \eea
where we have set $J_x + J_y=1$. At the Ising critical point 
($ \la = h-1=0$), one obtains using the continuum limit
\bea \chi_F \sim \int_{\frac{\pi}{L}}^{\frac{\pi(L-1)}{L}} dk~ 
\frac{\ga^2 k^2}{\ga^4 k^4} \sim L, \eea
where $\pi -k \to k$.
This is consistent with the scaling $\chi_F \sim L^{2/\nu -d}$ noting that the
Ising critical exponents $\nu=z=1$ and $d=1$ \ct{zanardi06} (see the right 
inset of Fig.~\ref{fig_fidelity1}). On the other hand, away from the critical 
point, i.e., $\la \gg L^{-1}$ when one can essentially assume $k$ to be 
continuous, we get 
\bea \chi_F \sim \int_0^{\infty} \frac{dk}{2\pi} \frac{\ga^2 k^2}{((\la
+\frac {k^2}{2})^2 + \ga^2 k^2)^2} \sim \la^{-1}, \eea
which is again in agreement with the proposed scaling $\chi_F \sim 
\la^{\nu d -2}$. 

\begin{figure}[htb]
\begin{center}
\includegraphics[height=1.8in,width=2.8in, angle = 0]{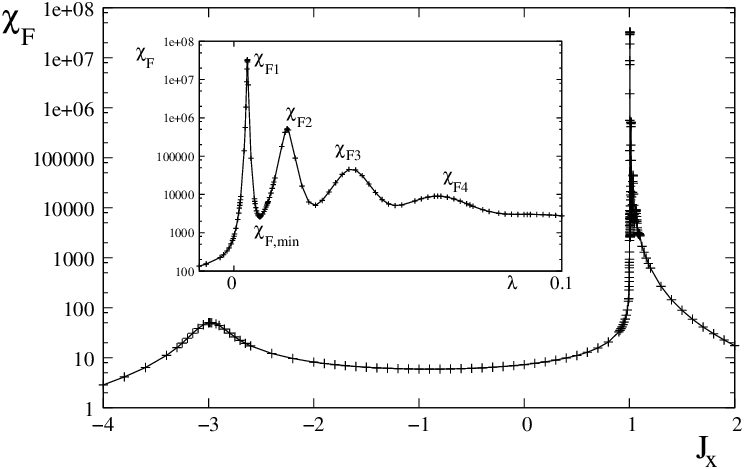}
\caption{The variation of $\chi_F$ with $J_x$, as obtained numerically, for a 
system size of $L = 100$. We have fixed $h = 2J_y = 2$. The peak near the 
Ising critical point $J_x = -3$ scales as $L$, whereas the peak near the 
MCP at $J_x = 1$ shows a $L^5$ divergence. Inset: The oscillatory behavior of 
the FS is a signature of the presence of the quasicritical
modes. Each of the maxima, denoted by $\chi_{Fj},~j=1,2,..$, scales as $L^5$. 
(After Mukherjee $et~al.$, 2010).} \label{fig_fidelity2} 
\end{center}
\end{figure}

While the FS close to a QCP satisfies the conventional scaling relations proposed
above, surprises emerge when the same is calculated close to a quantum
multicritical point. One finds remarkable features like oscillations in
FS and its anomalous scaling with the system size. To illustrate these,
let us consider the case $J_y=2h$ and calculate the overlap by changing the 
parameter $J_x$ so that the chain is at a critical point for $J_x = -3J_y$ 
and at a multicritical point (MCP) for $J_x=J_y$ \ct{mukherjee111} (see 
Fig.~\ref{fig_anisopath}). The FS ($\chi_F=(1/L)\sum_k ({\partial \theta_k}/
{\partial J_x})^2)$ is then found to be
\beq \chi_F = \frac{1}{L} \sum_k \frac{(J_y \sin k + 2J_y \sin 2k)^2}
{[(h+ (J_x+J_y)\cos k)^2 + ((J_x-J_y)\sin k)^2]^2}. \eeq
As expected from the scaling relation given above, one finds that $\chi_F 
\sim L$ at the Ising critical point ($h=-3J_y$). However, close to the quantum 
MCP ($J_x=J_y$), one finds a stronger divergence of the maximum value of FS 
given by $\chi_{\rm F}^{\rm max} \sim L^5$ (see Fig.~\ref{fig_fidelity2}); 
this cannot be explained using standard scaling 
relations because the critical exponents associated with the MCP are 
$\nu_{mc}=1/2$ and $z_{mc}=2$. Also, oscillations of the FS are observed 
close to the quantum MCP. This observation can be attributed to the existence 
of quasicritical points close to the MCP. For a finite chain, these 
quasicritical points are points of local minima of energy (also see 
Sec.~\ref{multicritical}) which lie on the FM side of the MCP. The 
exponents associated with these quasicritical points appear in the scaling 
relation of the maximum value of the FS, i.e., $\chi_F^{\rm max} \sim 
L^{2/\nu_{qc}-d}$; for the present model it can be easily shown that 
$\nu_{qc}=1/3$. As we move 
away from the MCP, the FS oscillates and attains a maximum whenever a 
quasicritical point is hit. As we approach the thermodynamic limit, these 
quasicritical points approach the MCP.

We now consider studies of the FS for a path defined
in a multidimensional parameter space \ct{zanardi07, venuti07} when more than 
one parameter is varied; such a situation arises in the transverse $XY$ model 
when the MCP is approached by simultaneously tuning the transverse field and
the anisotropy $\ga$ \ct{mukherjee111}.
Motivated by the geometrical interpretation of fidelity which is in fact the 
angle between two vectors in the Hilbert space, one can define a measure known
as the Fubini-Study distance, $d_{FS}$, given by \ct{wootters81,zanardi07}
\bea d_{FS} = \cos^{-1} F( \ket{\psi_0}, \ket{\psi_0 + \de\psi_0}). \eea 
Here $|\psi_0\rangle$ and $|\psi_0 + \de\psi_0\rangle$ represent two 
ground states infinitesimally spaced in the parameter space so that $F$ is 
close to unity, allowing for the expansion $d_{FS}^2 \simeq 2(1 - F)$. We 
consider a $M$-dimensional parameter space spanned by $\la_a$, 
$a=1,2,...,M$, with Hamiltonian given by $H = H_0 + \sum_a \la_a H_I^a$; 
the QCP is at $\la_i=0$ for all $i=1,2, \cdots,M$. Using a Taylor expansion, 
$\de \psi_0= \sum_a \partial_a \psi_0 d\la^a$
where $\partial_a \psi_0 \equiv \partial \psi_0 /\partial \la_a$,
one obtains $d^2_{FS}(\ket{\psi_0}, \ket{\psi_0 + \de\psi_0}) = L^d\sum_{a,b}
g_{ab}d\la_a d\la_b$. Consequently, the fidelity is given by
%\bea &~& F(\ket{\psi_0}, \ket{\psi_0 + \de\psi_0}) \simeq |1 + \langle\psi_0|
%\de\psi_0\rangle + \frac{1}{2}\langle\psi_0|\de^2\psi_0\rangle| \non \\ 
%&\simeq& 1 - \frac{1}{2}\langle\de\psi_0|\de\psi_0\rangle + 
%\frac{1}{2}|\langle\psi_0|\de\psi_0|^2. \label{geo_tensor} \eea
\beq F = 1 - \frac{L^d}{2}\sum_{ab} g_{ab}d\la_a d\la_b.
\label{ch1_fidelity_dist} \eeq
In (\ref{ch1_fidelity_dist}), $g_{ab}$ stands for the FS and is the real 
part of a complex geometric tensor given by \ct{venuti07} 
\bea Q_{ab} &=& \frac{1}{L^d}\left[\langle \partial_a \psi_0|\partial_b
\psi_0 \rangle - \langle \partial_a \psi_0|\psi_0 \rangle\langle \psi_0 |
\partial_b \psi_0\rangle\right] \non \\
&=& \frac{1}{L^d}\sum_{n \neq 0} \frac{\langle
\psi_0(\la)|\partial_a H|\psi_n(\la)\rangle \langle
\psi_n(\la)|\partial_b H|\psi_0(\la)\rangle}{[E_n - E_0]^2}. \non \\
& & \label{ch1_tensor} \eea
The imaginary part of $Q_{ab}$ gives the curvature 
two-form of the geometric phase. Using a similar scaling analysis of the 
connected correlation function defined in a $M$-dimensional space as shown 
above, we can readily show that $Q_{ab} \sim L^{2z+d -\De_a - \De_b}$ at
the QCP, where $\De_a$ is the scaling dimension of the operator 
$H_I^a={\partial H}/{\partial \la_a}|_{\la_i=0}$, for $i=1,2, \cdots,M$. 
Using the relevant or marginal nature of the driving operators, this scaling 
can be recast into the form $Q_{ab} \sim L^{1/\nu_a + 1/\nu_b -d}$, where 
$\nu_a$ is the correlation length exponent associated with the perturbation 
$\la_a$; the scaling reduces to the simple form $\chi_F \sim L^{2/\nu-d}$ for 
a path with $\la_a=\la_b=\la$. 
{In this connection, it has recently been shown that the 
geometric phase (GP) shows an interesting scaling behavior close to the QCP 
of the transverse $XY$ chain (\ref{eq_txy}) given by some of the quantum 
critical exponents \ct{zhu06,carollo05,pachos06,Zhu08,Hamma06,Cui06,quan09}; 
we present a note on these studies in Appendix \ref{sec_gp_scaling}.}

The FS approach has been implemented to identify the quantum phase 
transition from a gapless to a gapped phase in the Kitaev model \ct{yang08} 
when the interaction $J_{3}$ is varied; at the QCP at which the system enters 
the gapless phase (see Fig.~\ref{kitaev_phase}), $\chi_F \sim L^{1/2}$. This
can be explained noting the anisotropic nature of the associated critical 
point with $\nu_1=1/2, \nu_2=1$,$d=2$ and $m=1$\ct{mukherjee113}. Moreover, 
in the gapless phase $\chi_F \sim \ln L$, 
while no such scaling is observed inside the gapped phase \ct{gu09}.
The phase transition in the Kitaev model has also been studied using 
the fidelity per site approach \ct{zhao09}.

Finally, we recall the extended Ising model in (\ref{eq_mps_hamil}) with a
matrix product ground state. One can find the fidelity between two ground 
states with parameters $g_1 = g+ \de$ and $g_2 = g -\de$ for a spin chain 
of length $L$ \ct{zhou08,cozzini071}, given by
\beq F = \frac{[(1 + \sqrt{g_1 g_2})^L + (1 -\sqrt{g_1 g_2})^L]}{[\sqrt{
(1+g_1)^L + (1-g_1)^L}\sqrt{(1+g_2)^L +(1-g_2)^L}}, \label{eq_mps_fidelity}
\eeq
which shows that $F$ is symmetric in both $g$ and $\de$. The fidelity in 
(\ref{eq_mps_fidelity}) decays exponentially if the two states are in the same 
phase, but shows a strong oscillatory behavior with an exponentially decaying
envelope for two states in different phases \ct{zhou08}. At the QCP ($\la=g=0$),
the FS scales as $L$ which is consistent with the scaling $L^{2/\nu-d}$ 
with $\nu=d=1$; similarly, away from the QCP, $\chi_F \sim g^{-1}$ in
agreement with the scaling $\chi_F \sim \la^{\nu d -2}$
\ct{cozzini071}. In the thermodynamic limit \ct{rams11} for $|\de| \ll
|g| \ll 1$, the fidelity takes the form $F \simeq \exp(-\de^2 L/2|g|)$. 

 {We conclude with a brief note on random spin chains. The 
scaling of the FS can also be useful for determining the critical point and GM
singularities (see Sec. \ref{sec_griff+act}) in a disordered transverse $XY$ 
chain \ct{garnerone09}; this can be mapped to quasi-free fermions using the
JW transformation although the system is not reducible to decoupled 
$2 \times 2$ problems 
due to the loss of translational invariance. The results derived using the FS 
approach predicts the presence of Griffiths singular regions in the vicinity
of the Ising and anisotropic transition lines (see Fig. \ref{fig_xyphase}). 
Moreover, the scaling analysis of the FS shows a complete disappearance of the
anisotropic transition (for some disorder distributions) and the emergence of 
a GM phase; this is in congruence with the previous study of
the same model based on a mapping to Dirac fermions with random 
mass \ct{bunder99}.}

\bigskip

\noi {(ii) Fidelity per site and fidelity in the thermodynamic limit:}\\

The advantage of using the FS approach is that the parameter $\de $ 
gets factored out in the Taylor expansion rendering the FS dependent 
on $\la$ alone; however this approach fails to describe the Anderson 
orthogonality catastrophe. In an alternative approach 
\ct{zhou081,zhou08,zhou082}, the ground state fidelity per site, defined as 
${\cal F}(\la,\de) = \lim_{N \to \infty} F^{1/N}(\la,\de)$, $N=L^d$ or 
$\ln {\cal F} = \lim_{N \to \infty} \ln F(\la,\de)/N$, is calculated
in the large system limit with an arbitrary $\de$, in contrast to the FS 
approach in which $N$ is small and $\de \to 0$. As a result, the fidelity 
differs significantly from unity. It has been shown that ${\cal F}$ is 
able to detect QPTs in the transverse $XY$ chain and the Kitaev model on the 
hexagonal lattice. In particular, using the direct product form of the ground 
state in (\ref{eq_direct_product}) for an anisotropic transverse $XY$ chain, 
${\cal F}$ between two ground states with transverse fields $h$ and $h'$ is
given by 
\bea \ln {\cal F}(h,h') &=& \lim_{L \to \infty} ~\frac{1}{L} ~\sum_{k>0} ~\ln 
\left[\cos( \theta_k - \theta_k')\right] \non \\ 
&=& \frac{1}{2 \pi} \int_0^{\pi} ~\ln \left[\cos(\theta_k -\theta_k')\right] 
dk. \eea
Numerical studies show that the first derivative of ${\cal F}(h, h')$ shows 
a logarithmic divergence of the form $\partial {\cal F}(h,h')/\partial h \sim 
\kappa_1 \ln |h-h_c|$ as $h \to h_c=J_x+J_y=1$, in the thermodynamic limit. On 
the other hand, $(\partial {\cal F}(h,h')/\partial h)|_{h_m} \sim \kappa_2 
\ln L$, where $h_m$, as defined previously, is the effective critical point at 
which $\partial {\cal F}(h,h')/\partial h$ shows a pronounced maximum for 
large $L$. The ratio of the non-universal coefficients yields the exponent 
$\nu$; $\nu = |\kappa_1/ \kappa_2| \approx 1$ \ct{zhou082}.

As seen from Eq.~(\ref{def_fid}) and also the preceding discussion, the 
knowledge of $\chi_F$ should be sufficient to draw conclusions about the 
behavior of $F$ and the associated QCP for small system sizes and in the limit
$\de \to 0$ ($\de^2 L^d \chi_F/2 \ll 1$). However, in the thermodynamic 
limit ($L \to \infty$ at fixed $\de$), the expansion in 
Eq.~(\ref{def_fid}) up to the lowest order becomes insufficient. 
Recently, Rams and Damski \ct{rams11} have proposed a generic scaling relation 
 valid in the thermodynamic limit given by
\bea \ln{F(\la - \de, \la + \de)} \simeq -L^d|\de|^{\nu d} A
\left(\frac{\la}{|\de|}\right), \label{fidelity_generic} \eea
where $A$ is a scaling function; this relation interpolates between the 
fidelity susceptibility approach and the fidelity per site approach. In 
deriving the scaling relation in Eq.~(\ref{fidelity_generic}), it is assumed 
that the fidelity per site is well behaved in the limit $L \to \infty$, the 
QCP is determined by a single set of critical exponents, and $\nu d >2$ so 
that non-universal corrections are subleading \ct{gritsev10}. In particular, 
at the critical point $\la=0$, the fidelity, measured between the ground 
states at $+\de$ and $-\de$, is non-analytic in $\de$ and satisfies
the scaling $\ln{F} \sim -L^d |\de|^{\nu d}$. On the other hand, away from 
the QCP, i.e., for $|\de| \ll |\la| \ll 1$, the scaling gets modified
to $\ln{F} \sim -L^d \de^2 |\la|^{\nu d - 2}$. This scaling has been 
verified for an isolated quantum critical point using one-dimensional 
transverse Ising and $XY$ Hamiltonians \ct{rams11}. Moreover, near a QCP a 
cross-over has been observed from the thermodynamic limit ($L |\de|^{\nu} 
\gg 1 $) to the non-thermodynamic (small system) limit ($L |\de|^{\nu} \ll 1$) 
where the concept of fidelity susceptibility becomes useful. We note that 
Eq.~(\ref{fidelity_generic}) is an example of the Anderson orthogonality 
catastrophe \ct{anderson67} which states that the overlap of two states 
vanishes in the thermodynamic limit irrespective of their proximity to a QCP.
For the transverse Ising chain with $h \neq h_c$, it was found that
$F (g,\de) \simeq \exp \left(-L \de^2/16|h-h_c| \right)$
which reduces to the result derived above for the FS when the argument 
of the exponent is small and at the same time provides a new result when the 
lowest order Taylor expansion is insufficient; one finds no non-analyticity 
in $\de$ as we are away from the critical point. The scaling in
(\ref{fidelity_generic}) has recently been generalized to a massless Dirac
fermions and the two-dimensional Kitaev model \ct{mukherjee113}.

\bigskip 

\noi {(iii) Other fidelity measures:}\\

So far, our discussion has been limited to pure states.
The concept of fidelity has been generalized to mixed states through a 
measure known as the reduced fidelity \ct{uhlmann76,jozsa94} in which
the two ground states in a fidelity analysis are replaced with two states 
differing slightly in the driving parameter and describing only a local 
region $A$ of the system of interest. The reduced fidelity between two 
states described by reduced density matrices $\rho_A(\la)$ and 
$\rho_A(\la')$ for two parameters $\la$ and $\la'$ is defined as 
\beq F(\la,\la')= {\rm Tr} \sqrt{\sqrt{\rho_A(\la')}\rho_A(\la)
\sqrt{\rho_A(\la')}}. \label{eq_redfed} \eeq
The states under consideration describe a subsystem $A$; $\rho_A$, 
obtained by tracing out the rest of the system from the total density matrix, 
usually describes a mixed state. In contrast, for a pure state, $\rho_A = 
\ket{\psi}\bra{\psi}$ and $F(\la,\la')$ reduces to the modulus of the 
overlap $|\langle \psi (\la')|\psi(\la) \rangle|$. Denoting $\la'=\la+\de \la$
and taking the limit $\de \la \to 0$, we can define a FS
analogous to the pure state case as $\chi_F= -2 \lim_{\de \la \to 0} \ln 
F/(\de \la)^2$. In the context of a QPT, the reduced fidelity has been 
introduced
in connection to fidelity per site and renormalization flows \ct{zhou07}. 
Even though the reduced fidelity encodes local properties of a quantum system, 
surprisingly it is also able to capture the signature of both order-disorder
\ct{zindaric03,ma08,li08,ma10,paunkovic08,kwok08} and topological
\ct{eriksson09} QPTs. For a transverse Ising chain the susceptibility
associated with the reduced fidelity diverges logarithmically as $ \chi_F 
(h)\sim \ln^2 |h-h_c|$ as $h\to h_c=1$ which is a slower divergence in 
comparison to the global fidelity case \ct{eriksson09}. 

The notion of reduced fidelity for mixed states can be further generalized to 
finite temperatures using thermal mixed states $\rho= (1/{\cal Z}) \sum_n 
\exp (-\beta E_n) \ket{\psi_n} \bra{\psi_n}$, where the canonical partition 
function ${\cal Z} = \sum_n \exp (-\beta E_n)$ \ct{zanardi071}. To define the 
mixed state fidelity of two thermal states with small perturbations in the
inverse temperature $\beta$ and driving parameter $\la$, we rewrite 
(\ref{eq_redfed}) in the form 
\beq F(\beta_0,\la_0;\beta_1,\la_1) = {\rm Tr} \sqrt{\sqrt{\rho_0}\rho_1
\sqrt{\rho_0}}, \label{eq_fidthermal} \eeq
where $\rho_{\al}=\exp(-\beta_{\al} H(\la_{\al})) /{\cal Z} (\beta_{\al},
\la_{\al})$ with $\al=0,1$; we consider the limit $\beta_1=\beta+ \de \beta$ 
and $\la_1 = \la_0 + \de \la$ where perturbation theory is applicable. In the 
limit $\beta_0,\beta_1 \to \infty$, (\ref{eq_fidthermal}) reduces
to the ground state fidelity. When $\de \la =0$ and $\la_0= \la$, one can 
define the thermal fidelity which can be simplified to the form 
$$F_{\beta} (\beta_0,\beta_1,\la) = \frac{{\cal Z}\left(\frac{\beta_0 + 
\beta_1}{2},\la \right)}{{\sqrt{{\cal Z}(\beta_0,\la){\cal Z}
(\beta_1,\la)}}}.$$ $F_{\beta}$ can be related to well-known thermodynamic 
quantities \ct{zanardi07,you07,yang08,tzeng08}, e.g., the FS defined as 
$\chi_F = -2 \ln F/(\de \beta)^2|_{\de \beta \to 0}$ can be shown to be 
proportional to the specific heat, and has been useful for studying the 
classical phase transition at a finite temperature \ct{paunkovic081,quan09a}. 
On the other hand, $F(\la,\la+\de\la,\beta)$ defined at a fixed $\beta$ has 
been employed to detect the finite temperature signature of a QPT as occurs 
in the transverse Ising chain \ct{zanardi071,zanardi072}. It has been found 
that although there is a decay in the fidelity right at the QCP at low
temperatures, the sharpness diminishes with increasing temperatures and 
the decay is eventually wiped out at high temperatures.

\subsection{Quantum critical environment: decoherence and Loschmidt echo}
\label{lecho}

In recent years, there has been great interest in studies of decoherence,
namely, the quantum-classical transition by a reduction from a pure state to 
a mixed 
state \ct{zurek91,haroche98,joos03,zurek03}, specially from the viewpoint of 
quantum measurements and computations when a quantum system or qubit is coupled
to a macroscopic environment. In connection to the quantum-classical 
transition in quantum chaos, the concept of Loschmidt echo (LE) has been 
proposed to describe the hyper-sensitivity of the time evolution of the 
system to the perturbations experienced by the surrounding environment
\ct{peres95,karkuszewski02,cucchietti03,jalabert01}. 
Zurek and Paz \ct{zurek94}
argued that for a quantum system with a classically chaotic Hamiltonian, the 
rate at which the environment loses information about the initial state (e.g., 
the rate of production of von Neumann entropy calculated using the reduced 
density matrix of the system) is independent of the coupling strength between 
the system and the environment and is governed by the classical Lyapunov 
exponent for a wide window of the coupling strength.

The measure of the LE is the overlap between two states that evolve from the 
same initial state $\ket{\psi_0}$ under the influence of two Hamiltonians 
{$H_0$ and $H_0+V$, where $V$ is a perturbation}; the 
mathematical expression is
\beq L(t) = |\langle \psi_0 |e^{i(H_0 + V)t} e^{-iH_0 t}|\psi_0\rangle|^2.
\label{eq_deflsecho} \eeq
In some of the recent works, attempt has been made to connect these two fields 
by studying the behavior of the LE close to a QCP as a probe
to detect the quantum criticality. 
In this section, we will study the dynamics of a central spin coupled to a 
spin system (environment) and discuss that the decay of the Loschmidt echo can
indicate the proximity to a QPT in the surrounding spin chain \ct{quan06}.
In the process, we shall also point out the close connection between 
quantum fidelity and the echo.

\begin{figure}[h]
\begin{center}
$\begin{array}{cc}
\multicolumn{1}{l}{\mbox{\bf (a)}} &
	\multicolumn{1}{l}{\mbox{\bf (b)}} \\[-0.53cm]
\epsfxsize=1.5in
\epsfysize=1.5in
\epsffile{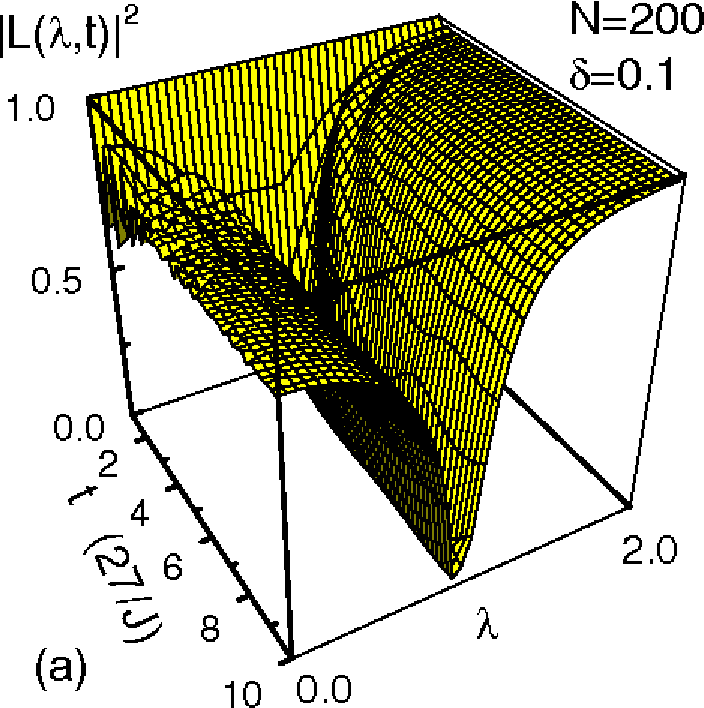} &
	\epsfxsize=1.5in
\epsfysize=1.5in
	\epsffile{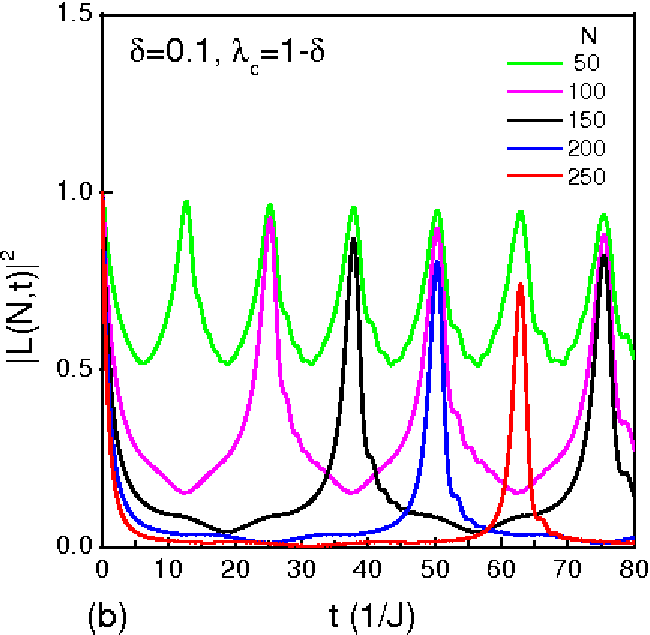} \\ [0.4cm]
%\mbox{\bf (aa)} & \mbox{\bf (bb)}
\end{array}$
\end{center}
\caption{(Color online) 
In the figure $\la=h$. Left panel: Three-dimensional ($3d$) diagram 
of the LE, $|L(\la, t)|^2$, as a function of $\la$ and $t$ for a system with
$N =200$. The valley around the critical point $\la_c=1$ indicates that the 
decay of the LE is enhanced by the QPT. Right panel: The cross section of the
$3d$ plot for different system sizes and $\la = \la_c -0.1=0.9$. It shows that
the period of the LE depends on the size of the surrounding spin chain.
(After Quan $et~al$, 2006).} \label{fig_lsecho}
\end{figure}

We consider the central spin model in which a central spin-1/2 (qubit) is coupled to the environment which is chosen to be
a transverse $XY$ chain in ({\ref{eq_txy}) with $N$ spins. 
Denoting the ground state and excited wave functions of the 
central spin as $\ket {g}$ and $\ket {e}$, respectively, one can write the 
Hamiltonian incorporating the interaction between the central spin and all 
the spins of the chain as \ct{quan06} 
\bea H ~=~ - ~\sum_{i=1}^{N-1} ~\left( J_x \si_i^x \si_{i+1}^x + J_y \si_i^y
\si_{i+1}^y \right) ~-~ h ~\sum_{i=1}^N ~\si_i^z ~-~ \de (J_x + J_y) ~
\sum_{i=1}^N ~\ket{e}\bra{e} \si_i^z, \label{eq_lsecho1} \eea
where the spin chain couples to only the excited state $\ket{e}$ of the central 
spin. One assumes that the central spin is in a pure state 
$\ket{\phi_s(t=0)} = c_e \ket {e} + c_g \ket g$ with $|c_e|^2+|c_g|^2=1$. If 
the spin chain is initially prepared in the ground state $\ket{\psi_0}$, then
the evolution of the spin chain gets split into two branches $\ket{ \psi_{\al}
(t)} = \exp (-i H_{\al}t) |\psi_0 \rangle$, with $\al=g,e$; the total wave 
function is $\psi_T = c_g \ket {g} \bigotimes \ket{ \psi_g (t)} + c_e \ket {e} 
\bigotimes \ket{ \psi_e (t)}$. We note that $H_{\al}$, for $\al =g$ and 
$e$, represents $XY$ chains with transverse fields $h$ and $h+
\de(J_x +J_y)$, respectively. Therefore the coupling between the central spin and the environmental spin chain provides two channels of evolution
of the spin chain with two different Hamiltonians. This leads to the decay of the Loschmidt echo 
defined as $L(J_x,J_y,\ga,t)=|\langle \psi_g(t)|\psi_e(t)\rangle|^2$.

Using the JW and Bogoliubov transformations (see 
Sec.~\ref{jorwigtr}), one can derive the exact analytical expression for 
the Ising case ($J_x=1$ and $J_y=0$),
\beq L(t) = \prod_{k>0} G_k = \prod_{k>0} \left[1 - \sin^2 (2 \al_k) \sin^2(
\om_k^e t) \right], \label{eq_lsecho2} \eeq
where $\al_k = [\theta_g^k - \theta_e^k]/2$ with $\theta_g^k= \arctan (-\sin 
k/(\cos k -h))$ and $\theta_e^k= \arctan (-\sin k/(\cos k -(h+\de))$, 
and $\om_k^e$ is the excitation energy for the transverse field $h+\de$. 
Defining a cut-off wave vector $k_c$, it has been shown \ct{quan06} that 
$L(t) \simeq \exp(-\Gamma t^2)$ as $h \to h_c=1$, where $\Gamma =\de^2 E(k_c)/
(1-h)^2$, $E(k_c) = 4 \pi^2 N_c(N_c + 1)(2N_c+1)/6N^2$, and $N_c$ is the 
integer closest to $n \pi k_c/(Na)$ with $a$ being the lattice spacing; 
hence one may expect an exponential decay 
of $L(t)$ in the vicinity of $h_c$ when $N$ is large. 
%On the other hand, if 
%$a \to 0$ and $N \to \infty$ such that the chain length $Na$ is constant, then
%$E(k_c) \sim 1/N^2$ so that numerator of $\Gamma$ vanishes. However, since 
%the denominator also tends to vanish as $h \to 1$, an exponential decay of 
%$L$ is still a possibility for a large finite system. 
Numerical studies for a 
finite system ($N =50$ to $250$ and $\de=0.1$) show (i) a highly enhanced decay
of $L$ at a fixed time around the critical point of the surrounding system, and
(ii) collapses and revivals of $L(t)$ as a function of time if $h+\de=1$,
the period of the revival being proportional to $N$, i.e., the size of the 
surrounding 
system as shown in Fig.~\ref{fig_lsecho}. The decay and revival of the 
echo is an indicator of a QPT. 
The present study has been generalized to the $XY$ spin chain \ct{yuan07,ou07},
% one finds that around the MCP, the LE is fixed unity at $h=1,\ga=0$ 
%(Fig.~{\ref{fig_xyphase}) while for small 
%$\ga$, LE shows a decay at short time and subsequent oscillations resulting
%from superposition of collapses and revivals. For higher $\ga$, one observes
%complete decay and revival as in the Ising case. 
where it has also been shown that
the ground state geometric phase of the central spin and its derivative with 
$h$ have a direct connection to the QPT of the surrounding system as discussed
in Sec.~\ref{fidelity}. The decay parameter $\gamma$ is found to scale as 
$1/N^{2z}$ which is verified close to the quantum MCP of the $XY$ spin chain
\ct{sharma12}.

%Recently, a direct observation of quantum criticality
%in Ising chains with antiferromagnetic couplings has been made using NMR 
%quantum simulators where the critical points of the transitions are detected 
%using the measurement of the Loschmidt echo \ct{zhang09}.
% We note that 
%Yi $et~al.$ \ct{yi06} studied the Hamiltonian in (\ref{eq_lsecho1}) 
%to calculate the Hahn
%echo \ct{hahn50} which measures the transverse relaxation or the dephasing 
%time and showed that the critical points of the surrounding chain are marked 
%by the extremal values of the Hahn echo.
Rossini $et~al$ \ct{rossini07}, studied a generalized central spin model in 
which the qubit interacts with a single spin of the environmental transverse 
Ising spin chain and it has been shown that the decay of the LE at short 
time is given by the Gaussian form
$\exp(-\Gamma t^2)$ where the decay rate $\Gamma$ depends on the symmetries 
of the phases around the critical point and the critical exponents.
For instance, for such systems with local coupling, it has also been reported 
that $\Gamma$ has a singularity in its first derivative as a function of 
the transverse field at the QCP \ct{rossini07}. 
In a subsequent work \ct{zhang09}, the LE has been used as a probe to
detect QPTs experimentally; at the same time, using a perturbative study in 
the short-time limit, the scaling relation $\Gamma \sim (\lambda)^{-2 z \nu}$ 
valid close to a QCP (at $\lambda=0$) has been proposed. 
%Here, $\nu$ and $z$ 
%are associated correlation length and dynamical exponents, respectively.
 In 
contrast to these studies where the coupling between the qubit and the 
environment is chosen to be weak, it has been shown that in
the limit of strong coupling the envelope of the echo becomes independent of 
the coupling strength which may arise due to a quantum phase transition in 
the surrounding \ct{cucchietti07a,cormick08}.
% Moreover the LE and the decoherence of the central spin has been studied 
% when the environmental transverse Ising spin chain is quenched across the 
% QCP by varying the transverse field linearly in time \ct{damski11}.

At this point, it is natural to seek a connection between the dynamical LE 
approach and the geometrical fidelity approach discussed in 
Sec.~\ref{fidelity}. In this regard, one defines a projected density of state 
function $D(\om;\la,\la') = \langle \psi_0(\la')|\de(\om -H(\la))
\ket{\psi_0 (\la)}$ which is related to the square of the overlap as $|\langle 
\psi_0(\la)|\psi_0(\la')\rangle|^2= 1 -\int_{E_1}^{\infty} D(\om) d\om$, where
$E_1$ is the first excited energy. One can further show that $L(\la,t) = |
\int_{-\infty}^{+\infty} D(\om) \exp(-i\om t) d\om|^2$. These relations show 
that if the ground states are significantly different, the overlap tends to 
vanish leading to a broader $D(\om)$ and hence a faster decay of the 
LE \ct{zanardi06}.

Let us now indicate how the LE can be shown to be related 
to the decoherence 
(decay of the off-diagonal terms) of the two-level reduced density matrix 
of the central spin. More precisely, one defines a quantity called purity $P= 
Tr_S ({\rho_S})^2$ where $\rho_S$ is the reduced density matrix of the
central spin 
%obtained through integrating out the surrounding spin chain 
%Variables from the total density matrix $\rho = \ket {\psi_T}\bra {\psi_T}$ 
\ct{cucchietti03}. One can show that $P = 1 - 2 |c_g c_e|^2 (1 -L(t))$ (in fact, $L(t)$ appears in the off-diagonal term
of $\rho_s$ \ct{damski11});
therefore the central spin transits from the initial pure state to a mixed state due to the 
enhanced decay of the LE around the QCP.

\section{Non-equilibrium dynamics across quantum critical points}
\label{sec_intrononeq}

In this section, we will review recent studies of non-equilibrium dynamics 
of transverse field spin systems. The non-equilibrium dynamics of a transverse
$XY$ spin chain was first investigated in a series of papers 
\ct{barouch70,barouch71,barouch711} where the time evolution of the model was 
studied in the presence of various time-dependent magnetic fields and the
non-ergodic behavior of the magnetization was pointed out. A similar result 
was also obtained in \ct{mazur69}. 

There is a recent upsurge in studies of non-equilibrium dynamics of a quantum 
system swept across a QCP. These studies are important to explore the 
universality associated with quantum critical dynamics. Moreover, recent 
experiments with ultracold atomic gases 
\ct{greiner02,sadler06,lewenstein06,bloch08} have stimulated numerous 
theoretical studies. The main properties of these atomic gases are low 
dissipation rates and phase coherence over a long time so that the dynamics 
is well described by the usual quantum evolution of a closed system.

In the subsequent sections, we shall discuss that when a quantum system
is driven across a QCP, the dynamics fails to be adiabatic however
slow be the change in parameter of the Hamiltonian be. This is due to the 
divergence of the characteristic time scale of the quantum system, namely,
the relaxation time close to the QCP. This non-adiabaticity results in 
the occurrence of defects in the final state of the quantum 
Hamiltonian. Interestingly, the defect density scales in a power-law fashion 
with the rate of quenching, when a parameter of the Hamiltonian is changed 
linearly in time, with the exponent given in terms of some of the critical 
exponents associated with the QCP. This implies that even after a 
non-equilibrium quench across a QCP, the scaling of defect density is given 
in terms of some of the equilibrium quantum critical exponents.
We shall discuss slow dynamics across quantum critical and multicritical 
points and shall comment on their relevance in adiabatic quantum 
dynamics. We shall also discuss how the quantum information theoretic 
measures discussed in Sec. 5 scale following such a quench. In all these 
studies, the transverse Ising/$XY$ spin chains and the two-dimensional 
Kitaev model play a crucial role because of their integrability and the direct
product form of the ground state.\footnote{{To be more 
precise, integrability means that the system has so many conserved
operators that the eigenstates of the Hamiltonian can be completely 
distinguished from each other by the eigenvalues of those operators.
States with different eigenvalues of those operators do not mix under
the dynamics. This makes such systems easier to study, but it also implies 
that such systems are not ergodic and do not equilibrate in the same way 
as non-integrable systems.}} {However, in Appendix 
\ref{app_ll}, we incorporate a brief discussion on the quenching of 
Tomonaga-Luttinger liquids, and we discuss some other interesting and related 
developments in Appendix \ref{kz_space} and \ref{sec_topology}.} Our focus 
will be mainly limited to systems at 
zero-temperature although we shall briefly comment on the role of 
quantum quenching at a finite temperature. 

We note that TIMs have been studied in the presence of a 
time-varying and periodic transverse \ct{acharyya94,vbanerjee95,das09} 
%and longitudinal field \ct{miyashita09} and 
and the dynamics of domain walls has been examined \ct{subrahmanyam03}. For a 
recent mean field study of hysteresis in a TIM in a time varying longitudinal 
field, we refer to \ct{miyashita09}. 

\subsection{Rapid quenching through a QCP}

Let us first discuss what happens when a parameter in the 
Hamiltonian of a system is suddenly changed so as to take the system across a 
QCP \ct{sengupta04}. We assume that the system is in its ground state before
the quench. Immediately after the quench, the system will be in the same 
state; however, it will then start evolving in time since it is no longer an 
eigenstate of the new Hamiltonian. Our aim is to understand how quantities 
such as the order parameter change under this time evolution. 

Das $et~al.$\ct{das06} studied quenching dynamics of an infinite 
range ferromagnetic spin-1/2 Ising model in a transverse field (see 
(\ref{haminfinite}) with $\si^x \to \si^z$, $\si^z \to \si^x$ and $\ga=0$) 
due to a sudden variation of the transverse field
across the QCP (at $h_c=J/2$) from an initial value $h_i \gg h_c$ to a final 
value $h_f$. The dynamics of the equal-time order parameter correlation 
function (EOC) (defined as $\left<(S_z)^2\right> /S^2$, where 
{$S_z = \sum_i \si_i^z$)} has been 
studied. Expressing the initial ground state of the system in terms of the 
eigenstates $\left|n\right>$ of the new Hamiltonian $ \left|\psi \right> = 
\sum_n c_n \left|n\right>$, one obtains
%by changing the transverse field $h$ suddenly. In this
%case, just after the quench, the ground state of the system can be expressed
%in terms of the eigenstates $\left|n\right>$ of the new Hamiltonian
%$H_f = -(J/4S) (S_{tot}^z)^2 - h_f S_{tot}^x$ as
%\bea \left|\psi \right> &=& \sum_n c_n \left|n\right>, \eea
%where $c_n$ denotes the overlap of the
%eigenstate $\left|n\right>$ with the old ground state $\left|\psi \right>$. 
%As the state of the system evolves in time, it is given by
%\bea \left|\psi(t) \right> &=& \sum_n c_n e^{-iE_n t} \left|n \right>, \eea
%where $E_n = \left<n\right| H_f \left|n\right>$ are the energy
%eigenvalues of the Hamiltonian $H_f$. The EOC can thus be written as
\beq \left<\psi(t) \right|(S_z)^2 /S^2 \left|\psi(t)\right> ~=~ \sum_{m,n} ~
c_n c_m ~\cos \left[\left(E_n-E_m \right)t \right] ~\left< m \right| (S_z)^2 /
S^2 \left| n \right>. \label{opcorr} \eeq
A numerical solution of (\ref{opcorr}) shows that the amplitude of 
oscillations is maximum when $h_f$ is near $h_c$ \ct{das06}.
%Here, we have quenched the transverse fields to $h_f/J \equiv \Gamma_f = %
%0.9,\,0.01\,\,
%{\rm and}\,\, 0.4$ starting from $h_i/J=2.0$. The oscillation amplitudes 
%of the EOC for $S=100$, as shown in Fig.~\ref{rapid_fig1}, are small for 
%$h_f=0.9$ and $0.01$, and much larger for $h_f=0.4$.
The long-time averaged value, given by
$O = \lim_{T\to \infty} \left< \left< (S_z)^2(t) \right > \right>_T / S^2
= \frac{1}{S^2} \sum_n c_n^2 \left<n\right|(S_z)^2 \left|n \right>$,
%for different $h_f$. Note that the long-time average depends on the
%product of the overlap of the state $\left|n\right>$ with the old ground
%state and the expectation of $(S_{tot}^z)^2$ in that state. We expect $O$ to 
%have a peak somewhere near the critical point where such an overlap is
%maximized. This is verified by explicit numerical computation of Eq.
%(\ref{longtime}) in Fig.~\ref{rapid_fig2} for several values of $S$ and
peaks around $h_f/J=0.25$ ($= h_c/2$), and the peak height decreases slowly 
with increasing $S$ (see Fig.~\ref{rapid_fig2}).

%\begin{figure}[htb]\centerline{\epsfig{figure=RL10033_fig15.ps,width=8.5cm}}
%\caption{Dynamics of $\left<(S_{tot}^z)^2\right>/S^2$ for $S=100$ after
%quenching the transverse field to different values $h_f/J = \Gamma_f$ from an
%initial field $h_i/J=2$. The oscillation amplitudes are small, as seen from the
%solid (red) and dotted (blue) curves corresponding to $h_f/J=0.9$ and $0.01$ 
%respectively, far away from the critical point $h_c/J=0.5$. The oscillation 
%is large in the ordered phase near the critical point as seen from the 
%dashed (black) curve $h_f/J=0.4$. Inset shows the. 
%(After Das $et~al.$, 2006).} 
%\label{rapid_fig1} \end{figure}

\begin{figure}[htb] \centerline{\epsfig{figure=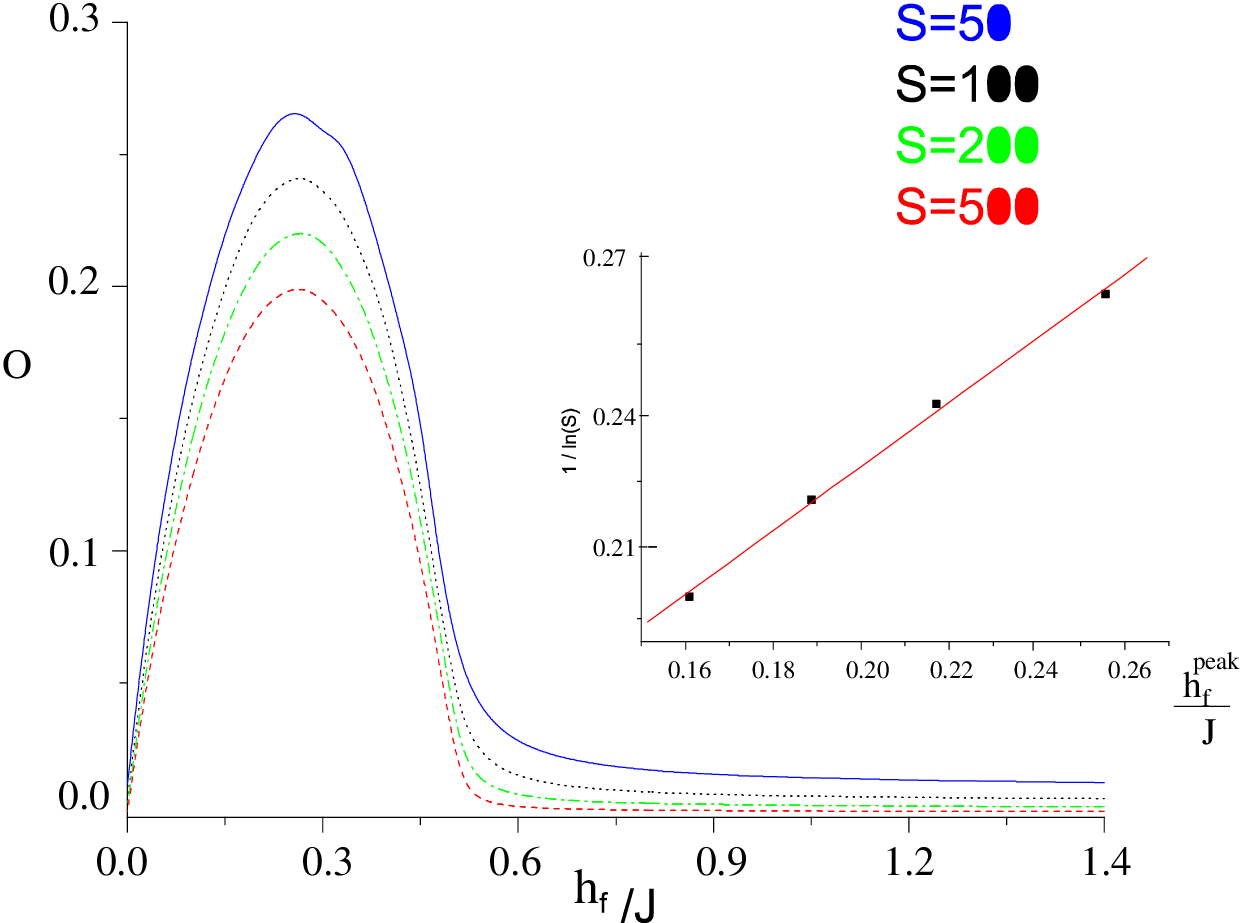,width=8.5cm}}
\caption{(Color online)
Plot of the long-time average $O$ as a function of $h_f/J$ for
different $S$. The solid (blue), dotted (black), dash-dotted (green) and the 
dashed (red) lines represent respectively the results for $S = 50$, $100$, 
$200$ and $500$. $O$ peaks around $h_f/J= 0.25$, and the peak value decreases 
with increasing $S$; $h_i/J=2$ for all the plots. Inset shows the plot of the 
maximum peak height $O_{\rm max}$ of the long-time 
average of the EOC as a function of $1/\ln(S)$. The straight line shows a
linear fit. (After Das $et~al.$, 2006).} \label{rapid_fig2} \end{figure}

%To understand the position and system size dependence of the
%peak in $O$, we now look at the thermodynamic (large system size)
%limit; in this model, this is given by the large $S$ and
To understand the above findings, we consider
the thermodynamic limit which is equivalent to the $S \to \infty$ 
(classical) limit for this model. One can study the classical equations 
of motion for ${\bf S} = S \left(\cos \phi \sin \theta, \sin \phi \sin 
\theta, \cos \theta \right)$
% rather of $\theta$ and $\phi$ for $h=h_f$,
% In the classical limit, we need to study the equations of motion for
%$\theta$ and $\phi$. To this end, we note that 
using the Lagrangian \ct{fradkin91}
$ L = - S ~\left[ 1-\cos \theta \right] ~d \phi /dt ~-~ H 
\left[\theta,\phi \right]$, with the initial condition 
$\theta=\pi/2 -\ep, \,\,\phi=\ep$, where $\ep$ is 
an arbitrarily small constant so that $S_x \approx S$ at $t=0$.
%This gives the equations of motion
%\bea \frac{d \theta}{dt} &=& h_f ~\sin \phi ~, \non \\
%\frac{d \phi}{dt} &=& -~\frac{J}{2} ~\cos \theta ~+~ h_f ~\cot \theta \cos 
%\phi. \label{eom1} \eea
%Eq.~(\ref{eom1}) has to be supplemented with the initial condition that
%$S_{tot}^x=S$ at $t=0$. This initial condition corresponds to $\theta=
%\pi/2,\,\,\phi=0$ which is, however, a fixed point of Eq.~(\ref{eom1}). 
%Therefore we will start from an initial condition which is very close to the 
%fixed point: $\theta=\pi/2 -\ep, \,\,\phi=\ep$, where $\ep$ is 
%an arbitrarily small constant. 
Since the motion occurs on a constant
energy surface after the quench has taken place, one gets
$h_f = (J/4) ~\cos^2 \theta ~+~ h_f ~ \sin \theta \cos \phi.$
%Using Eqs.~(\ref{eom1}) and (\ref{ec1}),
The equation of motion for $\theta$ then gives
%in closed form,
%\bea \frac{d \theta}{dt} =~ \frac{\sqrt{h_f^2 \sin^2 \left( \theta \right) -
%\left[ h_f -\frac{J}{4} \cos^2 \theta \right]^2}}{\sin \theta} ~\equiv f
%\left(\theta\right). \label{eom2} \eea
%It can be seen that the motion of $\theta$ is oscillatory and has classical
%turning points at $\theta_1 = \sin^{-1} \left(\left| 1-4h_f /J \right|\right)$
%and $\theta_2 = \pi/2$. One can now obtain $\left<(S_{tot}^z)^2\right>_T =
%\left<\cos^2 \theta \right>_T$ from Eq.~(\ref{eom2})$,
\bea \left<\cos^2 \theta \right>_T &=& {\mathcal N} /{\mathcal D}, 
\label{av1} \\
{\rm where} ~~~{\mathcal N} &=& \int_{\theta_1}^{\theta_2} d\theta ~
\frac{\cos^2 \theta}{f\left(\theta\right)} ~=~ 4 \sqrt{8 h_f \left( J-2 h_f 
\right)}/J, \non \\
{\rm and} ~~~{\mathcal D} &=& \int_{\theta_1}^{\theta_2} d\theta ~
\frac{1}{f \left(\theta\right)}, \non \eea
where $d \theta/dt = (\sqrt{h_f^2 \sin^2 \theta -
[ h_f -J/4 (\cos^2 \theta) ]^2})/\sin \theta$ $ ~\equiv f (\theta)$, and 
$\theta_1 = \sin^{-1} \left(\left| 1-4h_f /J \right|\right)$ and 
$\theta_2 = \pi/2$ are the classical turning points
obtained from the equation of motion of $\theta$.
Regularizing the end-point singularity in ${\mathcal D}$ at $\theta_2$,
%When trying to evaluate ${\mathcal D}$, we find that the integral has an
%end-point singularity at $\theta_2$; this can be regulated 
by a cut-off $\eta$ so that $\theta_2 = \pi/2 -\eta$, one finds that 
${\mathcal D} = -J \ln(\eta)/\sqrt{h_f \left(J-2h_f\right)/2}$. 
%The cut-off used here has a 
%physical meaning. To see this, note that the angles
%$(\theta,\phi)$ define the surface of a unit sphere of area $4\pi$.
%This surface, for a system with spin $S$, is also the phase space of the 
%system which has $2S+1$ quantum mechanical states. For large $S$, the area
%of the surface occupied by each quantum mechanical state is
%therefore $4\pi /(2S+1) \simeq 2\pi/S$. In other words, each quantum
%mechanical state has a linear dimension of order $1/\sqrt{S}$;
%this is how close we can get to a given point on the surface of the
%sphere. Note that this closeness is determined purely by quantum
%fluctuations and vanishes for $S\to \infty$. 
Note that $\eta$ is a measure of how close to the point 
$\theta=\pi/2$ we can get and must be of the order of $1/\sqrt{S}$.
%this determines the system-size dependence of
%$\left<\cos^2 \theta \right>_T$. Using Eq.~(\ref{av1}), 
One finally gets
\bea \left<\cos^2 \theta \right>_T &=& \frac{16 h_f \left(J -2 h_f \right)}{
J^2 \ln(S)}. \label{cr1} \eea
Equation (\ref{cr1}) 
%is one of the main results of this section. It 
demonstrates that the long-time average of the EOC must be
peaked at $h_f/J=0.25$ which agrees perfectly with the
quantum mechanical numerical analysis leading to Fig.~\ref{rapid_fig2}.
Moreover, it provides an analytical understanding of the $S$ (and
hence system size) dependence of the peak values of $h_f/J$. A
plot of the peak height of $O$ as a function of $1/\ln(S)$ indeed fits a
straight line as shown in the inset of Fig.~\ref{rapid_fig2}. 
%as shown in Fig.~\ref{rapid_fig3}. So we conclude that the peak
%in $O$ vanishes logarithmically with the system size $S$. Such a slow
%variation with $S$ shows that it might be experimentally possible to observe
%an experimental signature of a QCP for a possible realization of this model
%with ultracold atoms where $N \sim 10^5-10^6$ \ct{das06}.

The issue of thermalization after a rapid quench has been studied
considering the non-equilibrium dynamics of a one-dimensional
Ising model in a transverse field where the system is prepared in the 
ground state corresponding to $h_0$ \ct{rossini09,rossini10}. At $t=0$, the 
field is suddenly changed to $h$ so that the system is no longer in the 
ground state and hence relaxes in some fashion. The situation is equivalent 
to studying the temporal evolution of the correlation functions starting from 
a state which is not an eigenstate of the Hamiltonian \ct{igloi00}. The 
decay with time of the autocorrelation function defined
as $$C_{zz} (t) = \langle 0_i|\si_i^z(t)\si_i^z(0)|0_i\rangle = \langle 0_i| 
e^{iH_f t}\si_i^z e^{-iH_ft}\si_i^z|0_i\rangle,$$
where $|0_i\rangle$ is the initial ground state and $H_f$ is the final 
Hamiltonian, has been studied. As we have seen before in 
Sec.~\ref{jorwigtr}, $\si_i^x$ is non-local in terms of the fermions while 
$\si_i^z$ is quadratic in the fermions; this leads to completely different 
asymptotic behaviors of their autocorrelation functions. The order parameter 
autocorrelation function $C_{xx} (t)$ shows an effective thermal behavior, 
decaying exponentially in time even though the underlying Hamiltonian is 
integrable; the decay time depends on the initial and final values of the 
field. {In an integrable system, the states can be classified 
into a 
number of sectors defined by the eigenvalues of a large number of conserved
operators. States lying in different sectors do not mix under the dynamics;
this prevents the system from equilibrating as completely as non-integrable
systems}. The effective temperature can be deduced in two ways: from the decay 
time at finite temperatures, and from the excess energy $\langle 0_i | H_f|
0_i \rangle - E_{0f}$, 
where $E_{0f}$ is the ground state energy of the final Hamiltonian. The 
autocorrelation $C_{zz}(t)$, on the other hand, does not show thermalization
and decays only as a power-law with the exponent given by $1,3/2$ and $2$ for 
$g_f <0$,$=0$, and $>0$, respectively. The exponential decay of $C_{xx}(t)$ 
when the quench is in the FM region is shown in Fig.~\ref{fig_davide}. 
{This problem has been revisited from the point of view of 
the fluctuation-dissipation relations in a recent work \ct{foini11}}.

%Finally, we would like to mention some interesting work \ct{calabrese06} 
%in which the Hamiltonian of a system is suddenly quenched across a QCP 
%and the space-time dependent two-point correlation function is then 
%calculated under evolution by the new Hamiltonian. It was shown that the 
%asymptotic form of the correlation function can be found using boundary 
%conformal field theory.

A proposal has been made \ct{wichterich09} to exploit rapid 
quenching dynamics in spin chains for generating distant entanglement and 
quantum teleportation \ct{venuti071}.

The results discussed in this section can be tested in two kinds of
experimental systems. One class of systems are those with long-range
dipole-dipole interactions such as KH$_2$PO$_4$ or Dy(C$_2$H$_5$SO$_4$)$_3
$9H$_2$O (see Sec.~\ref{expt2}) which exhibit order-disorder transitions
driven by tunneling fields. The other class of systems are two-component
Bose-Einstein condensates where the inter-species interaction is
strong compared to the intra-species interaction; the relative strengths of
these interactions can be changed by tuning the system to be near a Feshbach
resonance as discussed for the $^{41}K-^{87}Rb$ system in
\ct{gordon99,micheli03,hines03,simoni03,cirac98}. The quench dynamics
that we have discussed can be realized by applying a radio frequency pulse 
to the system and suddenly changing the frequency of the pulse.

%\begin{figure}[htb] \centerline{\epsfig{figure=RL10033_fig17.ps,width=8.5cm}}
%\caption{ (After Das $et~al.$, 2006).} \label{rapid_fig3} \end{figure}

\begin{figure} \begin{center}
\includegraphics[height=2.0in]{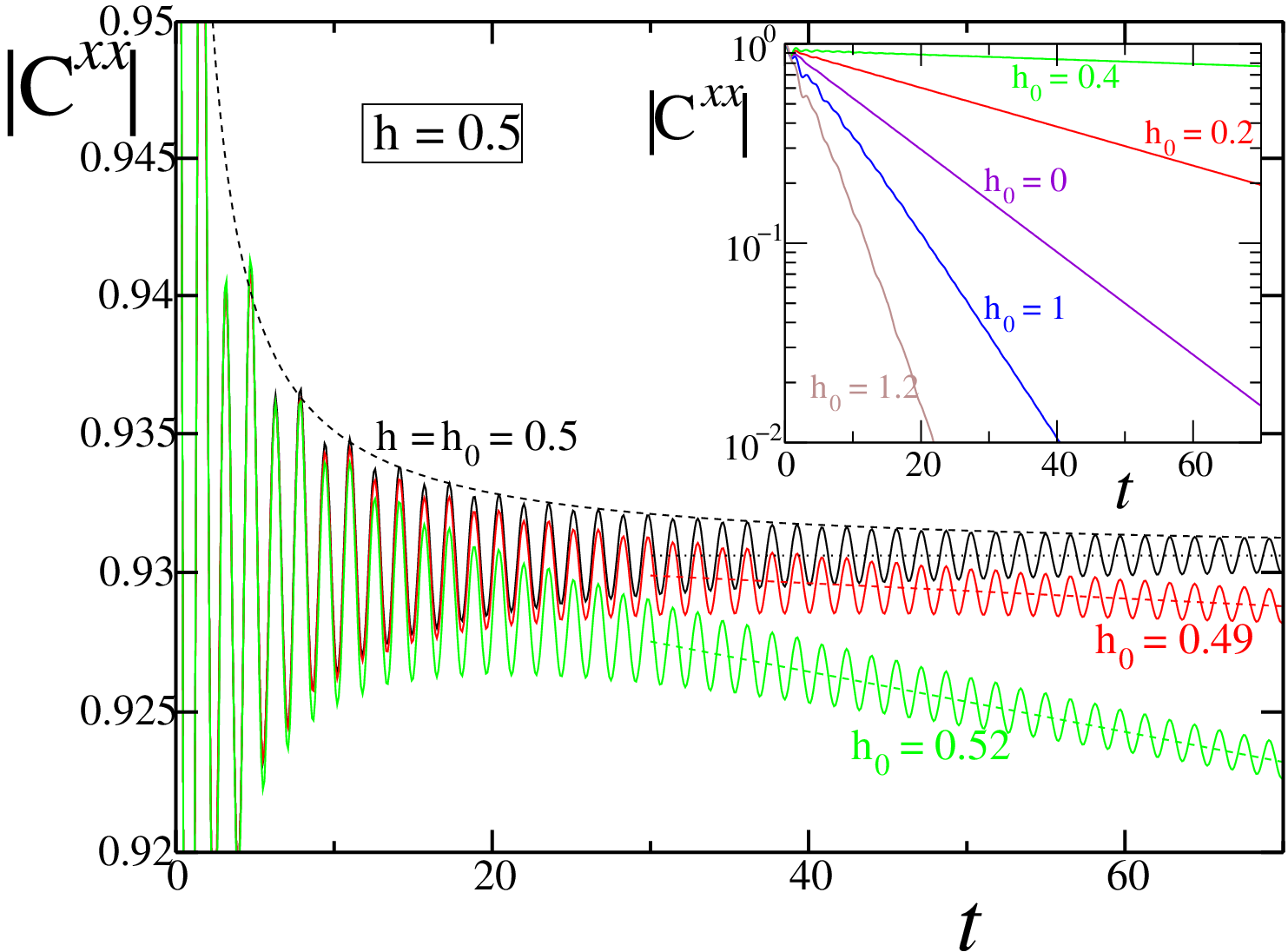} 
\caption{(Color online)
The exponential decay of correlators for different values of initial 
transverse field with final transverse field 
set equal to $0.5$. In the main frame, the solid curve corresponds to 
equilibrium: the horizontal dotted line indicates the asymptotic value of
$\langle \si^x \rangle^2$, while the dashed line denotes the power-law 
envelope $t^{-1}$. The inset shows data for strong quenches to show the 
exponential decay clearly. (After Rossini $et~al.$, 2009).} \label{fig_davide}
\end{center} \end{figure}

%\subsection{Rapid quenching and generalized susceptibilities}

%\ct{degrandi10}

\subsection{Defect generation: Kibble-Zurek (KZ) scaling}
\label{sec_kzs}

Recently a remarkable intersection of the fields of cosmology and condensed 
matter systems has caught the attention of physicists \ct{volovik03,kibble76}. 
In general, a symmetry breaking phase transition produces some defects. For 
example, when water freezes into ice below a critical temperature, the 
rotational symmetry is broken by the choice of the orientation of the crystal 
of ice. The
crystalline ordering sometimes fails especially in regions where one crystal 
meets another. In a similar spirit, Kibble in 1976 \ct{kibble76,kibble80} 
proposed that in the early universe too, symmetry breaking 
phase transitions might have led to the formation of topological defects, 
e.g., point-like monopoles, linear cosmic strings or planar domain walls. 

Later in 1985, Zurek pointed out that some of the low-temperature
phase transitions occurring in condensed matter systems can be used to mimic 
the formation of cosmic strings in the laboratory, for example, the vortices
formed in the normal to superfluid transition in liquid helium 
\ct{zurek85,zurek93,zurek96}. Zurek argued that in a continuous phase 
transition, a power-law relation exists between $\xi$ (the typical distance 
between two cosmic strings or defects) and 
the time taken by the system to pass through the phase transition. This 
power-law relation has been tested numerically \ct{laguna97} and 
experimentally in various systems, such as non-linear optical systems made up 
of liquid crystals \ct{chuang91,bowick94,ducci99}, annular Josephson junctions
\ct{monaco06}, and superfluids \ct{ruutu96}. Even though most of the
experimental data are in agreement with the theoretical predictions, the 
situation in liquid He$^4$ is not clear \ct{hendry94,dodd98}. 

The Kibble-Zurek scaling proposed for second order CPTs has recently been 
generalized to the quantum case independently by Zurek, Dorner and Zoller 
\ct{zurek05} and by Polkovnikov \ct{polkovnikov05}. To present the KZ 
argument extended to the quantum scenario, we note that according to the 
adiabatic theorem \ct{messiah99,suzuki05}, if the time scale associated with 
the minimum gap between the two states of a two-level system is sufficiently
larger than the rate at which a parameter of the two-level Hamiltonian is 
varied (called the adiabatic condition), the system remains in its initial 
state. If the gap between the two levels is smaller than the rate at which the 
parameter is varied, we have a non-adiabatic process which leads to 
excitations. 

In a driven quantum many-body system, the characteristic time scale is the 
relaxation time (inverse of the minimum gap) which diverges at the critical 
point. If the relaxation time of the system is smaller than the time scale on 
which the Hamiltonian is changed, the system is able to respond to the
changes in the Hamiltonian and it remains in the ground state. Clearly, the
adiabatic condition is satisfied when the system is far away from the critical 
point. Near the critical point, the relaxation time is large and the 
system is no longer able to follow the change in the Hamiltonian and hence 
non-adiabatic excitations appear. The time $\hat t$ at which the
behavior of the system changes from adiabatic to non-adiabatic can be
obtained as follows. Let us assume that the parameter $g$ is varied
such that it approaches the critical point $g_c$ as $t/\tau$, or $\la=
g-g_c = t/\tau$, such that the QCP occurs at $t=0$. As explained before, at 
$\hat t$, the relaxation time $\xi_{\tau}$ of the system is of the order of 
the time scale at which the Hamiltonian is changed; hence
\bea \left\vert \frac{\la}{\dot \la} \right\vert_{\hat t} = \hat 
t = \left . \xi_{\tau} \right \vert_{\hat t}. \label{hatt} \eea
This relaxation time diverges as $\xi_{\tau}\sim |g-g_c|^{-\nu z}$ which 
leads to
\bea \hat t &\sim& \left . (g-g_c)^{-\nu z} \right \vert_{\hat t}
\sim \left( \frac{\hat t} {\tau}\right)^{-\nu z}, \non \\
{\rm implying} ~~~{\hat t} &\sim& \tau^{\nu z/(\nu z+1)}. \label{hatt2} \eea
The times $\pm \hat t$ divide the entire evolution into three regions as 
shown in Fig.~\ref{fig_impulse}.

\begin{figure} 
\begin{center}
\includegraphics[height=2.0in]{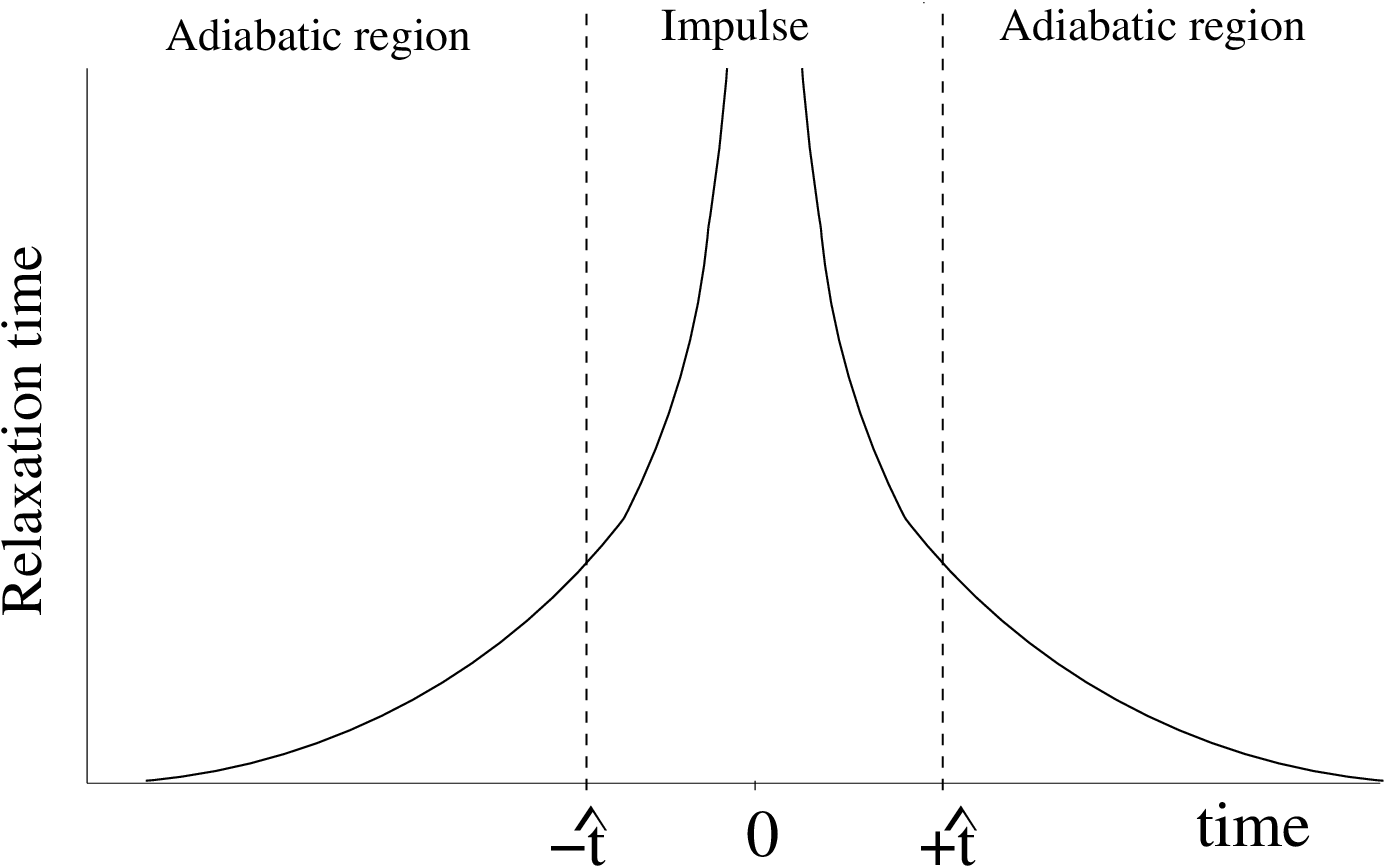}
\caption{Kibble-Zurek argument: the division of the entire time evolution into
three regions. For $t< |\hat t|$, the system is in the impulse region, whereas
for $t<-\hat t$ and $t>\hat t$, it is in the adiabatic region. (After 
Damski and Zurek, 2006).} \label{fig_impulse} 
\end{center}
\end{figure} 

According to the above argument, if the ground state wave function at 
$-\hat t$ is $\psi(-\hat t)$, the system does not evolve at all in the 
impulse region and hence remains in the same state $\psi(-\hat t)$ at time 
$+\hat t$ also \ct{damski05,damski06}; thus it loses track of the 
instantaneous ground state which leads to excitations. For large $\tau$ or 
slow variation of $g$, this argument is only valid for the low-energy modes,
while the high-energy modes do evolve adiabatically. Combining $\hat t \sim 
\xi_{\tau}$ (see (\ref{hatt})) and the healing length $\hat \xi \sim 
\xi_{\tau}^{1/z}$, and assuming that there is one defect per unit domain size 
of linear dimension $\hat \xi$, the density of defects $n$ is found to be
\bea n \sim \frac{1}{\hat \xi^d} \sim \tau^{-d\nu/(\nu z +1)},
\label{eq_kzm} \eea
where $d$ is the dimensionality of the system. Thus the Kibble-Zurek scaling 
(KZS) connects the equilibrium quantum critical exponents to the 
non-equilibrium dynamics and points towards universality even in the case of 
non-equilibrium dynamics through a QCP. The scaling in
(\ref{eq_kzm}) was independently derived in \ct{polkovnikov05},
using adiabatic perturbation theory, and the existence of an upper critical
dimension $d_u^c=2z+2/\nu$ was pointed out where the exponent
of the scaling saturates to 2 with possible logarithmic corrections to the 
scaling. For $d >d_u^c$, the KZS breaks down and one gets $n\sim 1/\tau^2$;
the main contribution to the defects then comes from the high-energy modes.

The KZS is also relevant to adiabatic transitions in a finite system 
of linear dimension $L$. In this case, the healing length $\hat \xi$ is 
limited by $L$ and one therefore gets a minimum value of $\tau =\tau_0 \sim 
L^{(1+\nu z)/\nu}$ such that for slower quenches ($ \tau \gg \tau_0)$, one can
achieve a perfect adiabatic transition \ct{zurek05,dziarmaga05}. It is in 
fact the finite energy gap at the QCP for a finite system that suppresses 
any excitation exponentially.

The KZS can also be arrived at using a phase space argument 
\ct{polkovnikov05}. Let us assume that the system is described by
the Hamiltonian $H(\la(t))$, and the time evolution of the parameter 
$\la(t)= \la_0 (t/\tau)$ takes the system through the critical point 
$\la=0$ at $t=0$. In the limit of large $\tau$, a non-vanishing 
probability of excitations requires the non-adiabaticity condition 
\ct{polkovnikov05,degrandi101} $\partial \De(\vec k, t)/\partial t \sim 
\De^2$, where $\De$ is the characteristic energy-scale near the QCP 
scaling as $\De \sim \la^{z\nu}$. Using $\partial \De(t)/\partial t=
[\partial \De(\la)/\partial \la)] \tau^{-1}$, the non-adiabaticity 
condition can be rewritten as $ \De^2 \sim \tau^{-1} \la^{z\nu -1}$. Near 
a QCP, we generally expect $\De \sim k^z$ and $\la \sim k^{1/\nu}$; the 
non-adiabaticity condition then
yields $k \sim \tau^{-\nu/(\nu z+1)}$ leading to a scaling of the gap
as $\De \sim \tau^{-z\nu/(z\nu+1)}$. The available phase space for 
quasiparticle excitations is given by $\Omega \sim k^d \sim \De^{d/z}\sim 
\tau^{-\nu d/(\nu z+1)}$ which is directly proportional to the defect density 
$n$, thus yielding the KZS. 

We will briefly discuss some interesting generalizations of the KZS to 
different quenching schemes in the following.

\noi (i) {\it A non-linear variation of the quenching parameter} 
\ct{sen08,barankov08}:
The quenching parameter $\la(t)$ varies as $\la_0 |t/
\tau|^{r}{\rm sign} (t)$, where $r$ denotes the power-law exponent and 
sign denotes the sign function. We consider the case when the critical point 
is crossed at time $t_0= 0$; the case $t_0\ne 0$ requires special attention 
\ct{sen08,mondal09}. Following the same line of arguments as in the linear 
case, we get $\hat t = \tau^{r \nu z/(1+ r \nu z)}$ which leads to
\bea n \sim \tau^{-r \nu d/(1+ r\nu z)}.
\label{eq_intrononlinearn} \eea
Thus the non-linearity exponent $r$ seems to renormalize the critical 
exponent $\nu$ to $r \nu$ which can also be seen employing the phase
space argument discussed above.

\noi (ii) {\it Quenching across a $(d-m)$-dimensional gapless surface} 
\ct{sengupta08,mondal08}: Let us consider a $d$-dimensional quantum system 
quenched across a $(d-m)$-dimensional critical hyperspace on which the energy 
gap vanishes. We then find that the phase space available for excitations 
gets modified to $\Omega \sim k^m$ which leads to a modified KZS given by
\bea n \sim \tau^{-m \nu/(1+\nu z)}. \label{eq_gapless1} \eea
For $m=d$ the standard KZS in ~(\ref{eq_kzm}) is retrieved.

\noi (iii) {\it Quenching through an anisotropic quantum critical point:}
The phase space argument can be extended to a $(d,m)$ anisotropic QCP 
\ct{dutta10,hikichi10} which is defined by $\De \sim k_i^{z_1}$ for $m$ 
momentum components while $\De \sim k_i^{z_2}$ for the 
remaining $d-m$ momentum components, with $z_2 >z_1$. For a non-linear 
quenching, one uses the relations $\De \sim \tau^{-\al\nu z/(1 + r \nu z)}$ 
and the phase space available for quasiparticle excitations, $\Omega 
= \prod_i k_i$. Since different momentum components scale differently with 
energy, one gets the scaling of the defect density given by \ct{hikichi10}
\bea n \sim \tau^{-[m+(d-m)z_1/z_2] \nu_1 \al/(1 + \al\nu_1 z_1)},
\label{eq_anisotropic} \eea
where $d=m$ is the standard KZS. The possibility of verifying 
the scaling relation in (\ref{eq_anisotropic}) for quenching through 
gapless Dirac points as observed in the low-energy properties of graphene 
\ct{castroneto09}, semi-Dirac \ct{banerjee09} and quadratic band crossing 
points \ct{sun09} has been proposed in \ct{dutta10}. The quench 
dynamics and scaling of defect density in monolayer and bilayer graphene
have also been studied in recent years \ct{dora10,dora10a}. 
The feasibility of studying graphene physics experimentally by loading 
ultracold fermionic atoms in a two-dimensional optical lattice has 
been reported \ct{lee09}.

{We have discussed above the KZS when the quenching parameter 
varies in time. One may wonder what happens when it varies in space. We defer 
this discussion to Appendix \ref{kz_space}. In Appendix \ref{sec_topology}, 
on the other hand, we discuss how topology influences the 
traditional KZS given in Eq.~(\ref{eq_kzm}).} 

We should note here that it is sometimes convenient to discuss the residual
energy per site, $e_r$, rather than the defect density $n$. The residual
energy is defined as the difference between the expectation value of the
final Hamiltonian in the state which is reached after the quench and
the ground state energy of the final Hamiltonian. Although $e_r$ and $n$
are often related to each other and scale in the same way with the 
quenching time $\tau$, there are situations where they scale differently;
this happens for disordered systems as we will discuss in 
Sec.~\ref{sec_quench_disorder}.

Contrary to the classical result, {the KZS at quantum 
critical points} is relatively new and has not been 
verified experimentally till now, although the possibility of verification in 
cold atom systems has been proposed \ct{sadler06,bloch08}. The formation of 
topological defects in the form of spin vortices in a quantum system following
a rapid quench across a QCP has already been reported \ct{sadler06} using
$^{87}$Rb spinor condensates. The Hamiltonian of $F=1$ $^{87}$Rb spinor gases 
has two terms (here $F$ is the sum of electronic and nuclear spin angular 
momenta), namely, $c_2 n^2 \langle {\vec J} \rangle^2 + q \langle J_z^2 
\rangle$. The first term (which is proportional to the square of the
density of atoms, $n$)
favors a FM phase with broken symmetry, whereas the second term (which depends
quadratically on a magnetic field applied in the $\hat z$ direction, $q \sim 
B_z^2$) favors a phase with $J_z = 0$; a quantum 
phase transition between the two phases occurs at some critical field. After 
preparing optically trapped BECs in the phase with $J_z = 0$ at a high field, 
the magnitude of the field is ramped down to a small value so that the FM
phase becomes favorable. The formation of small FM domains is observed 
throughout the condensate divided by unmagnetized domain walls as shown in
Fig.~\ref{fig_sadler}. The different domains correspond to different
orientations of $\langle {\vec J} \rangle$ in the $x-y$ plane; the domains 
grow with time which is indicated at the top of the figure in $ms$. This 
experiment {demonstrated for the first time} the formation of 
topological defects in a system undergoing a quantum phase transition. 

\begin{figure} 
\begin{center}
\includegraphics[height=3.2in]{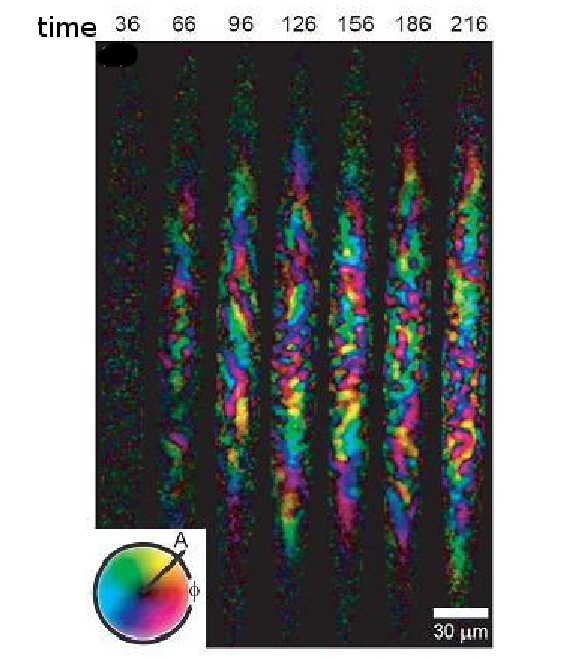}
\caption{(Color online) Ferromagnetic domains shown at different times after 
a quench (time in $ms$). The maximum brightness, shown by the color wheel on 
the left, corresponds to full magnetization of the condensate center. Near the
center of the condensate, one can see small magnetic domains divided by 
domain walls with zero magnetization (After Sadler $et~al.$, 2006).}
\label{fig_sadler} 
\end{center} 
\end{figure}

Following the prediction of the KZS, there have been a number of studies 
exploring the non-equilibrium dynamics of various quantum many-body systems 
\ct{saito07,uhlmann07,uhlmann10,uhlmann10a,bermudez09,bermudez10,polkovnikov08,fubini07,
mathey10} using various quenching protocol, for example, Rosen-Zener quenching 
\ct{klich07,rosen32}.
We limit our discussions mainly to studies related to transverse field spin 
systems, and refer to the review articles, 
\ct{dziarmaga09,schutzhold10,divakaran093,degrandi09,mondal091,gritsev10,
polkovnikov10,carr10}, for transverse field models as well as more generic 
quantum many-body systems. {We present a note on the 
quenching of Tomonaga-Luttinger liquids in Appendix \ref{app_ll}}.

\subsection{Adiabatic perturbation theory: Slow and Sudden Quenches}

Let us briefly discuss the adiabatic perturbation theory with which one can 
rederive the KZS \ct{polkovnikov05} given in (\ref{eq_kzm}) for a slow
quenching across a QCP. Following that we shall consider the limit of sudden 
quenching and review the concept of a generalized fidelity susceptibility 
that leads to the scaling of defects in the slow as well as the sudden limit. 

We consider a Hamiltonian $H(\la)=H_0+\la(t)H_I$ with a linear variation
of $\la$ which goes through a QCP at $t=0$. The wave function 
$|\psi(t)\rangle$ at any time $t$ can be expanded in its 
instantaneous (adiabatic) basis $|q(t)\rangle$ as 
$$|\psi(t)\rangle=\sum_q a_q(t) |q(t) \rangle.$$
The Schr\"odinger equation can be written as 
\bea i\frac{\partial a_p}{\partial t} + i \sum_q a_q(t) \langle p |
\frac{\partial}{\partial t} |q\rangle = \epsilon_p(t) a_p(t), 
\label{eq_aptschr} \eea
where $\epsilon_p(t)$ is the instantaneous eigenvalue in the 
basis $|p(t)\rangle$. Performing a unitary transformation
\bea a_p(t) = \tilde a_p(t) e^{-i\int^t \epsilon_p(\lambda(t))dt}
=\tilde a_p(\lambda)e^{-i \tau \int^{\lambda}\epsilon_p(\lambda) d\lambda},
\label{intro_unitrans} \eea
and assuming that the system is prepared in the ground state $|0\rangle$
of the initial Hamiltonian so that only the term with $q=0$ contributes
to the total sum in Eq. \ref{eq_aptschr}, the total density of excitations
at final time $t=+\infty$ is given by
\beq
n \approx \sum_{p\ne 0}|a_p (\infty)|^2 
\approx \sum_{p\ne 0}\left | \int_{-\infty}^{\infty} d \lambda \langle p
|\frac{\partial}{\partial \lambda}|0\rangle e^{i\tau \int^{\lambda}
(\epsilon_p(\lambda')-\epsilon_0(\lambda'))d\lambda'}\right |^2.
\label{eq_aptn} \eeq
Assuming that the system is translationally invariant, the above equation 
can be rewritten in momentum space $k$ replacing the summation
by an integration as follows:
\bea
n \approx \int \frac{d^d k}{(2\pi)^d} \left| \int_{-\infty}^{\infty}d\lambda 
\langle k|\frac{\partial}{\partial \lambda}|0\rangle 
e^{i\tau \int^{\lambda}(\epsilon_k(\lambda')-\epsilon_0(\lambda'))d\lambda'} 
\right|^2. \label{eq_aptdefects} \eea
Near a quantum critical point, the excitation energy can be written using 
a scaling function $F(x)$ as
\bea \epsilon_k-\epsilon_0 \sim \lambda^{z \nu} F(k/\lambda^{\nu}),
\label{eq_introaptscaling1} \eea
such that for $x \gg 1$, $F(x) \propto x^z$ and for $x \to 0$, $F(x) \to $ 
constant giving the correct scalings as required.
To analyze Eq.~(\ref{eq_aptdefects}), we introduce the following
scaling ansatz for the matrix element given by
\bea \langle k|\frac{d}{d\lambda}|0\rangle &\sim&\frac{1}{
\lambda}G(k/\lambda^{\nu}), \label{eq_introaptscaling2} \eea
where $G(x)$ is another scaling function. The form of this scaling function 
can be obtained by the following argument: for small $\lambda^{\nu}/k$, the 
matrix element should be independent of $\lambda$, i.e., $G(x)\propto 
x^{-1/\nu}$ for $x\gg 1$. Redefining $\lambda=\zeta \tau^{-1/(z\nu +1)}$
and $k=\mu \tau^{-\nu/(z\nu+1)}$, we get
\bea
n \sim \int \frac{d^dk}{(2\pi)^d}|\tilde a_k|^2 = 
\tau^{-\frac{d\nu}{z\nu+1}}\int \frac{d^d \mu}{(2\pi)^d} |\tilde a(\mu)|^2, \\
%\label{eq_aptscaling1}\\
{\rm where}~~ \tilde a(\mu) = \int_{\zeta_i}^{\zeta_f} \frac{d\zeta}{\zeta}
G(\frac{\mu}{\zeta^{\nu}}) \exp\left[i\int_{\zeta_i}^{\zeta_f} d\zeta_1 
\zeta_1^{z\nu}F(\mu/\zeta_1^{\nu}) \right].
%\label{eq_aptscaing2}
\eea

Thus we have rederived the KZS using adiabatic perturbation theory.
Let us study the convergence behavior of the above integrals carefully.
The integral over $\zeta$ is always convergent because of the fast oscillating 
function at large $\zeta$ and the absence of singularities at small $\zeta$.
We now need to check the convergence of the integral over $\mu$ at 
large $\mu$. If this integral converges,
one gets KZS, but if it does not, the defect density is dominated
by the high energy modes and is given by $\tau^{-2}$ thereby destroying the
KZ universality. Using explicit asymptotics of the scaling functions $F(x)$ 
and $G(x)$ at large $x$ $(\mu \gg \zeta^{\nu})$, it can be shown that
$$\alpha(\mu) \propto \frac{1}{\mu^{z+1/\nu}}.$$
This shows that there is a crossover from the low-dimensional universal 
KZS to a $\tau^{-2}$ scaling at a critical dimension
given by $d_c=2z+2/\nu$.
For a detailed discussion, we refer to \ct{polkovnikov05}.

So far our discussion has been limited to slow quenching across a QCP.
Let us now define a variation of a parameter $\la$ in some fashion starting 
from the critical point ($\la=0$) at $t=0$ to a final value $\la_f$ at time 
$t=t_f$; the system is initially in the ground state. If $|t_f| < |\hat t|$ 
(defined in (\ref{hatt})), we are dealing with a sudden quench in the sense that
the wave function is frozen in this region if we consider the low-energy modes 
only. For sudden quenching, the response of the system depends on the final 
value of $\la_f $ but not on the protocol of changing $\la$; this is in 
contrast to the slow quenching ($t_f > \hat t$) discussed so far 
for which the system is sensitive to the quenching protocol but is 
independent of $\la_f$. The question one would like to address is whether
there is a KZS for such a sudden quenching of parameters. 

The defect density and heat density ($Q$) following
a sudden quench have been studied in recent years \ct{gritsev10,degrandi101};
the heat density is the excess energy above the new ground state. The 
advantage of using the heat density is that it can be computed even for 
non-integrable systems. But for an integrable system with non-interacting 
quasiparticles, it is more useful to define the defect density, which is a 
measure of the density of excited quasiparticles generated in the system.
When $\la$ is suddenly increased from $\la = 0$ to its final value $\la_f$, 
all the momentum modes $k \lesssim k_0 \sim \la_f^{\nu}$ get excited with unit 
probability having excitation energy $\sim \la_f^{\nu z}$ for each mode. This 
gives the defect density $n \sim k_0^d \sim \la_f^{d\nu}$ and an excess energy 
density or heat density of the form $Q \sim \la_f^{\nu z} k_0^d \sim 
\la_f^{\nu(d+z)}$. The defect density is related to the probability of 
excitations, which in turn can be expressed in terms of the FS (see 
Sec.~\ref{fidelity}). From adiabatic perturbation theory 
\ct{gu09,degrandi102,degrandi101,gritsev10} one finds that
$n \sim \la_f^2 \chi_F \sim \la_f^{\nu d}$.
% this scaling 
%can also be derived by noticing that for a sudden quench of amplitude 
%$\la_f$, all the momentum modes $k_1 \lesssim \la_f^\nu$ get excited 
%with unit probability giving $n \sim \la_f^{\nu d}$. 

The concept of a generalized FS has been introduced in \ct{degrandi101}, 
using a generic quench from a QCP at time $t=0$ given by
\beq \la(t) = \de\frac{t^l}{l!} \Theta (t), \label{eq_genquench} \eeq
where $\de$ is a small parameter, and $\Theta$ is the step function. The 
case $l = 0$ denotes a rapid quench of amplitude $\de$, the case $l = 1$ 
implies a slow linear quench with a rate $\de$, and so on. In all these 
cases the limit $\de\to 0$ signifies a slow adiabatic time 
evolution. If the system is initially in the ground state, the transition 
probability to the instantaneous excited state as obtained from the adiabatic 
perturbation theory is given by 
\beq P_{ex} = \de^2 \sum_{n \ne 0}\frac{|\langle \psi_n | 
\frac{\partial H}{\partial \la} |\psi_0 \rangle|^2}{(E_n - E_0)^{2l + 2}} 
= \de^2 L^d \chi_{2l + 2}, \eeq
%which leads to a density of defect of the form
%\bea n_{ex} = \frac{1}{ L_{||}^m L_{\bot}^{d - m}} P_{ex} = \de^2 
%\chi_{\rm 2r + 2}. \eea
where we have defined a generalized fidelity susceptibility $\chi_m$ given by
\bea \chi_m = \frac{1}{L^d} \sum_{n \ne 0}\frac{|\langle \psi_n | 
\frac{\partial H}{\partial \la} |\psi_0 \rangle|^2}{(E_n - E_0)^m}.
\label{eq_gen_suscp} \eea
{}From (\ref{eq_gen_suscp}), one finds that $\chi_1$ stands for the 
specific heat density $\chi_E$, while $\chi_2$ is the fidelity 
susceptibility $\chi_F$; $\chi_4$, on the other hand, yields the excitation 
probability following a slow linear quench starting from a QCP.

In the same spirit as in (\ref{eq_fidcorr}), $\chi_m$ can also be 
expressed in terms of a time-dependent connected correlation function in 
imaginary time ($\bar \tau$) given by
\bea \chi_m = \frac{1}{L^d (m - 1)!} \int^\infty_0 d{\bar \tau} ~{\bar 
\tau}^{m - 1} \langle H_I(\bar\tau) H_I(0) \rangle_c. \label{eq_chil} \eea
Now, using $\la \sim L^{-1/\nu} $ and $\bar \tau \sim L^z$ in 
Eq. (\ref{eq_genquench}) leads to the scaling relation $L \sim 
\de^{-\nu/(1 + \nu z l)}$. One can further conclude (using the 
relevance or marginality of $H_I$) that $H_I \sim 
\la^{\nu z - 1} \sim \de^{(\nu z - 1)/(1 + l \nu z)}$,
and $\bar\tau \sim L^z \sim \de^{-\nu z/(1 + \nu z l)}$. 
Substituting for $L$, $H_I$ and $\bar \tau$ in (\ref{eq_chil}) 
with $m = 2l + 2$ one gets
\bea \chi_{ 2l + 2} &\sim& \de^{(\nu d -2 -2 \nu z l)/(1+\nu z l)} ~\sim~ 
L^{{2 zl +2/\nu -d}}, \non \\
P_{ex} &\sim& \de^2 L^{2/\nu + 2z l}, \label{eq_chil_scale} \eea
where we have assumed $d \nu < 2(1+z \nu l)$. The expression for $P_{ex}$ 
defined above is valid in the limit $\de \ll L^{1/\nu + zl}$ which ensures 
that $P_{ex} \ll 1$. One can therefore derive the scaling of the defect density
given by $n \sim \de^2 L^{(2/\nu -d +2 zl)}$ for $|\de|L^{1/\nu + zl} \gg 1$
and $n \sim \de^{d\nu/(z\nu l +1)}$ in the opposite limit. With $l=0$ and 
$\de = \la_f$, we retrieve the results for the sudden quench $n \sim 
\la_f^{\nu d}$. For $l=1$ and $\de =1/\tau$, we get back the scaling 
relation $n \sim \tau^{-\nu d/(\nu z +1)}$, while for $l >2$, the scaling of 
the defect density for non-linear quench is obtained. Similarly, one can 
derive the scaling of the heat density $Q \sim \la^{\nu z + \nu d}$ for $\de 
\gg L^{1/\nu + zr}$. We reiterate that the scaling relations presented above 
are valid as long as the corresponding exponents do not exceed 2. Otherwise 
contributions from short wavelength modes become dominant and hence the 
low-energy singularities associated with the critical point become 
subleading \ct{gritsev10, degrandi101}. Very recently,
a study of the influence of the geometric phase on the non-equilibrium dynamics
of a quantum many-body system has been reported \ct{tomka11}; it has been
shown that for fast driving the GP strongly affects transitions between levels, 
and the possibility of the emergence of a dynamical transition due to a 
competition between the geometrical and dynamical phases has been pointed out.

\subsection{Linear and non-linear slow evolution through critical points}

In this section and also in subsequent sections, we shall mainly focus on
slow quenching dynamics (defined through a rate of quenching) across critical 
and multicritical points.

Let us first illustrate the KZS taking the example of the $XY$ 
spin-1/2 chain in a transverse field to calculate the defect density produced 
when the spin chain is driven through critical points. Using the $2 \times 2$ 
form of the Hamiltonian (\ref{eq_ham2by2}), one use the Landau-Zener (LZ)
transition probability to calculate the probability of non-adiabatic 
transition and hence the defect density. We refer to Appendix \ref{app_lz} 
for a detailed discussion of the LZ transition formula.
 
If the transverse field of the $XY$ chain in (\ref{ham_xy}) is linearly 
quenched as $t/\tau$ \ct{cherng06}, where $t$ is varied from $-\infty$ to 
$+\infty$, 
the system crosses the two Ising critical lines ($h=\pm (J_x +J_y)$) shown in 
the phase diagram in Fig.~\ref{fig_xyphase}. For $J_y=0$ this reduces to the 
quenching of a transverse Ising chain \ct{dziarmaga05,damski05,polkovnikov05}.
We have shown in Sec.~\ref{jorwigtr} that the many-particle Hamiltonian 
decouples in Fourier space into $2 \times 2$ matrices $H_k$ for each mode $k$. 
%as given below
%\bea
%H_k(t)=-C_k^{\dg}\left[ \begin{array}{cc} h(t)+J\cos k & i\ga \sin k \\
%-i \ga \sin k & -h(t)-J\cos k \end{array} \right] C_k, \non \\
%&& \label{xy_hamfourier} \eea
%where $J= J_x+J_y$, the anisotropy $\ga=J_x-J_y$, and $h(t)$ is the 
%time-dependent transverse field. Clearly, the dynamics of each mode under a 
%temporal variation of the magnetic field can be treated independently.
%The full many-body evolution operator $U$ can be written as
%a tensor product of evolution operators $U_k$ acting in $(k,-k)$ 
%subspaces \ct{cherng06}. To obtain $U_k$, we consider the following 
%basis obtained from the vacuum $|0\rangle$ (where $c_k|0\rangle=0$):
%\bea |0\rangle, && |k,-k\rangle ~=~ c_k^{\dg}c_{-k}^{\dg}|0
%\rangle, \non \\
%|k\rangle &=& c_k^{\dg}|0\rangle, ~~~{\rm and}~~~ |-k\rangle = c_{-k}^{\dg}
%|0\rangle. \eea
%The latter two basis states $\ket k$ and $\ket {-k}$ are eigenstates of the 
%Hamiltonian $H_k$ as $$H_k(t)|\pm k\rangle = (h(t)+J\cos k)|\pm k\rangle.$$
%Thus, each of the $|\pm k\rangle$ states evolves in time with a
%phase factor, i.e., $|\pm k(t)\rangle = e^{-i\phi(t)}|\pm k\rangle$. 
%As seen from the Hamiltonian in (\ref{xy_hamfourier}), a mixing occurs
%between the other two states $|0\rangle$ and $|k,-k\rangle$.
Therefore, if the system starts in the state $\ket 0$, the dynamics takes 
place only between the two basis states $\ket 0$ and $\ket{k,-k}$ such that 
the time evolved state $\ket {\psi(t)}$ at time $t$ is a superposition 
given by $|\psi_k(t)\rangle=C_{1k}(t)|0\rangle+C_{2k}(t)|k,-k \rangle$. 
%The corresponding $2 \times 2$ evolution operator is denoted as 
%$S_k(t)$. Hence, the evolution operator $U_k$ is a $4\times 4$ matrix
%of the following block diagonal form
%\bea U_k=\left( \begin{array}{cc}
%S_k(t) & 0 \\
%0 & e^{-i\phi_k(t)}~I \end{array}\right), \eea
%where $I$ is the $2\times 2$ identity operator. The first
%block belongs to the $|0\rangle$ and $|k,-k\rangle$ sector while the
%second block corresponds to the states $|\pm k\rangle$. To describe
%$S_k(t)$, we project the Hamiltonian $H_k(t)$ onto the subspace
%$|0\rangle$ and $|k,-k\rangle$. 
We arrive at the following evolution equation for $C_{1k}(t)$ and $C_{2k}(t)$,
\bea i\left(\begin{array}{c}{\dot C}_{1k}\\ {\dot C}_{2k} 
\end{array} \right) ~=~ \left(\begin{array}{cc} h(t)+J\cos k & i \ga \sin k\\
-i\ga \sin k & -h(t)-J\cos k \end{array} \right) \left(\begin{array}{c}
C_{1k}\\C_{2k} \end{array}\right), \label{eq_transschor1} \eea
where $J = J_x + J_y$ and $\gamma=J_x-J_y$.

For $h= t/\tau$ for $-\infty < t <\infty$, the system starts in 
the state $|C_{1k}(-\infty)|^2=1$ and $|C_{2k}(-\infty)|^2=0$ and crosses the
two Ising critical lines at $t=\pm t_0=\pm \tau (J_x+J_y)$. The excitation 
energy of the system vanishes for the critical modes $k=0,\pi$, and the 
relaxation time diverges at $-t_0$ and $+t_0$, respectively; thus the system 
loses adiabaticity close to $t=-t_0$ (i.e., close to the QCP) no matter how 
slow the rate of variation of $h$ may be. The system starts from the ground 
state $|...\downarrow \downarrow \downarrow \downarrow \downarrow...\rangle$ 
(i.e., $\si_z=-1$ for all $i$)
which corresponds to the zero fermion state $|0\rangle$, whereas the final
state should be $|...\uparrow \uparrow \uparrow \uparrow\uparrow....\rangle$ 
if adiabaticity is followed throughout the quenching.
On the other hand, due to the reasons mentioned above, the system cannot
follow the instantaneous ground state close to the critical point and hence 
gets excited. The final state therefore has a structure such as
$|...\uparrow \uparrow \uparrow \uparrow\uparrow\downarrow \uparrow \uparrow
\uparrow \uparrow\downarrow \uparrow \uparrow \uparrow\uparrow....\rangle $,
where the density of down spins corresponds to defects. 

The Hamiltonian in (\ref{eq_transschor1}) resembles the 
LZ Hamiltonian as discussed in Appendix \ref{app_ll}. 
For convenience, let us first denote the off-diagonal elements of 
the matrix in (\ref{eq_transschor1}) by $\De_k$ and $\De_k^*$, and 
the diagonal elements by $\epsilon_k(t)$ and $-\epsilon_k(t)$. This equation 
then corresponds to the Schr\"odinger equation of a two-level system where 
two levels with energies $E_{1,2} =\pm \sqrt{\epsilon_k(t)^2+|\De_k|^2}$
approach each other linearly with a rate $1/\tau$. 
%This is the standard 
%Landau-Zener-Stueckelberg-Majorana problem 
%\ct{landau32,zener32,stuckelberg32,majorana32}
%provided that $\epsilon_k(t)$ is a linear function of time and $\De_k$
%is time independent as in (\ref{eq_transschor1}). 
Physically, the two energy levels $E_{1,2}$ approach each other 
from $t\to-\infty$ till some time $t^*$ where the energy gap is minimum 
($=2 |\De_k|$) which occurs when $\epsilon_k(t^*)=0$. For $t > t^*$, the 
energy levels start moving away from each other. If the system is initially 
($t\to-\infty$) prepared in the ground state of the two-level system, 
the LZ transition formula yields the probability of the system being in the
excited state at the final time $t\to+\infty$. 
{The Schr\"odinger
equation of such a two-level system can be solved by mapping it to 
the standard Weber equation with the above mentioned initial conditions} 
\ct{vitanov96,vitanov99,suzuki05,shevchenko10} (see Appendix \ref{app_lz}). 
The probability of excitations thus obtained at the final time is given by 
(see Eq.~(\ref{eq_lzformula}))
\beq p_k= e^{- 2\pi |\De_k|^2 /[d(\ep_k - (-\ep_k)) /dt]} = e^{-\pi\tau 
|\De_k|^2}, \label{eq_lzformula} \eeq
where we have used $d\ep_k/dt=1/\tau$.

Substituting $|\De_k|=\ga \sin k$ appropriately, the probability of 
excitations for each $k$ mode following the quenching in 
(\ref{eq_transschor1}) is given by (\ref{eq_lzformula}) and we get
\bea p_k = |C_{1k}(+\infty)|^2= e^{-\pi \tau \ga^2 \sin^2k}. 
\label{eq_pk} 
\eea
\begin{figure}
\begin{center}
\includegraphics[height=2.4in,angle=-90]{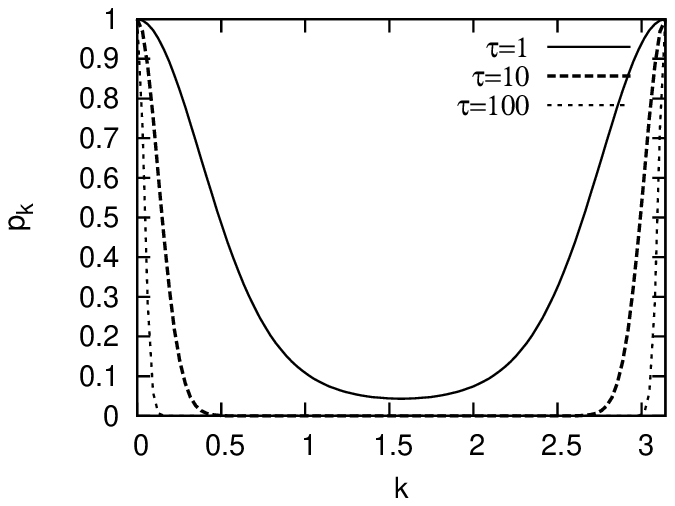}
\caption{Variation of probability of excitations $p_k$ with $k$ for different 
$\tau$ as given in (\ref{eq_pk}). For the modes $k=0$ and $\pi$, $p_k=1$ 
showing that these modes do not evolve. For large $\tau$, only modes close to 
these critical modes contribute to the defect.} 
\label{fig_pkvsk} 
\end{center}
\end{figure}
Fig.~\ref{fig_pkvsk} shows the variation of $p_k$ with $k$ for the transverse
field quenching. The critical modes at $k=0$ and $k=\pi$ for which the gap 
vanishes get excited with probability 1, while the modes close to the critical
modes have probability close to 1. The density of defects is obtained by 
integrating $p_k$ over all the modes in the Brillouin zone,
\bea n_h &=& \int^\pi_{-\pi} ~\frac{dk}{2\pi} ~p_k ~=~
\frac{1}{\pi}\int_0^{\pi} dk ~e^{-\pi \tau \ga^2 \sin^2 k} \non \\
&\simeq& \frac{1}{\pi}\left[ \int_0^{\infty} dk~ e^{-\pi \tau \ga^2 k^2} ~+~ 
\int_{-\infty}^0 dk~ e^{-\pi \tau \ga^2 (\pi-k)^2} \right] \non \\
&=& \frac{1}{\pi \ga \sqrt{\tau}}. \label{eq_transversen} \eea
The subscript $h$ in $n_h$ denotes the quenching of the transverse field.
In deriving (\ref{eq_transversen}), we have expanded $p_k$ around the 
critical modes $k=0$ and $\pi$ where the contributions to the defect density 
are peaked in the limit of large $\tau$; this allows us to extend the limits 
of integration to $\pm \infty$. The density of defects is the fraction 
of spins which could not accomplish the reversal, i.e., the density of
wrongly oriented spins in the final state. Noting that the critical exponents 
of the Ising transition are $\nu=z=1$, the $1/\sqrt{\tau}$ 
behavior in (\ref{eq_transversen}) for the density of defects is in perfect 
agreement with KZS noting that $d=1$ and $\nu=z=1$ for the Ising transition. 
The $n_h$ obtained by numerically integrating the 
Schr\"odinger equation also exhibits the predicted $1 /{\sqrt \tau}$ scaling. 
For a finite chain, one can naively argue that the $|\De_k|^2$ appearing in 
(\ref{eq_pk}) scales as $1/L^2$ at the QCP so that one arrives at the 
condition $\tau \gg L^2$ for the transition in a finite chain to be adiabatic.

%Fig.~\ref{fig_transversen}.

%If the transverse field $h$ is quenched in a non-linear fashion, $h(t)=
%|t/\tau|^{\al}sign(t)$, the quenching parameter does not vanish at the QCP. 
%The issue was resolved by linearizing the diagonal term of the two-level 
%Hamiltonian in Eq.~(\ref{xy_hamfourier}) around $t=t_0$, where the critical 
%point is reached at $t=t_0$ \ct{sen08,mondal09}. This
%reduces the problem to a linear quenching problem with an effective inverse 
%quenching rate $\tau_{\rm eff} = \tau/(\al g^{(\al-1)/\al})$;
%using the Landau-Zener formula one gets the defect density
%\beq n_h \sim \left[\al g^{(\al-1)/\al}/\tau \right]^{1/2}, \eeq
%which can be generalized to a $d$-dimensional system with critical exponents 
%$\nu$ and $z$, 
%\beq n_h \sim \left[\al g^{(\al-1)/\al}/\tau \right]^{d \nu/(\nu z +1)}. \eeq

\begin{figure}
\begin{center}
\includegraphics[height=2.0in]{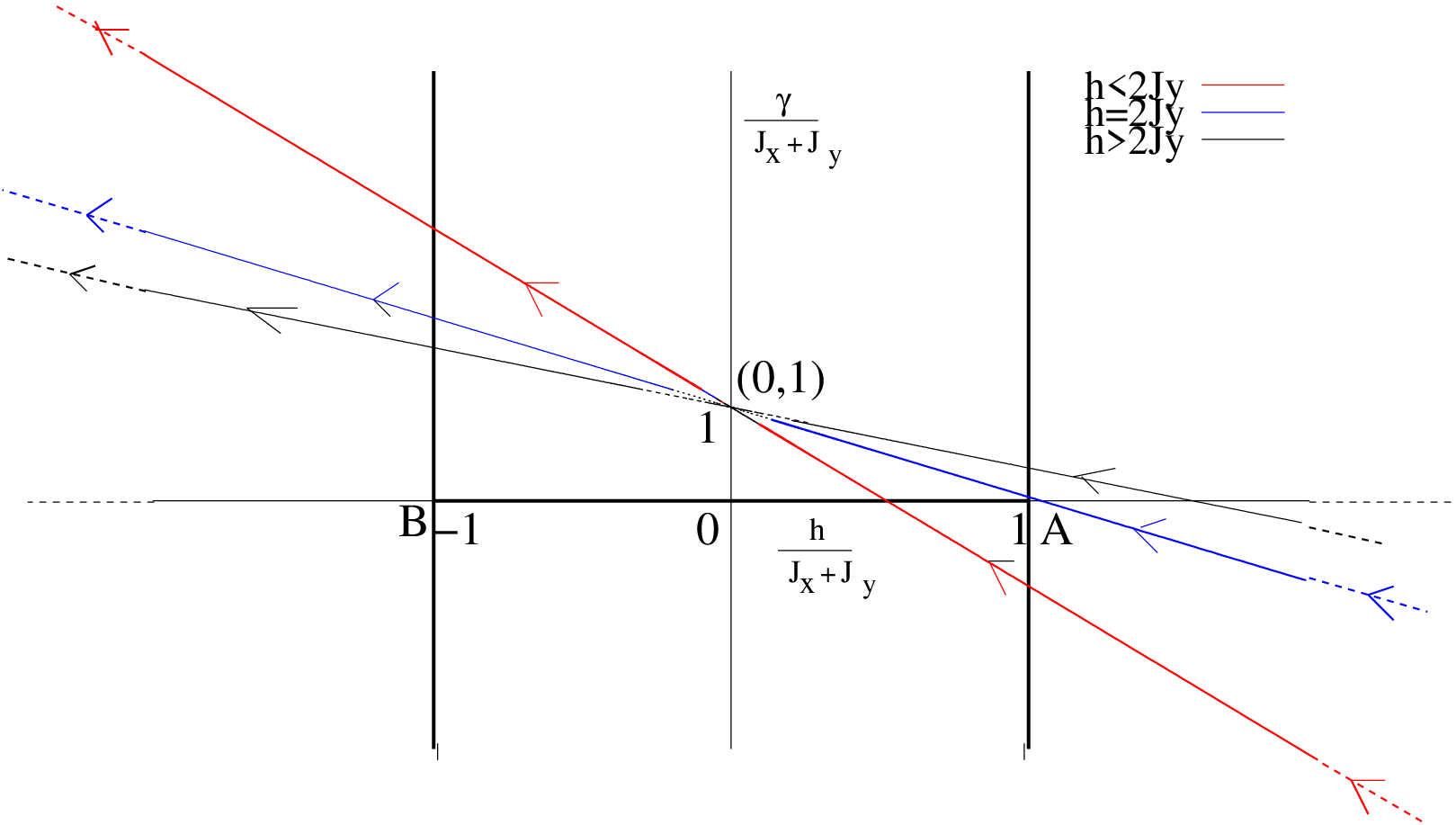}
\caption{(Color online)
The anisotropic quenching path for three different cases as discussed 
in the text: $h<2J_y,~h=2J_y$ and $h>2J_y$ shown in the phase diagram in Fig.
(\ref{fig_xyphase}). All the quenching paths cross the point $(0,1)$ at $t\to 
\pm \infty$.} \label{fig_anisopath} 
\end{center}
\end{figure}

In the anisotropic quenching scheme, $J_x = t/\tau$ goes from $-\infty$ to 
$+\infty$ with $J_y$ and $h$ fixed \ct{mukherjee07} (see 
Fig.~\ref{fig_anisopath}). If $h<2J_y$, the system crosses the two Ising 
critical lines ($h=\pm (J_x +J_y)$) and also the anisotropic transition line 
($J_x=J_y$); for $h > 2 J_y$ it crosses the two Ising critical lines only.
At a critical point on the anisotropic critical line, the critical mode with 
wave vector $k_0=\cos^{-1}(h/2J_y)$ (see (\ref{coska})) gets excited with
unit probability.
%The crossing of the anisotropic line is an additional feature of anisotropic 
%quenching as compared to transverse quenching. 
Performing a unitary transformation to a basis which are the eigenstates of 
the initial and final time Hamiltonians defined by \ct{mukherjee07}

$$ |e_{1k}\rangle = \sin (k/2) |0\rangle +i \cos (k/2) |k,-k \rangle, $$
$$|e_{2k}\rangle = \cos (k/2) |0\rangle -i \sin (k/2) |k,-k \rangle, $$
one can rewrite the reduced $2 \times 2$ Hamiltonian (\ref{eq_ham2by2}) $H_k$ 
in the form
\beq \left( \begin{array}{cc}
J_x+J_y\cos 2k+h \cos k & J_y \sin 2k + h \sin k\\
J_y \sin 2k + h \sin k& -(J_x+J_y\cos 2k+h \cos k)\\
\end{array} \right). \non
\label{rdm2} \eeq
In the process, the time dependence 
of the off-diagonal terms in (\ref{eq_transschor1}) gets shifted to only 
the diagonal terms of the transformed Hamiltonian so that the Landau-Zener 
expression becomes applicable. 
After expanding about all the three critical modes for $h<2J_y$, we get
\bea n_{jx} =n_1 + n_2 ~\simeq~ \frac{4J_y}{\pi \sqrt{\tau} ~
(4J_y^2 - h^2)}. \label{eq_anisojx} \eea
For $h>2J_y$, the total density of defects will be simply given by $n_{jx}/2$
as the system does not cross the anisotropic transition line. For $h=2J_y$, 
the system crosses the MCP during the evolution when (\ref{eq_anisojx}) 
is not valid; we will discuss this special case in the next section.

We recall the one-dimensional transverse Ising Hamiltonian in a longitudinal 
field (\ref{eq_t1d_long}) which is non-integrable for $h_L \neq 0$.
%given by
%\beq H = -J\sum_{<ij>} \si_i^x \si_i^x - h\sum_i \si_i^z - h_L\sum_i 
%\si_i^x; \label{eq_t1d_long} \eeq
% the presence of a finite $h_L$ renders the model non-integrable.
%(although the model is integrable in the limit $h_L \to 0$). 
For $h_L =0$, the system is at the QCP at $h_c=1$ if $J_x=1$, which is 
described by a 
Lorentz invariant theory (see Sec.~\ref{expcorr}) with $z=1$. In contrast, 
the correlation length exponent $\nu$ depends on how the theory is perturbed 
from the QCP. If the transverse field $h$ is perturbed then $\nu =1$. But if 
the longitudinal field $h_L$ is perturbed, the exponent $\nu_L=8/15$. This 
can be shown as follows. The scaling of the magnetization near the QCP, 
$\langle \si_z \rangle \sim (h-1)^{1/8}$, yields the scaling dimension of 
$\si_z$ to be $[\si_z] =1/8$. Consequently, we must have $[h_L] =15/8$ so 
that the change in the imaginary time action in the continuum limit, 
$\partial S = \int dx \int dt h_L \si_z(x,t)$, is dimensionless. Defining 
the correlation length $\xi_L$ corresponding to a perturbation
in the longitudinal field as $h_L \sim 1/\xi_L^{15/8}$ leads to $\xi_L\sim 
h_L^{-8/15}$ as $h_L \to 0$ (QCP); hence $\nu_L =8/15$. Therefore, if the 
critical point is crossed by linearly varying $h_L =t/\tau$, the defect 
density is expected to scale as $n \sim \tau^{-\nu_L d/(\nu_L z+1)} \sim 
\tau^{-8/23}$, since $d=z=1$ and $\nu_L=8/15$. If we quench through the 
QCP in a generic direction by varying both transverse and longitudinal fields 
as $h-1=h_L=t/\tau$, the defect density will have terms scaling as 
$\tau^{-1/2}$ and $\tau^{-8/23}$, respectively; the latter dominates in the 
limit $\tau \to \infty$. If one considers the Hamiltonian \ct{pollmann10}
\beq H= -\sum_i \left[ \si_i^x \si_{i+1}^x + \si_i^z + g(\cos \phi \si_i^z 
+ \sin \phi \si_i^x ) \right], \label{eq_gen_mag_field} \eeq
and $g$ is varied as $t/\tau$ across the QCP at $t=0$, the defect density $n$
is expected to scale as $\tau^{-1/2}$ if $\phi=0$ or $\pi$ but as 
$\tau^{-8/23}$ for any other value of $\phi$. This predictions are numerically
verified in \ct{pollmann10}.

Finally, let us comment on the non-linear quenching when the transverse field 
is quenched as $h(t)-J=|t/\tau|^{r} {\rm sign}(t)$ so that
the system crosses the Ising critical point at $t=0$. In this case, 
the probability cannot be evaluated by directly applying the LZ transition 
formula (\ref{eq_lzformula}) which is valid only for linear variation. 
However, using an appropriate scaling of the Schr\"odinger equation, it can 
be shown that the excitation probability $p_k= f(k^2 \tau^{2 r/(r +1)})$, i.e.,
it is only a function of the dimensionless combination $k^2 \tau^{2r/r+1}$; 
in the $\tau \to \infty$ limit, {one can then readily show 
that} $n_h \sim \tau^{-r/(r+1)}$ \ct{sen08} which is in agreement with 
(\ref{eq_intrononlinearn}). In contrast, if one employs the quenching scheme 
$h(t) =|t/\tau|^{r} {\rm sign}(t)$, the minimum energy gap that contributes 
maximally to the defect production does not occur at $t=0$, rather occurs at 
the instant $t=t_0$ such that $|t_0/\tau|^{r} {\rm sign} (t) + J \cos k =0$. 
Linearizing the Hamiltonian $H_k$ around $t=t_0$, we get
\beq H_k = \frac {t - t_0}{\tau_{\rm eff}} \si^z + k \si^x, 
\label{eq_linearized}\eeq
where we have expanded around the critical mode $k=\pi$ (with relabeling 
$\pi - k$ as $k$) 
and $\tau_{\rm eff} = \tau/r$ so that one can now use the LZ formula.

%\begin{figure}[htbp]
%\begin{center}
%\includegraphics[height=2.2in,width=1.8in]{rmp_transversequenching.eps}
%\caption{Variation of defect density in transverse quenching with the 
%quenching rate $1/\tau$ in log scale. The result confirms the Kibble-Zurek
%prediction.}
%\label{fig_transversen} \end{center} \end{figure}

\subsection{Quenching through multicritical points}
\label{multicritical}

As pointed out in the previous section, the density of defects given by
(\ref{eq_anisojx}) is not valid when the system is quenched through the 
multicritical point (MCP), implying that the standard KZS need not be valid
% as can be seen by substituting in
\ct{divakaran091}. Let us employ a quenching scheme in which both the 
transverse field and the anisotropy vary with time as $h-1=\ga=t/\tau$, so 
that the system hits the MCP at $t=0$. Using an appropriate unitary 
transformation and with $J_x +J_y=1$, we can recast the two-level problem 
into the form 
\bea H_k(t)=\left[ \begin{array}{cc} 
t/\tau + k^2 & -k^3 \\
-k^3 & -(t/\tau + k^2) \end{array} \right], \label{eq_mcp} \eea
where we have expanded close to the critical mode $k =\pi$. Applying the 
Landau-Zener formula, we find the defect density $n \sim \tau^{-1/6}$ which
is not consistent with (\ref{eq_kzm}) if one substitutes $d=1$, $z_{mc}=2$ 
and $\nu_{mc}=1/2$. Divakaran $et~al.$ \ct{divakaran091} proposed a 
generalized scaling 
$ n \sim \tau^{-d/(2z_{qc})}$, where $z_{qc}$ is determined by the 
off-diagonal term of the two-level Hamiltonian which scales as $k^{z_{qc}}$ 
close to the multicritical point; for the transverse $XY$ chain $z_{qc}=3$.

{Let us first qualitatively analyze the reason behind this 
anomaly before venturing into the technical details provided below. We have 
already discussed in the previous section that the low-energy modes (modes 
close to the critical mode) contribute to the density of defect in the limit 
of large $\tau$. It can be shown that the minimum gap point is not at the MCP, 
but rather at some special points close to it (on the ferromagnetic side); 
these are the
so-called quasicritical points. The exponents associated with the low-energy 
excitations above these quasicritical points in fact appear in the scaling of
the defect density which in turn becomes different from the predicted KZS.}

{Let us elaborate on the argument provided above.} Deng 
$et~al.$ \ct{deng09} attributed this anomalous KZS {(for the 
quenching scheme $h-1=\ga=t/\tau$, as discussed above)}, to the presence of 
some special quasicritical points mentioned above; they proposed a generic KZS
involving the exponents $\nu_{mc}$, $z_{mc}$ and $\nu_{qc} = \nu_{mc} z_{mc}/
z_{qc}$ associated with these quasicritical points \ct{deng09}. 
{For the transverse $XY$ spin chain, it can be shown that 
$z_{qc}= 1/\nu_{qc}=1/3$ using Eq.~(\ref{eq_mcp}); the energy gap is minimum 
at $t + k^2 =0$ (rather than at the MCP at $t=0$) where the excitations scale 
as $k^3$ yielding $z_{qc}=3$.}

%\begin{figure}[htbp]
%\begin{center}
%\includegraphics[height=2.6in,width=3.4in]{rmp_multicritical1.eps}
%\caption{$n$ vs $\tau$ obtained by numerically solving the Schr\"odinger 
%equation at the multicritical point with $h=10$ and $J_y=5$. The line has a 
%slope of $1/6$ confirming a different scaling for defect density for 
%quenching through a multicritical point.} \label{fig_multicn} \end{center}
%\end{figure}

{To illucidate the notion of these quasicritical points}, 
Mukherjee and Dutta, proposed a quenching scheme $h(t) = 1 + |\ga|^{r} 
{\rm sign} (t), r >0$ and $\ga = - t/\tau$, as $t$ goes from $-\infty$ to 
$+\infty$, which enables the system to hit the MCP along different directions 
characterized by the parameter $r$ \ct{mukherjee10} (see 
Fig.~\ref{fig_multicn2}). Here the case $r=1$, refers to the situation studied 
in \ct{deng09}.The Eq.~(\ref{eq_linearized}) turns out to be extremely useful 
in this case. One finds that $\tau_{\rm eff} = \tau k^{-2(r -1)/r}/r$ which 
leads to $\tau_{\rm eff}=\tau$ for $r=1$.
The higher order terms lead to faster decay of the defect density and hence 
linearization is a valid approximation for large $\tau$. For $r <1$, 
$\tau_{\rm eff} \to 0$ as $k \to 0$, rendering the linearization method 
inappropriate. The effective dynamical exponent, obtained via linearization 
method, $z_{qc} = 2/r + 1$, turns out to be a function of $r$ up to a critical 
value of $r = r_c = 2$, for $r > r_c$, it saturates at a constant value of 
$z_{qc} = z_{mc} = 2$; for $r=1$, we get $z_{qc}=3$ and $\nu_{qc}=1/3$. The 
value $\nu_{qc}=1/3$ as has already been used in explaining the scaling of the 
FS near the MCP in Sec.~\ref{fidelity}.

\begin{figure}
\begin{center}
\includegraphics[height=2.0in,width=3.0in,angle=0]{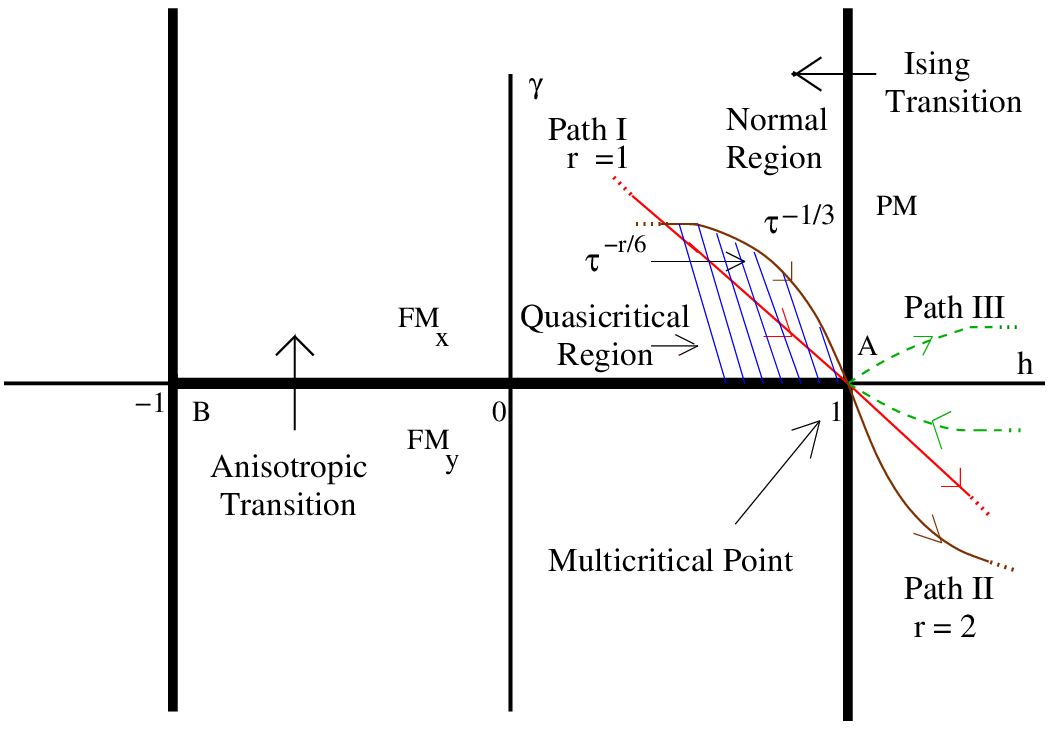}
\caption{(Color online) The phase diagram of the one-dimensional $XY$ model 
in a transverse field. The multicritical points are shown by points
$A$ ($h=1, \ga=0$) and $B$ ($h=-1, \ga=0$). Different quenching paths 
corresponding to different values of $r$ are shown; path I (path II) is 
for $r=1$ ($r=2$). Path III corresponds to a quenching scheme in which 
the system is always in the PM phase and touches the MCP at $t=0$. 
The shaded region corresponds to a continuously varying exponent for the defect
density, while for $r \geq 2$, $n$ scales as $n \sim \tau^{-1/3}$.
(After Mukherjee and Dutta, 2010).} \label{fig_multicn2} 
\end{center}
\end{figure}

Linearizing the equations of motion around a quasicritical point and using 
the Landau-Zener formula, it can be shown that for a path passing through the 
quasicritical region (i.e., for $r < 2$), the defect density varies 
continuously with $r$ as
\bea n \sim \tau^{-r/6} ~~~ \text{for ~$1<r<2$}. \label{eq_multi1} \eea
However, for $r \geq 2$, the system passes through a normal region, devoid 
of any quasicritical points, and hits the multicritical point vertically; 
this yields the same results as obtained by Divakaran $et~al.$, 2008, 
(Sec.~\ref{sec_gapless}) for a slow quench along the gapless line $h = J_x + 
J_y = 1$ (see Fig.~\ref{fig_xyphase}), given by
$ n \sim \tau^{-1/3} ~~~ \text{for $r \geq 2$}$.

Quenching along a special path (path III in Fig.~\ref{fig_multicn2}), characterized 
by $h(t) = 1 + |\ga|^{r}$ with $\ga = t/\tau$, which lies entirely in the 
PM region and touches the quantum multicritical point at $t = 0$ 
\ct{deng09,mukherjee10} leads to some unexpected results. For this path, one finds using linearization,
$ n \sim \tau^{-(4 - r)/(2(1 + r))} ~~~\text{for ~$r < 2$}, $
while for $r \geq 2$, the scaling $n \sim \tau^{-1/3}$ is retrieved. 
This was explained in terms of the shift in the center of the impulse 
region which leads to a dimensional shift ($d \to d+d_0$) \ct{deng09} in the 
KZS, where $d_0(r) = 2(2 - r)/r$ and varies continuously with $r$ for 
$r < 2$. Using adiabatic perturbation theory \ct{polkovnikov05}, a KZS for 
a generic MCP has been proposed \ct{deng09,dutta10} though the results for 
quenching through a MCP is still far from being fully understood and whether 
the scenario based on the assumption of the existence of quasicritical points
is a valid description close to a generic MCP is still an open question. A 
recent 
study of thermalization behavior in a transverse $XY$ chain following a sudden
quench towards the MCP has also pointed out the anomalous behavior close to 
it \ct{deng11}.

A similar behavior is observed for the FS (see Sec.~\ref{fidelity}) close 
to the MCP for a generic path with $\la_1 = h-1 = 
\ga^{r}$ ($\la_1 = \ga =0$ at the MCP) \ct{mukherjee111}. One finds that 
the maximum of $\chi_{\rm F}$ scales as $\chi_{\rm F}^{\rm max} \sim 
L^{2/\nu_{qc}-d}$, where $\nu_{qc} = (2/r +1)^{-1}$ for $r <2$, while for 
$r >2$, it saturates to $\chi_{\rm F}^{\rm max} \sim L^4$ which is the 
expected scaling of the $\chi_{\rm F}$ at the MCP when approached along the 
Ising critical line $\la_1=0$.

\subsection{Generalized quenching schemes}
\label{gen_quench}

In this section, we discuss defect generation in quenching schemes 
where the quenching parameter is varied in a non-monotonic fashion.

Let us first consider the effect of reversal of the quenching parameter right 
at the critical point \ct{divakaran09}. To study this, we choose the 
one-dimensional Kitaev model in (\ref{rmp_kitaev1d}) and quench the anisotropy 
parameter $J_- = J_1 - J_2$ linearly as {$J_- = -|t/\tau|$}, 
for $-\infty < t < 
\infty$. This implies that the spin chain initially prepared in the ground 
state is driven by increasing $J_-$ up to the quantum critical point at $t 
= 0$, after which $J_-$ is decreased at the same rate to its initial value.
The advantage of choosing the Kitaev model is that we can choose the parameter
variation in such a way that the minimum gap for all the modes occurs at the 
same time which makes the analytical calculations simpler. Solving the 
Schr\"odinger equation exactly with the reversal of the parameter (see 
(\ref{reverse_kitmat1})), the probability of excitations is found to be
\beq p_k (t \to \infty) ~=~ \frac{1}{4}(1-e^{-2\pi\al}) ~\left|\frac{\Gamma
(1-i\al/2)}{\Gamma(1+i\al/2)} ~+~i \frac{\Gamma(1/2-i\al/2)}{\Gamma
(1/2+i\al/2)} \right|^2, \label{reverse_eqpk} \eeq
where $\al=\tau \cos^2(k)/|\sin(k)|$ and $\Gamma(x)$ is the $\Gamma$-function. 
The density of defects obtained by 
numerically integrating this expression appears to be one-half of the defects 
generated in the conventional quenching without any reversal, i.e., for 
forward quenching ($-\infty <J_- <+\infty$) when $p_k$ can be derived by 
using the LZ formula appropriately in (\ref{reverse_kitmat1}).

We now consider a similar problem where the parameter $J_-$ is varied 
linearly up to the critical point at which the variation is stopped for some 
time $t_w$ \ct{divakaran10}. After this waiting time $t_w$, the variation is 
once again resumed in the forward direction. Although one cannot use the LZ 
formula in the
present case, the probability of excitations can be exactly obtained by 
considering the effect of time reversal on the Schr\"odinger equation which 
allows us to relate the time evolution from $t=0$ to $\infty$ to the evolution
from $t=0$ to $-\infty$. In the limit $t_w/\sqrt{\pi \tau} \ll 1$, the 
residual energy $e_r$ is found to be \ct{divakaran10}
\bea e_r 
%&=& \frac{1}{\pi \sqrt{\tau}} ~[~ 1 ~+~ e^{-4 (t_w + \frac{1}{2}
%\sqrt{\pi \tau})^2/(\pi \tau)}~] \non \\
\simeq \frac{0.32}{\sqrt{\tau}} ~[~ 1 ~+~ 0.37 ~e^{-2.3 t_w/\sqrt{\tau}}~].
\label{approx4} \eea
Interestingly, one finds that the residual energy decreases with increasing 
waiting time $t_w$.
%The decrease in residual energy as a result of waiting is also seen
%experimentally in single molecular magnets (SMM), but this decrease is more
%of a stretched exponential type \ct{hill06}. Similarities, if any, between
%SMM and the quenching problem is yet to be understood completely.
The analytical calculations for the two quenching schemes above are apparently
not applicable for a transverse $XY$ spin chain because the minimum gap for 
different modes occurs at different times given by the vanishing diagonal 
term in (\ref{eq_ham2by2}). However, it can be proved that in the limit 
$\tau \to \infty$, the time of occurrence of the minimum gap for all $k$ 
modes approaches the time at which the minimum of the critical mode occurs;
hence the above calculations valid for the Kitaev model should also hold 
good for the transverse $XY$ spin chain.

Mukherjee $et~al.$,2008, considered the effect of linearly varying the 
transverse field in a $XY$ chain from a large negative value to a large 
positive value and then bringing it back to its initial large negative value 
repeatedly $l$ times, in the process crossing the critical points many times 
\ct{mukherjee08}. Here, $l=1$ corresponds to a variation from $-\infty$ 
to $+\infty$ which is the usual case, while
bringing the system back to its initial value implies $l=2$. It has been shown
that for a general $l$, the probability of excitations varies as
\bea p_k(l) = \frac{1}{2}-\frac{(1-2e^{-\pi \tau \ga^2 
\sin^2k})^l}{2},~ \label{eq_pkrepeat} \eea
which is derived using the generalized form of the Landau-Zener formula for 
an arbitrary initial condition \ct{mukherjee08}. For $l=1$, we retrieve the 
excitation probability in (\ref{eq_pk})
for the transverse quenching across the Ising critical point.
%and assuming that cross terms $C_{1k appearing in the
%recursion relation of the probabilities vanish under coarse-graining.
The defect density for an arbitrary $l$ is given by
\bea n(l)=\frac{1}{2\pi\ga \sqrt{\tau}}\sum_{w=1}^{l+1}
\frac{l!}{w!(l-w)!} \frac{2^w}{\sqrt{w}}(-1)^{w+1}. \label{eq_repeatn} \eea
As $l \to \infty$, $p_k(l) \to 1/2$, i.e., after many oscillations, each spin 
has an equal probability of being in the up or down state. The expression for 
the defect density suggests that $n(l+1)<n(l)$ for small values of $\tau$ and 
odd $l$. This indicates some sort of corrective mechanism in the reverse path.
On the other hand, $n(l+1) \to n(l)$ for large $l$ (irrespective of whether 
it is even or odd) in the limit of large $\tau$.
One striking observation is that for odd $l$, the kink density $n(l)$ decreases 
monotonically with $\tau$, while for even $l$, $n(l)$ attains a maximum around
some characteristic value of $\tau$. In the limit $\tau \to \infty$, $n(l)$ 
decays as $1/\sqrt{\tau}$ for all $l$ in agreement with the KZS.
% as shown in Fig.~\ref{fig_repeatn}.
%\begin{figure}
%\includegraphics[height=2.3in]{fig_repeatn.eps}
%\caption{$n(l)$ vs $\tau$ for different $l$.$\overline n_4$ denotes the 
%defect density as obtained from the analytical expression in 
%Eq.~(\ref{eq_repeatn}).} \label{fig_repeatn} \end{figure}
 
Mukherjee and Dutta, 2009, studied the quenching dynamics of a 
transverse $XY$ spin chain using a time-dependent oscillatory 
transverse field $h$ given by $h(t) = h_0 \cos {\om t}$, with $|h_0| \gg J_x 
+ J_y = J$. They obtained the scaling of defects for a single crossing 
through the QCP ($h(t)=J$) for evolution from $t = -\pi/\om$ 
to $t = 0$, using linearized dynamical equations close to the gapless 
point. The excitation probability is given by
 $p_k=\exp(-\pi \ga^2 \sin^2k/\om \sqrt{h_0^2 -J^2 \cos^2k})$. In deriving 
$p_k$ through linearization, one defines an effective inverse quenching rate 
$\tau_{\rm eff} = 1/(2 \om \sqrt {h_0^2 - J^2 \cos^2 k})$; therefore the 
limits $\om \to 0$ and $h_0 \gg J$ refer to the slow quenching limit where the
KZS is applicable. For a single crossing, the defect density is found to
scale as $n \sim 1/\sqrt {\om}$.
On the other hand, for two successive crossings of the gapless point, i.e., 
with $0 \leq t < 2\pi/\om$, the interference effects were found to play a 
major role \ct{kayanuma93,kayanuma94,ashhab07,wubs05,kayanuma00}. The 
probability of non-adiabatic transition, $Q_{2k}$, is given by \ct{mukherjee09}
\bea Q_{2k} = 4p_k \left(1 - p_k\right) \sin ^2 {\left(\frac{2h_0 + J\pi 
\cos k}{\om} + \frac{\pi}{4}\right)}. \label{eq_repeat3} \eea
In the limit $\om \to 0$, $Q_{2k}$ oscillates rapidly; one can coarse grain 
over a small range of $k$ to obtain $Q_k \approx 2p_k (1-p_k)$ which is the 
excitation probability following a repeated linear quench \ct{mukherjee08} 
obtained by setting $l=2$ in (\ref{eq_pkrepeat}).

%\begin{figure}[htb]
%\includegraphics[height=2.0in,width=3.1in, angle = 0]{rmp_osc_fig1.ps}
%\caption{Graph showing the behavior of the excitation probability as a 
%function of number of full cycles ($N$) for $h_0 = 20, (J_x - J_y) = 0.005$, 
%$J = 1$ and $\om = 0.01$. The dashed line is obtained analytically for 
%$k = 81.9521^0$ with integral $2J\cos k /\om = 28$, while the solid line 
%is the analytical graph for $k = 25.8419^0$ with integral $2J\cos k /\om 
%= 180$. Numerical data points shown on the dashed line corresponding to 
%$k = 81.9521^0$ match exactly with the analytical results.} 
%\label{fig_repeat1} \end{figure}

Similar studies of multiple crossings \ct{ashhab07,wubs05} reveal the presence
of some special momentum modes, given by ${2 J\cos k}/{\om} = m$, $m$ being an
integer \ct{mukherjee09,ashhab07}, for which the non-adiabatic transition 
probabilities vary periodically with the number of full cycles; 
%(see Fig.~\ref{fig_repeat1}).
the other modes do not participate in the dynamics. The resultant kink density
and entropy density also show an oscillatory dependence on the number of full 
cycles. These studies have been generalized in recent years 
from the point of view of dynamical freezing \ct{das09} and using the 
Floquet approach \ct{russomanno12}.

% as shown in the Fig.~\ref{fig_repeat2}.

%\begin{figure}[htb]
%\includegraphics[height=2.0in,width=3.1in, angle = 0]{rmp_osc_fig2.ps}
%\caption{Graph showing the behavior of the kink density (solid line) and 
%entropy density (dashed line) as a function of number of full cycles for 
%$h_0 = 20, |J_x - J_y| = 0.005$, $J = 1$, and $\om = 0.01$, as 
%obtained numerically.} \label{fig_repeat2} \end{figure}

\subsection{Quenching through gapless phases}
\label{sec_gapless}

In the quenching processes discussed in previous sections, the initial and 
final phases are gapped while the spectrum is gapless at the QCP or quantum 
MCP that the system crosses. However, there are examples in which the system 
traverses along a gapless line or surface during the process of quenching
\ct{pellegrini08,sengupta08,mondal08,divakaran08}.

Quenching along an infinitely long gapless line can be achieved by varying 
the anisotropy $\ga=J_x-J_y=t/\tau$ along the Ising critical line $h=J_x+J_y$ 
(set $=1$) in a spin-1/2 $XY$ chain in a transverse field; see 
Fig.~\ref{fig_xyphase}. The two-level Hamiltonian gets modified to
\bea H_k(t)=\left[ \begin{array}{cc} kt/\tau & k^2 \\
k^2 & -kt/\tau \end{array} \right]. \label{eq_gapless} \eea
An application of the Landau-Zener formula leads to the scaling 
$ n \sim \tau^{-1/3}$.
Using adiabatic perturbation theory \ct{polkovnikov05}, a generalized scaling 
of the following form has been proposed as \ct{divakaran10}
$ n \sim \tau^{-d/(2a-z)},$ where $\ep_k=\ga(t)k^z+\de k^a$, where $z$ is the 
dynamical exponent associated with gapless line (i.e., Ising critical line in 
the present example) and $\de$ is a time-independent constant which depends 
on the other parameters of the Hamiltonian. The scaling relation presented 
here is applicable when the system is quenched across the MCP.

Recalling the phase diagram in Fig.~\ref{fig_dimerphase} of the transverse 
$XY$ spin chain with an alternating transverse field 
(\ref{eq_dimerham}), we find that the phase boundary given by $h^2 = 
\de^2+J^2$, with $\ga$ arbitrary and $J$ held fixed, defines a critical 
surface in the parameter space spanned by $h$, $\de$ and $\ga$. If $\ga$ is 
quenched linearly the system traverses a gapless line in the parameter 
space. Divakaran $et~al.$ \ct{divakaran08} mapped that problem into an 
equivalent two-level system and the scaling $n\sim \tau^{-1/3}$ was
again reproduced. When the transverse field $h$ and the alternating term 
$\de$ are quenched linearly with $\ga=0$, the system again traverses a 
gapless line and a scaling $n\sim \tau^{-1/2}$ was obtained. This 
observation brought in the concept of a dominant critical point \ct{deng08}; 
the exponents associated with the dominant critical point appear in the KZS in 
quenching across a gapless line in some situations. In the present case, the 
dominant critical point is the point $h=\de=0$ with critical exponents 
$\nu=z=1$.

Another situation arises when the system has a gapless line which has 
a finite length in parameter space. As pointed out earlier, this occurs
in the two-dimensional Kitaev model on a hexagonal lattice 
\ct{sengupta08} (see Fig.~\ref{kitaev_phase}). As $J_3$ is linearly 
quenched from $-\infty < t < \infty$ (i.e., $J_3 = t/\tau$), the system first 
passes through a gapped region, then through a gapless region, and finally 
ends in a gapped region. In the limit $\tau \to \infty$, using the
Landau-Zener transition formula in (\ref{eq_kitaev2by2}),
one finds the probability of ending up in the excited state 
%for the two-level subsystem with momenta $\pm \vec k$ 
given by
$p_{\vec k} = \exp(-2 \pi \tau [J_1 \sin ({\vec k} \cdot {\vec M}_1) ~-~ J_2
\sin ({\vec k} \cdot {\vec M}_2)]^2)$. In the limit $\tau \to \infty$ when 
only modes close to the gapless modes contribute, the defect density is found 
to scale as $1/\sqrt {\tau}$. This is consistent with the scaling relation 
in (\ref{eq_gapless1}) for quenching through a gapless 
surface since $d=2$ and $m =\nu=z=1$ in the present case.

Pellegrini $et~al.$ \ct{pellegrini08} looked at a similar finite region in a
gapless model given by the one-dimensional $XXZ$ chain with the Hamiltonian
\bea H=-\sum_i^{N-1}[\si_i^x\si_{i+1}^x+\si_i^y
\si_{i+1}^y+J_z(t)\si_i^z\si_{i+1}^z]. \eea
The low-lying excitations of the model have a gap $\De E
=4J(J_z-1)$ which closes for $J_z \to 1^+$. In the whole region $-1\leq J_z 
\leq 1$, the spectrum is gapless in the thermodynamic limit. For finite $N$, 
the gap vanishes as $1/N$ for $-1\leq J_z <1$, whereas it vanishes as $1/N^2$
at $J_z=1$. The point $J_z=-1$ corresponds to a BKT transition point
and for $J_z<-1$, the system is in the antiferromagnetic Neel phase with 
the energy gap being exponentially small in the system size. 

For evolution from the antiferromagnetic ground state with an initial value 
of the anisotropy $J_z^i\ll -1$ to the ferromagnetic region at a final value 
of $J_z^f \gg 1$ using a linear variation $J_z(t)=t/\tau$, the time-dependent 
density matrix renormalization group 
studies show that the residual energy follows a power law $1/\tau^{1/4}$;
this is in agreement with the Kibble-Zurek prediction since the minimum gap, 
relevant in the $\tau \to \infty$ limit, scale as $1/N^2$ and $N$ scales as $N \sim 
\tau^{\nu d/(\nu z+1)} \sim {\sqrt \tau}$. For a quench 
starting from $J_z=0$ and ending deep inside the Neel phase, the residual 
energy shows a power-law scaling with $\tau$ but with an exponent
$0.78 \pm 0.02$ which cannot be explained by a LZ treatment or the KZ argument.

The quenching dynamics of a $XY$ spin chain including multispin interactions
and a staggered transverse magnetic field \ct{titvinidze03,zvyagin06}, which 
shows a special QPT between two gapless phases involving doubling of the 
number of Fermi points \ct{fabrizio96,arita98}, has been studied 
\ct{chowdhury10}. It is found that for a linear quenching of the 
transverse field, the defect density in the final state decays exponentially 
with the inverse rate $\tau$ even after passage through gapless phases for 
models in which the interactions between the two sublattices are not symmetric.
This happens because the term that drives the QPT does not participate in 
the quench dynamics.

\subsection{Quenching of a disordered chain}

\label{sec_quench_disorder}

The quenching dynamics of a disordered transverse Ising chain cannot be mapped 
to decoupled $2 \times 2$ matrices due to the loss of translational invariance
\ct{dziarmaga06,caneva07}; hence the study of quenching dynamics becomes much 
more complicated. The disordered Hamiltonian considered in \ct{caneva07},
is that of a one-dimensional random transverse field Ising model defined by
$$H=-\sum_i J_{i}\si_i^x \si_{i+1}^x - \Gamma(t)\sum_i h_i\si_i^z.$$
The parameter $\Gamma(t)$ is quenched as $-t/\tau$ for $-\infty< t \leq 0$,
so that the Hamiltonian at $t=0$ reduces to a classical disordered Ising 
model. The random couplings $J_i, h_i$ are chosen from a uniform distribution 
between $0$ and $1$. The Schr\"odinger equation given in terms of the fermions
$c_i$ by time-dependent Bogoliubov theory is numerically solved \ct{rieger96}.
The numerically obtained scaling of the residual energy ($ e_r
$) and the defect density ($ n $) averaged over different 
realizations of randomness can be summarized as follows:
\bea n &\sim& (\log \tau)^{-2}, \non \\
e_r &\sim& (\log \tau)^{-3.4}. \eea
The logarithmic scaling behavior shows that $n$
decays very slowly with $\tau$ in the random case in contrast to the power-law
decay in pure systems. It is noteworthy that the scaling of the residual 
energy and defect density are different from each other; this is because
the defects are formed more easily near the bonds which are weaker (i.e., 
where $J_i$ is smaller), which implies that the energy cost of the defects 
(which is proportional to the values of $J_i$ at those bonds) goes to zero 
faster than the number of defects as $\tau$ becomes large. We also note that 
this logarithmic scaling is an artifact of the activated quantum dynamics of
a disordered chain discussed in Sec.~\ref{sec_griff+act}.

\subsection{Quenching with coupling to a bath}

We now discuss the quenching dynamics of a system which is weakly coupled to 
a thermal bath as studied in \ct{patane08,patane09}. The transverse Ising 
chain is initially in equilibrium with a bath whose temperature $T$ is kept 
constant during a linear quench of the transverse field, $h(t)= -t/\tau$. The 
quench of a parameter of the Hamiltonian takes 
the system through the quantum critical region at a finite $T$.
Assuming that the total density of excitations $n$ is obtained by adding up
the contributions from the Kibble-Zurek scaling at $T=0$ ($n_{kz}$) and the
thermally induced incoherent defects at finite $T$ ($n_{th}$), the
total defect density $n$ can be written as
\beq n ~=~ n_{kz} ~+~ n_{th}. \eeq
The decoupling of the thermal and Kibble-Zurek contributions is justified
by the assumption that the system is weakly coupled to the bath. Let us 
assume that the relaxation time $\xi_{\tau}$ for the excitations varies with 
the temperature $T$ as $\xi_{\tau}^{-1} \propto \al T^{\theta}$, where $\al$ 
is the system-bath coupling constant. Then the scaling of incoherent defects 
is given by
\bea n_{th} ~\propto ~\frac{\al}{\tau} ~T^{\theta+(d\nu+1)/(z\nu)}. \eea
For quenching times smaller than a critical $\tau=\tau_c$ (obtained by 
equating the KZ and thermal contributions), the KZ contribution dominates, 
whereas for quenching times larger than $\tau_c$, the incoherent thermal 
contribution is more important as the system spends more time within the 
quantum critical region.

In another series of works by Polkovnikov $et~al.$, the authors
studied a system which is prepared in an initial state in thermal
equilibrium at a temperature $T$ after which the dynamics is started
\ct{degrandi09,polkovnikov08,degrandi101,degrandi102}. Once the variation of 
the parameter of the Hamiltonian begins, no effect from the bath is 
considered. They show that at finite temperatures, the statistics of 
the low-energy quasiparticles plays an important role; 
{if one considers the quenching of free massive bosons and 
fermions\footnote{{These are actually two limits of the 
sine-Gordon model \ct{degrandi102} (see Appendix \ref{app_ll} and 
Eq.~(\ref{eq_SG})).}}, the number of quasiparticles excited to the momentum 
$k$ is different in the two cases depending on the statistics.}

The effect of the environment on the quenching dynamics of a spin chain has 
also been studied in recent years \ct{mostame07,damski09,cincio09}. Mostame 
$et~al.$, 2007, considered a global interaction between the transverse field 
of ({\ref{eq_tim1d}) 
%transverse Ising chain 
and the environment of the form $R \sum_i \si_i^z$, where $R$ is the Hermitian
operator describing the environment; the model is solvable due to the global 
nature of the interactions and it has been showed that the decoherence
increases the defect density. 

Cincio $et~al.$, 2009, studied the linear quenching of the transverse field 
of the Ising chain in (\ref{eq_tim1d}) coupled to a static environment through 
the interaction $V= -\sum_{m,n} \si_m^z V_{mn} \mu_n^z$, where the $\mu_n$'s
are the static spins describing the environment. If the dynamics starts from a
pure state when the system and the environment are uncoupled, it can be shown 
that the static environment generates a random transverse field in the spin 
Hamiltonian. The original problem of quench in a pure Ising system gets mapped 
to an average over quenches in an isolated Ising model in a random transverse 
field; the quenching dynamics then can be solved by numerically 
diagonalizing the Schr\"odinger equation. The KZS for the isolated system 
predicts that for an adiabatic evolution $\tau \sim L^2$, while in the model 
of decoherence, as is expected from the mapping to the dynamics of a 
disordered chain of size $L$ (compare with the result of the previous section),
$\tau \sim \exp (\sqrt L)$ which is non-polynomial in $L$; hence we have 
an upper bound on the system size that can be quenched adiabatically by 
coupling to a static bath. 
% To achieve the ground state of the pure Ising model, 
Assuming that influence of a random field is negligible, the dynamics becomes 
non-adiabatic when $t \sim \hat t$; KZS predicts a defect density scaling as 
$1/\sqrt{\tau}$. On the other, a perturbation expansion in the strength of the
disorder ($h_0$) predicts a defect density $\sim h_0^2$. We therefore get an 
additional condition for adiabatic evolution $1/{\sqrt \tau} \gg h_0^2$ or
$L \ll 1/h_0^2$.

%one additionally gets the condition for adiabatic evolution using the 
%perturbation in the disorder strength $\si$, namely, $\tau \si^2 \gg 1$ or 
%$L <1/\si$; for a pure chain the minimum $\tau$ scales as $L^2$.

%\begin{figure}
%\begin{center}
%\includegraphics[height=1.6in]{patane.eps}
%\n{}
%\end{center}
%\end{figure}
 
\subsection{Quenching and quantum information}

We have seen that slow and sudden quenches across a QCP lead to defects
in the final state of the quantum Hamiltonian. These defects influence the 
scaling behavior of different quantum information theoretic measures discussed 
in Sec. 5, e.g., the concurrence and negativity show an identical scaling 
relation as that of the defect density. We start the discussion focussing on 
the von Neumann entropy
density generated following a linear quench across the Ising critical point.

\subsubsection{Slow quenches}
\label{conneg}

\noi(i) {\it Decohered density matrix, von Neumann entropy density and 
correlation functions}

The quenching dynamics of the spin models considered so far is governed by the
Schr\"odinger equation and is unitary. Therefore, the final state following a 
quench must be a pure state. However, it is expected to have a fairly 
complicated local structure; {a sweep through a critical 
point generates defects which in turn lead to intrinsic decoherence and 
a local entropy which we discuss below \ct{cherng06}}.

%its local properties in the real space are 
%identical to those of a mixed state with a finite entropy \ct{cherng06}. In 
%other words, the final state can be viewed as a superposition of different 
%configurations of magnetically ordered domains.

Following an evolution from $-T <t <T$ of the transverse Ising chain, the 
final state is described by a density matrix of the product form $\rho = 
\bigotimes \rho_k$, where $\rho_k$ is given by
\bea \left[ \begin{array}{cc} p_k & q_k \\ 
q_k^* & 1 - p_k \end{array} \right]. \eea
with $p_k$ is the excitation probability and $q_k = C_{1k} (T) C_{2k}^* (T)$ 
(see the discussion around Eq.~(\ref{eq_transschor1})). In the 
thermodynamic limit, the correlation functions for any local observable are 
calculated by expressing them in terms of the $2 \times 2$ matrix of pair 
correlators and taking a trace over (i.e., integrating over $k$) the above
density matrix. The diagonal elements of the 
reduced $2 \times 2$ density matrix are smooth functions of $k$ and are 
independent of the total quenching time $T$ ($\gg \sqrt \tau$), whereas the 
off-diagonal terms are rapidly oscillating functions of both $k$ and $T$. In 
the limit $T \to \infty$, the phase of the off-diagonal term varies rapidly; 
as a result the integral of $q_k$ over $k$ vanishes.
%\begin{figure}[htb]
%\includegraphics[height=2.4in,width=3.1in]{rmp_aniso_fig1.ps}
%\caption{Variation of $p_k$ vs $k$ for $\tau = 5,\tau_{01},\tau_{02}$ and 0.01
%from bottom to top, where $p_k = 0.5$ at the two minima for $\tau = \tau_{01}$
%and $\tau_{02}$. We have fixed $J_y = 1$ and $h = 0.2$. For $\tau \gg 
%\tau_{01}, \tau_{02}$, the evolution is almost adiabatic except at $k = 0, ~
%\pi$ and $k_0$, where the system remains frozen in its initial state. For 
%very small $\tau$, $p_k$ is close to 1 for all $k$.}
%\end{figure}
\begin{figure}[htb]
\begin{center}
\includegraphics[height=2.4in,width=3.1in,angle=0]{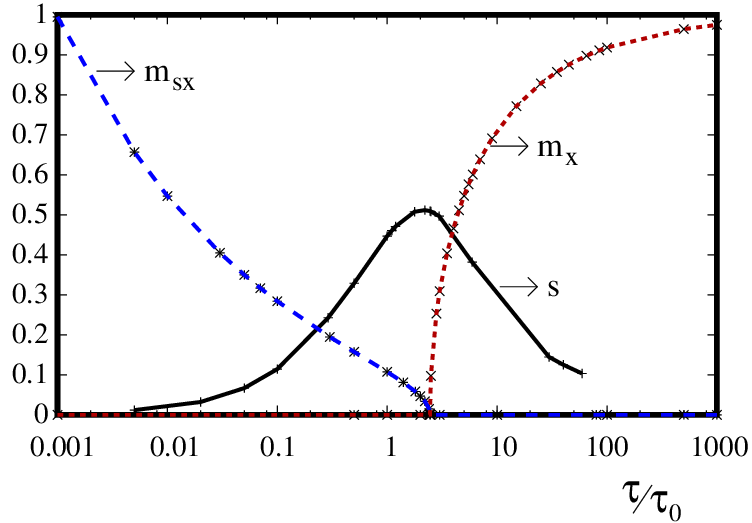}
\caption{(Color online)
Variation of von Neumann entropy density $s$, staggered magnetization
$m_{sx}$ and magnetization $m_x$ as a function of $\tau/\tau_{0}$, for 
$J_y = 1$ and $h = 0.2$. (After Mukherjee $et~al.$, 2007).} 
\label{fig_entropy1} 
\end{center} 
\end{figure}

The final state may therefore be viewed locally as a 
mixed state described by a decohered reduced density matrix $\rho_D$ given by
\bea \left[ \begin{array}{cc} p_k & 0 \\
0 & 1 - p_k \end{array} \right]. \eea
To quantify the amount of information lost in the decoherence process, we 
consider the von Neumann entropy density of the system, $s = -{\rm tr} 
(\rho_D \ln \rho_D)$, namely,
\beq s ~=~ -\int^\pi_0 \frac{dk}{\pi} ~[~p_k \ln (p_k) ~+~ (1 - p_k) \ln 
(1 - p_k) ~]. \eeq

The von Neumann entropy density shows an interesting dependence on the 
inverse quenching rate $\tau$ for both transverse and anisotropic quenching 
in a transverse $XY$ spin chain \ct{cherng06,mukherjee07}. For sufficiently 
small $\tau$, the final 
state is very close to the initial state, while for $\tau \to \infty$, the 
dynamics is nearly adiabatic and the system almost reaches the final ground 
state. In both these limits, the final state is almost pure and the entropy 
density tends to vanish; it attains the maximum value at an intermediate
value of $\tau \sim \tau_0$, for which $p_k=1/2$ for the wave vector $k=k_0$ 
corresponding to the critical mode.
Moreover, for the anisotropic quenching \ct{mukherjee07} in the limit $\tau 
\to 0$, the final state is antiferromagnetically ordered in the $\hat x$
direction with a non-zero staggered magnetization $m_{sx}$, and is 
ferromagnetically ordered ($m_x \neq 0$) for large $\tau$. The entropy density 
attains a maximum close to the value of $\tau$ for which the magnetic order 
of the final state changes from antiferromagnetic to ferromagnetic as 
shown in Fig.~\ref{fig_entropy1}.

Following a transverse quenching,
%of the transverse field $h = t/\tau$ as $t$ goes from $-\infty$ to $+\infty$,
we can calculate the correlation functions in the final state from the 
decohered density matrix using the notation introduced in 
Eqs.~(\ref{rhon}-\ref{matel}) \ct{sengupta09}.
We note that the correlation functions such as $\langle 
\si^{\pm}_i \si_{i+n}^z \rangle =0$ due to a $Z_2$ symmetry of the Hamiltonian. 
The non-zero correlation functions can be computed using the method developed 
in \ct{cherng06}, introducing a quantity $\al_n = \int_0^{\pi} 
({dk}/{\pi}) p_k \cos(nk) $; $ \al_n=0$ for odd $n$ as $p_k$ is invariant 
under $k \to \pi - k$. 
In terms of $\al_n$, the diagonal correlation functions are given by
\bea \langle \si_i^z \rangle &=& 1 ~-~ 2 \al_0~~~~{\rm and}~~~ \langle \si_i^z
\si_{i+n}^z \rangle =\langle \si_i^z \rangle^2 - 4 \al_n^2. \label{corr2} \eea
Thus, for any two spins separated by an odd number of lattice spacings, 
$\langle \si_i^z \si_{i+n}^z \rangle = \langle \si_i^z \rangle^2$
\ct{cherng06}. The off-diagonal correlators $\langle \si_i^a
\si_{i+n}^b \rangle$ (where $a$, $b$ can take the values $+$, $-$)
can also be computed in terms of $\al_n$.

For the correlation functions appearing in Eqs.~(\ref{matel}), we find that 
$\langle \si_i^{\pm} \si_{i+n}^{\pm} \rangle =b_1^n = 0$ for all $n$ since
these involve correlations between two fermionic
annihilation/creation operators and hence vanish. Further, $\langle
\si_i^{\pm} \si_{i+n}^{\mp} \rangle = b_2^n=0$ for all odd $n$ since these 
are odd under the transformation $\si^x_n \to (-1)^n \si^x_n, ~\si^y_n \to 
(-1)^n \si^y_n, ~\si^z_n \to \si^z_n$ which changes the signs of the couplings
$J_x$ and $J_y$ but leaves $p_k$ invariant. For even $n\le 6$, we find
\bea \langle \si_i^+ \si_{i+2}^- \rangle &=& \al_2 \langle \si_i^z \rangle,
\non \\
\langle \si_i^+ \si_{i+4}^- \rangle &=& (\al_4 \langle \si_i^z \rangle - 2 
\al_2^2) \langle \si_i^z \si_{i+2}^z \rangle, \non \\
\langle \si_i^+ \si_{i+6}^- \rangle &=& \Big[ \al_6 \langle \si_i^z 
\si_{i+2}^z \rangle +4\al_2(\al_2^2 +\al_4^2 -\al_4 \langle \si_i^z \rangle)
\Big] \times \non \\
&& \Big[ \langle \si_i^z \rangle [\langle \si_i^z \si_{i+2}^z \rangle -4 
(\al_2^2 + \al_4^2)] +16 \al_2^2 \al_4 \Big]. \non \\
& & \label{corr3} \eea
We will use these correlators to calculate the concurrence and negativity 
below. We note that the correlation function can also be calculated 
in a similar way when the transverse field of a transverse Ising chain is 
quenched from $-\infty$ to $0$ \ct{cincio07}.

For quenching through the gapless phase of the two-dimensional Kitaev model 
(see Sec.~\ref{sec_gapless}), it can be shown that the defect density is 
related to certain spin correlation functions. Let us consider a two-fermion 
operator 
\beq O_{\vec r} ~=~ i b_{\vec n} a_{\vec n + \vec r}, \eeq
where $b_{\vec n}$ and $a_{\vec n}$ are Majorana operators.
In terms of the spin operators, $O_{\vec r = \vec 0}$ is equal to
$\si_{j,l}^z \si_{j,l+1}^z$, i.e., a product of $\si^z$'s for two spins 
connected by a vertical bond. For $\vec r \ne \vec 0$, $O_{\vec r}$ can be 
written as a product of spin operators going from a $b$ site at $\vec n$
to an $a$ site at $\vec n + \vec r$ along with a string of $\si^z$'s in 
between. In momentum space, we have
\beq O_{\vec r} ~=~ \frac{4i}{N} ~\sum_{\vec k} ~[ b_{\vec k}^\dg a_{\vec k}
e^{i\vec k \cdot \vec r} ~-~ a_{\vec k}^\dg b_{\vec k} e^{-i\vec k \cdot \vec
r}], \eeq
where $N$ is the number of sites. We then find that on quenching $J_3$ from 
$- \infty$ to $\infty$ as $Jt/\tau$ (holding $J, ~J_1, ~J_2$ fixed), the 
expectation value of $O_{\vec r}$ in the final state is given by
\bea \langle O_{\vec r} \rangle &=& - ~\de_{\vec r, \vec 0} ~+~ \frac{2}{A} ~
\int ~d^2 \vec k ~ p_{\vec k} ~\cos (\vec k \cdot \vec r), \label{ovr} \eea
where the integral over momentum runs over half the Brillouin zone with
area $A = 4\pi^2 /(3\sqrt{3})$, and the excitation probability
$p_{\vec k} = e^{ - 2 \pi \tau ~[J_1 \sin ({\vec k} \cdot {\vec M}_1)- J_2
\sin (\vec k \cdot {\vec M}_2)]^2}$ as given by the LZ formula. For an 
adiabatic quench ($\tau \to \infty$), $p_{\vec k} = 0$ for all $\vec k$; 
Eq.~(\ref{ovr}) then shows that
only the on-site fermion correlation ($\vec r = \vec 0$) is non-zero. But if 
the quench is non-adiabatic, then longer-range correlations become non-zero; 
these have an interesting and highly non-anisotropic dependence on the 
separation $\vec r$ \ct{mondal08}.

\noi (ii) {\it Concurrence and negativity}

Sengupta and Sen \ct{sengupta09}, studied concurrence and 
negativity as measures of the
two-spin entanglement in the final state of the transverse Ising chain in 
(\ref{eq_tim1d}) following a linear quench of the transverse field $h = t/\tau$
as $t$ goes from $-\infty$ to $\infty$. As seen in the case of the von Neumann
entropy, the entanglement is expected to vanish in both the limits $\tau \to 
0$ and $\tau \to \infty$, when the system only retains information about the 
initial and final ground states, respectively. However, for a finite quenching
time $\tau$, the defects generated in the process of quenching across the QCP 
lead to a non-zero value of the two-spin entanglement \ct{sengupta09}. 
% We find that $\langle \si_i^{\pm} \si_{i+n}^{\pm} \rangle =b_1^n = 0
%$ for all $n$ since these involve correlations between two fermionic
%annihilation/creation operators and hence vanish. Further, $\langle
%\si_i^{\pm} \si_{i+n}^{\mp} \rangle = b_2^n=0$ for all odd $n$ since these 
%are odd under the transformation $\si^x_n \to (-1)^n \si^x_n, ~\si^y_n \to 
%(-1)^n \si^y_n, ~\si^z_n \to \si^z_n$ which changes the sign of the 
%cooperative term in Eq.~(\ref{eq_tim1d}) but leaves $p_k$ invariant.

%We consider the transverse Ising model with the Hamiltonian
%\beq H = \frac{J}{4} \sum_n [ 2 \si^x_n \si^x_{n+1} + h \si^z_n]. \label{h1} 
%\eeq
%We will set $J=1$, and measure all energies (times) in units of $J ~(1/J)$.
%Let us consider a linear quench of the transverse field given by $h(t) =
%t/\tau$, where $t$ goes from $-\infty$ to $\infty$. We will now compute the
%concurrence and negativity as measures of the two-spin entanglement
%generated by the quench \ct{sengupta09}. We note that the ground
%states of the initial and final Hamiltonians, at the beginning and
%end of such a quench process, do not posses any
%two-spin entanglement. Thus we expect that both for very fast ($\tau
%\to 0$) and very slow ($\tau \to \infty$) quenches, where the system
%retains information only about the initial and final ground states,
%the two-spin entanglement will vanish. Hence any finite entanglement
%obtained after such a quench with a finite rate $\tau$ must be generated
%by the non-adiabatic quench process and must therefore have contributions
%from excited states of the system.
To compute the concurrence and negativity, we use the expressions given in 
Eqs.~(\ref{rhon}-\ref{matel}) and the correlation functions given in 
Eqs.~(\ref{corr2}-\ref{corr3}). For $n\le 6$, we find
%(see Eq.~(\ref(conc1}) 
%and hence compute the concurrence and negativity. 
%The concurrence is defined in Eq.~(\ref{conc1}). In our case, 
(see (\ref{conc1})) that $\sqrt{\la_i^n}$ are given by $\sqrt{a_+^n 
a_-^n}$ (appearing twice), and $a_0^n \pm |b_2^n|$. Thus the spin chain has 
a non-zero concurrence given by
\bea C^n &=& {\rm max}\left\{0,2(|b_2^n|-\sqrt{a_+^n a_-^n})\right\}.
\label{conc2} \eea
To compute the negativity ${\cal N}^n$, we need to take a partial transpose
of $\rho^n$ with respect to the labels corresponding to the site $j=i+n$ in
(\ref{rhon}) \ct{peres96}. This interchanges $b_1^n \leftrightarrow
b_2^n$; the eigenvalues of the resultant matrix ${\bar \rho}^n$ are given
by ${\tilde \la}_0^n= a_0^n$ (appearing twice), and ${\tilde
\la}_{\pm}^n=(1/2) [ a_+^n + a_-^n \pm \sqrt{(a_+^n - a_-^n)^2 +
4 |b_2^n|^2}]$ of which only ${\tilde \la}_-^n$ can become
negative. This happens when $ |b_2^n| > \sqrt{a_+^n a_-^n}$ and yields 
{
\bea {\cal N}^n &=& {\rm max}\left\{0,{\tilde \la}_-^n\right\}.
\label{nega1} \eea}

\begin{figure} 
\begin{center}
%\rotatebox{0}{\includegraphics*[width=\linewidth]
%{RL10033_fig24.ps}}
\includegraphics[height=2.0in, width= 2.5in]{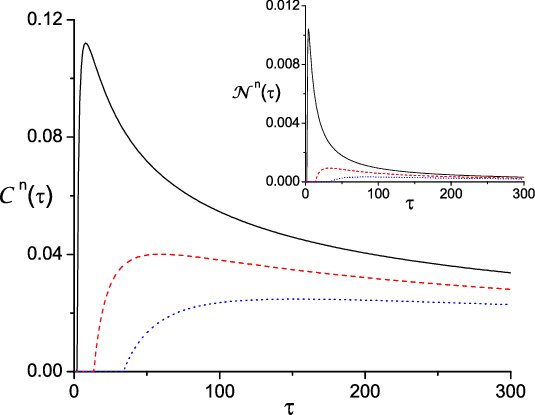}
\caption{(Color online) Plots of $C^n$ as a function of $\tau$ for $n=2$ 
(black solid line), $n=4$ (red dashed line) and $n=6$ (blue dotted line). The 
inset shows analogous plots for ${\cal N}^n$. (After Sengupta and Sen, 2009).} 
\label{entang_fig1} 
\end{center}
\end{figure}

Eqs.~(\ref{conc2}) and (\ref{nega1}) lead to several interesting findings 
\ct{sengupta09}. First, it can be easily shown that both $C^n$ and 
${\cal N}^n$ vanish when $n$ is odd, 
%$n$, $\langle \si_i^z \si_j^z \rangle
%= \langle \si_i^z \rangle^2$ and $\langle \si_i^+ \si_j^-
%\rangle = 0$. Thus all eigenvalues of $\rho^n (\si^y \otimes \si^y
%\rho^{n*} \si^y \otimes \si^y)$ are equal to $\al_0 (1- \al_0)$
%leading to $C^n =0$. All the eigenvalues of ${\bar \rho}$ are also
%positive; hence ${\cal N}^n=0$. Thus {\it 
i.e., the quench generates entanglement only between the even neighbor sites. 
Second, for large $\tau$ and $n \ll \sqrt{\tau}$, one finds that
%$\al_n$ scales similarly to the defect density: $\al_n \sim \sqrt{\al/\tau}$.
%Using this, we find from Eqs.~(\ref{corr2}-\ref{corr3}) that $b_2^n \sim
%\sqrt{\al/\tau}$ and $a_+^n a_-^n \sim \al/\tau$ which leads to
${ C}^n \sim \sqrt{1/\tau}$ and ${\cal N}^n \sim 1 /\tau$. Thus, for slow 
quenches, ${ C}^n$ (${\cal N}^n$) scales with the same (twice the) 
universal exponent as the defect density \ct{zurek05,polkovnikov05}.
%\ct{antunes99,polkovnikov05,polkovnikov08,sen08,mondal09,barankov08}.
%This result relating the generation of two-spin entanglement by a slow
%quench to defect production by such a process. 
Third, both the concurrence and the negativity become non-zero for a 
{\it finite critical quench rate $(\tau_c^n)^{-1}$} 
%(which is the solution of $|b_2^n|^2 = a_+^n a_-^n$) 
above which there is no entanglement between a site and its $n^{\rm th}$ 
neighbor. Numerical studies for $n=2,\,4\,{\rm and}\,6$ indicate
% Solving this equation
%numerically, we find $\tau_c^n =1.96,\,13.6\,{\rm and}\,33.8 $, for
%$n=2,\,4\,{\rm and}\,6$. [It turns out that for $\tau \le \tau_c^2$, the
%entanglement is entirely multipartite \ct{sengupta09}]. The numbers quoted
%above suggest that 
that $\tau_c^n$ scales as $n^2$ for large $n$. 
%This is
%consistent with the fact that the defect density scales as $1/\sqrt{\tau}$
%for large $\tau$; hence for $\tau \ll n^2$, there will typically be several
%defects between two sites separated by a $n$ which implies that
%will be no entanglement between them. 
Figure \ref{entang_fig1} shows plots 
of $C^n$ and ${\cal N}^n$ as functions of $\tau$ for $n=2,4,6$;
we see that ${ C}^n$ and ${\cal N}^n$
become non-zero between $\tau= \tau_c^n$ and $\infty$. Further, the
ratios ${ C}^4/{ C}^2$ or ${\cal N}^4/{\cal N}^2$ can
be selectively tuned between zero and 1 by tuning
$\tau$. The maximum values of both ${ C}^n$ and ${\cal N}^n$
decrease rapidly with $n$. For large $n \gg \sqrt{\tau}$, using the
properties of Toeplitz determinants used to compute the
spin correlators in these systems, it can be shown that $\langle \si_i^+
\si_{i+n}^- \rangle \sim \exp(-n/\sqrt{\tau})$ \ct{cherng06}. Thus one
expects the entanglement to vanish exponentially for $n \gg \sqrt{\tau}$.

\bigskip

\noi (iii) {\it Entanglement entropy}

\medskip

The defects generated by a quantum quench lead to a non-zero 
entanglement entropy between two parts of the final state of the system. When 
a one-dimensional chain is quenched linearly across the QCP with $z=1$, the 
KZS provides the scaling of the healing length given by $\hat \xi \sim 
\tau^{\nu/(\nu +1)}$ which also determines the entanglement entropy. For an 
open chain, one gets the entanglement entropy (see 
Sec.~\ref{e_entropy}) $S_{\rm quench} = c \nu/6(\nu+1) \log_2 \tau$. For 
quenching of a transverse Ising chain by linearly varying a generic
magnetic field (see (\ref{eq_gen_mag_field})), one gets $\nu = 1$ for 
$\phi=0$ and $\pi$, and $\nu = 8/15$ otherwise, while the central charge 
$c=1/2$ \ct{pollmann10}.

Cincio $et~al.$, 2007, studied the linear quenching of a transverse Ising 
chain in (\ref{eq_tim1d}) with $h(t)=t/\tau$, where $t$ changes from $-\infty$ 
to $0$ so 
that the system is expected to reach the Ising ferromagnetic ground state.
%They showed that following the quench a characteristic length scale $\xi = 
%\sqrt{\tau}$ develops. This length scale $\xi$, which appears in the KZ 
%scaling of the defect density, also determines the entanglement entropy
% due to the quench\ct{cincio07}. 
The entanglement entropy of a block of $l$ spins which is coupled to the 
rest of the system saturates to $S_l = 1/6 \log_2 \hat \xi$
which is in complete agreement with the argument provided above. However, this
saturation is found to occur when the block size $l \gg l_0 =\sqrt {\tau} \ln 
\tau$. This secondary length scale $l_0$ is argued to 
develop as a result of a dephasing process which is in fact the dynamical 
equivalent of the classical phase-ordering process \ct{bray02}. 

The geometric phase \ct{basu10}, quantum discord \ct{nag11} and generalized 
entanglement \ct{deng081} in the final state of a quantum $XY$ chain following 
a quench across the QCP or the quantum MCP have been studied in recent years. 
Cincio $et~al.$ \ct{cincio07} calculated the fidelity between the final state 
reached through quenching the transverse field of the transverse field Ising 
model as $h=t/\tau$, and the ground state of the final Hamiltonian at $t=0$, 
given by $F = \prod_{k >0} (1-p_k)$, where $p_k$ is the excitation probability 
calculated in previous sections. In the limit $\tau \gg 1$, one finds that
\beq \ln F ~\simeq~ \frac L{\pi} \int_0^{\pi} dk \ln \left(1-p_k \right) ~
\sim~ -1.3 ~n ~L \label{eq_fid_quench} \eeq
for a chain of length $L$.
In deriving the above equation, one assumes that $N \gg \hat \xi$ and uses 
the excitation probability $p_k = \exp(-2 \pi \tau k^2)$ and the density
of excitations $n$. Eq.~(\ref{eq_fid_quench}) shows that the 
fidelity decays exponentially with the system size $L$ over a length scale 
determined by the healing length $\hat \xi \sim 1/n$.

Let us now address the question of dynamical generation of decoherence; this 
has been studied by measuring the LE within the central spin model introduced 
in Sec.~ \ref{lecho}. Damski $et~al.$, studied the decoherence of a single 
qubit (spin-1/2) coupled 
to all the spins of a transverse Ising chain with the transverse field being 
quenched linearly across the QCP as $t/\tau$\ct{damski11}. At the initial 
time $t \to - \infty$, the qubit is 
chosen to be in a pure state which is a superposition of up and down states, 
and the spin chain is in the ground state. By calculating the reduced density 
matrix of the qubit as a function of time, it has been shown that the 
decoherence is dramatically increased (i.e., the modulus of the off-diagonal 
term of the reduced density matrix shows a sharp decay) when the surrounding 
spin chain is close to the QCP. Moreover the logarithmic of the LE or the decoherence
factor that appears in the off-diagonal term of the reduced density matrix is
found to scale in an identical fashion to that of the defect density when the environmental
spin chain is quenched across isolated critical \ct{damski11} and multicritical points; the scaling is not identical when the spin chain is quenched across 
a gapless critical line (for example for quenching along the Ising critical 
line through the MCP of Fig.~(\ref{fig_xyphase})) \ct{nag12}. Whether
the decoherence factor satisfies a universal scaling relation for such a 
non-equilibrium quenching of the environment is still an
open question. It has also been observed that a sudden quenching of the 
environment increases the decoherence of the central spin 
\ct{mukherjee121}.

\subsubsection{Sudden quenches}

The time evolution of the entanglement 
entropy in a quantum spin chain following a sudden quench of a parameter $\la$
from an initial value $\la_i$ to a final value $\la_f$, where the initial 
state is the ground state for the parameter $\la_i$ and the final Hamiltonian 
is critical has been studied recently \ct{calabrese05,calabrese06}. 
If the Hamiltonian is held fixed after the quench, it was found 
using both CFT and exactly solvable models that
\bea S_L(t) &\sim& t ~~~~{\rm for}~~~ t \leq t^*, \non \\
&\sim& L ~~~~{\rm for}~~~ t \geq t^*, 
\label{eq_ent_global}\eea
where $t^*$ is of the order of $L/2$. A similar crossover is also seen for 
quenching to a non-critical final Hamiltonian as seen from numerical 
studies of the transverse Ising chain (with initial and final transverse field 
both greater than $h_c$) \ct{calabrese05}. This can be understood as follows.
Following the sudden quench, pairs of entangled quasiparticles are produced at
different points of the chain and propagate in opposite directions with a 
velocity which is set equal to 1. The entanglement entropy receives a 
contribution from those pairs for which one quasiparticle lies inside the 
block while the other lies outside; hence the entropy increases linearly in 
time till it saturates when all the pairs which started inside the block 
reach its boundary.

However, following a slow quench in the generic transverse Ising Hamiltonian 
in (\ref{eq_gen_mag_field}) (with $\phi=0$) from $g=g_i$ to $g_f = -g_i$ 
through the critical point at $g=0$, if the system is allowed to evolve with 
the final Hamiltonian, one finds that the entanglement entropy $S$ oscillates 
as a function of time around a linearly increasing mean 
(Fig.~\ref{fig_ent_osc}) \ct{pollmann10}. This can be explained in terms of 
the Loschmidt echo (see Sec.~\ref{lecho}) for an integrable system which can 
be written as a product of two-level systems parametrized by the wave vector 
$k$ (e.g., for the transverse Ising and $XY$ spin chains). Let us assume 
that in one of the subsystems, the probabilities of being in the ground 
and excited states following the quench are $1-p_k$ and 
$p_k$, respectively; these states evolve in time with the final Hamiltonian. 
Assuming the energy difference between these two levels to be $\De E_k$, one 
can calculate the overlap between the state $\ket {\psi_k(0)}$ immediately 
after the quench and the state $\ket {\psi_k(t)}$ after an additional time 
$t$, given by
\bea \langle \psi_k(0)| \psi_k(t) \rangle = \left(\begin{array}{cc}
\sqrt p_k & \sqrt{1-p_k} \end{array} \right) \left(\begin{array}{c}
\sqrt p_k \\ \sqrt {1-p_k} e^{-i\De E_k t} \end{array}\right). \non \\
\eea
Hence, the overlap for the total wave function is
\bea |\langle \psi(0)|\psi(t)\rangle|^2 &=& \bigotimes_k |\langle \psi_k(0)| 
\psi_k(t) \rangle|^2 \non \\
&=& \exp \left[ -L \int \frac{dk}{2\pi} \log [1 -4 p_k (1-p_k)\sin^2 (
\frac{\De E_k t}{2})] \right]. \non \label{eq_ent_overlap} \eea
The mode for which $p_k=1/2$, gives rise to the cusp in the above equation 
whenever $\sin^2( {\De E_k t}/2)=1$. This oscillation does not occur if the 
system is non-integrable ($\phi \neq 0$ or $\pi$ in (\ref{eq_gen_mag_field})).

\begin{figure}
\begin{center}
\includegraphics[height=2.0in, width= 2.5in]{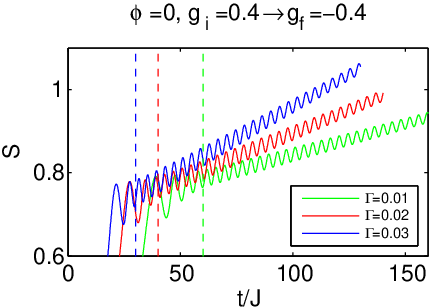}
\caption{(Color online)
Entanglement entropy as a function of time for a sweep of the Hamiltonian in
(\ref{eq_gen_mag_field}) with $\phi=0$ from $g_i = 0.4$ to $g_f =- 0.4$. The
dashed lines indicate the time at which the final value of $g_f = -0.4$ has 
been reached, and the Hamiltonian remains unchanged thereafter. Here, 
$\Gamma$ is the inverse rate $1/\tau$. (After Pollmann $et~al$, 2010).}
\label{fig_ent_osc} 
\end{center}
\end{figure}

Venuti and Zanardi studied a sudden quench of small amplitude around a QCP and 
showed that a generic observable does not equilibrate in the sense that its 
distribution is not peaked around the mean \ct{venuti10}. 
Moreover the distribution function 
is found to assume a universal, bistable form rather than the generic Gaussian 
form. The essential point is that the overlap $c_n = \langle \psi_n|\psi_0 
\rangle$, where $\psi_0$ is the initial state and $\psi_n$ is the eigenstate 
of the final Hamiltonian with eigenvalue $E_n$, decays very rapidly 
($c(E)\sim E^{-1/(z\nu)}$) near the QCP; this in turn implies a power-law 
double-peaked distribution. This prediction has been verified for the quantum 
ANNNI chain in (\ref{annni1}) with $\kappa<1$. 

\subsection{Local quenching processes}
\label{sec_localquench}
Another interesting problem going in parallel with global quenches is local 
quenches \ct{abraham71}. As the name suggests, in local quenches, a parameter
of the Hamiltonian $H_0$ is changed only locally, say at site $r$ or in a 
small region $R$. The system which is prepared in the ground state $|\psi_0 
\rangle$ of $H_0$ is now in the excited state corresponding to the locally 
disturbed Hamiltonian $H$. One can then look at the evolution of the
average values of various quantities like the entanglement 
entropy \ct{eisler07,calabrese07,eisler08}, correlation
function and magnetization \ct{calabrese07,divakaran11}, Loschmidt echo and 
light cone effects \ct{stephan11}. Experimentally, this type of situation 
corresponds to the $X$-ray absorption problem in metals where the creation of 
a hole plays the role of a local defect which is then removed by filling a
conduction electron.

Eisler $et~al.$ 
numerically studied the evolution of the entanglement entropy $S$ in a critical
$XX$ chain expressed in terms of non-interacting fermions hopping between 
neighboring sites when the hopping parameter is changed at a particular site
\ct{eisler07}. 
In global quenches, $S$ exhibits the scaling given in (\ref{eq_ent_global}).
In contrast, for local quenches where the defect is at the center of a 
system of size $L$, $S$ does not change till time $t=L/2$ after which there is 
a sudden jump followed by a slow relaxation towards the value in the 
homogeneous system of size $L$ as shown in Fig.~\ref{local1}. They 
looked at the effect of various defect positions and defect strengths on 
the evolution of $S$. These observations were analytically justified using 
conformal field theory by Calabrese and Cardy \ct{calabrese07}; they also 
looked at the evolution of the correlation function and magnetization after 
such a local quench in a system consisting of two disconnected (but 
individually critical) parts for $t<0$
which are suddenly joined together at $t=0$. They showed that the correlation 
function, like the entanglement entropy, shows a horizon effect, i.e., it 
remains in its initial value till some time $t$, which depends upon the defect 
position. After this time $t$, the correlation function shifts towards 
its equilibrium value corresponding to the homogeneous system. 
%The expressions for correlation function depend on the relative position of
%the defect with respect to the reference points. 

Let us consider a local quench of a critical transverse Ising model where the 
interaction term connecting two sites at the center of the chain is suddenly 
increased from zero to the homogeneous value $J_x$ ($=h_c$) (see 
(\ref{eq_tim1d})). If the two sites between which the correlations are measured
are on the same side of the defect (where the local quench is performed) 
such that $r_2$ is the distance of the defect from the closer reference point 
and $r_1$ is the distance from the farther point, with $r_2<r_1$, then the 
correlation function $\langle \psi_0|\sigma^x(r_1,t)\sigma^x(r_2,t)| \psi_0
\rangle$ for $t<r_2$ behaves as
\bea C_t(r_1,r_2) &=& \left( \frac{(r_1+r_2)^2}{4 r_1 r_2 (r_1 - r_2)^2} 
\right)^x F\left(\frac{4r_1r_2}{(r_1+r_2)^2}\right), \non \\
\eea
where $x$ is the scaling dimension of the operator $\sigma^x$, and the 
function $F$ has the form
\bea F(\eta)=\frac{\sqrt{1+\sqrt{\eta}}-\sqrt{1-\sqrt{\eta}}}{\sqrt{2}}. \eea
Clearly, the correlations for $t<r_2$ do not depend on the time since the 
signal traveling at unit velocity from the defect located outside the two 
reference points will not arrive at the closer reference point till $t=r_2$. 
As the signal reaches the first reference point, the correlation changes to
\beq C_t(r_1,r_2) = \left( \frac{(r_1+r_2)(r_2+t)\ep} {(r_1-r_2)(r_1-t)4r_1
(t^2-r_2^2)} \right)^x ~F \left( \frac{2r_1(r_2+t)}{(r_1+r_2)(r_1+t)}\right),
\eeq
where $\ep$ is the regularization parameter used as a damping factor
in conformal field theory calculations.
Finally, at $t=r_1$, the second reference point also receives the
signal from the defect and the system recognizes the presence of the enlarged 
system. The correlation saturates to the equilibrium value corresponding to 
the homogeneous system of size $L$ and is given by $C_t(r_1,r_2)\propto 
|r_1-r_2|^{-2x}$. The expression for $C_t$ is different when the 
reference points are located on opposite sides of the defect.

\begin{figure}
\begin{center}
\includegraphics[height=3.0in, angle=-90]{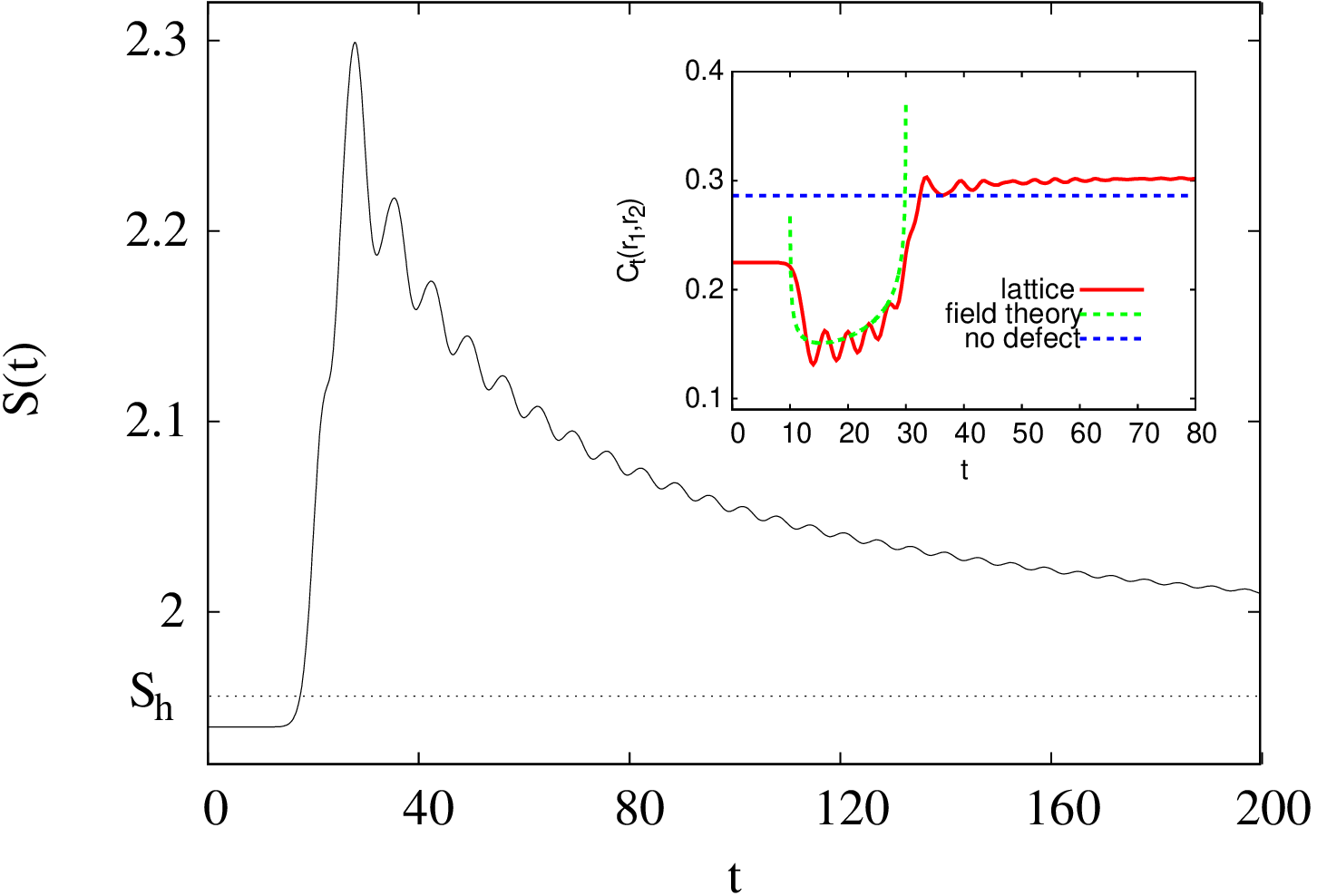}
\caption{(Color online)
Time evolution of entanglement entropy for a subsystem of L=40 sites
with a central defect. A sudden jump is followed by a slow relaxation towards
the homogeneous value $S_h$ (After \ct{eisler07}). Inset shows the time 
evolution of the correlation function after a local quench comparing the 
numerical and field theory results (After \ct{divakaran11}).}
\label{local1} 
\end{center}
\end{figure}

These results have been numerically verified by Divakaran $et~al.$
\ct{divakaran11}. They numerically studied the evolution of the
magnetization and correlation functions of the transverse Ising model in the 
critical system where the conformal theory predictions were successfully 
verified and also discussed some results when the final Hamiltonian is in 
the ferromagnetic and PM phases. The numerically obtained results along with 
the conformal field theory results for the evolution of correlation functions 
is shown in the inset of Fig.~\ref{local1} for comparison.
%The evolution of magnetization function is discussed in detail when the
%final Hamiltonian is ferromagnetic, critical and paramagnetic using the 
%framework of quasiparticle picture \ct{sachdev97,calabrese071}. 
The non-zero correlations can be explained using the quasiparticle picture 
\ct{sachdev97,calabrese071}; in this framework, kink like excitations (a pair 
of quasiparticles with opposite momenta) are created at the defect which 
move semi-classically with a momentum dependent velocity.
% and result in the reduction of the order parameter in the system. 
Since these quasiparticles are emitted only at the site of defect, they
are quantum entangled causing correlations between the two reference points.
This is contrary to the picture of global quenches where the quasiparticles
are emitted at all the sites \ct{rossini09,rossini10,igloi11}. 
%Only those quasiparticles are entangled which are within the correlation 
%length resulting to relaxation of the correlations.

\subsection{Thermalization}

In the major part of this section, we have studied what happens immediately 
after a parameter in the Hamiltonian of a many-body system is varied so as 
to take it across a QCP. We typically find that starting
with the ground state of the initial Hamiltonian $H_i$, we end in a state 
which is {\it not} the ground state of the final Hamiltonian $H_f$. For 
instance, for a slow quench, the final state reached differs from the 
ground state of $H_f$ by a finite density of defects.

One can then ask: what would now happen if the Hamiltonian is held fixed
at $H_f$ and the many-body system is allowed to evolve further in time?
The problem of ergodicity and thermalization following a quantum quench 
\ct{igloi00,rigol08,patane08,iucci09,iucci10,canovi10} has been of great 
interest following the experimental observations with cold atoms 
\ct{sadler06,greiner02}. We will not discuss this important problem here, 
and we refer the reader to recent review articles 
\ct{polkovnikov10,dziarmaga09}.

%$\chi_F}$ diverges as $J_z \to \pm J$. We note that the fidelity 
%susceptibility is able to detect even the KT transition.

\subsection{KZ mechanism and adiabatic evolution}

The defect generation following a passage through a QCP is relevant in the 
context of adiabatic quantum computation \ct{farhi01}, where the adiabatic 
evolution of a system from an initial state to a computationally non-trivial 
state is explored. When a QCP is encountered during the evolution, the 
defects are inevitably generated according to the KZS. Considering a 
non-linear quenching of the driving parameter $\la \simeq \la_0 |t/T|^{r} 
{\rm sign} (t)$, where $-T <t < T$ and $T$ is the characteristic 
passage time which is held fixed, we note that a large number of excitations 
are produced for both $r \to 0$ and $r \to \infty$. 
{To minimize defect production, one therefore needs to find 
an optimized value of $r$ and check whether this optimal value is determined 
by some universal properties of the associated QCP.}

%a compromise 
%between being as slow as possible near the QCP and being not very fast away 
%from it. 
Using adiabatic perturbation theory for the above mentioned quenching scheme, 
Barankov and Polkovnikov \ct{barankov08} found that the defect density 
$n \sim (r \de)^{d 
r \nu/(r z \nu +1)}$ where $\de = 1 /(T \De_0)$ and $\De_0$ is 
the lowest excitation energy at $\la = \la_0$, i.e., the final value of the 
coupling at $t = T$. We note that an identical exponent for the scaling of
the defect density was also derived by Sen $et~al.$ \ct{sen08} (see 
Sec. \ref{sec_kzs}), although the dependence on the parameter $\de$ through 
$r \de$ was not included. Optimizing $n$ with respect to $r$, one finds 
that minimum excitation occurs at $r_{\rm opt} \de \simeq 
\exp(-r_{\rm opt} z \nu)$. For a transverse Ising chain with $z=\nu=1$, one 
finds $n \simeq [A(r) \de]^{r/(r+1)}$ where $A(r)$ scales linearly 
with $r$; hence the optimal condition gets modified to $r_{\rm opt} 
\simeq -\ln[(\de/C) \ln (C/\de)]$ with the non-universal constant $C=14.7$ 
\ct{barankov08}.
In the context of a power-law sweep of the confining potential of a 
trapped system, the above study has been further generalized to the case 
where the quenching potential varies both in space and time as $\la (x,t) 
\simeq \la_0(t)|x|^w$, where the QCP is at $x=t=0$; the amplitude $\la_0(t)
\sim |t/\tau|^{r}$ \ct{collura10} for $-T <t <T$ (see also 
Sec.~\ref{kz_space}). It has been shown that the mean excitation density 
follows an algebraic law as a function of the sweeping rate with an exponent 
that depends on the space-time properties of the potential.

In relation to the generalized quenching schemes (see Sec.~\ref{gen_quench}), 
the concept of a quench echo has been introduced to test the adiabaticity of 
quantum dynamics \ct{quan10}. In this approach, one calculates the fidelity 
between the initial ground state and the final evolved state following a slow 
quench in which the final Hamiltonian of the quantum system happens to be the 
same as the initial one. For example, for a transverse Ising chain of finite 
length, one can define a quenching protocol for the transverse field $h$,
\bea h(t) &=& h_0 ~-~ \frac t {\tau} ~~~{\rm for}~~~ 0 < t < (h_0 -h_T) \tau, 
\non \\
&=& 2h_T ~-~ h_0 ~+~ \frac t {\tau} ~~~{\rm for}~~~ (h_0 -h_T) \tau < t <
2 (h_0 -h_T) \tau. \eea
Namely, the system is initially prepared in the ground state of $h_0$. Then 
$h$ is first ramped in the forward direction from $h=h_0$ to $h=h_T$, and is 
then reversed from $h_T$ to $h_0$, where $h_T$ is the turnaround point. The 
idea is that if $\tau$ is so large that the system is protected against 
excitations due to a finite gap at $h=h_c=1$ (for $J_x=1$), both forward and 
reverse 
evolutions are adiabatic, and the fidelity between the initial and final state 
is close to unity. The advantage of the quench echo approach is that no prior 
knowledge of the eigenstates or eigenvalues of the Hamiltonian is necessary.

In another approach, the adiabaticity is tested locally in time during the 
evolution and the local speed is adjusted accordingly so that the system stays 
close to the instantaneous ground state though the energy gap may be unknown 
\ct{nehrkorn11}. Similarly, for a local Hamiltonian like the transverse Ising 
chain, the linear evolution from an initial to a final Hamiltonian can be 
replaced by a series of straight-line interpolations, along which
the fundamental energy gap is always greater than a constant independent of 
the system size; this enables a very efficient and adiabatic preparation of 
a highly entangled ground state \ct{schaller08}. We note that the role of
quenching protocol has also been studied in \ct{eckstein10}.

\subsection{A brief note on quantum annealing}

In this section, we will include a brief note on the optimization method 
known as quantum annealing {which is also related to the 
adiabatic evolution}; for an extensive review we refer to 
\ct{das08,santoro06}. In this method, one attempts to find the ground state 
of a random classical Hamiltonian (or a multivariate cost function viewed as 
the same). To achieve this, a (non-commuting) quantum kinetic term (e.g., a 
transverse field term in a classical random Ising Hamiltonian) is added. The 
kinetic term is initially set at a value much higher than all the interaction 
terms so that the initial ground state is trivially is known. The kinetic 
term is then reduced to zero slowly following some protocol. If the dynamics 
is completely adiabatic, one is expected to reach the ground state
of the classical random Hamiltonian at the final time. Of course, one 
assumes that there is no crossing of energy levels with the ground state in the
course of evolution. The quantum annealing is meaningful for finding the 
ground state of classical Ising SK spin glass where, as mentioned already, 
ergodicity is broken at $T=0$ and barriers between neighboring local minima 
become proportional to system size. We have already mentioned that quantum
fluctuations may induce tunneling between these minima and the ergodicity may 
get restored \ct{ray89}. However, in the presence of disorder extended states 
in a quantum system may get localized \ct{lee85}.

{Let us clarify quantitatively in what sense one believes 
that quantum annealing may be more efficient than thermal annealing. 
%We have already mentioned that quantum fluctuations may
%induce tunneling between different minima of a rugged free energy landscape and the ergodicity may get
%restored \ct{ray89}. 
In fact, if two local minima of the free energy landscape are
separated by a barrier of height $l$ and width $w$, the
classical escape rate at a finite temperature $T$ is given by $\exp(-l/T)$. 
On the other hand, the quantum tunneling
probability induced by the transverse field $h$ is approximately given by $\exp(-({\sqrt l w})/h)$, ensuring the clear
advantage of quantum tunneling when $ w << {\sqrt l}$ \ct{das05,smelyanskiy12}.} 
Recent numerical studies clearly demonstrate that quantum annealing
can occasionally help reaching the ground state of a complex glassy system 
much faster than could be done using thermal annealing. Following the 
pioneering demonstrations, theoretically in the references 
\ct{amara93,finnila94, kadowaki98,farhi01,santoro02}, and experimentally
by Brooke $et~al.$, 1999 and Aeppli and Rosenbaum, 2005, the quantum annealing 
technique has now emerged as a successful technique for optimization of 
complex cost functions {(see Sec.~\ref {expt5})}. (For a 
mathematical discussion on the bounds and 
schedule strategies for quantum annealing see the ref. \ct{Morita98}.)

However, it needs to be mentioned that recent studies show
that the possibility of Anderson localization \ct{altshuler09}, and also of
first order phase transitions \ct{young10,farhi09} with exponentially
small gaps \ct{amin09,jorg08} may lead to an exponential complexity
in quantum adiabatic algorithms. {These studies raise some
concerns regarding the prospect of quantum annealing in
searching for efficient solutions (bounded in time by a polynomial
in the size of the problem) of computationally hard problems for which
classical algorithms take a time which is exponential in
the problem size. However the typical problems of quantum
annealing, arising due to the exponential decay (in the problem size)
of the minimum energy gap and the consequent failure of the adiabatic theorem, 
may not pose a serious threat to the efficiency of quantum annealing. It has 
recently been demonstrated that a suitable design of the tunable quantum 
driving term in the Hamiltonian may remove the difficulty \ct{seki12}.} 
%We also note that
%optimal control theory may lead to drastic improvement of the
%performance of quantum information analysis \ct{caneva09}.

%However, the questions of localization
%of these extended states due to disorder in barrier heights and
%widths remains still to be investigated(cf., Lee and Ramakrishnan,
%1985)

%However, it needs to be mentioned that recent studies show that possibility of
%Anderson localization , and also of first order phase 
%transitions \ct{young10,farhi09} with exponentially small gaps
%\ct{amin09,jorg08} may lead to an exponential complexity in quantum 
%adiabatic algorithm. These studies however raise concerns whether quantum 
%annealing can solve truly hard problems in polynomial time.
%We note that optimal control theory may also lead to drastic improvement of 
%the performance of quantum information studies \ct{caneva09}.
%\section{Adiabatic evolution through a QCP: quantum annealing}

%\subsection{Theoretical summary (\it {BKC})$\sim 1 p$}

\section{Experimental realizations of transverse field Ising systems}
\label{expt}

\subsection{Singlet ground state magnets}
\label{expt1}

In a crystalline lattice, the crystal fields often take a form such 
that the ground states of magnetic ions 
are singlets separated from the higher lying spin multiplets. 
Sufficiently strong magnetic dipolar or exchange couplings between different 
sites can mix the singlets with the multiplets enough so as to induce 
non-zero expectation values for the magnetic dipoles at individual sites. 
Examples of this phenomenon are generally found among metals and insulators 
based on rare earths. Such metals were a subject of extensive research in the 
1960's and 1970's, and the book by MacKintosh and Jensen \ct{jensen91} 
gives an excellent account of both the data as well as mean field approaches. 
{ With respect to insulating materials, LiTbF$_4$ emerged as a model system 
in the 1970's. It is an example of an ideal Ising dipolar coupled ferromagnet;}
here the long-range nature of the dipolar interactions leads to an upper 
critical dimension $d_u^c$ \ct{chaikin95} (at which mean field theory with 
logarithmic corrections calculable using RG methods becomes 
an exact description of thermal phase transitions) of three rather than four. 

\begin{figure}
\begin{center}
\includegraphics[height=2.8in]{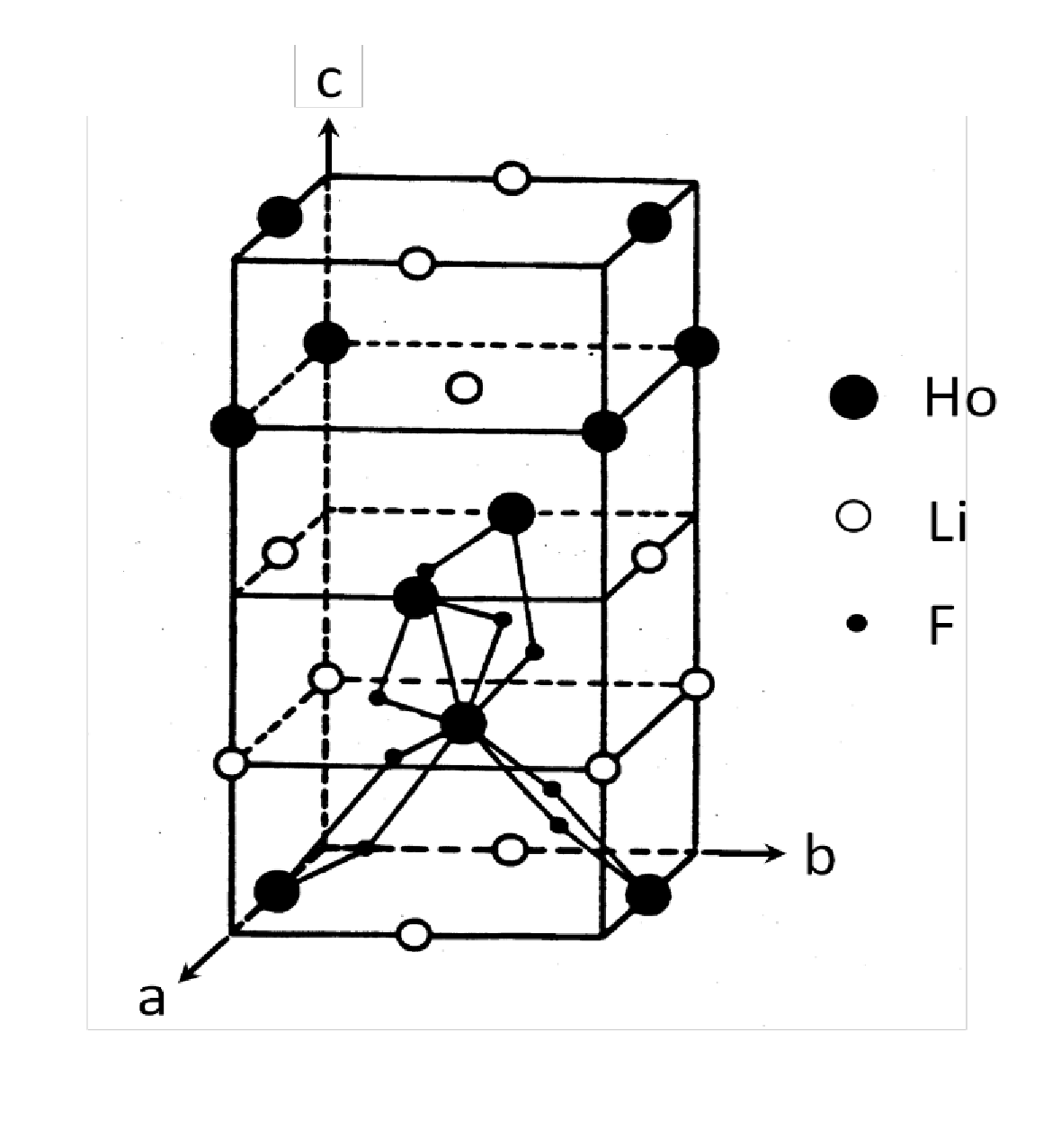}
\caption{{Crystal structure for LiHoF$_4$, one of the series of rare earth 
tetrafluorides that illustrates the physics of three dimensional Ising
model in a transverse field.}} \label{GA1} 
\end{center}
\end{figure}

\begin{figure}[h]
\begin{center}
%\begin{flushleft}
\includegraphics[height=2.6in, width=2.6in]{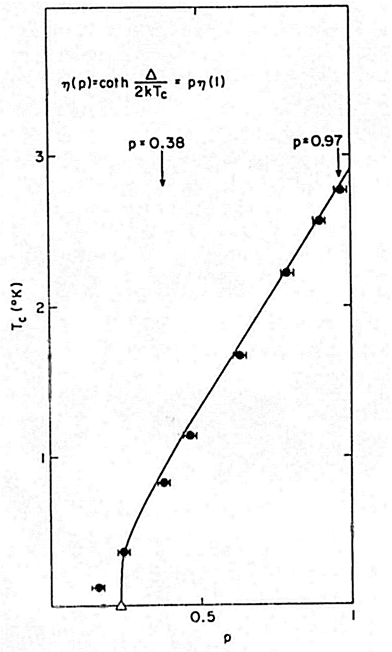}
%\end{flushleft}
\caption{(Color online)
{Phase diagram for the dipolar-coupled Ising ferromagnet LiTbF$_4$ as a 
function of substitution of magnetic Tb with non-magnetic Y ( denoted by $p$).} 
(After Youngblood $et~al.$,1982.)} \label{GA2} 
\end{center}
\end{figure}

LiTbF$_4$ belongs to the LiREF$_4$ isostructural series of ionic salts. 
Li and F carry valences $+1$ and $-1$ respectively, leaving RE, which can be 
any rare earth atom or the nonmagnetic element Y, with a valence of $+3$. 
Figure \ref{GA1} shows the body-centered tetragonal crystal 
structure for the family. The underlying LiF$_4$$^{-3}$ lattice together 
with the number of $f$-electrons (i.e., the position in the RE row of 
elements) determine the magnetic crystal field levels for the RE ions. 
Because the RE ions have compact $f$ orbitals with little site-to-site overlap,
the dominant magnetic interactions are dipolar. The early work of Luttinger 
and Tisza \ct{luttinger46} 
then shows that the ground state (ferromagnetic or antiferromagnetic) for 
pure compounds in this series of salts is entirely determined by the identity 
of the rare earth ion, leading to, for example, ferromagnetism with a Curie 
temperature $T_c=1.53$ K for RE=Ho, with a non-Kramers doublet ground state, 
and {antiferromagnetism with a Neel temperature $T_N \sim 0.4K $} 
for RE=Er \ct{beauvillain77,mennenga84}.
The lowest lying crystal field states for RE=Tb are actually two singlets, 
separated by a splitting $\De=1$K. There is a strong Ising character to 
the dipole matrix elements connecting them, 
and because the dipolar interactions between ions are actually larger than 
$\De$, the outcome is again a ferromagnet, but with the more convenient 
$T_c$ of $3.5$K. {Figure \ref{GA2} \ct{youngblood82}
shows what is most relevant for this review: upon substitution of Y for Tb at 
a fraction $1-p$ of the RE sites to produce LiTb$_{p}$Y$_{1-p}$F$_4$, $T_c$ 
does not follow the classical mean field prediction $T_c(p)=p T_c(p=1)$.} 
Instead, it is well-described by the mean field form taking single site 
quantum mixing into account, with a quantum critical point 
%at $p=p_c=?$ 
where the dipolar interaction ceases to be strong enough to yield the 
quantum mixing needed to produce magnetic moments. {The 
reason for the deviation of $T_c(p)$ from this form, represented by the solid 
line, below the quantum critical concentration $p\sim 0.2$ in Fig. 28 is the 
nuclear hyperfine interaction which continues to mix the singlets and thereby 
restores the dipole moments on the Tb ions. Such effects are discussed in 
more detail in the context of LiHoF$_4$ below.}
%Another paradigmatic result, 
%which we will discuss in more detail below in the context of LiHoF$_4$, 
%is that for $p$ close to $p_c$, the QPT is avoided, 
%due to the intervention of the nuclear hyperfine interactions which 
%impose dipolar transitions between the electronic singlets.

A key feature of transverse field Ising systems is the mode softening 
which occurs on approaching the quantum critical point {where 
fluctuations become pronounced and long-ranged}. Figure \ref{GA3} shows
this phenomenon in the form of inelastic neutron scattering data for 
LiTb$_{0.38}$Y$_{0.62}$F$_4$. The observed magnetic excitation has the required
dispersion, but there is both considerable damping of the excitation and strong
quasielastic scattering. There is also 
no trace of the excitation for the pure compound LiTbF$_4$, and all the 
spectral weight seems to be concentrated in quasielastic scattering. 
%\begin{widetext}

\begin{figure}
\begin{center}
\includegraphics[height=3.0in,width=6.0in]{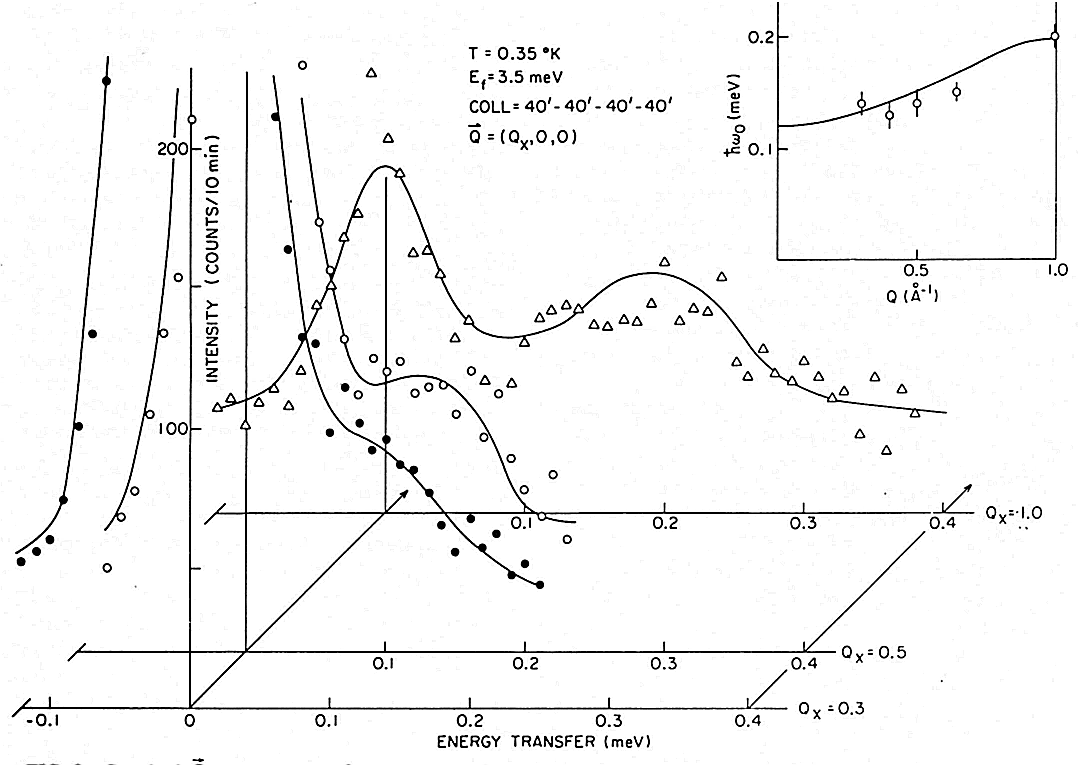}
\caption{ {Magnetic excitons in LiTb$_p$Y$_{1-p}$F$_4$ at $p=0.38$
in the ferromagnetic phase; solid line corresponds to the theoretical line 
shape fitted to the data. Inset compares the theoretical dispersion curve for 
excitons (solid line) and fitted value of the exciton energies.} (After 
Youngblood $et~al.$,1982.)} \label{GA3} 
\end{center}
\end{figure}
%\end{widetext}

\subsection{Order/disorder transitions in hydrogen-bonded and other 
ferro/antiferroelectric systems}
\label{expt2}

In hydrogen-bonded systems, it is quite common for there to be two possible 
sites which any H ion can occupy. An electric field could bias the ions 
towards particular site occupancy patterns, and this field could be 
internally generated via the electric charges of the protons themselves. 
We therefore have a situation very much analogous to that for the singlet 
ground state magnets discussed above, and indeed much of the same physics is 
observed. The ferroelectricity of potassium dihydrogen phosphate 
(KH$_2$PO$_4$) and isomorphous systems and their order-disorder transition 
behavior were satisfactorily explained employing transverse Ising models
\ct{blinc72,stinchcombe73}. The hydrogen ion (proton) in the O-H-O bond 
linking the PO$_4$
tetrahedra on the three-dimensional lattice is in a symmetric double well 
potential provided by the pair of oxygen ions at each lattice
site. An asymmetric position of the proton in any of these wells would
induce a tiny electrical dipole moment and these local dipole moments
at different sites would interact (ferroelectrically in the case of
KH$_2$PO$_4$) and induce order at low temperatures in the
absence of any quantum tunneling across the double wells. However, since 
the oxygen-oxygen double well at each site has a finite width and finite 
height of the barrier separating the individual wells, quantum tunneling of 
the proton can take place (with a probability dependent on the proton mass 
and the well structure or lattice constant), {thereby inducing 
quantum relaxation that thwarts the development of order}.

In a transverse Ising model of such ferro- or antiferroelectric
systems, the degenerate (double well minima) proton states at each
site can be represented by Ising spin states ($\si_i^z = \pm 1$), 
interacting with each other through long-range dipole-dipole interactions
$J$. The quantum tunneling field $h$ (which couples to
$\si_i^x$, inducing transitions from one Ising spin state to the
other) is determined by the mass of the proton or deuteron (in the
case of KD$_2$PO$_4$) or the pressure induced changes in the lattice
constants. For low values of the tunneling field $h$, the dipolar
interactions $J$ induce order (ferro or antiferro type, depending on
the sign of $J$), while if the value of the field is higher than a
certain threshold (which depends on
the magnitude of $J$ and the lattice structure), one gets a
disordered (para-electric) phase. The resulting analysis of the
static and dynamic behavior across such order-disorder transitions
of the ferro-/antiferroelectric systems have been very satisfactory
and well-accepted in the literature (see e.g., Blinc and Zeks, 1987).
Solid state mixtures of ferroelectric and antiferroelectric components 
induce glassy behavior, where the tunneling of protons dominates the low 
temperature relaxation and can be related directly to 
the quantum mechanical action \ct{feng96}.

\begin{figure}
\begin{center}
\includegraphics[height=1.5in]{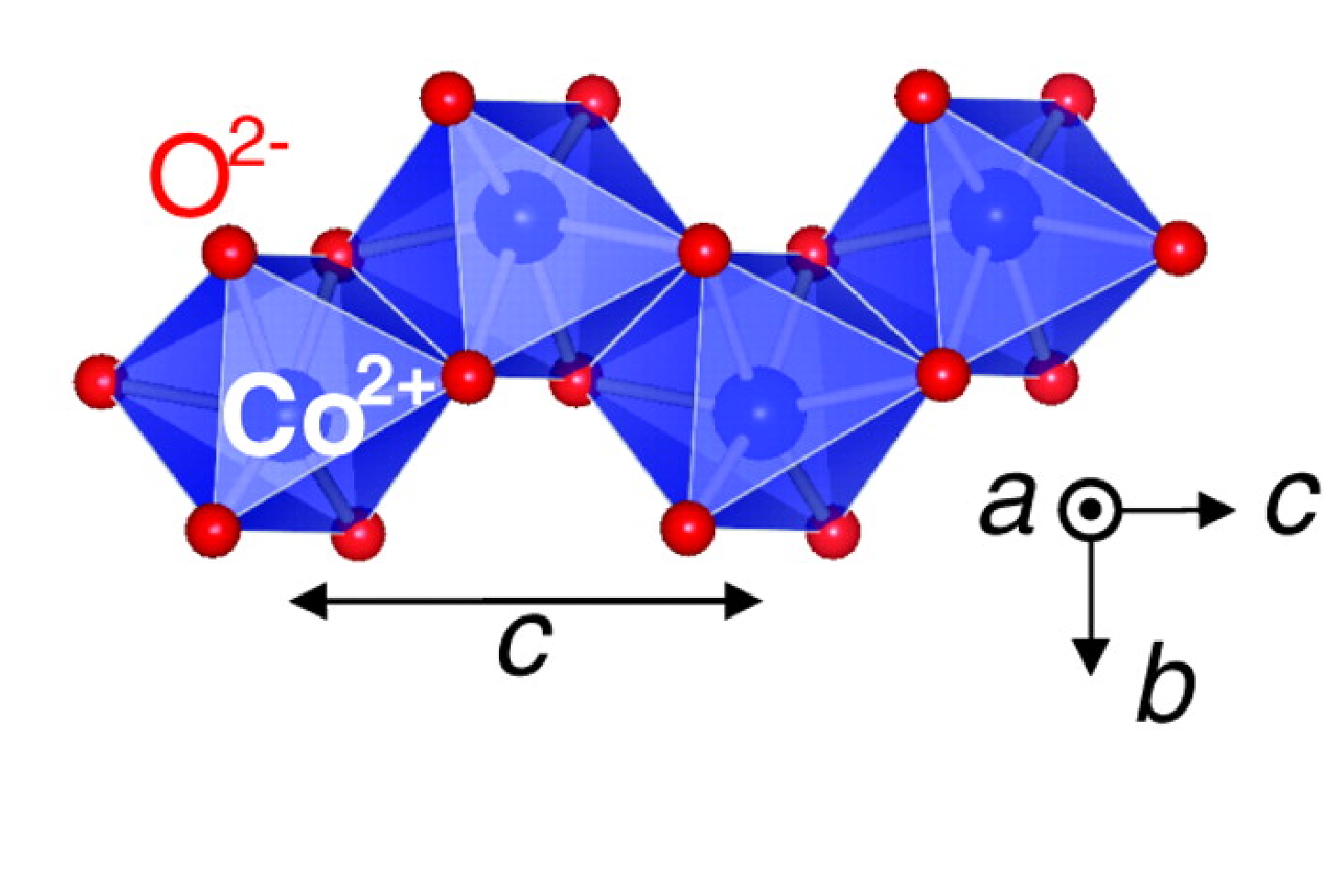}
\caption{(Color online) Structure of CoNb$_2$O$_6$.} \label{GA4} 
\end{center}
\end{figure}

\subsection{Low-dimensional magnetic realizations of the transverse field 
Ising model}
\label{expt3}

Because of the analytic tractability of one-dimensional systems, there is a 
long history of experiments on magnetic chain compounds, and there has been a
correspondingly large body of reviews. Even so, it is only within the last year 
that a definitive experiment on the one-dimensional transverse field Ising 
model has been performed \ct{coldea10}. The material for this study is 
CoNb$_2$O$_6$, whose crystal structure is illustrated in Fig.~\ref{GA4}. 
The building blocks are ferromagnetic zig-zag chains of edge sharing CoO$_6$ 
octahedra; this is in contrast to the linear chains of apex 
sharing NiO$_6$ octahedra in the model $S=1$ antiferromagnet, YBaNiO$_5$ 
\ct{guangyong07} which are 
arranged so as to be coupled antiferromagnetically. In an external field 
transverse to the ``easy" spin axis, the ordering along the easy axis is 
destroyed at a QPT due to delocalized domain walls, 
sometimes referred to as kinks or solitons. The associated magnetic Bragg 
reflections, resulting from the full three-dimensional order, disappear at the
quantum critical field $h=5.5 T$. There will be a related mode softening, where
the frequency of the zone center excitation vanishes at the quantum critical 
field. The zone center is where three-dimensional order appears among the 
weakly coupled chains; elsewhere, the softening is incomplete. Fig.~ 
\ref{GA5} displays the corresponding 
inelastic neutron scattering data for transverse fields near, at and above the 
critical transverse field. The dominant feature in the ordered state is a 
domain wall pair continuum split into discrete levels 
due to the longitudinal field $h_L$ derived from neighboring chains, which 
results in an attraction between domain walls rising linearly with their
separation (because the number of spins misoriented relative to the molecular 
field in the ordered state is equal to the separation between walls); this 
is very similar to the confining potential for quarks in high energy physics.
As the quantum critical point is approached and $h_L\to 0$, an exact 
scattering matrix analysis predicts that 
a scaling regime should be entered where the Lie algebra of $E_8$ accounts for 
eight bound states with energies such that the ratio of the lowest two are 
given by the golden mean \ct{zamolodchikov89} (see also Sec.~\ref{expcorr}).

\begin{figure}
\begin{center}
\includegraphics[height=2.4in]{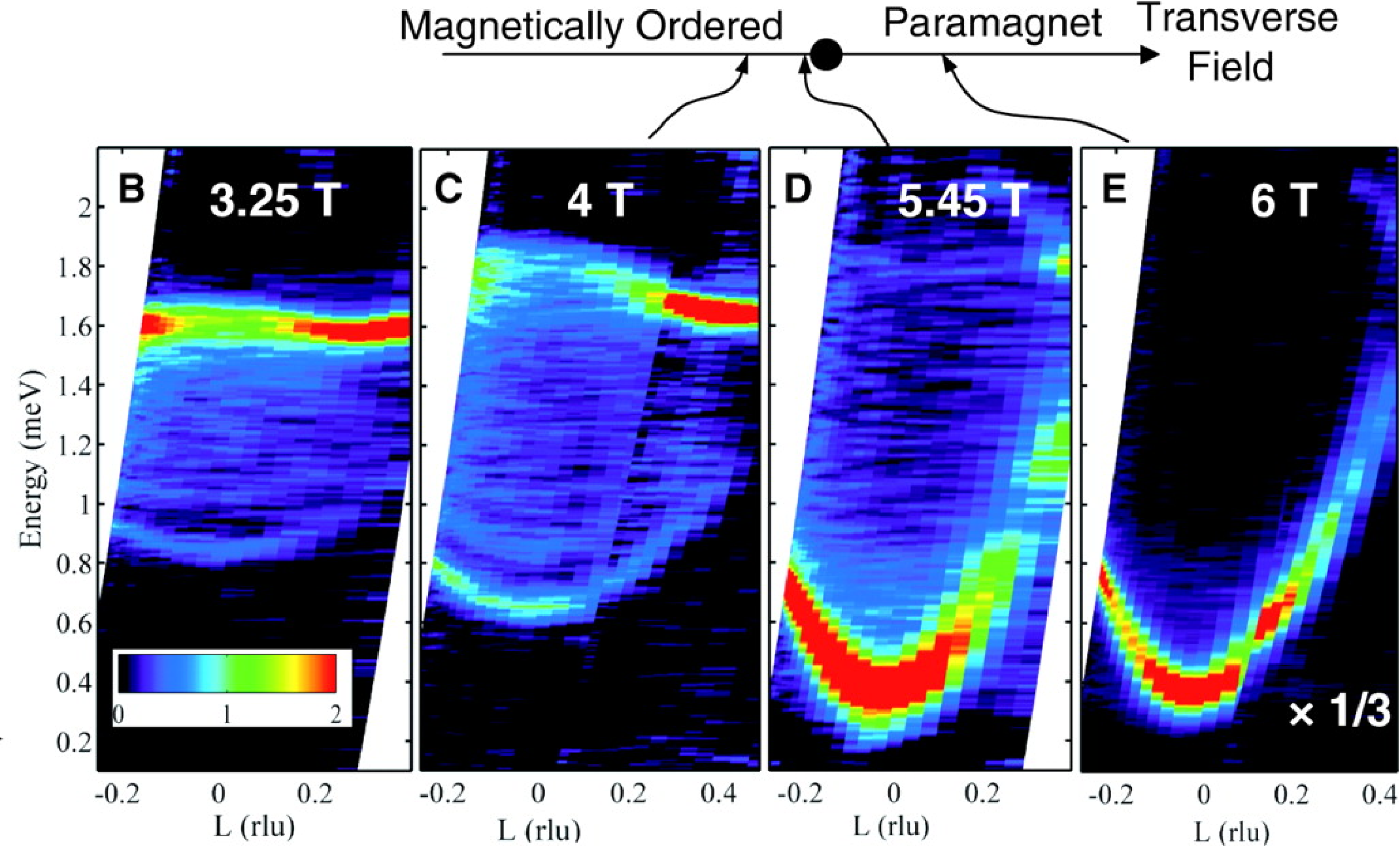}
\caption{(Color online) Magnetic spectra in CoNb$_2$O$_6$ for transverse fields
above and below QCP. In the ordered phase [(B) and (C)], excitations form a 
continuum due to scattering by pairs of kinks; in the paramagnetic phase (E), 
a single dominant sharp mode occurs, due to scattering by a spin-flip 
quasiparticle. Near the critical field (D), the two types of spectra tend to 
merge into one another. (After Coldea $et~al.$, 2010).} \label{GA5} 
\end{center}
\end{figure} 

\begin{figure}
\begin{center}
\includegraphics[height=2.4in]{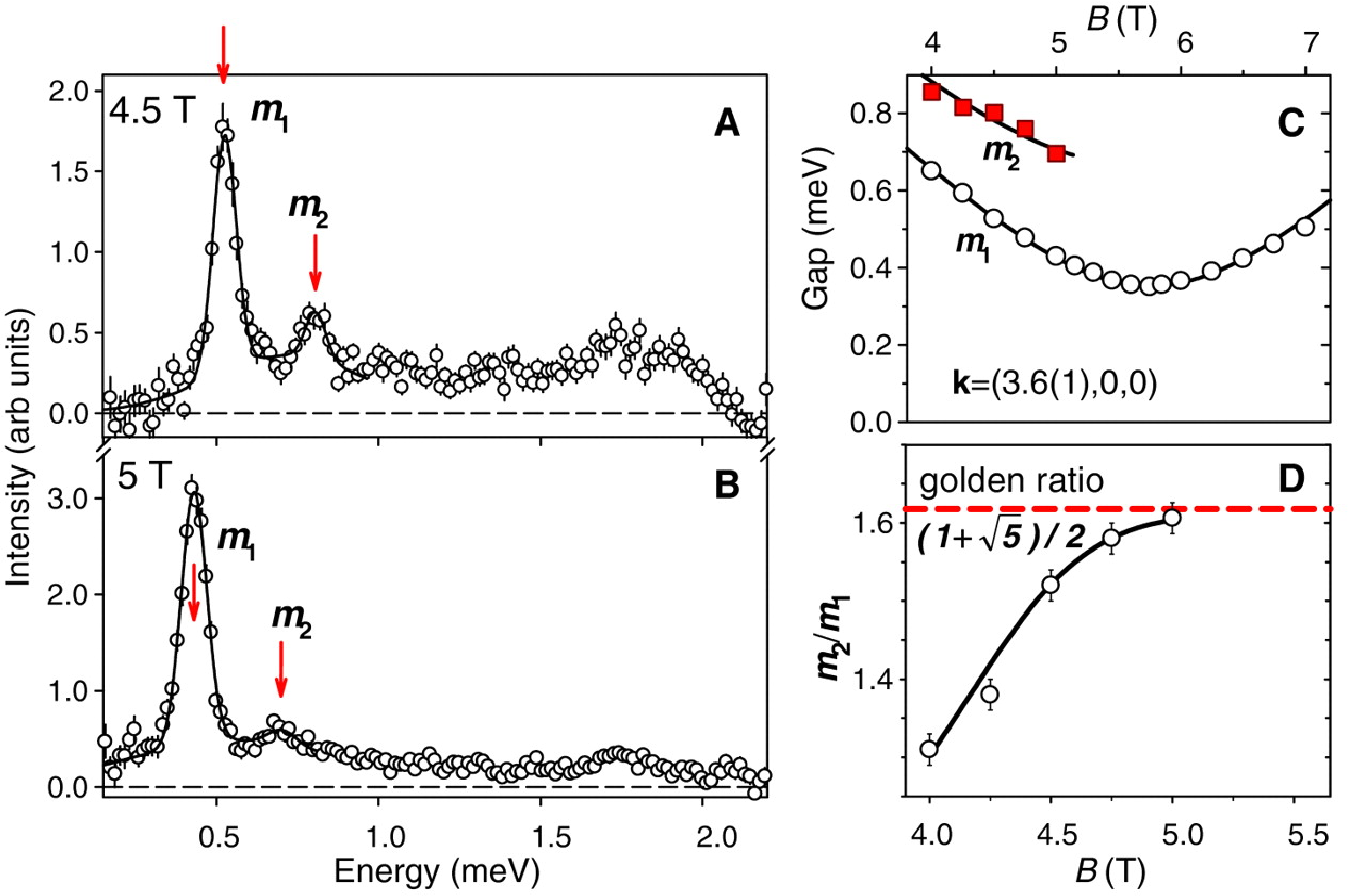}
\caption{(Color online) Scaling regime for QCP in CoNb$_2$O$_6$. (A) and (B): 
Energy scan at zone centers at $4.5$ and $5T$ with two peaks $m_1$ and $m_2$. 
(C) Softening of two energy gaps near the critical field. (D) The ratio 
$m_2/m_1$ approaches the $E_8$ golden ratio just below the critical field. 
(After Coldea $et~al.$, 2010).} \label{GA6} 
\end{center}
\end{figure}

Experimental results for 
CoNb$_2$O$_6$, reproduced in Fig.~\ref{GA6}, bear out this prediction. For 
$h>h_c$, the spectrum is dominated by single spin flips instead of domain 
wall pairs, with the result that the continuum is replaced by a dispersing 
magnetic excitation (see Fig.~\ref{GA5}). The experiments on CoNb$_2$O$_6$ are
important not only because of the many exact theoretical results tested so 
far, but also because they suggest new avenues for research. {In 
particular, we look forward to experiments dealing with magnetic and 
non-magnetic impurities substituted for Co to parse the effects of disorder 
and chemical doping on the statics and dynamics of transverse Ising models
in reduced dimensions.} Furthermore, pressure could regulate the 
interchain coupling, thus modifying the bound state spectra.

\subsection{Ising doublets in external magnetic fields in three-dimensions}
\label{expt4}

The system for which the most data are available is LiHoF$_4$, from the 
LiREF$_4$ family already introduced in Sec.~\ref{expt1}. The ground state term 
is a non-Kramers doublet with strong anisotropy, implying effective 
Ising spins {with angular momentum $J=8$}
along the tetragonal $c$-axis. External magnetic fields can mix this term with
the excited states at energies of $\sim 1$ meV and above. Of particular 
interest are laboratory fields applied perpendicular to the $c$-axis, which 
split the degenerate ground state into two terms with $\langle J_z \rangle=0$.
{The splitting of the ground state doublet in a transverse field, 
$\Gamma$,} is even in the external field because it is generated via mixing, 
implying that to lowest order a transverse Ising model representation of the 
system with an effective transverse field $h \sim \Gamma \sim H_t^2$. 
{Here $H_t$ is the laboratory magnetic field applied transverse to 
the Ising axis}. Figure \ref{GA7} A displays the field-temperature phase 
diagram \ct{bitko96,ronnow05} for LiHoF$_4$, with a boundary 
between Ising ferromagnetism and paramagnetism terminating at the classical 
phase transition for $H_t$=0 and a quantum critical point at $H_t=5T$. The 
dashed line corresponds to the mean field expression for the boundary where 
the nuclear spins ($I=7/2$) are ignored. The solid line takes into account
the nuclear hyperfine coupling, and provides a good description of the data; 
the bulge setting in below $0.7 K$ means that the nuclear spins inhibit the 
electronic quantum fluctuations, i.e., produce composite electro-nuclear 
degrees of freedom with effectively higher and more classical spin, and 
therefore {raise the critical field from $4T$ to $5T$}. 
This correction notwithstanding, 
LiHoF$_4$, with its crystalline and chemical perfection, represents an 
excellent venue for a quantitative examination of quantum critical phenomena, 
in the same way that it and more prominently LiTbF$_4$ previously served as 
testbeds for the RG theory of classical critical behavior.
Figure \ref{GA8} shows the divergence of the bulk susceptibility as the 
quantum critical point is approached along thermal and field trajectories. 

\begin{figure}
\begin{center}
\includegraphics[height=2.8in]{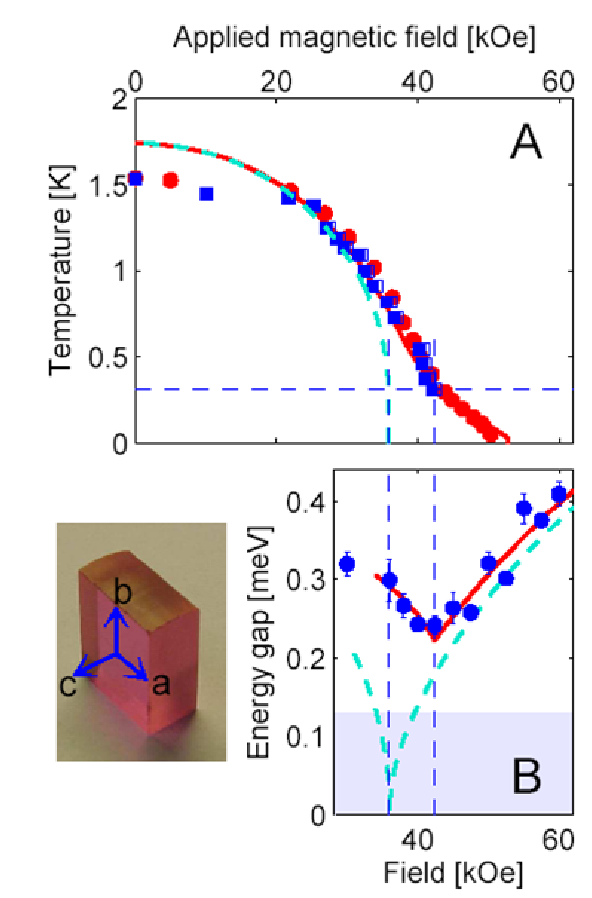}
\caption{(Color online) Phase diagram of LiHoF$_4$ shown in (A). In (B), we 
show the field dependence of the lowest excitation energy. (After 
Ronnow $et~al.$, 2005.)} \label{GA7} 
\end{center}
\end{figure}

\begin{figure}
\begin{center}
\includegraphics[height=3.4in]{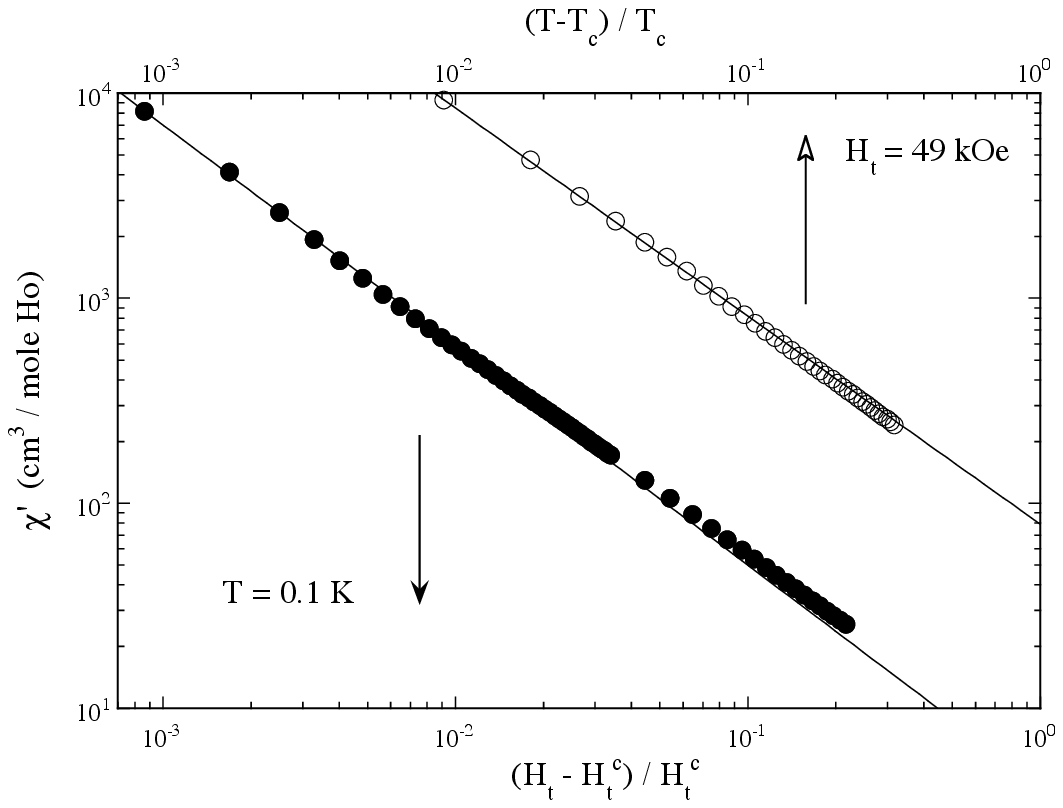}
\caption{Divergence of magnetic susceptibility on approaching QCP in ${\rm Li 
Ho F_4}$. (After Bitko {\it et~al.}, 1996).} \label{GA8} 
\end{center}
\end{figure}

The important result is that they are both described by exponents 
indistinguishable from unity, which is what is also anticipated based on 
mean field theory which should be valid given that $d^u_c$ is already $3$ for 
the classical system, and is reduced by $z$, the dynamical critical exponent 
which will be a positive number, for the quantum phase transition. 
Inelastic neutron scattering measurements show that the nuclear spins also 
affect the quantum critical dynamics strongly \ct{ronnow05,ronnow07}.
In particular, the magnetic excitation which dominates the inelastic neutron 
scattering data in Fig.~\ref{GA9} undergoes incomplete softening at the 
quantum phase transition. What accounts for the diverging susceptibilities 
seen in the bulk measurements (Fig.~\ref{GA8}) is then the quasielastic peak 
which appears near zero frequency in the images of Fig.~\ref{GA9} is due to 
the mixing of electronic and nuclear degrees of freedom.

\begin{figure}
\begin{center}
\includegraphics[height=2.8in]{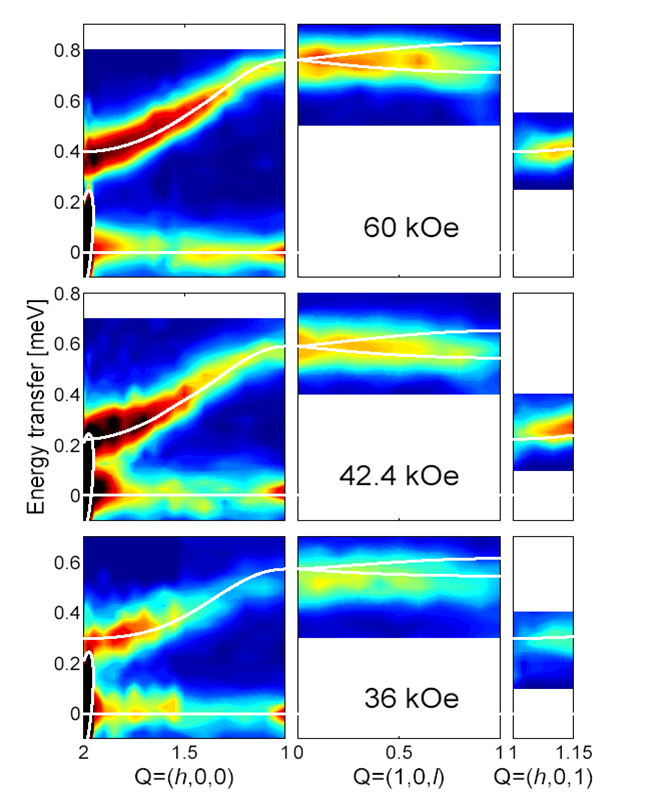}
\caption{(Color online)
Incomplete mode softening on passing through QPT in LiHoF$_4$. Figure shows
pseudocolor representation
of the inelastic neutron scattering intensity of LiHoF$_4$ at $T=
0.31$ K observed along the reciprocal space trace $(2,0,0) \to (1,0,0)
\to (1,0,1) \to (1.15,0,1)$. (After Ronnow $et~al.$, 2005).} \label{GA9} 
\end{center}
\end{figure}

\begin{figure}
\begin{center}
\includegraphics[height=1.8in]{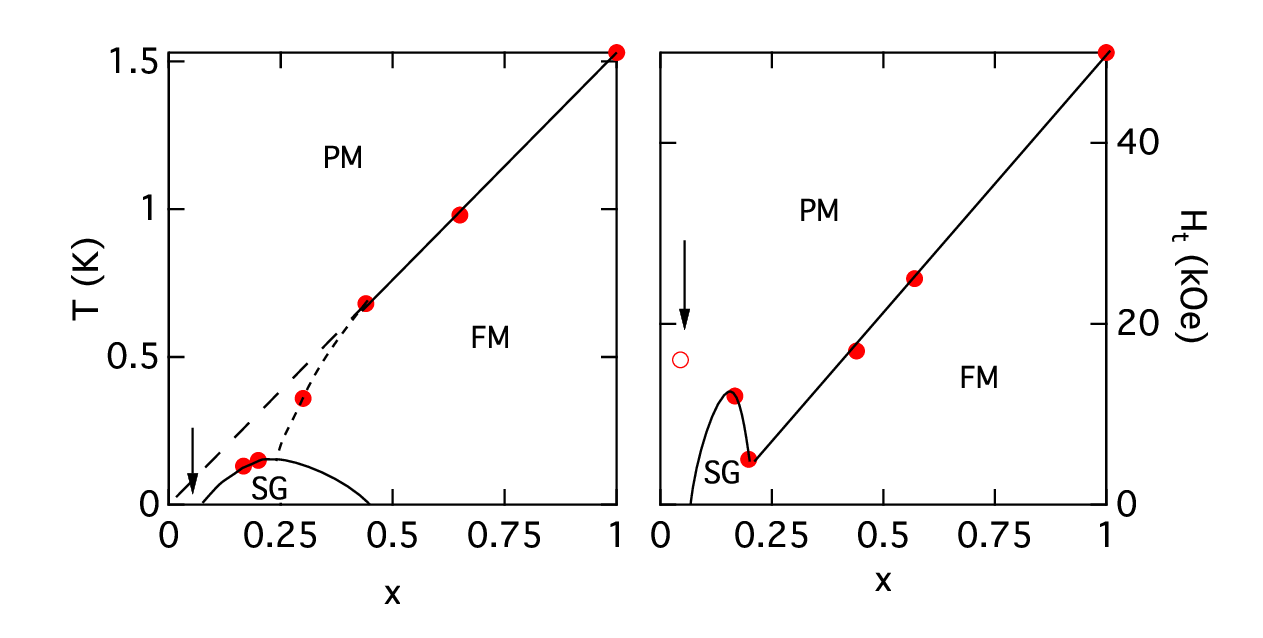}
\caption{{Dilution of the dipolar-coupled, Ising ferromagnet 
LiHo$_{x}$Y$_{1-x}$F$_4$. Left panel: magnetic phases in the $x-T$ plane. 
Arrow denotes spin liquid ``antiglass'' phase. Right panel: Magnetic phases 
in $x-H_t$ plane where $H_t$ is the field applied transverse to the Ising 
axis. Open circle shows the peak in susceptibility for the antiglass.} 
(After Ancona-Torres $et~al$, 2008)}. \label{GA_new} 
\end{center}
\end{figure}

\subsection{Disorder and the transverse field Ising model}
\label{expt5}

The LiTb$_x$Y$_{1-x}$F$_4$ and K(H$_x$D$_{1-x}$)$_2$PO$_4$ series of compounds
are naturally disordered materials where the mean transverse fields are 
scanned by changing the composition variables $x$. Because of localization and
random field effects naturally associated with disorder, it would obviously be 
useful to scan the transverse field and disorder separately. This is 
achieved for LiHo$_x$Y$_{1-x}$F$_4$ in a transverse external field. As the Ho 
sublattice is diluted by non-magnetic Y, the fact that the dipolar interaction
is ferromagnetic between Ising spins with separation vector parallel to the 
Ising axis and antiferromagnetic between spins separated by a vector 
orthogonal to the Ising axis becomes highly relevant in that it will 
introduce numerous antiferromagnetically correlated spins into the (classical)
ground state of the system. Furthermore, a very long time ago, Aharony and 
Stephens predicted that this would lead to a spin glass ground state for any 
$x>0$. Initial experiments \ct{reich86,reich90} focused on the zero field 
phase diagram, {shown in Fig.~\ref{GA_new} along with its analogue
in transverse field \ct{torres08}}. For $x>0.4$, the dominant feature is a
para-ferromagnetic transition located along the straight line $T_c=xT_c(x=1)$
given by the most simple effective medium theory. For $x<0.4$, the transition
is clearly below this line, and the signatures of ferromagnetism 
{apparently disappear} entirely below $x \sim 0.2$, where 
there is first a low-temperature state displaying {glassy} 
behavior, which persists at least to $x \sim 0.1$.
We note parenthetically that there has been some controversy \ct{jonsson07}
surrounding this spin glass state, but this has died away in view of the 
agreement between the two groups \ct{torres08,wu91,quilliam08} which probed it 
using low fields and low frequencies; the seemingly different results were 
obtained under the very different experimental circumstances of dc SQUID 
magnetometry performed with rapid sweep rates to high fields.

For $x<0.1$, there are reports of both continued spin glass behavior 
\ct{quilliam08,quilliam07} as well as 
a novel antiglass state \ct{reich87}, so named because a reduction in the
temperature lead to a reduction of the width of the distribution of barriers 
to relaxation rather than the conventional increase. The antiglass state is 
marked by strongly entangled spins, analogous to the Bhatt-Lee state 
\ct{bhatt82} found 
below the insulator-metal transition in doped silicon, and indeed, 
a decimation calculation \ct{ghosh03} which ignores nuclear spin degrees of 
freedom but takes into account both the diagonal and off-diagonal dipolar 
couplings between Ho$^{3+}$ ions (the latter providing the necessary quantum 
mixing), gives a remarkably good description of the magnetic susceptibility 
in the limit of small frequencies. An additional astonishing effect 
(Fig.~\ref{GA10}) is spectral hole burning \ct{ghosh02}, the acoustic 
frequency analog to optical experiments where the frequency($f$)-dependent 
small
signal response $\chi (f)$ is measured in the presence of a large pump signal 
at fixed frequency $f_o$. Very sharp minima at $f_o$ are induced in $\chi (f)$,
indicating the decoupling of low frequency magnetic excitations from each 
other; these experiments motivated a successful search for similar behavior 
in the classic geometrically frustrated system Gadolinium Gallium Garnet 
(${\rm Gd_3 Ga_5 O_{12}}$ or GGG) \ct{ghosh08}.

\begin{figure}
\begin{center}
\includegraphics[height=2.8in]{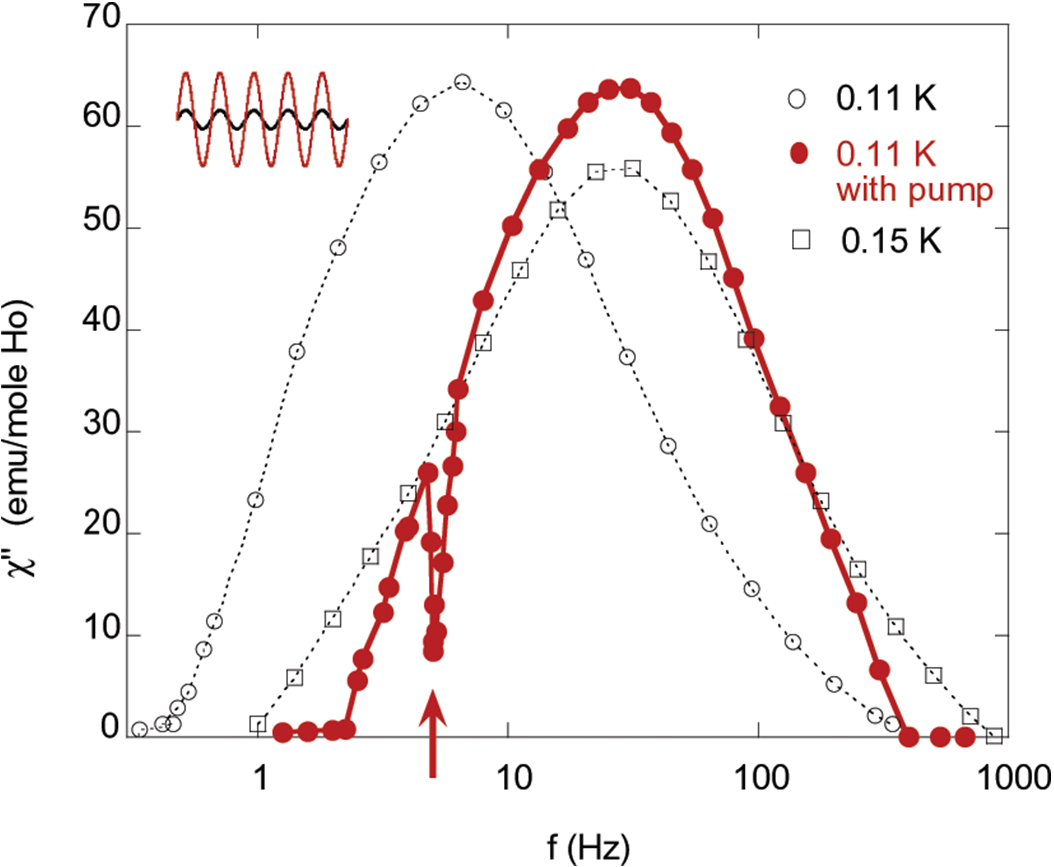}
\caption{(Color online)
The presence of a sharp minimum is indicative of spectral hole 
burning in ${\rm LiRE_xY_{1-x}F_4}$ with RE=Ho and $x=0.046$. (After Ghosh 
$et~al.$, 2002).} \label{GA10} 
\end{center}
\end{figure} 

The antiglass regime remains a highly active area of contemporary research, 
both for the fundamental reason that it seems an excellent venue for probing 
non-equilibrium quantum effects in macroscopic systems of strongly coupled 
degrees of freedom, but also on account of the divergent claims based on 
different experiments. The differing claims are made for samples of nominally 
the same composition $(x=0.046)$, but mounted in radically different ways in 
the measuring cryostat. The discrepancy is real and interesting: trivial 
aspects of thermalization cannot account for the differing results because 
in both sets of measurements, cooling the cryostats towards their base 
temperatures does result in monotonic changes in properties. It is worth 
adding that the different measuring apparatuses (superconducting SQUID and 
ac induction coils) used by the two groups involved similar measuring 
fields and frequencies, and are therefore unlikely to be responsible 
for the different results. For the lowest concentrations, the magnetic 
behavior is dominated by single ions and the occasional pairs of ions, and 
SQUID magnetometry shows the hysteresis loops characteristic of single 
molecule magnets, including quantum tunneling effects \ct{giraud01}.
Transverse fields $H_t$ applied to LiHo$_x$Y$_{1-x}$F$_4$ lead to three 
important families of phenomena.

%\begin{enumerate}

%\item 

\noi 1. The transverse field induces a random field via the off-diagonal terms 
of the dipolar interaction, which have non-zero expectation values due to the 
lower local symmetries \ct{tabei06,schechter081} resulting from random 
substitution of Y for Ho. This 
is manifested in a hardening of the (longitudinal field-induced) magnetization
loops on application of the transverse field for $T$ near $T_c(x)$ 
\ct{silevitch10}, as well as different critical exponents for the
longitudinal susceptibility as a function of transverse field for $T=T_c(x)$ 
in the cases where $x=0.44$ and $x=1$, respectively \ct{silevitch07}. In 
particular, the divergence of the longitudinal susceptibility at the classical
critical point is defined by the exponent $1/\de$ in 
$$\chi'=\frac{dM_z}{dh_z} (H_t)|_{T=T_c(H_t=0)}\sim \frac{1}{|H_t|^{1/\de}}.$$
The experiment 
%(Fig.~\ref{fig_ga12}) 
shows that for ordered, pure LiHoF$_4$, 
$1/\de=2$, while for the disordered $x=0.46$ compound $\de=1$, where all 
of the data in the PM phase are described by the form 
\beq \chi'= \frac{C}{\al'\nu_B |H_t| +(T - T_c) + \ga T}. \label{eq_ga1} \eeq

%\begin{figure}[htbp]
%\begin{center} 
 %\end{center}
%\includegraphics[height=4.0in,width=3.4in]{RL10033_fig36a.eps}
%\end{figure}
%\begin{figure}[htbp]
%\begin{center} \includegraphics[height=4.0in,width=3.4in]{RL10033_fig36b.eps}
%\caption{Left panel: Top: The schematic picture of quantum annealing. Bottom:
%The phase diagram of the disordered Ising magnet ${\rm LiHo_{0.44}Y_{0.56}F_4}$
%in a transverse field. The quantum (red) and classical (blue) annealing 
%protocols provide different pathways for the free-energy minima. Right panel: 
%Spectroscopy of the material at points A to D after quantum and classical 
%computations. (After Brooke $et~al.$, 1999).} \label{fig_ga12} \end{center} 
%\end{figure}

\begin{figure}[h]
\begin{center}
$\begin{array}{cc}
\multicolumn{1}{l}{\mbox{\bf (a)}} &
	\multicolumn{1}{l}{\mbox{\bf (b)}} \\[-0.53cm]
\epsfxsize=1.5in
\epsfysize=2.5in
\epsffile{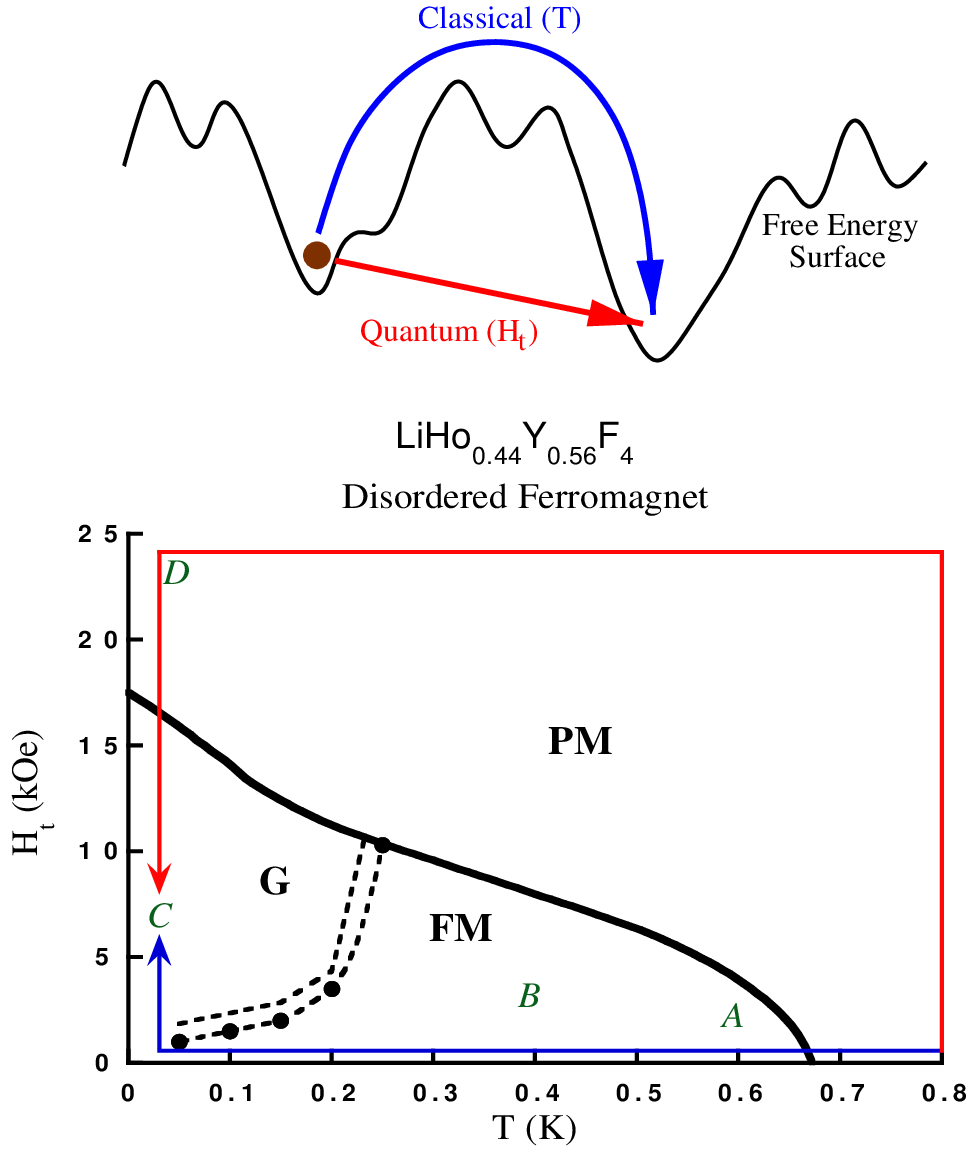} &
	\epsfxsize=2.0in
\epsfysize=3.0in
	\epsffile{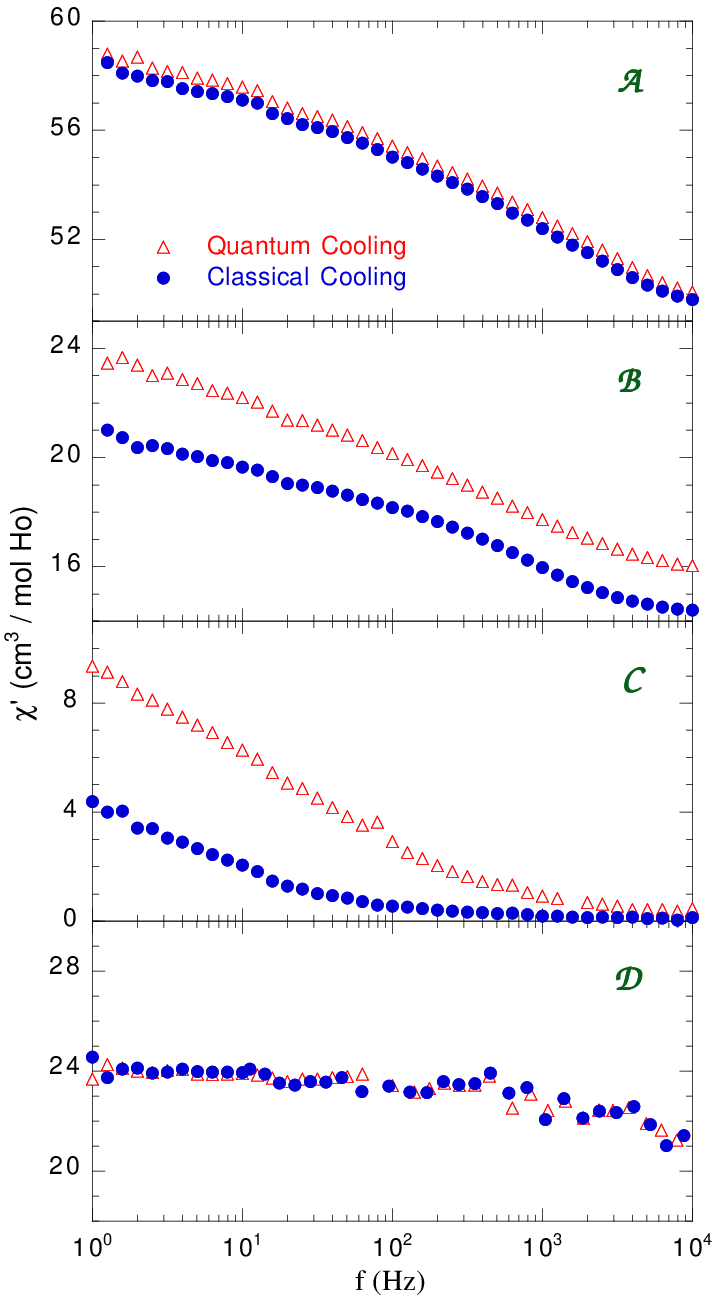} \\ [0.4cm]
%\mbox{\bf (aa)} & \mbox{\bf (bb)}
\end{array}$
\end{center}
\caption{(Color online)
Left panel: Top: The schematic picture of quantum annealing. Bottom:
The phase diagram of the disordered Ising magnet ${\rm LiHo_{0.44}Y_{0.56}F_4}$
in a transverse field. The quantum (red) and classical (blue) annealing 
protocols provide different pathways for the free-energy minima. Right panel: 
Spectroscopy of the material at points A to D after quantum and classical 
computations. (After Brooke $et~al.$, 1999).} \label{fig_ga12} \end{figure}

The power law seen at $T_c$ is remarkable enough, but the accuracy of 
(\ref{eq_ga1}) is even more so, especially because it is non-analytic as 
$H_t\to 0$, encapsulating the singular nature of the magnetic response above 
$T_c$, and the corresponding experiments represent a particularly clear 
manifestation of the Griffiths phenomenon, where singular behavior inherited 
from rare fluctuations towards higher $x$ is visible in a thermodynamic 
quantity. In spite of the simplicity of the data and phenomenology for small 
$H_t$ for $T$ near $T_c$, there is no analytical theory accounting for the 
observed effects, although much less finely grained numerical data for the 
phase diagram are consistent with (\ref{eq_ga1}), which entails a linear 
approach of the critical temperature towards $T_c$ as $H_t \to 0$ \ct{tabei06}.

%\item

\noi 2. For low temperatures, the principal effect of the transverse field 
$H_t$ is to facilitate quantum fluctuations. The dc magnetization experiments 
\ct{silevitch10} show a monotonic narrowing of the magnetic hysteresis loops
with increasing $H_t$, and ac susceptibility data \ct{brooke01} show a
speed-up of the magnetic response well described by a WKB model of domain wall 
tunneling, where the transverse field controls the mass of the domain walls 
($M^*$) which play the role of the particles in the model. Pinning sites 
provide confining potential barriers, whose magnitudes are established 
independently from the Arrhenius behavior observed in the thermal activation 
regime at higher temperatures. $M^*$ was measured to be of order $ 10^4 m_e$ 
(where $m_e$ is the electron mass), 
consistent with coherent tunneling of wall segments containing 10 spins and 
the magnitude of the magnetic dipole coupling between Ho$^{3+}$ moments. 
The control of the tunneling by an external transverse field enabled the 
first experimental exploration of quantum annealing (or `adiabatic quantum 
computation') in 1999 \ct{brooke99}. The computational problem was that of 
finding the low-lying states of the disordered magnet with the impurity 
configuration of the particular $x=0.44$ sample used. The readout was achieved 
simply by measuring the frequency-dependent magnetic susceptibility 
$\chi(\om)$ along the Ising axis, and different trajectories in the $H_t-T$ 
plane were used as `programs' for finding optimal states. The experiments 
(see Fig.~\ref{fig_ga12}) established that quantum and thermal protocols give 
rise to states characterized by visibly different properties. The principal 
effect seems to be that reaching 
low $H_t$ and $T$ via the quantum route of reducing $H_t$ results 
in a state with more rapid small amplitude fluctuations than that reached via 
a thermal route. Indeed, quantum annealing moves $\chi(\om)$ towards higher 
frequencies in such a way that the log $\om$ behavior of $\chi'$, 
corresponding to a marginally stable ferromagnetic state, is apparent. 
Because more rapid small amplitude fluctuations are associated with sharper 
free energy minima, the experiments show that quantum annealing uncovers less 
gentle minima than thermal annealing. In addition, if there is some scale 
invariant distribution of minima curvatures, as suggested by the log $\om$ 
decay of $\chi'$, quantum annealing has proven its value by virtue of 
revealing this behavior over easily reachable frequencies. Notwithstanding 
this result, the question of which state has lower free energy has yet to be 
answered by careful thermodynamic measurements.

%\item 

\noi 3. The phase boundaries in the $H_t -T$ plane, which bound a ferromagnetic 
phase for $x>x_c\sim 0.2$ and a spin glass-like phase for $x<x_c$, do not scale 
as simply as one might expect. In particular, neither the laboratory quantum 
critical field $H_t^c$ nor the effective transverse field $\De$ scale as 
$T_c$. There is even a minimum in the critical field at $x_c$, which 
represents evidence for the very different nature of the spin glass-like and 
disordered ferromagnetic ground states. Also, as we pass from the pure to the
disordered samples with $x\sim 0.5$, the shape of the phase boundary in the 
$H_t-T$ plane also changes from being highly curved near $T=T_C(x)$ to much 
more linear. In the spin glass regime, the phase boundary between the 
PM and spin-glass states is established by looking for the vanishing of 
the power law $\al$ in the fluctuation spectrum $\chi(\om)\sim 
\om^\al$ \ct{wu91}. This is correlated with the (analysis-free)
observation of the onset of a plateau in $\chi(\om)$ for suitably low, 
fixed $\om$, as a function of $H_t$, and it yields a very well-defined 
phase boundary. There was also an examination \ct{wu93,torres08} 
of the non-linear susceptibility along this phase boundary, and its behavior 
was consistent with divergence near $T_g(H_t=0)$. The associated 
critical exponent, however, was anomalously small, and as $T \to 0$ and $H_t 
\to H_t^c$ it becomes smaller as the singularity is washed out, even as the 
signal for the phase boundary becomes sharper in $\chi(\om)$. There is 
also a theoretical puzzle coming from the knowledge that $H_t$ also induces 
internal random fields, and should, according to classical theory, eliminate 
phase transitions at finite $H_t$. The resolution for this as well as an 
explanation of the various peculiar observations is likely to lie in the fact
that the classical theory does not consider a quantum glass with entangled 
spin pairs and the associated distribution of level crossing resonances. 
The configurational entropy of the glass state can be due to a near degeneracy
of the quantum pairing of spins rather than the relative orientation of 
classical spins; we look forward to a theory of this approach to provide 
a better understanding of the experiments.

%\end{enumerate}

\section{Concluding remarks}

In this review, we have discussed quantum phase transitions in transverse 
field Ising and $XY$ models in one and higher dimensions. Quantum phase 
transitions for pure models as well as models with random interactions or 
random fields have been discussed. As mentioned earlier, these models 
were introduced in the early 1960's in the context of order-disorder 
ferroelectric systems. Our review focuses on the salient issues for which 
these models continue to be important and interesting even fifty years after 
their inception. 

On the theoretical side, the integrability of the transverse field Ising/$XY$ 
models in one dimension has provided an ideal testing ground for field 
theoretical and information theoretical studies. Numerous predictions of 
theoretical studies have been verified experimentally in recent years.
\ct{coldea10}. The one-dimensional models have been 
extremely useful in studying information theoretic measures like concurrence, 
entanglement entropy, fidelity and fidelity susceptibility, and also the 
scaling of the defect density generated by quantum critical and multicritical 
quenches, namely, the Kibble-Zurek scaling. These models have played a 
crucial role in the development of quantum annealing techniques and adiabatic
quantum algorithms. The equilibration or thermalization following quantum 
quenches have been studied using variants of transverse field models. 

An advantageous feature of transverse field Ising models is the 
quantum-classical mapping which renders these models ideally suited for 
quantum Monte Carlo studies for $d>1$. One can study the QPTs in these models, 
both pure and random, by investigating the thermal phase transitions of 
equivalent classical Ising models in higher dimensions (see for example, 
Sec.~\ref{sec_qdis}). We have also discussed Griffiths-McCoy
singularities and activated quantum critical dynamics associated with 
low-dimensional random quantum transitions. 
%Recently, a proposal has been made to study non-equilibrium quantum dynamics 
%using an imaginary time quantum Monte Carlo method \ct{grandi11}.

On the experimental side, the discovery of ${\rm LiHoF_4}$ systems and their 
disordered versions has opened up new vistas in the studies of transverse field
Ising models. The experiments have established the existence of both quantum 
phase transitions and quantum glassy states. Unique phenomena like ``spectral 
hole burning" and the existence of ``antiglass states" and ``entangled 
quantum states of magnetic dipoles" have emerged from these experimental 
studies. More importantly, we have discussed the possibility of realizing 
quantum annealing in these systems$^{*}$. We have also pointed out some 
experimental observations which do not have any theoretical explanations to 
date. In short, the transverse field models have played a crucial role in 
 {the understanding of 
quantum phase transitions}, and we believe that they will continue 
to remain important for future theoretical and experimental research.

{$^{*}$ Note added:} {After this review was originally 
prepared, a realization of the traverse field Ising model for an eight-qubit 
chain using tunable Josephson junctions was implemented. This represents a 
significant breakthrough because it also permits experimental examination 
of quantum annealing (adiabatic quantum computation) with 
unprecedented detail \ct{johnson11}}.

%\subsection{Experimental summary ({\it Aeppli $\sim$ 5 pp})}

%\noi Experiment: \ct{brooke99,aeppli05}

\section*{Acknowledgments}
We thank M. Acharyya, C. Ancona-Torres, J. Axe, G. Baskaran, D. P. Belanger, 
J. K. Bhattacharjee, D. Bitko, C. Broholm, J. Brooke, E. Bucher, A. 
Chakrabarti, A. Chandra, Y. Chen, R. Chitra, R. Coldea, P. Coleman, A. Das, 
W. DeGottardi, J. F. DiTusa,
B. Ellman, C. D. Frost, S. Ghosh, J. Griffin, H. J. Guggenheim, J.-I. Inoue, 
T. Ito, J. Jensen, M. Kenzelmann, C. Kraemer, H. R. Krishnamurthy, D. F. 
McMorrow, S. Mondal, V. Mukherjee, T. Nag, K. Oka, R. Parthasarathy, S. Pati, 
A. Polkovnikov, S. Ramasesha, R. Ramazasvilli, S. Rao, P. Ray, D. H. Reich, H. 
Rieger, H. M. Ronnow, D. Samanta, G. Santoro, A. Schr\"oder, P. Sen, K. 
Sengupta, R. Shankar, S. Sharma, D. Silevitch, R. R. P. Singh, Y.-A. Soh, R. 
B. Stinchcombe, H. Takagi, M. Thakurathi,
S. Vishveshwara, W. Wu, G. Xu, and P. Youngblood 
for collaborations and discussions on the many works presented in this review.
We specially thank V. Mukherjee for immense help in preparing the manuscript. 
AD acknowledges CSIR, India, for financial support, UD acknowledges AV 
Humboldt foundation for financial support and CSIR, India, for research 
associateship, DS acknowledges DST, India for Project SR/S2/JCB-44/2010,
TR acknowledges DOE BES Grant No. DE-FG02-99ER45789, and GA thanks the 
Wolfson Foundation and the UK ESPRC for support.

\appendix

\section{Large spin limits: transverse XY spin chain}

\label{app_largespin}

{We have already discussed the exact solution of the spin-1/2 
transverse $XY$ model (\ref{eq_txy}) in Sec.~\ref{jorwigtr}. In this Appendix, 
we shall study the model in the large spin limit ($S \to \infty$). This method 
sometimes turns out to be useful for a given quantum spin Hamiltonian.} In 
this limit, the spins become classical objects 
whose components commute with each other. Quantum fluctuations can then be 
studied by going to the first order in the small parameter $1/S$ using the 
Holstein-Primakoff transformation which maps the spin operators at site $n$ 
to the raising and lowering operators of a simple harmonic oscillator at the 
same site \ct{holstein40,anderson52,kubo52}. To illustrate this, let us 
consider the Hamiltonian of the $XY$ spin chain in a transverse magnetic field,
\beq H=-\sum_n [J_x ~S_n^x S_{n+1}^x + J_y ~S_n^y S_{n+1}^y + h S S_n^z], 
\label{ham5} \eeq
where ${\vec S}_n^2 = S(S+1)$; we have introduced a factor of $S$ in
the last term in Eq. (\ref{ham5}) so that all the terms scale as $S^2$ in
the limit $S \to \infty$. Let us assume that $|J_x| \ge |J_y|$. We can ensure 
that $J_x, ~h ~\ge ~0$ by performing unitary transformations if necessary. In 
the limit $S \to \infty$, let us assume that the spins point in the direction 
given by $(\theta,\phi)$ in polar coordinates, namely, $(S^x,S^y,S^z) = S(\sin 
\theta \cos \phi, \sin \theta \sin \phi, \cos \theta)$ at all sites $n$. 
Eq.~(\ref{ham5}) then implies that the energy per site, $e_0$, is given by 
\beq e_0 ~=~ - S^2 (J_x \sin^2 \theta \cos^2 \phi + J_y \sin^2 \theta \sin^2 
\phi + h \cos \theta). \eeq
Minimizing this as a function of $\theta$ and $\phi$ gives $\phi = 0$ if $J_x 
> |J_y|$. Further, we find that $\theta = 0$ if $h \ge 2J_x$ while $\theta =
\cos^{-1} (h/2J_x)$ lies between 0 and $\pi/2$ if $h < 2J_x$. We will
now consider these two cases in turn.

\noi (i) For $h \ge 2J_x \ge 2|J_y|$, the spins point along the $z$ direction.
The Holstein-Primakoff transformation then takes the form
\bea S_n^z &=& S ~-~ a_n^\dg a_n, \non \\
S_n^x ~+~ i S_n^y &=& (2S ~-~ a_n^\dg a_n)^{1/2} ~a_n, \non \\
S_n^x ~-~ i S_n^y &=& a_n^\dg (2S ~-~ a_n^\dg a_n)^{1/2}. 
\label{holstein1} \eea
In the large $S$ limit, we expand the square roots in Eq. (\ref{holstein1})
to the lowest order to obtain
\bea S_n^x &=& (S/2)^{1/2} ~(a_n ~+~ a_n^\dg), \non \\
S_n^y &=& -i ~(S/2)^{1/2} ~(a_n ~-~ a_n^\dg). \eea
Keeping terms up to order $S$, the Hamiltonian in Eq. (\ref{ham5}) takes the
form
\bea H ~=~ E_0 ~+~ S ~\sum_n ~\left[ h a_n^\dg a_n ~-~ \frac{J_x + J_y}{2} 
(a_n^\dg a_{n+1} + a_{n+1}^\dg a_n) ~-~ \frac{J_x - J_y}{2} (a_n a_{n+1} + 
a_{n+1}^\dg a_n^\dg) \right], \label{ham6} \eea
where $E_0 = - NS^2 h$ is the classical ground state energy which is of order 
$S^2$. Here $N$ denotes the number of sites and we assume periodic boundary 
conditions; we will eventually take the limit $N \to \infty$. The remaining 
terms in (\ref{ham6}) can be analyzed by Fourier transforming from $a_n$ to 
$a_k = (1/\sqrt{N}) \sum_n a_n e^{-ikn}$, where $-\pi < k < \pi$. These 
terms then take the form
\bea H ~=~ S ~\sum_{0 < k < \pi} ~\left[(a_k^\dg a_k + a_{-k}^\dg a_{-k})
(h - (J_x + J_y) \cos k) ~+~ (a_k a_{-k} + a_{-k}^\dg a_k^\dg ) (J_y - J_x) 
\cos k) \right]. \eea
This can be diagonalized using a bosonic Bogoliubov transformation of the
form $b_k = \cosh \al ~a_k + \sinh \al ~a_{-k}^\dg$ and $b_{-k} = \cosh 
\al ~a_{-k} + \sinh \al ~a_k^\dg$. We then obtain
\bea H &=& \sum_{0 < k < \pi} ~\om_k ~(b_k^\dg b_k + b_{-k}^\dg 
b_{-k}), \non \\
\om_k &=& h S ~[(1 - \frac{2J_x}{h} \cos k)(1 - \frac{2J_y}{h} \cos k)]^{1/2}.
\eea
We see that the excitation spectrum is gapped unless $h=2J_x$ in which
case the gap vanishes at $k=0$.
% which signals the Ising transition of a transverse Ising
%chain discussed before.

\noi (ii) For $2J_x \ge h \ge 0$ and $J_x \ge |J_y|$, the spins lie in the 
$z-x$ plane and point in a direction which makes an angle $\theta = \cos^{-1}
(h/2J_x)$ with the $z$ axis. The Holstein-Primakoff transformation then takes 
the form
\bea S_n^z \cos \theta + S_n^x \sin \theta &=& S ~-~ a_n^\dg a_n, \non \\
S_n^x \cos \theta - S_n^z \sin \theta + i S_n^y &=& (2S ~-~ a_n^\dg a_n)^{1/2}~
a_n, \non \\
S_n^x \cos \theta - S_n^z \sin \theta - i S_n^y &=& a_n^\dg (2S ~-~ a_n^\dg 
a_n)^{1/2}. \label{holstein2} \eea
Expanding the square roots in Eq. (\ref{holstein2}) to the lowest order and
keeping terms up to order $S$, the Hamiltonian in Eq. (\ref{ham5}) takes the
form
\bea H &=& E_0 ~+~ S ~\sum_n ~[ (h \cos \theta + 2 J_x \sin^2 \theta) 
a_n^\dg a_n ~-~ \frac{J_x \cos^2 \theta + J_y}{2} (a_n^\dg a_{n+1} + 
a_{n+1}^\dg a_n) \non \\
& & ~~~~~~~~~~~~~~~~~~~-~ \frac{J_x \cos^2 \theta - J_y}{2} (a_n a_{n+1} + 
a_{n+1}^\dg a_n^\dg) ], \label{ham7} \eea 
where $E_0 = - NS^2 [J_x + h^2/(4J_x)]$ is the classical ground state energy.
After doing Fourier and Bogoliubov transformations, the remaining terms in 
(\ref{ham7}) take the form
\bea H &=& \sum_{0 < k < \pi} ~\om_k ~(b_k^\dg b_k + b_{-k}^\dg 
b_{-k}), \non \\
\om_k &=& 2J_x S ~[(1 - \frac{h^2}{4J_x^2} \cos k)(1 - \frac{J_y}{J_x} \cos 
k)]^{1/2}. \eea
The excitation spectrum is gapped unless $J_y =J_x$ (or $-J_x$) in which
case the gap vanishes at $k=0$ (or $\pi$). 

Putting together the above conditions for gaplessness of the spectrum, we find
the following critical lines in terms of the parameters $J_x, ~J_y$ and $h$. 
Let us scale these parameters simultaneously so as to make $J_x + J_y = 1$. 
Then in terms of the variables 
$\ga = J_x - J_y$ and $h$, we find that the critical lines correspond to: 
(i) $-2 \le h \le 2$ and $\ga = 0$, (ii) $h \ge 1$ and $\ga = \pm (h-1)$, and 
(iii) $h \le -1$ and $\ga = \pm (h+1)$. Here (i) corresponds to the anisotropic
phase transition, while (ii) and (iii) correspond to the Ising transition.
These may be compared with the critical lines of the spin-1/2 model shown 
in Fig. \ref{fig_xyphase}. We note that the large-spin method has been useful
study other quantum models like the quantum ANNNI chain \ct{sen90,dsen91} and 
the Kitaev model \ct{baskaran08}.

\section{Derivation of a matrix product Hamiltonian}
\label{app_mps}

In this Appendix, we will show how the Hamiltonian in Eq.~(\ref{eq_mps_hamil})
can be derived \ct{wolf06} given that its ground state has a matrix product 
form with two $2 \times 2$ matrices $A_1 = \left( \begin{array}{cc} 
0 & 0 \\ 
1 & 1 \end{array} \right)$ and $A_2 = \left( \begin{array}{cc}
1 & g \\ 
0 & 0 \end{array} \right)$. Note that since there is a matrix $X = \left( 
\begin{array}{cc}
0 & g \\
1 & 0 \end{array} \right)$ such that $X^{-1} A_1 X = A_2$ and $X^{-1} A_2 X = 
A_1$, we expect the Hamiltonian $H$ to be $Z_2$ symmetric. To derive $H$, we
observe that the matrix product structure involving $A_1$ and $A_2$ is such
that if the values of $\si^z$ at two next-nearest-neighbor sites $i-1$ and 
$i+1$ are given, this uniquely fixes the state at the site $i$ which lies in 
between. There are four possibilities as follows.

\noi (i) $\si^z_{i-1} = +1$ and $\si^z_{i+1} = +1$ goes with a normalized
state at site $i$ given by 
\beq |\psi_{i,++} \rangle ~=~ \frac{1}{\sqrt{1+g^2}} ~\left( \begin{array}{c}
1 \\
g \end{array} \right). \eeq

\noi (ii) $\si^z_{i-1} = +1$ and $\si^z_{i+1} = -1$ goes with a 
state at $i$ given by 
\beq |\psi_{i,+-} \rangle ~=~ \frac{1}{\sqrt{2}} ~\left( \begin{array}{c} 
1 \\
1 \end{array} \right). \eeq

\noi (iii) $\si^z_{i-1} = -1$ and $\si^z_{i+1} = +1$ goes with a 
state at $i$ given by 
\beq |\psi_{i,-+} \rangle ~=~ \frac{1}{\sqrt{2}} ~\left( \begin{array}{c} 
1 \\
1 \end{array} \right). \eeq

\noi (iv) $\si^z_{i-1} = -1$ and $\si^z_{i+1} = -1$ goes with a 
state at $i$ given by 
\beq |\psi_{i,--} \rangle ~=~ \frac{1}{\sqrt{1+g^2}} ~\left( \begin{array}{c}
g \\
1 \end{array} \right). \eeq

Now, a state $|\psi_i \rangle$ will be the ground state (with zero energy) of 
a Hamiltonian $h_i$ if we choose $h_i$ to be the projection operator to the
state orthogonal to $|\psi_i \rangle$, i.e., if $h_i = I - |\psi_i \rangle
\langle \psi_i |$. Hence the four possibilities listed above show that
the Hamiltonian $h_i$ should be chosen to be
\bea h_i &=& ~~\frac{I + \si^z_{i-1}}{2} ~\frac{I + \si^z_{i+1}}{2} ~(I -
|\psi_{i,++} \rangle \langle \psi_{i,++} |) \non \\
&& +~ \frac{I - \si^z_{i-1}}{2} ~\frac{I + \si^z_{i+1}}{2} ~(I -
|\psi_{i,-+} \rangle \langle \psi_{i,-+} |) \non \\
&& +~ \frac{I + \si^z_{i-1}}{2} ~\frac{I - \si^z_{i+1}}{2} ~(I -
|\psi_{i,+-} \rangle \langle \psi_{i,+-} |) \non \\
&& +~ \frac{I - \si^z_{i-1}}{2} ~\frac{I - \si^z_{i+1}}{2} ~(I -
|\psi_{i,--} \rangle \langle \psi_{i,--} |) \eea
in order to make the matrix product state the ground state of $h_i$ for
all possible values of $\si^z_{i-1}$ and $\si^z_{i+1}$. This gives 
\bea h_i ~=~ \frac{1}{4(1+g^2)} ~\left[ (g^2 -1) ~(\si^z_{i-1} \si^z_i + 
\si^z_i \si^z_{i+1}) ~-~ (1+g)^2 ~\si^x_i ~+~ (g-1)^2 \si^z_{i-1} \si^x_i 
\si^z_{i+1} ~+~ 2 (1+g^2) \right]. \eea
This in turn implies that the matrix product state is the ground state
(with zero energy) of $H = \sum_i h_i$. We see that this is identical to 
Eq.~(\ref{eq_mps_hamil}) up to an overall factor of $4(1+g^2)$ and a constant 
$- \sum_i 2 (1+g^2)$ which is the ground state energy of 
Eq.~(\ref{eq_mps_hamil}).

\section{From Jordan-Wigner to bosonization}
\label{jorwigbos}

Before discussing bosonization, let us briefly consider systems of 
interacting fermions in one dimension. It turns out that many such 
systems are described by Tomonaga-Luttinger liquid (TLL) theory 
\ct{haldane80,haldane81a,haldane81b,mahan00,schulz95,schulz00,gogolin98,
vondelft98,rao01,giamarchi04,giuliani05} in contrast to interacting fermionic 
systems in three dimensions which are usually well described by Fermi liquid 
(FL) theory. Let us assume for the moment that we are considering gapless
systems in which no symmetry is spontaneously broken; for instance, we are
not considering charge/spin density waves and superconductors. Then the main 
differences between FLs and TLLs are as follows. 

In a FL, the low-energy excitations are quasiparticles which are 
fermions; they do not have a one-to-one relationship between energy 
and momentum. There are also low-energy bosonic excitations consisting
of particle-hole pairs which can be thought of as sound modes.
The one-particle momentum distribution function $n({\vec k})$, obtained by 
Fourier transforming the one-particle equal-time correlation function, has a 
finite discontinuity at the Fermi surface; this is called the quasiparticle 
renormalization factor $z_{\vec k}$, and it lies in the range $[0,1]$ (for
a non-interacting system, $z_{\vec k} = 1$). Finally, the various correlation 
functions decay asymptotically at long distances as power laws, where the 
powers are independent of the strength of the interactions and are
therefore universal. 

In a TLL, on the other hand, there are no quasiparticle excitations. All
the low-energy excitations consist of particle-hole pairs. The one-particle
renormalization factor $z_{\vec k}$ therefore vanishes, and the momentum 
distribution function $n({\vec k})$ has no discontinuity at the Fermi 
surface (in one dimension this usually consists of two points), although
it has a cusp of the form 
\beq n (k) ~=~ n (k_F ) ~+~ c~{\rm sign} (k-k_F ) ~\vert k- k_F \vert^\beta ~,
\label{beta} \eeq
where ${\rm sign} (x) \equiv 1$ if $x>0$, $-1$ if $x<0$, and $0$ if $x=0$, and
$c$ is some constant. The exponent $\beta$ is a positive number whose value 
depends on the strength of the interactions. For a non-interacting system, 
$\beta =0$ and $n(k)$ has a discontinuity at $k_F$, but for $\beta > 0$ there
is no such discontinuity. Finally, the various correlation functions 
decay asymptotically at long distances as power laws, but the powers now
depend on the strength of the interactions and are therefore not universal.

We now turn to bosonization. This is a very useful field theoretic technique 
which can be used to study a variety of quantum systems in one dimension 
including TLLs 
\ct{schotte69,mattis74,luther75,coleman75,mandelstam75,affleck89,
schulz95,schulz00,gogolin98,vondelft98,rao01,giamarchi04,giuliani05}.
Bosonization allows us to map fermionic operators into bosonic ones, and then 
use whichever set of operators is easier to compute with. This technique is 
most directly applied to translation invariant and gapless systems of fermions 
in which the low-energy excitations have a linear energy-momentum dispersion 
($z=1$). As we will see below, bosonization can also be used to study 
various perturbations from gapless systems of the above type, and it can be 
applied to systems such as a chain of spin-1/2 objects which can be mapped 
to fermions using the JW transformation.

Although the dispersion relation is generally not linear for all the modes of
a given system, it may happen that the low-energy and long-wavelength modes 
can be studied using bosonization. For a fermionic system in one dimension, 
these modes are usually the ones lying close to the two Fermi points with 
momenta $\pm k_F$ respectively. One can then define right- and left-moving 
second quantized fields $\psi_R$ and $\psi_L$, which vary slowly on the 
length scale $1/k_F$, as
\beq \psi (x) ~=~ \psi_R (x) ~e^{ik_F x} ~+~ \psi_L (x)~ e^{-ik_F x}. 
\label{slow} \eeq
Assuming that the system is on a line of length $L$ with periodic boundary 
conditions, we define the Fourier expansion of the fermionic fields as
\bea \psi_R (x) &=& \frac{1}{\sqrt L} ~\sum_{k=-\infty}^{\infty} ~c_{R,k}~
e^{ikx}, \non \\
\psi_L (x) &=& \frac{1}{\sqrt L} ~\sum_{k=-\infty}^{\infty} ~c_{L,k} ~e^{-ikx},
\label{psiRL} \eea
where $k=2\pi n_k /L$ is now being measured with respect to $\pm k_F$ (so 
that the Fermi momentum is at $k = 0$), and $n_k = 0, \pm 1, \pm 2, \cdots$. 
Further,
\beq \{ c_{\nu,k}, c_{\nu^\prime, k^\prime} \} ~=~ 0 ~ \quad {\rm and}
\quad \{ c_{\nu,k}, c_{\nu^\prime, k^\prime}^\dg \} ~=~ \de_{\nu 
\nu^\prime} ~\de_{k k^\prime}, \eeq
where $\nu = R,L$.

Next, let us consider a model of bosons with the right- and left-moving fields
being denoted by $\phi_R$ and $\phi_L$ respectively. These fields have the 
Fourier expansion
\bea \phi_R (x) &=& \frac{i}{2 {\sqrt \pi}}~ \sum_{q>0} ~\frac{e^{- \al 
q/2}}{\sqrt n_q} ~[b_{R,q} ~e^{iqx} - b_{R,q}^\dg ~e^{-iqx}], \non \\
\phi_L (x) &=& -~\frac{i}{2 {\sqrt \pi}}~ \sum_{q>0} ~\frac{e^{- \al 
q/2}}{\sqrt n_q} ~[b_{L,q} ~e^{-iqx} - b_{L,q}^\dg ~e^{iqx}], \non \\
& & \label{phiRL} \eea
where $q=2\pi n_q /L$, and $n_q = 0, \pm 1, \pm 2, \cdots$. Further,
\beq [ b_{\nu,q}, b_{\nu^\prime, q^\prime} ] ~=~ 0 ~ \quad {\rm and} \quad 
[ b_{\nu,q}, b_{\nu^\prime, q^\prime}^\dg ] ~=~ \de_{\nu \nu^\prime} ~
\de_{q q^\prime}, \eeq
where $\nu = R,L$. In (\ref{phiRL}), we have introduced a cut-off $\al$
in order to ensure that the contributions from high-momentum modes do not 
produce divergences when computing various commutators and correlation 
functions; $\al$ will often be of the same order as $1/k_F$. [In writing 
(\ref{phiRL}), we have ignored the zero modes. The interested reader may 
find discussions of these in the references on bosonization 
\ct{haldane80,haldane81a,haldane81b,mahan00,schulz95,schulz00,gogolin98,
vondelft98,rao01,giamarchi04,giuliani05}].

The bosonization formulae allow us to write the bosonic operators in terms 
of the fermionic ones and vice versa. For instance, we find that the 
creation/annihilation operators in momentum space are related as
\bea b_{\nu,q}^\dg ~&=&~ \frac{1}{\sqrt n_q} ~\sum_{k=-\infty}^\infty ~
c_{\nu,k+q}^\dg c_{\nu,k}, \non \\
b_{\nu,q} ~&=&~ \frac{1}{\sqrt n_q} ~\sum_{k=-\infty}^\infty ~
c_{\nu,k-q}^\dg c_{\nu,k}, \label{bos1} \eea
where $q = 2\pi n_q/L$ and $\nu = R,L$ as before. Eq.~(\ref{bos1}) 
shows that a bosonic excitation is generally a superposition of an infinite 
number of fermionic particle-hole pairs. On the other hand, the fermionic 
operators can be written as exponentials of the bosonic operators as follows:
\bea \psi_R (x) &=& \frac{1}{\sqrt {2\pi \al}} ~\eta_R ~ e^{-i2 {\sqrt \pi}
\phi_R (x)}, \non \\
\psi_L (x) &=& \frac{1}{\sqrt {2\pi \al}} ~\eta_L ~ e^{i2 {\sqrt \pi} 
\phi_L (x)}. \label{fer1} \eea
The unitary operators $\eta_R$ and $\eta_L$ are called Klein factors, and they
are required to ensure that the field operators given in (\ref{fer1}) 
anticommute with each other at two different spatial points.

It is convenient to define the two bosonic fields
\bea \phi ~&=&~ \phi_R ~+~ \phi_L, \non \\
\theta ~&=&~ -~ \phi_R ~+~ \phi_L. \label{bose} \eea
Eq.~(\ref{bos1}) then implies that 
\beq \psi_R^\dg \psi_R ~+~ \psi_L^\dg \psi_L ~=~ - ~\frac{1}{\sqrt 
\pi} ~\frac{\partial \phi}{\partial x}. \eeq
The fermionic density is given by
%\bea & & \rho ~-~ \rho_0 ~=~ \psi^\dg \psi \non \\
%& & = \psi_R^\dg \psi_R + \psi_L^\dg \psi_L + e^{-i2k_F x} 
%\psi_R^\dg \psi_L + e^{i2k_F x} \psi_L^\dg \psi_R, \non \\
%& & \label{density1} \eea
\beq \rho ~-~ \rho_0 ~=~ \psi^\dg \psi ~=~ \psi_R^\dg \psi_R + \psi_L^\dg 
\psi_L + e^{-i2k_F x} \psi_R^\dg \psi_L + e^{i2k_F x} \psi_L^\dg \psi_R, 
\label{density1} \eeq
where $\rho_0$ is the background density, and we have written the density
fluctuation around $\rho_0$ in terms of fermionic fields. Ignoring the last 
two terms in (\ref{density1}) which vary rapidly (since their length 
scale of variation is $1/k_F$), we see that 
%\bea \rho - \rho_0 &=& -~\frac{1}{\sqrt \pi} ~\frac{\partial \phi}{\partial 
%x} ~+~ \frac{1}{\sqrt {2\pi \al}} ~[\eta_R^\dg \eta_L ~e^{i 2 {\sqrt 
%\pi} \phi - i 2 k_F x} \non \\ 
%& & ~~~~~~~~~~~~~~~+~ \eta_L^\dg \eta_R ~e^{-i 2 {\sqrt \pi} \phi + i2k_F x}].
%\label{density2} \eea
\beq \rho - \rho_0 ~=~ -~\frac{1}{\sqrt \pi} ~\frac{\partial \phi}{\partial 
x} ~+~ \frac{1}{\sqrt {2\pi \al}} ~[\eta_R^\dg \eta_L ~e^{i 2 {\sqrt 
\pi} \phi - i 2 k_F x} ~+~ \eta_L^\dg \eta_R ~e^{-i 2 {\sqrt \pi} \phi + 
i2k_F x}]. \label{density2} \eeq

The formalism outlined so far does not explicitly refer to any particular
dynamics of the bosonic or fermionic system. Let us now introduce some 
dynamics by specifying appropriate Hamiltonians. We first consider a system 
of non-interacting fermions; let us assume that the velocity of the modes 
described by $\psi_{R,L}$ is given by $(dE_k /dk)_{k=k_F} = - (dE_k /
dk)_{k=-k_F} = v_F$, where $v_F$ is called the Fermi velocity. The 
Hamiltonian for such a system is given by
\beq H_0 ~=~ -v_F \int ~dx ~[\psi_R^\dg (x) i \partial_x \psi_R (x) ~-~
\psi_L^\dg (x) i \partial_x \psi_L (x)], \label{hamf} \eeq
where $\partial_x \equiv \partial /\partial x$. Let us now add to this
a short-range density-density interaction of the form
\beq V ~=~ \frac{1}{2} ~\int ~dx ~[ 2g_2 ~\rho_R (x) \rho_L (x) ~+~g_4 ~
( \rho_R^2 (x) ~+~ \rho_L^2 (x))], \label{vf} \eeq
where $\rho_R = \psi_R^\dg \psi_R$ and $\rho_L = \psi_L^\dg \psi_L$.
Physically, we may expect an interaction term like $g \rho^2$, so that
$g_2 = g_4 =g$. However, we will allow $g_2$ to differ from $g_4$ for the sake
of generality. Then the Hamiltonians for the corresponding bosonic systems are
as follows. The non-interacting Hamiltonian in (\ref{hamf}) maps to 
\beq H ~=~ \frac{v_F}{2} ~\int ~dx ~[(\partial_x \theta)^2 ~+~ (\partial_x 
\phi)^2], \label{hamb} \eeq
while the Hamiltonian for the interacting fermion system $H=H_0 + V$ maps to 
the bosonic Hamiltonian 
\beq H ~=~ \frac{v}{2} ~\int ~dx ~[~ K (\partial_x \theta)^2 ~+~ \frac{1}{K} ~
(\partial_x \phi)^2 ~], \label{hamb2} \eeq
where 
\bea v ~&=&~ \Bigl[ ~(~ v_F + \frac{g_4}{2\pi} - \frac{g_2}{2\pi} ~)~
(~ v_F + \frac{g_4}{2\pi} + \frac{g_2}{2\pi} ~) ~\Bigr]^{1/2}, \non \\
K ~&=&~ \Bigl[ ~(~ v_F + \frac{g_4}{2\pi} - \frac{g_2}{2\pi}~) ~/~ (~v_F +
\frac{g_4}{2\pi} + \frac{g_2}{2\pi} ~) ~\Bigr]^{1/2}. \non \\
& & \eea
This describes bosonic modes with an energy $E_q = v |q|$.
Note that the interaction has changed the velocity from $v_F$ to $v$,
and has also given rise to a parameter $K$ which is called
the Luttinger parameter. We observe here that this analysis breaks down
if $g_2 > 2 \pi v_F + g_4$ since $K$ and $v$ then become imaginary numbers.

An important use of bosonization is that one can easily compute certain 
correlation functions using the bosonic theory which would be very difficult 
to compute in an interacting fermionic theory. Eq.~(\ref{phiRL}) can be 
used to show that the equal-time correlation functions of the fields $\phi$ 
and $\theta$ are given by
\bea \langle 0 \vert \phi (x) \phi (0) \vert 0 \rangle &=& -~\frac{K}{2\pi}~
\ln ~\Bigl( \frac{|x|}{\al} \Bigr), \non \\
\langle 0 \vert \theta (x) \theta (0) \vert 0 \rangle &=& -~ 
\frac{1}{2\pi K}~ \ln ~\Bigl( \frac{|x|}{\al} \Bigr), \label{corrfn1} \eea
for $|x| \gg \al$. (In (\ref{corrfn1}), we have not written the time 
coordinates for the fields since these are equal). Let us now consider an 
exponential operator of the form
\beq O_{m,n} ~=~ e^{i2 {\sqrt \pi} (m \phi + n \theta)}. \label{omn} \eeq
Eq.~(\ref{corrfn1}) can then be used to derive the following result 
for the equal-time correlation function
\beq \langle 0 \vert O_{m,n} (x) O_{m^\prime, n^\prime}^\dg (0) \vert 0 
\rangle ~\sim ~\de_{mm^\prime} \de_{nn^\prime} ~\Bigl( 
\frac{\al}{|x|} \Bigr)^{2(m^2 K + n^2 /K)}, \label{corromn} \eeq
where $K$ is the Luttinger parameter. Note that the correlation function 
decays as a power law, but the power law depends on $K$ and is therefore not 
universal. In the language of the RG, the scaling dimension
of $O_{m,n}$ can be read off from (\ref{corromn}) as $m^2 K + n^2 /K$.

We can now study a quantum spin chain using bosonization. Let us consider 
the $XXZ$ spin-$1/2$ chain in a magnetic field which is described by the 
Hamiltonian 
%\bea H &=& \sum_{n=1}^N ~[~ \frac{J}{2} ~(\si_n^+ \si_{n+1}^- + \si_n^- 
%\si_{n+1}^+) \non \\
%& & ~~~~~~~~+~ \frac{J_z}{4} \si_n^z \si_{n+1}^z ~-~ \frac{h}{2} \si_n^z ~], 
%\label{xxzh} \eea 
\beq H ~=~ \sum_{n=1}^{N-1} ~[~ \frac{J}{2} ~(\si_n^+ \si_{n+1}^- + \si_n^- 
\si_{n+1}^+) ~+~ \frac{J_z}{4} \si_n^z \si_{n+1}^z] ~-~ \frac{h}{2} 
\sum_{n=1}^N ~\si_n^z, \label{xxzh} \eeq
where $\si_n^+= (1/2) (\si_n^x + i \si_n^y)$ and $\si_n^- = (1/2)(\si_n^x - i 
\si_n^y)$ are the spin raising and lowering 
operators, $h$ denotes the strength of the magnetic field, and we assume that
$J > 0$. (We will eventually be interested in the thermodynamic limit $N \to
\infty$). Note that the model has a $U(1)$ invariance, namely, rotations about
the $z$ axis. When $J_z=J$ and $h=0$, the $U(1)$ invariance is enhanced to an 
$SU(2)$ invariance, because at this point the model can be written as $H =
(J/4) \sum_n {\vec \si}_n \cdot {\vec \si}_{n+1}$. Although the model 
in (\ref{xxzh}) can be exactly solved using the Bethe ansatz \ct{baxter82}, 
and one has the explicit result that the model is gapless for a certain
range of values of $J_z /J$ and $h/J$ (see \ct{cabra98}), it is not easy to 
explicitly compute correlation functions in that approach. We will therefore
use bosonization to study this model.

%We first use the to map the spin model to a model 
%of spinless fermions. We map an $\uparrow$ spin or a $\downarrow$ spin at any 
%site to the presence or absence respectively of a spinless fermion at that 
%site. This can be done by introducing a fermion annihilation operator $c_n$ 
%at each site, and writing the spin at that site as
%\bea \si_n^z &=& 2 c_n^\dg c_n -1 = 2 \rho_n - 1 \non \\
%\si_n^- &=& c_n ~e^{i\pi \sum_j \rho_j}, \label{jorwig} \eea 
%where the sum over $j$ runs from the left end of the chain up to the 
%$(n-1)^{\rm th}$ site (we are assuming open boundary conditions for 
%convenience), and $\rho_n =0$ or $1$ is the fermion occupation number at site 
%$n$. The expression for $\si_n^+$ can be obtained by taking the Hermitian 
%conjugate of $\si_n^-$. The string factor in the definition of $\si_n^-$ is 
%necessary to ensure the correct anticommutation relations between the fermionic
%operators, namely, $\{ c_m, c_n^\dg \} = \de_{mn}$ and $\{ c_m, c_n \} = 0$.

Using the JW transformations (\ref{jw1}), (\ref{xxzh}) can be written as
%\bea H &=& \sum_n ~[~-~ \frac{J}{2} ~(c_n^\dg c_{n+1} +H.c.) \non \\
%& & ~~~~~~~~+~ J_z ~c_n^\dg c_n c_{n+1}^\dg c_{n+1} ~-~ h ~c_n^\dg c_n], \eea
\beq H ~=~ \sum_n ~\left[-~ \frac{J}{2} ~(c_n^\dg c_{n+1} +H.c.) ~+~ J_z ~
c_n^\dg c_n c_{n+1}^\dg c_{n+1} ~-~ h ~c_n^\dg c_n \right], \eeq
where we have ignored some constants.
We see that the spin-flip operators $S_n^\pm$ lead to hopping terms 
in the fermionic Hamiltonian, whereas the $S^z_n S^z_{n+1}$ interaction 
term leads to an interaction between fermion densities on adjacent sites.

Let us first consider the non-interacting case given by $J_z =0$. By Fourier 
transforming the fermions, $c_k = \sum_n c_n e^{-ikna}/\sqrt{N}$, where 
$a$ is the lattice spacing and the momentum $k$ lies in the first Brillouin 
zone $-\pi /a < k \le \pi /a$, we find that the Hamiltonian is given by 
\bea H &=& \sum_k ~\om_k ~c_k^\dg c_k, \non \\
{\rm where} ~~~\om_k &=& -J ~\cos (ka) ~-~ h. \eea
The ground state of this system is one in which all the single-particle
states with $\om_k <0$ are occupied, and all the states with 
$\om_k >0$ are empty. If we set the magnetic field $h=0$, the 
magnetization per site $m \equiv \sum_n \si_n^z /N$ will be zero in the
ground state; equivalently, in the fermionic language, the ground state is 
half-filled. Thus, for $m=0$, the Fermi points lie at $ka = \pm \pi/2$ where 
$\om_k =0$; hence $k_F a = \pi/2$. Let us now add the magnetic field term. In 
the fermionic language, this is equivalent to adding a chemical potential 
term which couples to $\rho_n$. Now the 
ground state no longer has $m=0$ and the fermion model is no longer 
half-filled. The Fermi points are then given by $\pm k_F$, where 
\beq k_F a ~=~ \frac{\pi}{2} ~( m ~+~ 1). \eeq
It turns out that this relation between $k_F$ (which governs the oscillations 
in the correlation functions as discussed below) and the magnetization $m$ 
continues to hold even if we turn on the interaction $J_z$, although the 
simple picture of the ground state (with states being occupied below some 
energy and empty above some energy) no longer holds in that case.

In the linearized approximation, the modes near the two Fermi points have 
the velocities $d\om_k / dk = \pm v$, where
$v$ is some function of $J$, $J_z$ and $h$. Let us define the continuum
fermionic field $\psi (x) = c_n$ where $x=na$. We introduce the slowly 
varying fermionic fields $\psi_R$ and $\psi_L$ as defined in 
(\ref{slow}), and then bosonize these fields. The spin fields 
can be written in terms of either the fermionic or the bosonic fields. For 
instance, $\si^z$ is given by the fermion density as in (\ref{jw1})
which then has a bosonized form given in (\ref{density2}). Similarly, 
\bea \si^+ (x) ~&=&~ (-1)^{x/a} ~[e^{-ik_Fx} \psi_R^\dg (x)+ e^{ik_Fx} 
\psi_L^\dg (x)] \non \\
&& \times ~[e^{i\pi \int_{-\infty}^x dx^\prime (\psi^\dg (x^\prime ) 
\psi (x^{\prime} ) + 1/2a)} + H.c. ], \eea
%\beq \si^+ (x) ~=~ (-1)^{x/a} ~[e^{-ik_Fx} \psi_R^\dg (x)+ e^{ik_Fx} 
%\psi_L^\dg (x)] ~ [e^{i\pi \int_{-\infty}^x dx^\prime (\psi^\dg (x^\prime ) 
%\psi (x^{\prime} ) + 1/2a)} + H.c. ], \eeq
where $(-1)^{x/a} = \pm 1$ since $x/a$ is an integer. This can now be written
entirely in the bosonic language; the term in the exponential is given by
%\bea & & \int_{-\infty}^x dx^\prime ~\psi^\dg (x^\prime) \psi (x^\prime )
%\non \\
%& & = -~\frac{1}{\sqrt{\pi}} \int_{-\infty}^x dx^\prime ~\frac{\partial 
%\phi}{\partial x^\prime} ~=~ -~\frac{1}{\sqrt{\pi}} ~[\phi_R (x) + \phi_L(x)],
%\eea
\beq \int_{-\infty}^x dx^\prime ~\psi^\dg (x^\prime) \psi (x^\prime ) ~=~ -
\frac{1}{\sqrt{\pi}} \int_{-\infty}^x dx^\prime ~\frac{\partial \phi}{\partial
x^\prime} ~=~ -\frac{1}{\sqrt{\pi}} ~[\phi_R (x) + \phi_L(x)], \eeq
where we have ignored the contribution from the lower limit at $x^\prime = 
-\infty$.

We can now use these bosonic expressions and Eq.~(\ref{corromn}) 
to compute the two-spin, equal-time correlation functions $G^{ab}(x) \equiv 
<0| \si^a (x) \si^b (0)|0>$. We find that
\bea G^{zz} (x) &=& m^2 ~+~ \frac{c_1}{x^2} ~+~ c_2 ~\frac{\cos (2k_F x)}{
x^{2K}}, \non \\
G^{+-}(x) + G^{-+}(x) &=& c_3 ~\frac{(-1)^{x/a}}{x^{1/(2K)}} ~+~ c_4 ~
\frac{(-1)^{x/a}~\cos (2k_F x)}{x^{2K+1/(2K)}}, \label{corrfn2} \eea
where $c_1, \cdots, c_4$ are some constants. The parameter $K$ in 
(\ref{corrfn2}) is a function of $J_z /J$ and $h/J$ as discussed in 
\ct{cabra98}. For $h=0$, $K$ is given by the expression 
\beq \frac{1}{K} ~=~ 1 ~+~ \frac{2}{\pi} \sin^{-1} \left(\frac{J_z}{J} \right). 
\label{kpara} \eeq
Note that at the $SU(2)$ invariant and isotropic antiferromagnetic point, $J_z
=J$ and $h=0$, we have $m=0$, $k_F a= \pi/2$ and $K=1/2$, and the two 
correlations $G^{zz}$ and $G^{+-}$
have the same forms, as expected; since the terms which decay the slowest do 
so as $1/x$, we conclude that $\si^x$, $\si^y$ and $\si^z$ all have the
scaling dimension equal to $1/2$ at the isotropic point. Two other points of 
interest are given by $J_z=0$ where $K=1$ (non-interacting fermions), 
and $J_z \to - J$ (the isotropic ferromagnetic point) where $K \to \infty$. 
In general, $J_z > (<) ~0$ corresponds to fermions with repulsive (attractive)
interactions and $K < (>) ~1$ respectively.

We can compute the exponent $\beta$ appearing in the momentum distribution
function $n(k)$ in (\ref{beta}). In terms of the parameter $K$, we 
find that $\beta = (1-K)^2 /(2K)$. This is always positive except for
the special case of non-interacting fermions where $K=1$.

In addition to providing a convenient way of computing correlation functions, 
bosonization also allows us to study the effects of small perturbations.
We will consider a few examples here.

A physically important perturbation is a dimerizing term
\beq V ~=~ \de ~\sum_n ~(-1)^n ~[~ \frac{J}{2} ~(\si_n^+ \si_{n+1}^- + \si_n^-
\si_{n+1}^+) ~+~ \frac{J_z}{4} \si_n^z \si_{n+1}^z ~], \eeq
where $\de$ is the strength of the perturbation. Upon bosonizing and
ignoring terms which oscillate on the scale of the lattice spacing $a$
or $1/k_F$ (and which therefore integrate to zero on length scales much 
larger than $a$), we find that the slowly varying terms are proportional to 
$e^{\pm i2{\sqrt \pi} \phi}$. Eq.~(\ref{corromn}) then implies that 
the scaling dimension of these 
terms is $K$. Namely, under an RG flow, the effective length-scale dependent 
dimerization parameter $\de (L)$ satisfies the equation 
\beq \frac{d\de}{d \ln L} ~=~ (2-K) \de \eeq
to first order in $\de$. Hence this perturbation is relevant, irrelevant and 
marginal for $K < 2$, $>2$ and $=2$ respectively. If $K<2$, a small value of 
$\de$ at the microscopic length scale $a$ will grow to be of order 1 at a 
length scale $L_0$ given by $L_0/a \sim 1/(\de(a))^{1/(2-K)}$. This implies 
that the two-point correlation function will decay exponentially at large 
separations with a correlation length of order $L_0$. To put it differently,
this perturbation produces an energy gap in the system which scales 
with $\de (a)$ as
\beq \De E ~\sim ~(\de(a))^{1/(2-K)}. \eeq
For an isotropic antiferromagnet with $K=1/2$, we see that $\De E \sim 
\de^{2/3}$. This kind of phenomenon occurs in spin-Peierls systems such 
as $\rm CuGeO_3$; below a transition temperature $T_{sp}$, such a system goes 
into a dimerized phase which has a gap \ct{boucher96}.

Another interesting perturbation is to consider the spin-1/2 $XYZ$ chain
where the $x$ and $y$ couplings are slightly different from each other.
(This is a generalization of the ``anisotropic transition" discussed in
Sec.~\ref{jorwigtr} where we had taken $J_z = 0$ and therefore $K$ was equal
to 1). Namely, we consider the Hamiltonian
\beq H ~=~ \frac{1}{4} \sum_n ~[ J_x S_n^x S_{n+1}^x ~+~ J_y S_n^y S_{n+1}^y ~
+~ J_z S_n^z S_{n+1}^z ]. \label{xyz} \eeq
Let us denote $J= (J_x +J_y)/2$, and $\ga = (J_x -J_y)/(J_x +J_y)$.
The JW transformation turns this Hamiltonian into
\bea H ~=~ \sum_n ~\left[ -\frac{J}{2} ~\{ c_n^\dg c_{n+1} +H.c. + \ga 
(c_n^\dg c_{n+1}^\dg - c_n c_{n+1})\} ~+~ J_z ~c_n^\dg c_n c_{n+1}^\dg 
c_{n+1} \right]. \eea
Upon bosonizing and ignoring terms which oscillate on the scale of the lattice
spacing, we find that the slowly varying terms are proportional to $e^{\pm i
2{\sqrt \pi} \theta}$. Eq.~(\ref{corromn}) then implies that the 
scaling dimension of these terms is $1/K$. Using RG arguments similar to 
those given above, we conclude that this perturbation will lead to the 
opening of an energy gap which scales as
\beq \De E ~\sim ~\ga^{1/(2-1/K)}. \eeq

Another important perturbation is a weak coupling between two isotropic 
antiferromagnetic spin-$1/2$ chains (with the spin variables in the two chains
being denoted by ${\vec \si}^{(1)}_n$ and ${\vec \si}^{(2)}_n$) given by
\beq V ~=~ J' ~\sum_n ~{\vec \si}^{(1)}_n \cdot {\vec \si}^{(2)}_n ~. \eeq
We saw above that for a single isotropic chain, the scaling dimension of the 
spin operators is equal to $1/2$; hence the perturbation ${\vec \si}^{(1)} 
\cdot {\vec \si}^{(2)}$ has the scaling dimension $1$. Hence this perturbation
is relevant, and it produces an energy gap which scales as $\De E \sim J'$. 
This has been confirmed by numerical calculations \ct{pati97}.

Finally, let us consider a spin-1/2 chain with both nearest- and 
next-nearest-neighbor isotropic antiferromagnetic interactions governed 
by the Hamiltonian 
\beq H ~=~ J_1 \sum_n ~{\vec \si}_n \cdot {\vec \si}_{n+1} ~+~ J_2 \sum_n ~
{\vec \si}_n \cdot {\vec \si}_{n+2}. \eeq
This is called a frustrated system since the classical ground states of the 
$J_1$ and $J_2$ pieces of the Hamiltonian are in conflict with each other; 
the first term has a classical ground state of the Neel form $\uparrow
\downarrow \uparrow \downarrow \cdots$ while the second term has a period-4 
ground state of the form $\uparrow \uparrow \downarrow \downarrow \cdots$.
Since the spin-1/2 chain is a quantum system, the effect of frustration is 
not as easy to visualize here as in a classical spin system. A field theoretic
study indicates that the model makes a transition from a gapless phase 
to a gapped and spontaneously dimerized phase when $J_2/J_1 \gtrsim 1/6$
\ct{haldane82}. Numerical calculations show that the 
spin-1/2 chain is gapless and in a Neel state for $0 < J_2/J_1 \lesssim 
0.241$, gapped and in a spontaneously dimerized Neel state for $0.241 
\lesssim J_2/J_1 < 1/2$, and gapped and in a spiral state for $1/2 < J_2/J_1$ 
\ct{okamoto92,bursill94,bursill95,chitra95,white96,schollwock96,kumar10}. Here,
the words Neel and spiral refer to ground states in which the structure factor
$S(q)$, given by the Fourier transform of the equal-time two-spin correlation 
function $\langle 0 | {\vec \si}_0 \cdot {\vec \si}_n | 0 \rangle$, has peaks 
at $q=\pi$ and $q < \pi$ respectively. 
The point $J_2/J_1 = 1/2$ is called the Majumdar-Ghosh point; the ground
states are doubly degenerate and exactly solvable there. The two ground 
states are given by a product of singlets in which each spin on an even 
numbered sites forms a singlet with its neighbor either to its left or to
its right \ct{majumdar69}. An interesting perturbation to consider now
is to study what happens when $J_2$ crosses the critical value $J_{2c} \simeq
0.241$. This turns out to be a marginal perturbation (i.e., the scaling 
dimension is 2), so that in the RG equation one has to go to the second 
order in the perturbing parameter $J_2 - J_{2c}$. One finds eventually that 
for $J_2 > J_{2c}$, an energy gap opens up with an essential singularity of 
the form $\De E \sim \exp [- c/(J_2 - J_{2c})]$, where $c$ is a constant 
\ct{okamoto92}. 

%\section{Next appendix}

\section{Scaling of the geometric phase close to a QCP}
\label{sec_gp_scaling}

{When the Hamiltonian of a system changes adiabatically in 
time in the parameter space, then in addition to the usual dynamical phase, 
there is a phase that arises purely due to the geometry of the parameter 
space. This is known as the Pancharatnam-Berry phase or geometric phase (GP) 
\ct{pancharatnam56,berry84}; this phase is non-zero only if there is a 
degeneracy inside the circuit enclosed. At a QCP, the energy gap between the 
ground and the first excited state vanishes, i.e., there is a degeneracy 
which is expected to lead to a non-zero GP. In this section, we shall briefly 
note the scaling of the GP (which is also the imaginary part of the geometric 
tensor defined in Sec.~\ref{fidelity}) close to the QCP of a transverse 
$XY$ spin chain \ct{zhu06,carollo05,pachos06,Zhu08,Hamma06,Cui06,quan09}.}

The GP is estimated introducing a new family of Hamiltonians which is 
generated by rotating every spin of Hamiltonian (\ref{eq_txy}) by an angle 
$\phi$ around the $\hat z$ direction. The direct product ground state in 
(\ref{eq_direct_product}) then takes the form
\beq |\psi_0 (\ga,h) \rangle = \bigotimes_{k>0} ( \cos {\theta_k} |0\rangle + 
i e^{2i\phi} \sin {\theta_k}|k, -k\rangle), \label{eq_gsxy} \eeq
where $\theta_k$ is defined in (\ref{eq_def_theta}); the energy spectrum 
and hence the critical exponents remain unaltered under this rotation.
The GP of the ground state is given by $\beta_g = (\pi/L) \sum_{k>0} (1-\cos 
2 \theta_k)$. Carollo and Pachos \ct{carollo05} 
studied the behavior of the GP close to the anisotropic critical point in 
a transverse $XY$ chain (Fig.~\ref{fig_xyphase}) and showed that a 
non-contractible geometric phase difference between the ground state and the 
first excited state exists when the Hamiltonian encounters a critical point 
while passing through an adiabatic cycle\footnote{{The 
geometric phase depends on the path over 
which the Hamiltonian is traversed. If one considers a loop and tries to find 
the net GP accumulated, then there can be two cases. If the enclosed region 
does not include any critical point, the loop can be continuously deformed 
to a point and hence produces a zero GP. On the other hand, if a critical 
point is enclosed in the loop, then one finds a finite GP even when the loop is
contracted to a point. The GP is non-contractible in the sense that even when 
one deforms the loop, it always encloses the critical point and thus has a
associated non-zero phase associated with the critical point.}}. In 
subsequent work \ct{zhu06}, it was shown that the ground state GP
is non-analytic along the Ising critical line; the derivative $d 
\beta_g/d h$ shows a peak at the Ising critical point when plotted as a 
function of $h$ and diverges logarithmically with the chain length. From the 
scaling relations $d\beta_g/dh \sim \kappa_1 \ln L + C_1$ at the effective 
critical point $h_m(L)$ where the peak occurs for large $L$, and $d
\beta_g/dh \sim \kappa_2 \ln |h -h_c| + C_2 $ in the thermodynamic limit as 
$h \to h_c= J_x + J_y=1$ (where $C_1$ and $C_2$ are non-universal constants), 
the critical exponent $\nu$ can be obtained from the relation $\nu = \kappa_2/
\kappa_1 = 1$. On the anisotropic critical line ($\ga=0$), the GP is 
given by $\beta_g = 2\pi - 2 \cos^{-1} (h)$ in the thermodynamic limit, and 
$d \beta_g/d h \sim (1-h)^{-1/2}$ for $h \to 1^-$ which provides the correct
exponent $\nu_{mc}=1/2$, associated with the MCP. Close to the multicritical 
point, 
the derivative of the GP shows an oscillatory behavior due to the presence 
of quasicritical points as seen in the FS and exhibits a power-law divergence 
with the system size with an exponent that depends on the path of 
{approach to} 
the MCP \ct{patra11}. Recently a NMR interferometry study which exploits the 
geometric phase as a probe of quantum transitions of the ground state in a 
spin-1/2 transverse $XY$ models has been reported \ct{peng10}. 

\section{Landau-Zener tunneling}
\label{app_lz}

We present here an outline of the calculational details leading to an exact 
solution for the conventional Landau-Zener-Stueckelberg-Majorana-Stuckelberg 
problem \ct{landau32,zener32,stuckelberg32,majorana32}. Let us first consider 
a general Hamiltonian in the basis $\ket 1$ and $\ket 2$ as shown below 
\ct{zener32,damski06,suzuki05}
\bea H(t)~=~ \left[ \begin{array}{cc} \epsilon_1(t) & \Delta(t) \\
\Delta^*(t) & \epsilon_2(t) \end{array} \right], \non \eea
where $\epsilon_1$ and $\epsilon_2$ are the two bare energy levels (diagonal 
elements) varying as $\sim t/2\tau$ and $-t/2\tau$ respectively. If $\ket 
{\psi(t)}=c_1(t)\ket 1+c_2(t)\ket 2$, then the Schr\"odinger equation for 
$c_1(t)$ and $c_2(t)$ is
\bea i\frac{d}{dt} \left( \begin{array}{c} 
c_1 \\ c_2 \end{array} \right) ~~=
\left[ \begin{array}{cc} \epsilon_1(t) & \Delta(t) \\
\Delta^*(t) & \epsilon_2(t) \end{array} \right] \left( \begin{array}{c}
c_1 \\ c_2
\end{array} \right). \label{eq_appschr} \eea
{The wave function $\psi(t)$ can be analytically obtained 
at any instant $t$ when $\epsilon_1$ and $\epsilon_2$ vary linearly in time 
and $\Delta$ is time-independent. If the off-diagonal term is time-dependent, 
in some cases one can use a unitary transformation to shift the time-dependence
entirely to the diagonal terms to find an exact solution
\ct{mukherjee07}.} We now redefine $c_1(t)$ and $c_2(t)$ as follows
\bea c_1(t) = \tilde c_1(t) e^{-i \int_{-\infty}^{t}\epsilon_1(t')dt'} ~~~
\rm{and}~~~ c_2(t) = \tilde c_2(t) e^{-i \int_{-\infty}^{t}\epsilon_2(t')dt'}.
\non \eea
The Schr\"odinger equation for $\tilde c_2$ is
\bea i\frac{\partial}{\partial t}\tilde c_2(t)= \Delta^{*} \tilde 
c_1(t) e^{-i\int_{-\infty}^{t}(\epsilon_1(t')-\epsilon_2(t'))dt'}.
\label{eq_appschr2} \eea
Replacing $\tilde c_2(t)$ by $U_2(t)$ defined as
\beq \tilde c_2(t) = e^{-\frac{i}{2}\int_{-\infty}^t
(\epsilon_1-\epsilon_2) dt'} U_2(t), \eeq
followed by a change of variable given by 
$$z=\frac{t}{\sqrt{\tau}}e^{-i\pi/4},$$
the Schr\"odinger equation can be re-written as
\beq \frac{\partial^2 }{\partial z^2}U_2(z)+(m+\frac{1}{2}-
\frac{z^2}{4})U_2(z)=0, \eeq
\noindent where $m=i\Delta^2\tau$ (where $\Delta^2 = \Delta \Delta^{*}$) and $\epsilon_1-\epsilon_2=t/\tau$. 
By all these transformations, we are able to recast the Schr\"odinger 
equation \ref{eq_appschr} in the form of the Weber differential equation
whose solutions are linear combination of well known Weber functions
$D_{-m-1}(iz)$ and $D_{-m-1}(-iz)$ $i.e.,$
$U_2(z)=aD_{-m-1}(iz)+bD_{-m-1}(-iz)=\tilde{c}_2(t) e^{\frac{it^2}{4\tau}}$,
or, going back to the notation of $\tilde c_1(t)$ and $\tilde c_2(t)$,
\bea \ket{\psi(t)}&=&\frac{i}{\Delta}[\partial_t- \frac{it}{2\tau}] ~[
aD_{-m-1}(iz)+bD_{-m-1}(-iz)]~\ket 1 \non \\
&& +~ [aD_{-m-1}(iz)+bD_{-m-1}(-iz)] ~\ket 2. \label{psit} \eea
The initial condition demands that at $t\to-\infty$, $\ket{\psi(t)}\sim 
\ket1$ which implies $\tilde c_2(-\infty)$=0. This forces $\tilde {c}_2(t)$ or
$U_2(z)$ to be a function of only $D_{-m-1}(-iz)$ as $D_{-m-1}(-iz)$ goes
to zero at $t \to -\infty$ but $D_{-m-1}(iz)$ does not. This can be shown as
follows. The asymptotic forms of the Weber functions are
\bea \lim_{z \to +\infty} D_n(z) \sim e^{-\frac{1}{4}z^2}z^n -
\frac{\sqrt{2\pi}}{\Gamma(-n)}e^{n\pi ir} e^{\frac{1}4 z^2}z^{-n-1} ~~~~{\rm 
for}~~~ \frac{\pi r}{4}<arg(z)<\frac{5\pi r}{4}, \label{eq_appasymptote1} \eea
\noindent where $r$ is either $1$ or $-1$, and
\bea \lim_{z \to +\infty} D_n(z)\sim e^{-\frac{1}{4}z^2} z^n ~~~~{\rm for}~~~ 
|arg(z)|< \frac{3\pi}{4}. \label{eq_appasymptote2} \eea

The $t \to -\infty ~(z \to -\infty)$ form of the
two Weber functions $D_{-m-1}(iz)$ and $D_{-m-1}(-iz)$ is
\bea \lim_{z\to -\infty}D_{-m-1}(iz)
&=& \lim_{R\to +\infty} D_{-m-1}(R e^{-i3\pi/4}) \non \\
&\sim&
\frac{\sqrt{2\pi}}{\Gamma(m+1)}e^{i7\pi m/4} e^{iR^2/4} R^{m+1}.\non \\
\label{eq_appapprx1} \eea
and,
\bea \lim_{z \to -\infty}D_{-m-1}(-iz) ~=~ \lim_{R \to +\infty}(R e^{i\pi/4})~
=~ 0. \eea
where $R=|t|/\sqrt{\tau}$.

Since at $t \to -\infty$, $\tilde c_2=0$, and $D_{-m-1}(iz)$ does not
go to zero, we must have $a=0$ whereas the form of $b$ can be obtained by 
these asymptotic forms along with the initial condition as follows:
Putting $a=0$, we have
$$\tilde c_2(t)=e^{-\frac{it^2}{4\tau}} b D_{-m-1}(-iz).$$
Expanding Eq. (\ref{eq_appschr2}) after substituting $\tilde c_2(t)$
as given above, we get
\bea \Delta \tilde c_1(t) e^{-\frac{it^2}{2\tau}} &=&
\frac{bt}{2\tau}e^{-\frac{it^2}{4\tau}}
\left[e^{-\frac{iR^2}{4}}R^{-k-1}e^{-\frac{\pi(k+1)i}{4}}\right]\non \\
&& +~ b i e^{-\frac{it^2}{4\tau}}\frac{dR}{dt}
\left[\frac{d}{dR}e^{-iR^2/4}R^{-k-1}e^{-\frac{\pi(k+1)i}{4}}\right].\non 
\eea
Neglecting the second term and substituting $c_1(-\infty)=1$ gives
$$b=\Delta\sqrt{\tau}e^{-\frac{\pi}{4}\Delta^2\tau}.$$

Once we know the value of b with the given initial conditions,
it is easy to obtain the excitation probability at $t \to +\infty$
which is simply given by $p=|c_1|^2=1-|c_2|^2$, where
\bea |c_2(t \to +\infty)|^2 = |b~ \lim_{z \to +\infty}
D_{-m-1}(-iz)|^2. \eea
Substituting the expression of $b$ obtained along with the asymptotic form
of $D_{-m-1}(-iz)$ when $z\to +\infty$ which is the same as given in Eq. 
(\ref{eq_appapprx1}), we get
\bea p=e^{-2\pi \tau \Delta^2}. \eea
This can be generalized to finite coupling duration \ct{vitanov96}, and 
the LZ transition times have been calculated in diabatic and adiabatic limits 
\ct{vitanov99}. We note that this problem has recently 
been extended to various other situations \ct{damski06,chakrabarti10} and 
to many particle systems \ct{altland08,altland09}.

\section{KZ mechanism in space}
\label{kz_space}

{In Sec.~\ref{sec_intrononeq}, we have considered different 
quenching schemes (except in Sec.~\ref{sec_localquench}) 
where the quenching parameter is homogeneous 
throughout the system.} This is to some extent an idealized situation; for 
example, in ultracold atom gases confined in magnetic/optical traps, the 
trapping potential results in an inhomogeneous density of atoms and the 
critical point depends on the local density of atoms. In this context, the 
notion of a symmetry breaking (inhomogeneous) QPT in space has been 
introduced \ct{damski09,zurek08,platini07,collura09,damski07}. For a classical 
phase transition, spatial inhomogeneities influence the critical behavior if 
the resulting perturbation is relevant in the sense of the RG; 
for a review, see \ct{igloi93}. Also, the critical behavior of a 
Bose condensate in a trapped potential has recently been studied, and a 
power-law scaling of the correlation length with the trap size has been 
proposed \ct{campostrini09}.
 
Let us define the driving parameter $\la(\vec r)$ as follows. The part of the 
system where $\la(\vec r)>0$ is in a different phase than the region in which 
$\la(\vec r)<0$; a phase transition between the two phases occurs at the 
critical surface $\la (\vec r)=0$. We consider the variation of $\la$ along 
the $\hat x$ direction and linearize near the critical point $x_c$ ($\la(x_c)
=0$); this gives $\la \simeq \la_0 (x-x_c)/\al_x$, 
where $1/ \al_x = (\partial \la/\partial x)_{x_c}$. 
The system is in the symmetry broken phase for $x < x_c$ while in the 
symmetric phase for $x > x_c$. Within the local density approximation (LDA) 
which assumes that the system is locally uniform, the
order parameter vanishes as $(x_c -x)^{\beta}$ as $x \to x_c^-$. However, 
the LDA breaks down in the vicinity of $x_c$ due to the diverging healing 
length $\hat \xi \sim |x - x_c|^{-\nu}$ which is the only 
characteristic length as $x \to x_c$. We now encounter a situation 
analogous to that which appeared in deriving (\ref{hatt}); in a 
similar spirit we argue that if $\la(x)$ is varied linearly through $x = x_c$,
the LDA must break down when the local correlation length $\xi \sim 
[(x -x_c)/\al_x]^{-\nu}$ is of the order of the distance from the critical 
point $|x-x_c|$. We define a length scale $\hat x$ where the healing length 
$\hat \xi$ becomes of the order of the length scale associated with the 
spatial driving $\la/(d\la/dx)$ at which the adiabatic to non-adiabatic 
transition occurs, given by $\hat x \sim \al_x^{\nu/(\nu+1)}$. The above 
equation is very similar to that of $\hat t$ in (\ref{hatt2}) showing a 
striking resemblance between the temporal and spatial quenches.

The KZ mechanism in space predicts that the order parameter 
is frozen in $x$ in the impulse region given by $ -(x-x_c) <\hat x < 
x -x_c$; it penetrates into the disordered symmetric phase up to the 
length $\hat x$ which is given in terms of the quantum critical exponent 
$\nu$, i.e., the QCP is rounded off on the scale $\hat x$. Consequently, the 
energy gap is finite at $x=x_c$ and scales as $\hat x^{-z} \sim \al_x^{-\nu z/
(\nu+ 1)}$ which vanishes only in the limit $\al_x \to \infty$. 
%One can therefore define a threshold $\hat \tau = \De^{-1}/\hat x \sim 
%\al_x^{\nu(z-1)/(1+\nu)}$ such that for $. 
The scaling can be easily generalized to non-linear quenching by renormalizing 
$\nu \to r \nu$ (see Sec.~\ref{sec_kzs}). This prediction has been verified 
for an Ising chain in an inhomogeneous transverse field $h(x)-1 = x/\al_x$ 
with an open boundary condition; the gap was found to scale as 
$\al_x^{-1/2}$ and $\hat x \sim \al_x^{1/2}$ which is expected since 
$\nu=z=1$ \ct{platini07,zurek08}. On the other hand, when the MCP of the
transverse $XY$ chain (Fig.~\ref{fig_xyphase}) is approached along a linear 
path varying both $h-1$ and $\ga$, the gap scales as $\al_x^{-3/4}$ because 
the quasicritical exponents $\nu_q =1/3$ and $z_q =3$ \ct{dziarmaga10b}.

The situation has been further generalized to study an inhomogeneous phase 
transition that takes place both in space and time 
\ct{dziarmaga10a,dziarmaga10b}; here the driving parameter can be linearized 
close to the QCP as $\la(x,t) = (x - v t)/\al_x$, where $\al_x$ and $v$ denote 
the inhomogeneity and the speed of the critical front $\la(x,t)=0$, 
respectively. At a fixed point in space, 
this represents a homogeneous temporal quenching discussed in previous 
sections with $\tau = \al_x/v$. The part of the system for which $x < v t$ is
in the symmetry broken phase and the orientation of the order parameter cannot
propagate with velocity exceeding the critical speed $\hat v =\hat \xi /\hat 
t$; with $\hat t \sim \tau^{\nu z/(z \nu +1)}$ and $\hat \xi \sim \tau^{\nu /
(z \nu +1)}$, one finds $\hat v \sim \tau^{\nu (1-z)/(z \nu +1)}$ or, 
equivalently, $\hat v \sim \al_x^{\nu (1-z)/(z \nu + 1)}$ as $\tau = \al_x/v$.
If $v < \hat v$, one expects the final state to have fewer excitations than 
predicted by the KZS. One can equivalently find an optimal rate which scales 
with the inhomogeneity parameter $\al_x$ as $\tau_{\rm opt}= \sim 
\al_x^{(z\nu +1)/(1+\nu)}$; for $\tau \gg \tau_{\rm opt}$,
inhomogeneities become relevant and fewer defects are produced. In the 
homogeneous limit, $\al_x \to \infty$ and hence $\tau_{\rm opt} \to \infty$. 
For a transverse $XY$ chain, the $\hat x$ and $\hat \tau$ has been 
derived analytically for the Ising transition as
well as the transition across the MCP ($z=2$) \ct{dziarmaga10a,dziarmaga10b}.

\section{Effect of topology on quantum quenching}

\label{sec_topology}

In certain models, it turns out that topology can have an important effect on 
defect production by quenching through a QCP. We will briefly discuss two
such models here both of which have the form of a two-legged ladder.

Bermudez $et~al.$, 2009, 
considered a two-legged ladder with spinless electrons 
hopping between nearby sites on both legs. In addition, there is a magnetic
flux through each plaquette; as this flux is varied over one period (given
by the flux quantum), there is one point where the system becomes gapless. 
With open boundary conditions, the model
contains states which are localized either in the bulk or at the ends. 
If the system is initialized so that only a bulk-localized state is
occupied by an electron, a quantum quench (which consists of slowly varying 
the magnetic flux over one period) leads to a defect density given by the
KZ power-law $\tau^{-1/2}$. But if the system is initialized so as to
occupy only an end-localized state, a quantum quench leads to a defect
density which was numerically found to scale as $\tau^{-1.35}$. So far 
there is no analytical understanding of this unusual power law \ct{bermudez09}.
{In another paper, Bermudez $et~al.$, 2010, showed that when 
a Majorana chain with open boundary conditions is quenched across a QCP, an 
initial Majorana edge state gives rise to a density of excitations which scales
as $\tau^0$, i.e., the density does not go to zero as $\tau \to \infty$.
This again differs from the KZ power-law. The reason for this anomalous
behavior is that at the QCP, the edge state gets completely delocalized and
mixes with all the bulk excitations \ct{bermudez10}.}

Sen and Vishveshwara, 2010, considered a two-legged ladder version of the 
Kitaev model in which there are spin-1/2's at each site which are coupled to 
nearest neighbors on both legs. The model has a 
$Z_2$-valued invariant on each plaquette; hence the system decouples into 
$2^N$ sectors for a system with $N$ sites. Denoting the coupling between
the two legs by $J_3$, it turns out that $J_3 = 0$ is a QCP for every sector.
The effect of quenching $J_3 = t/\tau$ from $-\infty$ to $\infty$ was studied 
in a number of sectors in which the $Z_2$ invariants form a periodic pattern.
It was found that in one class of sectors, the quenching leads to a defect 
density or residual energy which scales as $\tau^{-1/2}$. However, in 
another class of sectors, these quantities were found to scale as 
$\tau^{-2/3}$. This can be analytically understood as follows. In the 
first class of sectors, close to the QCP the system decouples into 
subsystems (parametrized by a wave number $k$ as usual) in which two states 
lie close to zero energy and are directly coupled by terms proportional to
$J_3$. A Landau-Zener derivation of the excitation probability followed by an 
integration over $k$ then leads to the standard $\tau^{-1/2}$ scaling of 
the defect density since $d=\nu = z =1$. However, in the second class of 
sectors, close to the QCP the system decouples into a set of subsystems in 
which two states again lie close to zero energy. However, there is now no
direct coupling between these two states; rather, they are coupled through
some intermediate high-energy states. Eliminating the high-energy states using
second order perturbation theory, one obtains an effective coupling between 
the two low-energy states which is of the order of $J_3^2 = t^2/\tau^2$. Hence
the quenching in such sectors effectively has the non-linear form discussed 
around Eq.~(\ref{eq_intrononlinearn}), and one gets a defect density 
scaling as $\tau^{-2/3}$ since $r = 2$ \ct{sen10}. These arguments were 
developed further in Thakurathi $et~al.$, 2012, which considered
a spin-1/2 chain with equal $xx$ and $yy$ couplings and a periodically 
varying magnetic field in the $\hat z$ direction with period $2q$,
where $q$ is a positive integer. When the strength of the magnetic field 
is linearly quenched at a rate $1/\tau$ across a quantum critical point, it 
was found that the density of defects produced scales as $1/\tau^{q/(q+1)}$
\ct{sen11}. 
This behavior could be understood by mapping the low-energy physics of the 
model to a set of fermionic two-level systems in which the two states are 
coupled at $q$-th order in perturbation theory. Hence these systems effectively
undergo a non-linear quench in which a parameter in the Hamiltonian varies 
with time as $(t/\tau)^q$; this explains the defect scaling of the form 
$1/\tau^{q/(q+1)}$ since $d=\nu = z =1$ and $r=q$ (see 
Eq.~(\ref{eq_intrononlinearn})).

\section{Studies of Tomonaga-Luttinger liquids}
\label{app_ll}
In this appendix, we will briefly mention some recent studies of 
Tomonaga-Luttinger liquids (TLLs) from 
the perspective of quenching dynamics and quantum information. The loading of 
interacting bosons onto a one-dimensional optical lattice provides an example 
of quenching in a TLL \ct{grandi08}. Suppose that the periodic potential
of the optical lattice is changed in time as $V(x,t)=V(t) \cos (2\pi x/a)$, 
where the lattice spacing $a$ is commensurate with the bosonic density. The 
potential couples the right and left moving particles leading to a mass 
term. The action in terms of the bosonic field variables $\phi(x,t)$ can be 
written in the sine-Gordon form
\beq S= \int dx dt \left [ \frac{1}{v} \left(\frac{\partial \phi}{\partial t}
\right)^2 - v \left (\frac{\partial \phi}{\partial x} \right)^2 + V(t) \cos 
(2 \sqrt{\pi K \phi}) \right], \label{eq_SG}\eeq
where $K$ and $v$ are as defined in Appendix \ref{jorwigbos}. The Luttinger 
parameter $K$ is governed by the interaction strength; $K \to 1$ is the limit 
of strongly repulsive interactions (note that bosons with very strongly 
repulsive interactions are equivalent to weakly interacting fermions in
one dimension), while $K \to \infty$ is the weakly interacting limit. The 
scaling dimension of the cosine term can be read off from the equal-time 
correlation function $ \langle \cos(2 \sqrt{\pi K \phi(x,0)})\cos (2 
\sqrt{\pi K \phi(0,0)}) \rangle\sim 1/|x|^{2K}$ leading to the scaling dimension 
$[\cos(2 \sqrt{\pi K \phi(x,0)})] =K$. Therefore the mass term generated by 
the cosine term satisfies the relation $V(t) \sim m^{2-K}$ so that the 
correlation length $\xi \sim 1/m \sim V(t)^{1/(2-K)}$; hence the correlation 
length exponent is $\nu = 1/(2-K)$.
If one considers a linear ramp $V(t) = t/\tau$, with $d=z=1$, the number of 
excitations will be given by the KZS in (\ref{eq_kzm}), $n \sim 
\tau^{-1/(3-K)}$ for $0<K<2$. For $K>2$, the cosine term is irrelevant and 
one gets $n \sim \tau^{-1}$ as there are contributions from the high-energy 
modes also \ct{grandi08,degrandi09}. $K=2$ is the BKT transition point.

A quench of the interaction strength in a TLL from zero to a finite value 
has been studied in \ct{dora11}. It was shown that the residual 
energy goes as $\ln \tau/\tau^2$ in the limit $\tau 
\to \infty$. Further, the space-time dependent one-particle density matrix 
and the time-dependent momentum distribution exhibit crossovers between 
sudden quench, adiabatic and Fermi liquid behaviors as functions of the 
spatial (or momentum) coordinate, time and $\tau$.

The ground state FS of a one-dimensional gapless model 
has also been studied using the TLL Hamiltonian in (\ref{hamb2}) 
\ct{yang08,fjaerstad08}. The fidelity between two ground states with 
Luttinger parameters $K$ and $K'$, respectively, is found to be $F(K,K') = 
\prod_{k\neq 0} 2/(\sqrt{K/K'}+ \sqrt{K'/K})$, so that the fidelity is unity 
when $K \to K'$; one can then show that for a large system, the fidelity 
decays very fast as $K$ separates from $K'$. For the $XXZ$ chain 
in (\ref{xxzh}) with $h=0$, the parameter $K$ is related to the 
$zz$ coupling $J_z$ through the relation (\ref{kpara}); one then finds the 
FS to be $\chi_F = 1/[4(\pi -\arccos (J_z))^2 (1-J_z^2)]$ \ct{yang08}. 
Therefore, the fidelity susceptibility is able to detect the BKT transition 
at $J_z=1$.

\bibliography{review.bib}

\end{document}